\documentclass[pdflatex,sn-basic]{sn-jnl}

\usepackage{graphicx}%
\usepackage{lipsum}%
\usepackage{ulem}%
\usepackage{multirow}%
\usepackage{amsmath,amssymb,amsfonts}%
\usepackage{amsthm}%
\usepackage{mathrsfs}%
\usepackage[title]{appendix}%
\usepackage[dvipsnames]{xcolor}
\usepackage{textcomp}%
\usepackage{manyfoot}%
\usepackage{booktabs}%
\usepackage{algorithm}%
\usepackage{algorithmicx}%
\usepackage{algpseudocode}%
\usepackage{listings}%

\usepackage{aas_macros}
\renewcommand{\aap}{A\&A\ }

\renewcommand{\apjl}{ApJL\ }
\renewcommand{\apjs}{ApJS\ }

\usepackage{comment}

\newcommand{\kms}{km\,s$^{-1}$}

\begin{document}

\title[Magnetic activity in cool stars]{Magnetic activity in cool stars: manifestations and relevance to exoplanets}


\author*[1]{\fnm{Emre} \sur{I\c{s}{\i}k}}\email{isik@mps.mpg.de}

\author[2]{\fnm{Adriana} \sur{Valio}}\email{adriana.valio@mackenzie.br}

\author[3]{\fnm{Antoine} \sur{Strugarek}}\email{antoine.strugarek@cea.fr}

\author[4]{\fnm{Silva} \sur{J\"arvinen}}\email{sjarvinen@aip.de}

\author[5]{\fnm{Kriszti\'an} \sur{Vida}}\email{vidakris@konkoly.hu}

\author[6]{\fnm{Andrea} \sur{Buccino}}\email{abuccino@iafe.uba.ar}

\author[7,8,9]{\fnm{Kosuke} \sur{Namekata}}\email{namekata@kusastro.kyoto-u.ac.jp}

\author[10]{\fnm{Thomas} \sur{Hackman}}\email{thomas.hackman@helsinki.fi}

\author[4]{\fnm{Juli\'an} \sur{Alvarado-Gomez}}\email{julian.alvarado-gomez@aip.de}

\author[11,12]{\fnm{Dibyendu} \sur{Nandy}}\email{dnandi@iiserkol.ac.in}

\author[13,14]{\fnm{Alison} O. \sur{Farrish}}\email{alison.o.farrish@nasa.gov}

\author[4,15]{\fnm{Katja} \sur{Poppenh\"ager}}\email{kpoppenhaeger@aip.de}

\author[16]{\fnm{Pedro} \sur{Figueira}}\email{pedro.figueira@iaa.es}

\affil*[1]{\orgname{Max-Planck-Institut f\"ur Sonnensystemforschung}, \orgaddress{\street{Justus-von-Liebig-Weg 3}, \city{G\"ottingen}, \postcode{37077}, \country{Germany}}}

\affil[2]{\orgdiv{Centro de Rádio Astronomia e Astrofísica Mackenzie}, \orgname{Universidade Presbiteriana Mackenzie}, \orgaddress{\street{Rua da Consolação, 930}, \city{São Paulo}, \postcode{01302-907}, \state{SP}, \country{Brazil}}}

\affil[3]{Universit\'e Paris-Saclay, Universit\'e Paris Cit\'e, CEA, CNRS, AIM, 91191, Gif-sur-Yvette, France}

\affil[4]{\orgname{Leibniz-Institut für Astrophysik Potsdam (AIP)}, \orgaddress{\street{An der Sternwarte 16}, \city{Potsdam}, \postcode{14482}, \country{Germany}}}

\affil[5]{HUN-REN Research Centre for Astronomy and Earth Sciences, Konkoly Observatory, MTA Centre of Excellence, Konkoly Thege Miklós út 15-17., H-1121 Budapest, Hungary}

\affil[6]{\orgname{Instituto de Astronomía y Física del Espacio (IAFE, UBA-CONICET)}, \orgname{Departamento de Física, FCEN, UBA}, \orgaddress{Ciudad Universitaria}, \city{CABA}, \postcode{1428}, \country{Argentina}}

\affil[7]{\orgname{The Catholic University of America}, \orgaddress{\street{620 Michigan Avenue, N.E.}, \city{Washington}, \state{DC}, \postcode{20064}, \country{USA}}} 

\affil[8]{\orgname{The Hakubi Center for Advanced Research, Kyoto University}, \orgaddress{\street{Yoshida-Honmachi, Sakyo-ku}, \city{Kyoto}, \postcode{606-8501}, \country{Japan}}} 

\affil[9]{\orgname{Department of Physics, Kyoto University}, \orgaddress{\street{Kitashirakawa-Oiwake-cho, Sakyo-ku}, \city{Kyoto}, \postcode{606-8502}, \country{Japan}}} 

\affil[10]{\orgdiv{Department of Physics}, 
\orgaddress{\street{P.O. Box 64}, 
\postcode{FI-00014 University of Helsinki}, \country{Finland}}}

\affil[11]{\orgdiv{Center of Excellence in Space Sciences India}, \orgname{Indian Institute of Science Education and Research Kolkata}, \city{Mohanpur 741246, West Bengal}, \country{India}}

\affil[12]{\orgdiv{Department of Physical Sciences}, \orgname{Indian Institute of Science Education and Research Kolkata}, \city{Mohanpur 741246, West Bengal}, \country{India}}

\affil[13]{\orgname{Heliophysics Science Division, NASA Goddard Space Flight Center}, \orgaddress{\street{8800 Greenbelt Road}, \city{Greenbelt}, \state{MD}, \postcode{20771}, \country{USA}}} 

\affil[14]{\orgname{George Mason University}, \orgaddress{\street{4400 University Dr}, \city{Fairfax}, \state{VA}, \postcode{22030}, \country{USA}}} 

\affil[15]{\orgdiv{Institut f\"ur Physik und Astronomie}, \orgname{Universit\"at Potsdam}, \orgaddress{\street{Karl-Liebknecht-Str.\ 24/25}, \city{Potsdam}, \postcode{14476}, \country{Germany}}}

\affil[16]{\orgdiv{Instituto de Astrof\'{i}sica de Andaluc\'{i}a-CSIC}, \orgaddress{\street{Glorieta de la Astronom\'{i}a s/n}, \city{Granada}, \postcode{18008}, \country{Spain}}}

\abstract{
Beyond its intrinsic astrophysical significance, understanding stellar magnetic 
activity is central to exoplanet science in two complementary ways. On the one hand, stellar magnetic
activity constitutes the dynamic inner astrospheric boundary condition that governs the space environment
of orbiting planets: from direct magnetic star-planet coupling, to energetic-particle
bombardment, radiation-driven atmospheric photoionization, evolution of planetary atmospheres and habitability. On the other hand, stellar activity is the primary obstacle to the unambiguous detection and characterisation of exoplanets, as magnetically driven stellar variability imprints correlated quasi-periodic signals across the time series of planet-detection and characterisation diagnostics. In cool stars, magnetic fields generated by magnetohydrodynamic dynamos emerge at the
photosphere as bipolar magnetic regions, drive chromospheric and coronal heating,
modulate stellar irradiance and plasma wind, power transient events such as flares and coronal mass ejections. The spatial scales involved range
from individual flux tubes to global coronal field configurations, and the temporal scales of associated dynamical phenomena range from minutes to decades (magnetic cycles) and beyond, requiring observational
and theoretical tools of correspondingly wide scope. We review observational manifestations and physical models of magnetic activity in
stars harbouring outer convective envelopes, with the exoplanet community as the intended, primary audience. We develop the solar-stellar connection through the ``Sun in Time'' framework and a sequence of solar analogues that serve as evolutionary
snapshots of a solar-mass star over several gigayears. We survey photospheric,
chromospheric, and coronal activity diagnostics across timescales together with the
forward-modelling tools that translate surface field distributions into signals
at or above the level of exoplanet detection. Empirical rotation-activity 
relationships and their physical interpretation are examined in all three 
atmospheric layers. Surface reconstruction techniques -- Doppler imaging, 
Zeeman-Doppler imaging, Zeeman broadening, and transit spot mapping -- are 
assessed for their diagnostic reach and limitations. The evolution of magnetism in solar-like stars is discussed as a context to habitability conditions and as a window to other worlds. The chapter closes with
an account of how stellar magnetism sculpts the astrospheric environment and affects close-in exoplanets, followed by a synthesis of outstanding issues and an outlook on future observational and theoretical prospects that have interdisciplinary implications across stellar and exoplanetary sciences.
}

\keywords{keyword1, Keyword2, Keyword3, Keyword4}

\maketitle
\tableofcontents

\section{Introduction}
\label{sec:intro}

Cool stars shape their environment through magnetic activity in ways that have profound implications for (exo)planets. Stellar magnetic fields drive
chromospheric and coronal heating, modulate the radiation and particle flux received by planets, and power eruptive events such as flares and coronal mass ejections (CMEs). Magnetically mediated activity therefore bridges stars and the planets they host across the vast interplanetary scales within astrospheres \citep{Nandy2023}. Thus understanding stellar magnetic processes that shape exoplanetary space environments, their atmospheres, and by extension their habitability, has emerged as a critical domain of interdisciplinary research. 

On a completely different dimension, stellar activity imprints systematic signals on the radial-velocity and photometric time series used for exoplanet detection and atmospheric characterization. To understand, predict, and ultimately correct for these effects, a quantitative grasp of stellar magnetic activity across all relevant timescales and stellar types
is indispensable.

Here we review the observational manifestations and theoretical understanding of magnetic activity in cool stars -- those with outer convection zones -- and the
physical models used to interpret them, with the exoplanet community in mind as a primary audience. This contribution is part of a series in this volume that progressively follows the influence of stellar magnetism from the stellar interior outward through the wind and astrosphere to the exoplanetary environment. Those topics are addressed in the companion chapters and are touched on here only to the extent needed to
provide continuity. Our focus here is on stars: their photospheric, chromospheric, and coronal activity, their evolution, and the activity diagnostics that probe them.

Several principles guide our approach throughout. First, we aim to be explicit
about the limits of our current knowledge, not only about what has been established
but about where the foundations remain uncertain. Second, we put emphasis on careful
comparisons between stars of genuinely similar properties: a mid-K-type star is not a
solar analog, and conflating the two leads to systematic errors that propagate
into exoplanet science \citep[cf.][]{Vidotto21LRSP}. Third, we
emphasize empirical relationships and forward models that can be directly applied
to stellar characterization, rather than reviewing dynamo physics for its own sake
\citep{Schussler2025book}. Where possible, we provide the data and
scaling relations underlying key diagnostics, so that this chapter can serve as a 
practical reference alongside a conceptual one.

The chapter is organized as follows. Section~\ref{sec:sun} uses the Sun as a
well-resolved laboratory to define the observables and physical processes that
must be inferred for distant stars, establishing the reference frame for the rest
of the chapter. Section~\ref{sec:diagnostics} surveys the principal diagnostics of
stellar magnetic activity -- from rotational modulation and transient phenomena to
long-term activity cycles -- and what they reveal about the underlying field
configurations and their variability. Section~\ref{sec:rotation-activity} examines
the empirical rotation--activity relationships that link stellar spin to magnetic
output across the cool-star parameter space, a critical ingredient for predicting
activity levels in exoplanet host stars. Section~\ref{sec:mapping} addresses the
techniques used to map the surface magnetic field directly, including Zeeman
broadening, Zeeman--Doppler imaging, and forward modeling of photometric and
spectroscopic variability, together with their limitations in the context of
exoplanet characterization. Section~\ref{sec:impact} bridges to the companion
chapters by examining how surface and coronal magnetic field configurations
sculpt the stellar environment and drive the radiative and particle inputs
experienced by close-in exoplanets. Section~\ref{sec:outlook} closes with a
synthesis of open questions and an outlook on how near-future facilities and
theoretical developments may resolve them.

\section{Solar-stellar magnetism and its evolution}

\label{sec:sun}

To the solar physicist, the Sun is an end in itself; to the stellar astrophysicist, it is a means. The former explores the richness of dynamic phenomena in a high-dimensional data space. The latter navigates a high-dimensional parameter space, treating the Sun variously as a single data point, an inactive template, or a laboratory for modeling observables -- all in service of understanding other stars. To the (exo)planetary scientist the Sun and stars are sources of exotic external forcing and sometimes, noise! 

What can be directly measured on the Sun must be inferred in the case of distant stars, making the Sun an indispensable reference for studies of stellar magnetism, variability, and exoplanetary environments \citep{Engvold+19book}. The better we understand solar magnetism, the more we can learn about other stellar systems \citep{Brun+Browning17}. Yet this solar analogy becomes increasingly limited as one moves away from the Sun in the stellar parameter space. Effective temperature, rotation rate (itself correlated with age), and metallicity are the primary parameters governing magnetic activity patterns in cool stars. The first two are often combined into the Rossby number,
\begin{equation}
    Ro := P_{\rm rot}/\tau_c,
    \label{eq:Ro}
\end{equation}
which captures the combined effect of the rotation period, $P_{\rm rot}$ and the convective 
turnover time $\tau_c$, required for a convective parcel to traverse the convection zone. Since $\tau_c$ scales with the fractional depth of the convection zone and varies with effective temperature, $T_{\rm eff}$, along the main sequence, the Rossby number elegantly unifies two factors that enhance hydromagnetic dynamo action: faster rotation and thicker convection zones both drive stronger magnetic activity \citep{Kapyla+23}. Indeed the stellar Rossby number is related to the dynamo number ($N_D$) -- which directly governs the efficiency of the stellar dynamo mechanism; specifically, $N_D$ $\sim$ $1/{R_o}^2$ \citep{Nandy2004}. Stars with different rotation rates and convection zone properties display diverse magnetic output, which in turn is manifest in different levels of forcing of exoplanetary environments \citep{Gupta2023}. 

\subsection{Global activity patterns of solar-stellar magnetism}
\label{ssec:solar}
Magnetic fields are produced by a magnetohydrodynamic (MHD) dynamo mechanism in stellar interiors \citep{CharbonneauSokoloff2023} from where they buoyantly emerge \citep{Kumar2019,Weber2023} in the form of bipolar magnetic regions (BMRs), also called active regions, the larger ones involving sunspots \citep{vanDriel-Gesztelyi15}. The average orientation of BMR polarities is antisymmetric about the equator (Hale's polarity rule). The line joining the opposite polarities of a BMR generally shows a small tilt angle with respect to the local latitudinal circle of a few degrees, so that the leading polarity (in the sense of rotation) has a slightly lower latitude than the follower polarity, though there is a large scatter around the mean tilt angle at a given latitude (Joy's law). Hale's law is the primary evidence for the existence of a toroidal field in the convection zone. Joy's law is essential if the poloidal field is to be regenerated by emerging BMRs, as is the case for Babcock--Leighton-type dynamo scenario \citep{Pal2023,Cameron23}.

Extending this framework to solar-mass stars at
various rotation rates, \citet{Zhang_Zhebin+24} showed that the
rotation-dependent shift of emergence to higher latitudes with larger tilt
angles \citep{Isik18} deposits poloidal
flux closer to the $\pm55^\circ$ latitudes of peak $\Omega$-effect
efficiency, thereby shortening magnetic cycles in faster rotators. The model
also predicts a parity transition from quadrupole toward dipole as stars
spin down, consistent with ZDI observations of large-scale field topology
(Sect.~\ref{sssec:zdi}).

A complementary line of three-dimensional and kinematic Babcock--Leighton
dynamo modelling reaches the same qualitative conclusion through a
geometric argument: when rapidly rotating stars are forced to deposit
flux at high latitudes -- whether by Coriolis-deflected parallel rise of
flux tubes or by larger Joy's-law tilts -- the cross-equatorial
cancellation of the leading polarity becomes inefficient and the
solution settles into a quadrupolar configuration, independent of the
initial parity \citep{Karak+14,Vashishth+23,Vashishth+26}. The transition
between low-latitude (radial rise) and high-latitude (rise parallel to 
rotation axis)
emergence regimes thus sets the equilibrium parity of the global field,
linking flux-emergence statistics directly to the large-scale dynamo
mode probed by ZDI (Sect.~\ref{sssec:zdi}).

Magnetic flux loops leading to BMRs appear to be rising buoyantly from the convection zone into the photosphere, and are affected to some extent by convective flows \citep{Weber2023}. Following emergence, large-scale flows (mainly in the form of differential rotation, meridional flow, and turbulent convection in supergranular scales) transport and disperse magnetic flux across the solar photosphere \citep{Yeates2023}, shaping the large-scale magnetic field of the Sun throughout the outer atmosphere and the heliosphere. It is important to note that the internal dynamo, its variability, and its evolution are the ultimate sources of magnetic forcing of the heliosphere and astrospheres \citep{Dash2023}.

Flux emergence shows a hierarchy of structures from small to large scales on the Sun. It occurs in episodic bursts in time, shaping monthly to annual-scale variability. This is associated with the tendency of active regions to emerge in the vicinity of recent magnetic flux emergence, called active nests or complexes \citep{Castenmiller86,Brouwer90}. While automated statistical techniques for the quantification of nesting have recently been implemented \citep{Csaszar25,Karapinar26}, the physics underlying the nesting tendency of emerging flux is not well understood, though a few possible mechanisms relating it to magnetic instabilities and dynamo processes were suggested \citep{Usoskin07,Raphaldini23}. Such nests can be important drivers of eruptive outer-atmospheric phenomena \citep{Finley25}. They can also be responsible for strong variability on solar-type stars with near-solar rotation periods but more active than the Sun \citep{Isik20}, especially if the nesting degree increases with the activity level. 150 years of sunspot group data from Greenwich and Kislovodsk sunspot catalogs recently analysed by \citet{Karapinar26} showed such a correlation. Active nests and their possible association with latitudinal differential rotation on solar-like stars are also under investigation \citep{Ozavci18,Breton24}. 

Alongside the spatial organization, there are also distinct temporal patterns in solar activity. Active regions emerge at the surface in a cyclic manner, with the emergence frequency modulated over 9--11 years (the so-called \textit{activity cycle}), with opposite polarity orientations about the equator \citep{Hathaway15}. The emergence and the subsequent surface transport of magnetic flux are important drivers of the solar dynamo process \citep{Cameron23}, in which the poloidal and toroidal field reservoirs interchange magnetic energy (in interaction with the kinetic energy deposit). This energy exchange occurs quasi-periodically, with an average of 22 years, the so-called \textit{magnetic cycle} \citep{CharbonneauSokoloff2023}. Beyond the decadal solar cycle timescale, supra-decadal envelopes in solar magnetic activity are observed which have centennial to millennium scale fluctuations \citep{Saha2025, Pal2023}, including episodes of grand minima in activity (with very few sunspots) and grand maxima (with stronger than average activity levels); see e.g., the review by \cite{Usoskin2023}; notably these long-term variations are higher in amplitude than decadal scale fluctuations in the amplitude of the sunspot cycle. 

A less-appreciated but physically important aspect of the Sun's magnetic activity cycle concerns the magnetic parity of the large-scale field -- specifically, whether the solar cycle operates in a dipolar (antisymmetric across the equator) or quadrupolar (symmetric) configuration. \cite{Hazra2019} examined the origin of parity changes in the solar cycle through dynamo modeling, demonstrating that the Sun's large scale magnetism can transition between dipolar and quadrupolar states under appropriate conditions -- and can remain in a specific state for multiple cycles. While the current Sun is predominantly in the dipolar parity state, observations indicate that the Sun does display mixed parity states over the course of a sunspot cycle, with some studies indicating that during the Maunder minimum (the last known grand minima) -- the Sun may have switched to a quadrupolar parity state (\cite{Passos2014} and references therein). This transition from dipolar to quadupolar parity in the large scale magnetism of solar-like stars has cascading influence on the location and structuring of the heliospheric (and astrospheric) current sheet \citep{Smith2001}, stellar winds \citep{Reville2022} and propagation of stellar CMEs -- which modulate the space environments of close-in and far-out exoplanetary systems.
 
We note that magnetic activity in solar-like stars are expected to display such global scale structuring and fluctuations across decadal to millennium timescales, along with the expected (relatively slower) long-term evolution in activity as the star ages. 

\subsection{The Sun in time: magnetic history and future of solar-type stars}
\label{ssec:sun-in-time}
The Sun's current magnetic activity level represents only a fraction of what it once was. The foundational observational framework for reconstructing the history of a solar-like star comes from the ``Sun in Time'' program, which studied solar analogs spanning ages of approximately $0.1-7$ Gyr using multi-wavelength observational diagnostics from X-ray through ultraviolet wavelengths \citep{Guinan03,Ribas2005,Gudel2007}. The results demonstrate that the coronal X-ray and EUV emissions of the young main-sequence Sun were approximately 100–1000 times stronger than those of the present day Sun, while the transition region and chromospheric FUV–UV emissions were 20–60 and 10–20 times stronger than the current Sun, respectively. 

These enhanced chromospheric and coronal radiative flux at early times trace directly to the young Sun's faster rotation \citep{Nandy2007}; magnetic activity in solar-type stars is governed by a rotation-driven dynamo, such that high-energy radiation from solar-like main-sequence stars decays over time as a result of stellar spin-down \citep{Skumanich1972} -- with X-ray luminosity declining with stellar age \citep{Nandy2007, Lin2015}. This has direct connection with the efficiency and evolution of the magnetohydrodynamic (MHD) dynamo mechanism operating within stars; we shall revisit this theoretical basis of long-term stellar activity evolution later in this section. 

Given the observational evidence of stronger coronal activity that solar-like stars exhibit early in their life, one expects magnetically driven flares and CMEs to be more intense and frequent in young stars or faster rotating stars. While observations of stellar flares conform to this expectation \citep{Maehara2012}, observing CMEs is rather challenging and only a few confirmed detections exist \citep{Callingham2025}; thus, with limited statistics, it is too early to comment on the evolution of stellar CME activity with age. The wind properties of the young Sun and stars in their youth are expected to be more extreme. To understand the past and future evolution of the Sun -- including its magnetism, radiation, flares and wind -- we rely on information from other solar-like stars at different relative ages to the current Sun \citep{Nandy2007, Ribas2005}.

Reconstructing the solar wind's evolution in time through the stellar proxy approach reveals that a break in wind behavior occurs at approximately 2 Gyr \citep{vanSaders2016}, where a sharp decline in coronal temperatures results in a steep decay in mass loss rates for older, slowly rotating stars \citep{Fionnagain2018}. This regime change has important dynamical consequences. The wind is responsible for stellar spin-down through angular momentum loss via magnetized winds, this decline in mass-loss rates for older stars explains the anomalously high rotation rates observed in middle-aged stars, a phenomenon that suggests a breakdown of stellar gyrochronology, 
which is a technique that derives ages of stars using only their rotation periods and colors \citep{2007ApJ...669.1167B}.

One compelling recent development in our understanding of the Sun's magnetic future concerns a fundamental physical transition in dynamo behavior at approximately solar age. \cite{Tripathi2021} demonstrates that at about the age of the Sun, the magnetic field generation mechanism in solar-like stars becomes subcritical or less efficient, allowing stars to exist in two distinct activity states -- a low-activity mode and an active mode -- such that a middle-aged star can often switch to the low-activity mode resulting in drastically reduced angular momentum losses via magnetized stellar winds (see also the discussion in Sect.~\ref{sssec:zdivszeeman}. This physical picture provides a unified explanation for three previously puzzling observations -- the breakdown of gyrochronology at intermediate stellar ages \citep{vanSaders2016}, the relatively low activity of the Sun compared to other solar-like stars \citep{Reinhold2020} and the occurrence of Maunder-like grand minima episodes in solar-stellar activity \citep{Usoskin2014}. 
Observational support for the underlying magnetic transition has since been 
provided by \citet{Metcalfe25}, who used direct spectropolarimetric 
measurements of large-scale field strengths across a sample of F-to-K type 
stars to demonstrate an abrupt decrease in both wind-braking torque and 
large-scale dipole field as the Rossby number approaches a critical value 
near ${\rm Ro}_\odot$ -- interpreted as evidence for the collapse of the 
global stellar dynamo (see Sect.~\ref{sssec:zdivszeeman} for a detailed 
discussion).

In what they call a stellar mid-life crisis, \cite{Tripathi2021} propose that Sun-like stars undergo a mid-life transition where their magnetic field generation weakens, disrupting predictable spin-down rates, which may explain why stellar gyrochronology fails beyond middle age and why the Sun has experienced quiet periods such as the Maunder Minimum. This subcritical dynamo framework suggests that the future of the Sun's magnetic activity is not simply a smooth continuation of the current declining trend in long-term activity. Instead, the Sun may periodically transition between active and quiescent phases before sliding down into a magnetically inactive future.

On billion-year timescales, stellar evolution models predict that the Sun will leave the main sequence in about another 4 Gyr, evolving from its current G2V classification toward the Red Giant phase, and thereafter through a planetary nebula phase before ending up as a white dwarf. The interplay between this stellar evolution and secular trend in dynamo activity, and the latter's long-term behavior as suggested by \citet{Tripathi2021}, will have profound implications for the long-term habitability of Earth and the space environment experienced by solar system's planets \citep{Nandy2021, Vidotto21LRSP}. This understanding is expected to be a guiding light in interpreting coupled star-planet evolution in exoplanetary systems.

\subsection{Solar analogs as evolutionary snapshots of a Sun-like star}
\label{ssec:solar-analogs}

Below we present a curated discussion of stars drawn from various phases of the life of a near-solar-mass star. A summary of stellar parameters is given in Table~\ref{tab:sun-in-time}. With the exception of $\iota$~Hor and $\tau$~Cet, the selected stars have temperatures and metallicities very close to solar values, making them solar analogs. $\iota$~Hor is some 450~K hotter than the Sun and somewhat more metal-rich, here chosen mainly owing to a dedicated recent study reviewed in Sect.~\ref{sssec:iotaHor}, in addition to its representative state of its age and rotation period. 
In spite of its low temperature ($<450$~K cooler than the Sun) and low 
metallicity, $\tau$~Cet was chosen for its age older and much slower than 
the Sun, making it a well-studied nearby star in a grand-minimum state. 

\begin{table}[h]
\caption{Parameters of selected solar-type stars}\label{tab:sun-in-time}%
\begin{tabular}{@{}lllllllll@{}}
\toprule
Name & Mass [$M_\odot$] & Radius [$R_\odot$] & $T_{\rm eff}$ & $\log g$ & Age [Myr] & $P_{\rm rot}$ [d] & [Fe/H] & Ref. \\
\midrule
EK Dra    & 1.04 & 1.07 & 5770 & 4.4 & 27 & 2.61 & 0.03 & 1 \\
$\chi^1$ Ori & 1.03   & 1.05\footnotemark[1]  & 5882 & 4.34 & 300 & 4.83 & -0.036 & 2 \\
$\iota$ Hor & 1.21 & 1.16\footnotemark[2]  & 6207 & 4.53 & 480 & 7.73 & 0.180 & 3 \\
HD 56124  & 1.02 & 1.00\footnotemark[2] & 5848 & 4.46 & 3880 & 18 & -0.02 & 4 \\
18 Sco & 1.02 & 1.01\footnotemark[2]  & 5808 & 4.44 & 4300 & 23 & 0.041 & 4 \\
Sun & 1.00 & 1.00\footnotemark[1] & 5777 & 4.44 & 4600 & 25 & 0.00 & - \\
$\tau$ Cet & 0.69 & 0.793\footnotemark[1] & 5320 & 4.48 & 8000? & 48 & -0.50 & 5 \\
\botrule
\end{tabular}
\footnotetext{(1) \citet{Senavci21}, (2) \citet{2025A&A...693A.269B}, 
  (3) \citet{2025A&A...704A..68A}, 
  (4) \citet{Kochukhov20}; \citet{2023ApJ...958...57D}, 
  (5) \citet{2023AJ....166..123K}}
\footnotetext[1]{Radius from \citet{2023AJ....166..123K} (interferometry).}
\footnotetext[2]{Radius from isochrone fit; see reference in column Ref.}
\end{table}

\subsubsection{EK Draconis}
\label{sssec:ekdra}
EK\,Draconis (HD\,129333) is probably the best-studied and the most active analogue of the young Sun. A detailed analysis of fundamental parameters and abundances by \citet{Senavci21} indicates that it is a pre-main sequence star approaching ZAMS, at an age of $27^{+11}_{-8}$ Myr, whereas the Li abundance indicates 70 Myr \citep{GorgeiEKDra}. \citet{Senavci21} located EK\,Dra on the $\log g$--$T_{\rm eff}$ plane with evolutionary tracks, to show that it lied very close to the Sun in terms of its internal structure. This does not imply that the Sun passed through the same configuration as EK\,Dra during its early evolution, but rather that EK\,Dra is a young solar analogue suitable for studying how the rotation rate (along with its gradients) \textit{alone} affects stellar magnetism, as all the other quantities are very near the solar values. 

As recently discussed in \citet{GorgeiEKDra},
the photometric behaviour of this star can be traced back to 1891. Albeit the earlier observations are not very dense, when combined with newer data
\citep[e.g.][]{2018A&A...620A.162J}
the photometric record covers more than a century. The first about 50 years of observations show that the star got brighter whereas the past $\sim$64 years have a fading trend. Still, even with this long data set, it is not possible to distinguish if there is a long term cycle or just a trend. In any case, the star clearly shows variability in a time scale of a century that may be similar to the solar Gleissberg cycle. According to the analysis of
\citet{GorgeiEKDra},
the whole dataset shows a 10.7--12.1 years long cycle and the more densely covered years reveal also an additional 7.3--8.2 years long signal.
Although the newer photometry allowed to study the spot evolution of EK\,Dra to some extent
\citep[e.g.][]{2005A&A...440..735J},
the TESS light curves with continuous coverage over several rotation periods are much more suitable for detailed studies of spot evolution and can already give a rough estimate for differential rotation.

Though the first Doppler map of EK\,Dra is based on spectra obtained in 1995
\citep{1998A&A...330..685S},
the next data set was obtained more than five years later. Since then, as summarised in
\citet{2018A&A...620A.162J}
or in
\citet{Senavci21},
more Doppler imaging or Zeeman--Doppler imaging maps have been published from time to time, but usually with long gaps in between them. Apart from the one by
\citet{2018A&A...620A.162J}
that was obtained with very high resolution mode of $R=230\,000$, all maps show coexisting high-latitude and low-latitude spots. Interestingly, a high-latitude spot is always the more dominant feature and from time to time even covering the polar region. 

Numerical simulations of magnetic flux emergence and transport on EK Dra offered new ways to interpret observed variability patterns. \citet{Senavci21} used the FEAT model \citep{Isik18} to synthesise Doppler images with imposed differential rotation. In these simulations, near-polar spots were maintained by bipolar regions with emergence latitudes and tilt angles determined by flux-tube emergence dynamics and the surface redistribution by horizontal flux transport. Interestingly, near-equatorial spots not existing in the input simulations were recovered as artefacts of the less visible southern-hemisphere activity in synthetic Doppler images, indicating room for improvement in the imaging technique (see Sect.~\ref{sssec:di}). 

The magnetic field of EK\,Dra was mapped only a few times and all maps did fall within the same activity cycle showing no signs of a polarity reversal.
Unfortunately, no further Zeeman--Doppler imaging of EK\,Dra exists since those of
\citet{2016A&A...593A..35R,2017MNRAS.465.2076W} and although results indicate fast evolution of the field topology, polarity reversal remains unobserved.
So far the most dense spectroscopic monitoring of EK\,Dra was performed by
\citet{GorgeiEKDra}.
They obtained over 900 spectra within February 2021 and July 2024 allowing construction of 13 stellar surface maps including multiple consecutive maps. Each of these maps reveal multiple mid- and  also low-latitude spots in line with the  results by
\citet{2018A&A...620A.162J}.
The consecutive maps provided for the first time the possibility to obtain a reliable measure for the differential rotation that resulted in a solar-type one. 

Flaring activity of EK\,Dra has been known for a long time
\citep[e.g.][]{1999ApJ...513L..53A}.
Again, TESS photometry has provided a significant improvement in flare statistics for this star
\citep[e.g.][]{2022ApJ...926L...5N,GorgeiEKDra}.
Both studies reported similar flare frequency distributions. The latter study also compared longitudinal distribution of the flares with the rotational phase and with the spotted longitudes finding no correlations.

\subsubsection{$\chi^1$ Orionis}

$\chi^1$\,Orionis (HD\,39587) is in a detached binary system with an M-type companion \citep{1978PASP...90..330L,1992PASP..104..101I,2002PASP..114..224H}. It is also one of the solar type stars flagged as variable by
\citet{Baliunas1995ApJ...438..269B}.
Although chromospheric emission of this star showed clear variability, a clear cyclic behaviour was not detected. Later,
\citet{2018RAA....18...76B}
included more data to the analysis and detected three cycles with period changes over the years (main $\sim$25--20 years and two weaker $\sim$16 -- 10 years each).

$\chi^1$\,Ori does not rotate fast enough for mapping its temperature of brightness but Zeeman--Doppler imaging of this target has been performed for various epochs by
\citet{2016A&A...593A..35R,2022A&A...659A..71W}.
The former study revealed magnetic polarity reversed twice within 4.8 years, indicating that the star might have a very short magnetic cycle. The latter study produced a map with same polarity configuration as the last map from the former study -- however, the maps have a significant gap in time by six years. Based on all available data, 
\citet{2022A&A...659A..71W}
estimated that polarity reversals could happen with an interval between 1.9 to 3.6 years making the magnetic cycle to be between 3.8 and 7.2 years. Further monitoring of the target is required in order to confirm the cycle, but the existing data implies that $\chi^1$\,Ori is one of the targets that may allow the characterization of a magnetic cycle on a reasonably short time.

\subsubsection{$\iota$ Horologii}
\label{sec:Horologii}

The current knowledge of the young and very active planet-hosting solar-type star $\iota$\,Hor has been recently summarized by
\citet[][see also Sect.~\ref{sssec:iotaHor}]{2025A&A...704A..68A}.
 This star does not only show cyclic chromospheric and coronal activity behavior\citep{2010ApJ...723L.213M,2018MNRAS.473.4326A}
 but it also displays one of the shortest and best-sampled X-ray cycles to date, with a length of 1.6 year  \citep{2013A&A...553L...6S,2019A&A...631A..45S}.
The long-term evolution of the S-index is, however, a superposition of two periodicities ($P_1=1.49$ year and $P_2=1.09$ year), whereas the star undergoes a full magnetic cycle over roughly 2.1 years. This suggests a 2:1 correspondence for the shorter chromospheric activity cycle, mirroring the
relationship observed in the Sun. On the other hand, the coronal activity cycle traced by the X-ray emission could be linked to the longer magnetic variability timescale identified in the butterfly diagrams of the radial magnetic field.
 \citet{2025A&A...704A..68A}
have presented the first estimates of the large-scale flow properties on a star other than the Sun
and identified possible poleward and equatorward drift speeds for different field polarities (see Sect.~\ref{sssec:iotaHor}).

\subsubsection{HD\,56124}
\label{sssec:hd56124}
HD\,56124 is regularly included into a sample of stars in stellar magnetism studies
\citep[e.g.][]{2014MNRAS.441.2361V, See19}. Although the star does not show strong chromospheric activity,
\citet{Kochukhov20}
managed to determine a consistent and formally significant $\langle B \rangle$ of about 0.22\,kG from the Narval spectra corresponding to four different observing epochs using the Zeeman broadening method. The mean magnetic field they obtained for HD\,56124 is formally compatible with the solar average field strength inferred with 
their method.

The first large scale magnetic field maps of HD\,56124 were published only recently
\citep{2025A&A...693A.269B}. 
The maps reveal that the star features a predominantly poloidal, dipolar, and axisymmetric field. Further, the maps reveal two evident polarity reversals. From the existing information, the authors estimated that the star experiences a magnetic cycle characterised by a timescale of 3--4 years between polarity reversals.

\subsubsection{18 Scorpii}

18\,Scorpii (HD\,146233) is the closest approximation to the Sun among bright stars. The similarity of stellar parameters of 18\,Sco to those of the Sun was reported by
\citet{1997ApJ...482L..89P}
and a good summary of current knowledge of the star is given in
\citet{2023ApJ...958...57D}.
However, several detailed studies have reported a slightly higher Li abundance and faster rotation than that of the Sun. This indicates that 18\,Sco is slightly younger -- around 3.4--3.7 Gyr old -- than the Sun.

The seven years long chromospheric activity cycle of 18\,Sco was reported by
\citet{2007AJ....133.2206H}.
Although the stellar parameters of the star are very similar to those of the Sun, the activity cycle is clearly shorter. However,
\citet{2007AJ....133.2206H}
also showed that during that cycle, the brightness of the star behaves in a similar manner to that of the Sun, that is, the photometry of 18\,Sco varies directly with its activity cycle. A compilation of long-term monitoring of the chromospheric activity of the star, spanning four decades,  confirmed the main cycle to be $6.95\pm0.60$ years long
\citep{2023ApJ...958...57D}.
However, the longer time span allowed also to detect a secondary cycle which is $14.91\pm2.67$ years long.

\citet{2008MNRAS.388...80P}
published the magnetic field topology of 18\,Sco. The reconstructions revealed a dominantly poloidal geometry with negligible toroidal component, which is reminiscent of the large-scale solar magnetic geometry.
\citet{BCoolSnapshots} 
analysed about 60 spectropolarimetric observations of 18\,Sco but only a few resulted in marginal or definite detections of a polarized signature. The maximum measured value of longitudinal magnetic field was only a few gauss. Later,
\citet{2023ApJ...958...57D}
reanalysed the previous data as well as some newer observations. As in the previous analysis, in the majority of the observations, no Zeeman signature was detected. The detected signatures allowed to get a picture of expected polarity and relate it with the chromospheric activity cycle, revealing that the $\sim$15 years long cycle detected in 18\,Sco is analogous to the 22 years long solar magnetic polarity cycle.

\subsubsection{$\tau$ Ceti }

$\tau$\,Ceti (HD\,10700), albeit inactive \citep[e.g.][]{2021A&A...646A..77G}, has been extensively studied due to having stellar parameters very close to those of the Sun and for being extremely bright. While it was previously thought to host planets near its habitable zone
\citep{2017AJ....154..135F}, the recent study of \cite{2025A&A...700A.174F} showed that these are attributed to stellar activity. The same study identified a long-term trend in radial velocity that can be attributed to a stellar activity cycle. Once corrected for this effect and that of activity in the rotational time scale, the star shows the remarkable long-term precision of 40\,cm/s, the lowest value obtained to date on any star.

Furthermore, it has a debris disk that spans approximately 10--50 au
\citep{2016ApJ...828..113M}
and has inclination of $35^{\circ}\pm10^{\circ}$
\citep{2014MNRAS.444.2665L}.
The nearly pole-on orientation along with other revised stellar parameters have been reported by
\citet{2023AJ....166..123K}.

\citet{Baliunas1995ApJ...438..269B} flagged the chromospheric activity of $\tau$\,Cet to be probably flat.
\citet{2022AJ....163..183B} compiled a 52 years long data set for this star and confirmed the flat activity.

Information about the magnetic field strength of $\tau$\,Cet was obtained for the first time only recently by
\citet{2023ApJ...948L...6M}
who reported, assuming dipole morphology, $B_{\rm d}=-0.77\pm0.31$\,G and a mean longitudinal magnetic field $\langle B_{\rm z} \rangle = -0.37\pm0.08$\,G.

\subsection{Rotation--magnetism relationship}
\label{ssec:rotmag}

The rotation rate is the most fundamental observational 
indicator of magnetic activity, 
because with increasing rotation rate, stars with similar 
internal structure exhibit stronger magnetic activity 
\citep{SchrijverZwaan00}. In the context of stellar evolution, stars lose angular momentum by magnetic activity, making them spin down as they age, leading to weaker dynamos \citep[see][for a review]{Isik23}. 

Early measurements by \citet{Saar96} indicated a power-law dependence between the Rossby number and magnetic flux density, implying $\langle B\rangle_{\rm obs}\propto\omega^{1.7}$ for solar-type stars. An indirect estimate combining X-ray flux, Ca II HK flux relationships, and the solar field - Ca II K relationship yielded an empirical exponent of $0.9\pm 0.3$ \citep{SchrijverZwaan00}, motivating early models to assume linear scaling with rotation rate.

\citet[][hereafter K20]{Kochukhov20} provided high-precision measurements using Zeeman intensification for solar-type stars. For the Sun, they measured $0.18^{+0.11}_{-0.05}$~kG from HARPS Sun-as-a-star spectra, likely dominated by small-scale dynamo (SSD) fields and network fields. Using the same method for other stars provides a homogeneous dataset showing smooth scaling of mean magnetic fields with rotation. 

\citet{Reiners22} presented a comprehensive analysis revealing a clear dependence between average magnetic field and Rossby number over more than three orders of magnitude in Ro. The rotation-magnetic field relation exhibits a break between saturated (Ro$<0.13$) and non-saturated groups, with surface-average magnetic fields showing tight correlation with stellar rotation and X-ray luminosity, demonstrating that the dynamo itself depends on rotation rate.

\begin{figure}
    \centering
    \includegraphics[width=\linewidth]{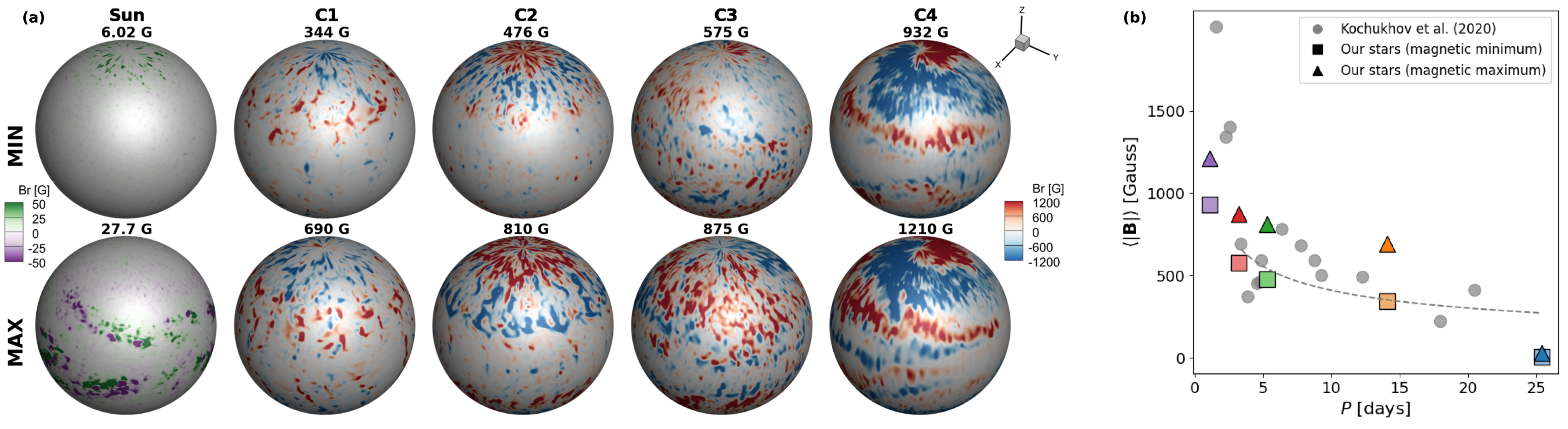}
    \caption{(a) Radial magnetic field distributions on the surfaces of increasingly faster-rotating solar-type stars according to 3D global MHD models; (b) the mean unsigned field strength as a function of the rotation period for models in (a) compared to mean fields of stars with near-solar parameters measured by \citet{Kochukhov20}. Figure adapted from \citet{Chen25}.}
    \label{fig:Chen25}
\end{figure}
\citet{Chen25} recently modeled the coronal emission of solar-type stars at different rotation rates using 3D magnetohydrodynamic simulations. They prescribed surface magnetic boundary conditions using dynamo-generated radial field maps from \citet{Viviani18}, scaled to match observed field strengths, and computed magnetic field distributions for stars at both activity minimum and maximum phases. Figure~\ref{fig:Chen25} shows mean magnetic field strengths ranging from ~400 G at minimum to ~1200 G at maximum for their fastest rotator ($P_{\rm rot}\sim1.6$~d), overlaid on the K20 observational data. Their gray dashed line power-law fit to the non-saturated stars from \citet{Kochukhov20} provides the observational constraint on the rotation-magnetism relationship. Chen et al. demonstrated that the activity cycle phase can introduce scatter of factors of 2-3 in the mean field strength at a given rotation rate, an important consideration when comparing observations taken at different cycle phases.

Building on observational constraints, \citet{Isik26} used the Flux 
Emergence And Transport (FEAT) framework to constrain how emerging 
magnetic flux must scale with rotation to reproduce stellar 
fields measured by \citep{Kochukhov20}. The model decomposes the total 
unsigned mean field into three components:
\begin{eqnarray}
    \langle B_{\rm tot}\rangle &=& \langle B_{\rm SSD}\rangle + 
    0.6\langle B_{\rm SSE\odot}\rangle\omega^p + \langle B_{\rm SFT}(p)\rangle,
    \label{eq:Btot}
\end{eqnarray}
where $\langle B_{\rm SSD}\rangle \simeq 180$~G is the rotation-independent 
small-scale dynamo field (solar measurement by K20), $\omega$ is the 
rotation rate in solar units, $\langle B_{\rm SSE\odot}\rangle$ 
represents small-scale bipolar (including ephemeral) regions, and 
$\langle B_{\rm SFT}(p)\rangle$ is the active-region-driven field 
processed by surface flux transport, with rotation-dependence 
controlled by the flux-injection power index $p$.

\begin{figure}
    \centering
    \includegraphics[width=0.45\linewidth]{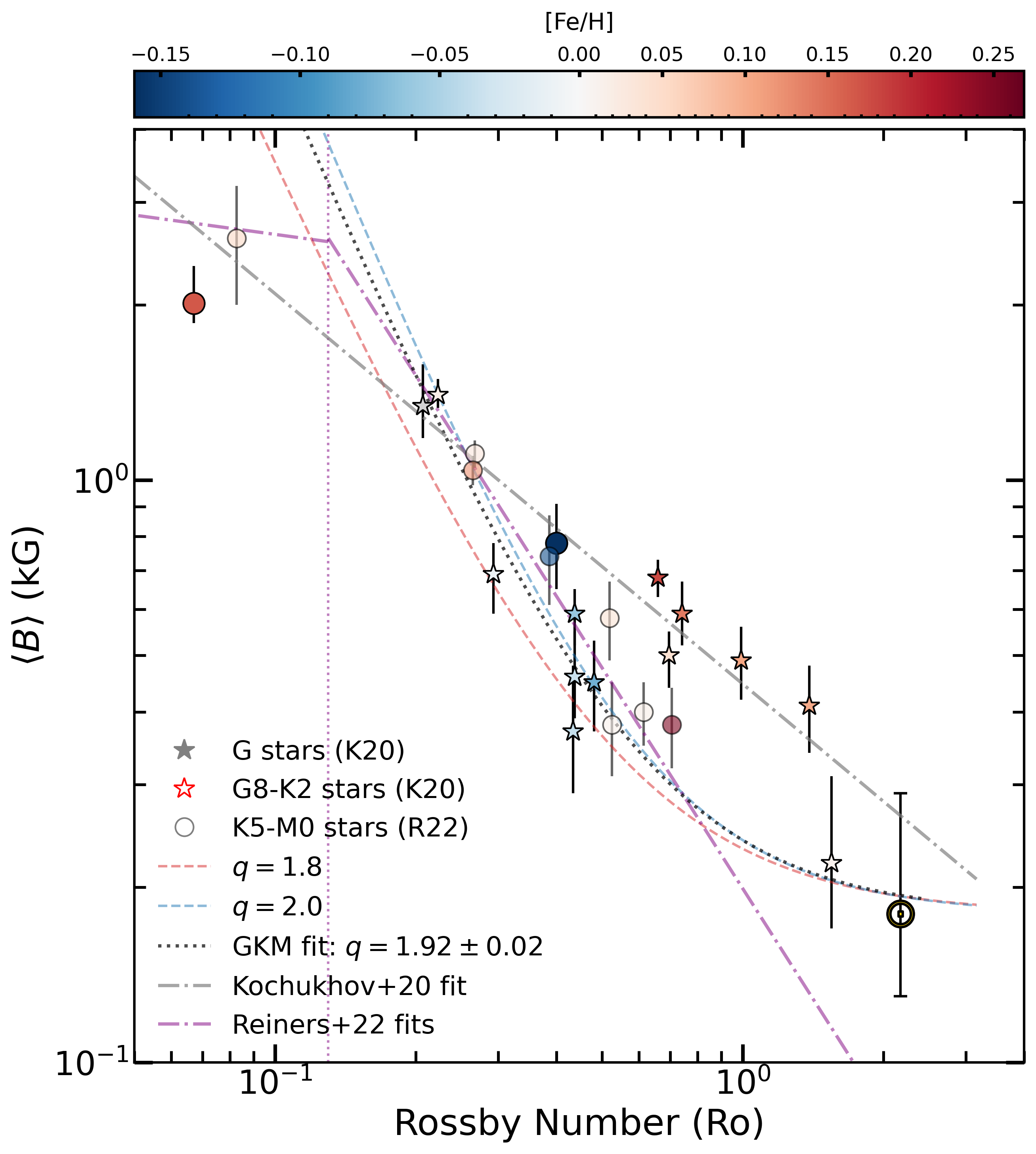}
    \includegraphics[width=0.45\linewidth]{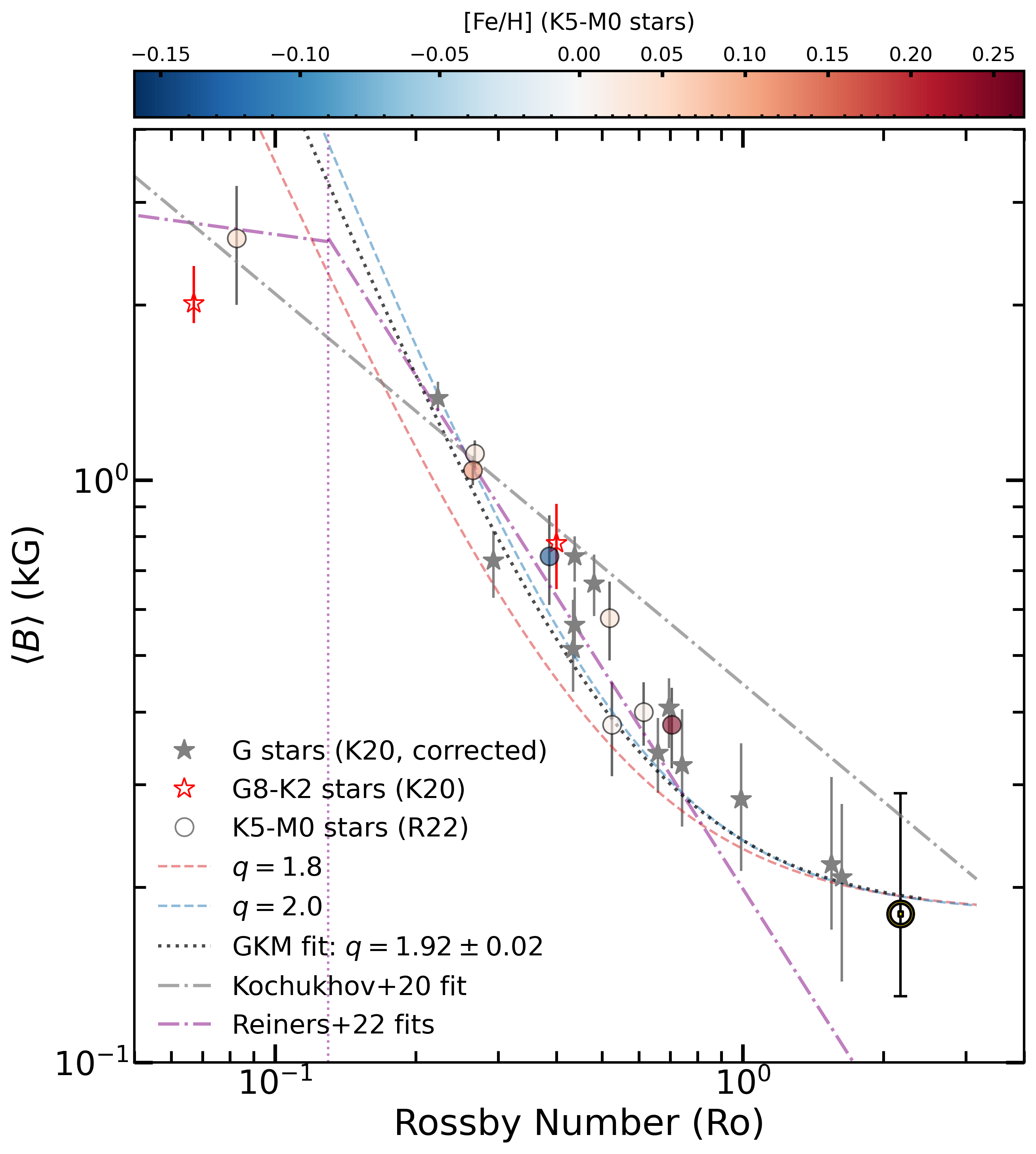}
    \caption{Mean unsigned surface magnetic field as a function of 
    Rossby number for solar-type stars from \citet{Kochukhov20} 
    (filled symbols) and late-K to early-M stars from \citet{Reiners22} 
    (open symbols), compared to FEAT model curves for flux-injection 
    exponents $p=1.0$--$2.0$. Left: \citet{Kochukhov20} measurements 
    as observed. Right: G-star measurements corrected for stellar 
    metallicity and effective temperature to solar values; the 
    \citet{Reiners22} subsample is shown uncorrected in both panels. Figure adapted from \citet{Isik26}.}
    \label{fig:isik26}
\end{figure}

A key diagnostic step was examining systematic deviations between 
model predictions and the \citet{Kochukhov20} measurements. These 
deviations correlate strongly with stellar 
metallicity ($r=0.83$) and effective temperature ($r=-0.76$), with a 
combined correlation coefficient of 0.90. Enhanced metallicity 
increases spectral line opacity and dynamo efficiency through deeper 
convection zone onset, similar to cooler stars \citep{Karoff18,See21,See23,See24}. 
\citet{Pezzotti25} independently found that metal-poor stars follow 
steeper slopes than metal-rich stars in the unsaturated regime of the 
rotation dependence of X-ray luminosity, supporting this picture. The 
effect of applying metallicity and temperature corrections 
to the \citet{Kochukhov20} G-star sample is illustrated in 
Fig.~\ref{fig:isik26}: the uncorrected data (left panel) show 
substantial scatter about the model curves, which is 
markedly reduced after correction (right panel). Correcting for these 
systematic effects reveals that magnetic flux emergence rates must scale 
steeply with rotation: $p \approx 1.9$, in the range $1.8$--$2.0$. This 
is significantly steeper than the linear dependence 
assumed in earlier work \citep{Isik18} and exceeds previous indirect
estimates.

The model reveals a transition from SSD-dominated fields 
($\sim 92$\% for the Sun) to active-region dominance (up to 82\% 
for $P_{\rm rot}\sim 3$~d, $p=2.0$). Estimated flux 
emergence rates increase from $4.6\times 10^{24}$~Mx/yr at solar 
rotation to about $6\times 10^{26}$~Mx/yr at $P_{\rm rot}\sim 3$~d 
for $p=2.0$, more than two orders of magnitude. The results imply that 
sample homogeneity in fundamental stellar parameters is 
essential for deriving unbiased rotation-magnetism relationships, 
independent of the measurement technique.

\subsubsection{Large-scale versus total field: complementary diagnostics}
\label{sssec:zdivszeeman}

Spectropolarimetric studies based on Stokes V measurements probe different field components than total unsigned field measurements. \citet{Metcalfe25} found that the dipole component from Zeeman-Doppler Imaging (ZDI) exhibits a concave relationship with Rossby number, rapidly declining to zero near the solar value. They determined a critical Rossby number ${\rm Ro}_{\rm crit}=1.014\pm 0.026~{\rm Ro_\odot}$, marking the transition to weakened magnetic braking and establishing an approximate dividing line between stars with activity cycles and those with flat activity. The suggested interpretation is that 
the global dynamo can be considered a weakly nonlinear system near a supercritical Hopf bifurcation, with dynamo number $D\sim Ro^{-2}$ as the control parameter \citep{Cameron17}. Below critical $D$ (when Ro is sufficiently large), the global dynamo is not excited. 

At this point, a reconciliation of two different observables is in order. On the one hand, direct Zeeman broadening observations \citep{Kochukhov20,Reiners22} and MHD-based modelling of their target variable \citep{Chen25,Isik26}, both probe surface-averaged total unsigned magnetic field $\langle B\rangle$ (including SSD fields and mixed-polarity structures that cancel in Stokes V). On the other hand, the Stokes-V-based observations \citep{2025MNRAS.542.1318S} measure the signed field with considerably different responses to field geometry, polarity cancellation, and spatial scale \citep[see][for their forward modelling]{Vidotto16,2019MNRAS.483.5246L}. The latter technique captures large-scale organized field components and their scaling with the rotation rate, with empirical corrections estimated by \citet{See19}. 

Within the framework proposed by \citet{Isik26}, a shutdown of the large-scale dynamo (as observed by \citet{Metcalfe25} for $Ro\gtrsim Ro_\odot$) would not imply vanishing $\langle B\rangle$, but rather convergence toward the rotation-independent basal SSD field (100-200~G). This explains the complementary behaviour: 

\begin{enumerate}
\item Towards fast rotators, both the total unsigned field and the large-scale organized field increase with decreasing $Ro$ as the global dynamo becomes more efficient. 
\item In solar-like rotators ($Ro\sim Ro_\odot$), the large-scale dipole approaches criticality and declines \citep{Metcalfe25}, while the total unsigned field converges to SSD-dominated basal level maintained by small-scale turbulent dynamo. 
\item Metallicity effects are present in both diagnostics, with metal-rich stars lying above mean trends \citep[see also][for effects on SSD]{Witzke23}. 
\end{enumerate}

\section{Magnetic activity diagnostics across time scales}
\label{sec:diagnostics}


Magnetic activity, driven by the emergence of dynamo-generated magnetic flux from the stellar interior, manifests itself in a complex set of phenomena across the stellar atmosphere. These stellar activity features are highly dynamic, exhibiting variations across a vast range of time-scales, from impulsive events like flares (seconds to minutes) to long-term modulations characteristic of stellar activity cycles (years to decades). To systematically study this activity, a diverse array of observational diagnostics across the electromagnetic spectrum is required to probe the different layers of the stellar atmosphere, from the cool photosphere to the hot corona.

\subsection{Rotational modulation of activity indicators}
\label{ssec:rotmod}


Rotational modulation of stellar activity indicators arises because surface inhomogeneities such as starspots, plages, and prominences are carried across the visible hemisphere by stellar rotation. These features alter chromospheric and coronal diagnostics  -- including Ca {\sc ii} H\&K, H$\alpha$, and photometric brightness -- producing quasi-periodic signals that trace the rotation period. The resulting modulation provides a valuable probe of the distribution of active regions.

The most common manifestation of solar-type activity is dark, cool starspots, which lead to photometric variability from rotational to cycle timescales. In the case of solar-like activity levels, bright faculae become more dominant than spots in cycle variability  \citep[][ see also Sect.~\ref{sssec:photos_diag}]{Solanki13}.
Starspots cause periodical brightness changes in the optical domain, with amplitudes ranging from one hundredth to two–three tenths of a stellar magnitude. This modulation (with some limitations) can be used to reconstruct the map of the spotted surface -- for details see Section \ref {sect:modelling-photometry}. The modulation occurs with the star's axial rotation, that in the case of quiet M dwarfs -- generally the slowest rotating main-sequence stars -- typically does not exceed 2 \kms, while on active stars, such as UV Cet-type flare stars, can be noticeably faster. For 40\% of single UV Cet-type stars, rotation rates are close to or somewhat higher than 10 \kms . Stellar rotation generally evolves from fast rotation in young objects to very slow rotation in old stars \citep{johnstoneStellarWindsPlanetary2021}, although this deceleration can be slowed by binary systems through tidal interaction.

The rotational modulation of the H$\alpha$  line  occurs because chromospheric active regions are distributed non-uniformly across the stellar surface. As the star rotates, these bright regions move into and out of the observer's field of view, causing the integrated H$\alpha$ flux to vary periodically. 
On the Sun, photospheric and chromospheric activity are often closely linked, and similar connections have been identified in other stars, including the BY Dra–type star EY Dra \citep{Korhonen2010AN....331..772K}, the T Tauri star TWA 6 \citep{Skelly2008MNRAS.385..708S}, the K-dwarf LQ Hya \citep{Frasca2008A&A...481..229F}, and RS CVn-type binaries such as RT Lacertae \citep{Frasca2002A&A...388..298F}.
However, H$\alpha$ emission is not always directly associated with photospheric spots. For example, \citet{Biazzo2009A&A...499..579B} did not detect any H$\alpha$ rotational modulation in SAO 51891, despite chromospheric variability in the Ca \textsc{ii} HK, IRT, and H$\varepsilon$ lines along with the $V$-band modulation. In this case, any H$\alpha$ modulation may be masked by other forms of activity, such as microflares.

The Calcium {\sc ii} H and K emission lines, typically quantified by the Mount Wilson $S$ index or the $R'_{HK}$ ratio (the ratio of the emission in the core of the Ca {\sc ii} H \& K lines to the total bolometric emission of the star), serve as fundamental indicators of stellar chromospheric activity, reflecting the non-uniform distribution of magnetic active regions  on the star's surface. Monitoring the intensity of these lines shows rotational modulation with periods typically measured in days, which provides a key method for directly estimating the axial rotation periods of stars, especially slow rotators where spectroscopic methods are ineffective. The long-term stability of the modulation phases indicates the persistence of asymmetric active region distributions along the stellar longitude \citep[][see also Sect.~\ref{ssec:solar}]{1983ApJ...275..752B}.

Stellar flares can also show modulation  with stellar rotation, provided the time scale of the flare is long enough for the star to complete a significant portion of its rotation \citep[CD-36 3202;][]{biczUnveilingSpectacular24hour2024}, or even multiple rotations \citep[V405 And;][]{vidaPhotosphericChromosphericActivity2009}.
These rare long-duration events can be used to estimate the latitude of the flares and the flare location in relation to starspots.
The broader question
of how flare rate and energy relate to surface spot distributions, and
of what active-region property actually drives flare productivity, is
taken up in Sect.~\ref{ssec:flare}.


\subsection{Transient phenomena: flares, prominences, CMEs}
\label{ssec:transients}

Stellar atmospheres are also subject to impulsive, eruptive phenomena that release stored magnetic energy on timescales of minutes to hours, and in some cases up to days. The most prominent of these events are stellar flares, which produce significant perturbations to the radiative environment (Section \ref{ssec:flare}). In addition, recent observations have increasingly revealed disturbances in the plasma environment, including filament and prominence eruptions as well as coronal mass ejections (CMEs) (Section \ref{ssec:eruptions}). 
These intense bursts of high-energy radiation and energetic particles can substantially modify the quasi-steady field surrounding stars, and their relative impact on exoplanets has therefore become a subject of growing interest. While Chapter~2 discusses transient aspects in detail, here we focus on the observable properties of these events and their dependence on stellar type, rotation, and age.

\subsubsection{Stellar flares}\label{ssec:flare}
Stellar flares are observed across the electromagnetic waves from radio to X-rays, and thought to occur as a result of the rapid release of magnetic energy in the stellar atmosphere through magnetic reconnection in the coronae \citep[e.g.,][for review]{Benz2017LRSP,2024LRSP...21....1K}. 
Early observational works focused primarily on nearby active stars, such as young stars, dMe flare stars, and close binary systems, for which large events could be detected repeatedly \citep[e.g.,][]{1991ApJ...378..725H,Gudel2004A&ARv,2007LRSP....4....3G,2005ApJ...621..398O}. 
These studies established the basic phenomenology of stellar flares, with close relationship with solar flares (see Chapter~2 for details). 
These early studies also found that such active stars frequently produce large flares called ``superflares'', exceeding the maximum energy ever observed for solar flares ($\sim 10^{32}$ erg). 
One of the open questions has been whether such extreme events could be understood within the framework of solar flare models.

\begin{figure*}[!t]
\centering
\includegraphics[trim=0.0cm 0.0cm 0.0cm 0.0cm, clip=true, width=0.99\textwidth]{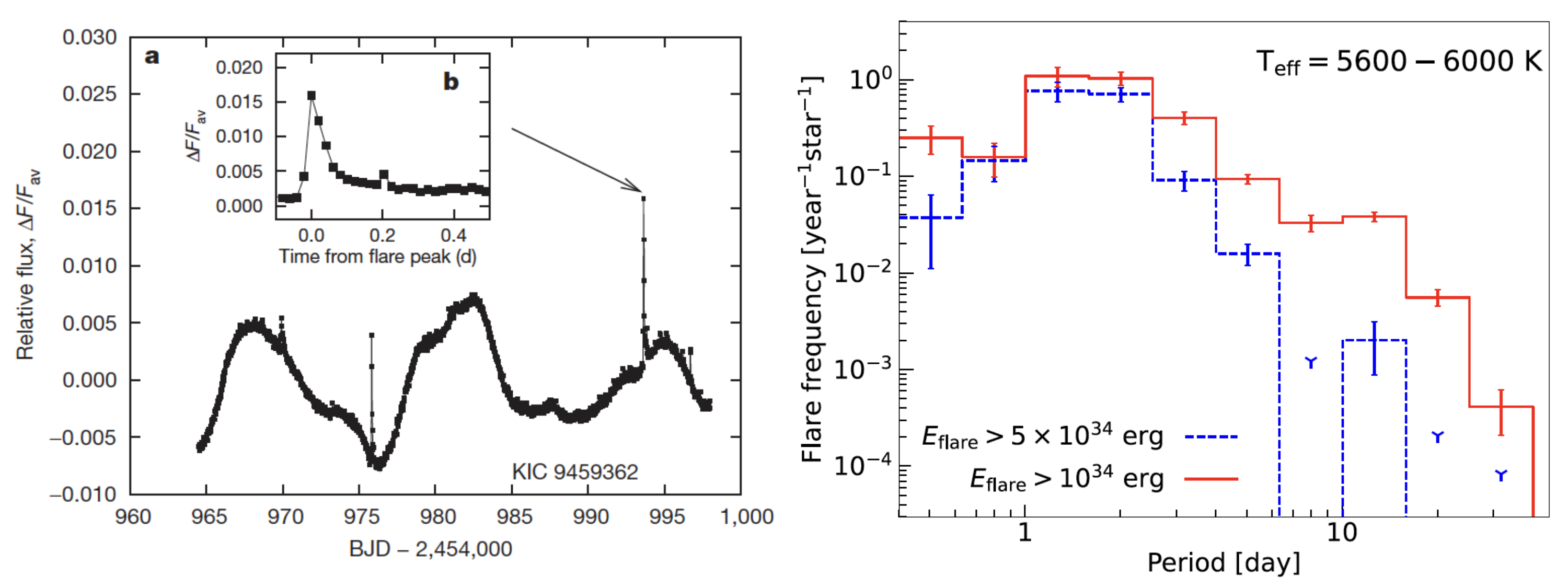}
\caption{Superflares on a solar-type star observed by Kepler Space Telescope \citep[left panel, adapted from][]{2012Natur.485..478M}. Flare frequency as a function stellar rotation period for solar-type stars with temperature of 5500--6000 K \citep[right panel, adapted from][]{2020arXiv201102117O}. }\label{Fig:Kepler_Flare}
\end{figure*}

A major advance came with space-based time-domain photometry, particularly from \textit{Kepler} \citep{borucki2010}, which transformed flare studies into a statistical discipline. 
Large homogeneous samples by \textit{Kepler} made it possible to quantify flare frequency-energy distributions, which generally follow solar-like power laws, and to establish their dependence on stellar type, rotation, and age. 
Figure \ref{Fig:Kepler_Flare} shows an example of stellar superflares on a solar-type star detected by \textit{Kepler} \citep{2012Natur.485..478M}.
Together with cluster studies and rotation analyses, a broad evolutionary picture has been established in which the flare occurrence rate and maximum energy decline as stars spin down with age 
\citep[see Figure \ref{Fig:Kepler_Flare}; e.g.,][]{2012Natur.485..478M,2013ApJS..209....5S,2016ApJ...829...23D,2019ApJ...871..241D,2019ApJ...876...58N,2019A&A...622A.133I,2020AJ....160..219F,2021A&A...645A..42I,2020arXiv201102117O}. 
This is important in an exoplanetary context, as it demonstrates that the transient radiative, and likely also particle, environment around planets evolves strongly over stellar lifetimes (e.g., represented by the sample of stars in Sect.~\ref{ssec:solar-analogs}). In a solar context, \textit{Kepler} also revealed that even slowly rotating, mature Sun-like stars are capable of producing superflares \citep[e.g.,][]{2012Natur.485..478M,2013ApJS..209....5S,2019ApJ...876...58N,2020arXiv201102117O,2024Sci...386.1301V}. 
This finding has attracted growing attention in the context of the modern Sun’s activity, as well as its possible implications for the human civilization and space exploration. 

The reconciliation of stellar superflare statistics with the
expectations for solar-type stars has recently been advanced on the
solar side by \citet{Krivova+26}, who used the \citet{Kazachenko+17}
ribbon-to-energy scaling (to be discussed below) to extrapolate from 
the largest historical
sunspot groups -- the 1859 Carrington spot and the 1947 great
group -- and found that such exceptionally complex regions could
plausibly have powered flares of up to a few $\times 10^{34}$\,erg.
This places the once-per-century stellar superflare rate inferred from
Kepler statistics \citep{2024Sci...386.1301V} at the rare-but-physical
extreme tail of, rather than beyond, the empirical solar distribution.

A long-standing question in the stellar context is the extent to
which flare productivity correlates with surface spot distributions,
mirroring the well-established connection between flares and active
regions on the Sun. Disc-integrated photometric analyses have
produced a mixed picture. Using Kepler light curves,
\citet{Roettenbacher2018ApJ...868....3R} found that the most
energetic flares showed no correlation with starspots, while weaker
flares tend to occur when starspots are visible. In a larger TESS
sample, \citet{Zhang2025arXiv251201051Z} found no evidence for a
correlation between flare rate and spot occurrence, and combined
Doppler imaging plus TESS photometry of the K2V dwarf PW~And
likewise indicated a lack of correlation between spot occupancy and
flare occurrence \citep{Lee+26}. In contrast, transit-mapping studies
that probe smaller spatial scales along the transit chord do reveal
positive correlations: \citet{Araujo2021} found a positive
correlation between the total area of starspots and the energy of
superflares in Kepler-411 (Sect.~\ref{sec:transit-mapping}).

On the Sun, where individual active regions are spatially resolved,
two statistical studies converge on a clear conclusion:
disc-integrated active-region magnetic flux is a poor predictor of
flare productivity, whereas the small-scale complexity of the flux
that actually reconnects is what matters. \citet{Dhakal+24}
demonstrated that the length of the strong-gradient polarity
inversion line (SgPIL) is the strongest photospheric predictor of
flare productivity among 20 ARs of contrasting sunspot areas,
yielding a correlation coefficient of $r=0.78$ with the flare index,
compared to only $r=0.14$ for the total unsigned magnetic flux.
Super-productive ARs required SgPIL lengths exceeding 50~Mm, and
magnetic flux emergence was found to be insufficient on its own
unless it drove convergence of opposite-polarity nonconjugate pairs.
A much larger dataset was introduced by \citet{Kazachenko+17}, based on
3137 SDO-era flares of GOES class C1.0 and above, reached a
complementary conclusion from a different observable: the GOES peak
X-ray flux is uncorrelated with the AR unsigned magnetic flux but
strongly correlated with the flare-ribbon reconnection flux,
$I_{X,\mathrm{peak}}\propto \Phi_{\mathrm{ribbon}}^{1.5}$.
Taken together, these solar results explain why disc-integrated
stellar studies have produced inconclusive spot--flare correlations:
total spot coverage does not trace the small-scale magnetic
complexity that actually drives flare energetics.

These results naturally lead to a central question addressed: how strong XUV radiation and plasma environment are associated with flares?
Recent work has increasingly shifted toward coordinated multi-wavelength observations aimed at understanding how flare energy is distrubuted among the optical--UV--X-ray bands, as well as among radiative--thermal--nonthermal--kinetic energy ranges. 
Optical surveys with \textit{Kepler}/TESS are now being complemented by HST, \textit{XMM-Newton}, \textit{Chandra}, \textit{NICER}, and radio facilities \citep[e.g.,][]{2005ApJ...621..398O,2021ApJ...911L..25M,2023ApJ...951...33T,2024ApJ...961...23N,2024ApJ...961..189N,2025ApJ...978...81K,2026ApJ...996...13O}. Some studies suggest relatively simple scaling relations between optical and X-ray flare energies \citep[e.g.,][]{2022AA...667L...9S,2024ApJ...961...23N}, whereas others have shown the unexpected large fraction of near-ultraviolet continuum in M-dwarf mega-flares \citep{2025ApJ...978...81K} or the lack of the solar-like \textit{Neupert} effect  \citep{2023ApJ...951...33T,2025ApJ...978...81K,2026ApJ...996...13O}. This is important in an exoplanetary context, because atmospheric ionization, photochemistry, and heating depend not only on the total flare energy, but also on the energy distribution in wavelength and its variation in time. 
A key observational goal for the coming years is therefore to extend simultaneous optical--UV--X-ray observations to a wider range of stellar types and activity levels.

\subsubsection{Prominence eruptions and coronal mass ejections}\label{ssec:eruptions}

In the Sun, large flares are often accompanied by large-scale plasma eruptions, i.e., prominence/filament eruptions and Coronal Mass Ejections (CMEs).
Compared with stellar flares, eruptive mass-loss phenomena have proven much more difficult to establish observationally on other stars. 
One must infer mass motions indirectly from spectral line asymmetries, transient absorption, coronal dimming, or radio signatures from a single pixel \citep[see,][for review]{2017IAUS..328..243O,2022arXiv221105506N,2022SerAJ.205....1L,2024Univ...10..313V,2025LRSP...22....2V}. 
Much of the early work focused on optical spectroscopy, particularly Balmer lines, where blueshifted emissions/absorptions are interpreted as evidence for erupting cool prominence/filament material \citep[e.g.,][see Chapter~2 for the further references]{2016A&A...590A..11V,2022NatAs...6..241N}. 
These studies demonstrated that eruptions with projected velocities of a few hundred km~s$^{-1}$ are not uncommon in active stars, and that some events reach substantially higher velocities \citep[e.g.,][]{1990A&A...238..249H}, suggesting that at least a subset of stellar eruptions may escape as CMEs rather than remaining confined.

In recent years, the observational picture has expanded beyond Doppler signatures alone. X-ray and EUV dimming events, analogous to solar coronal dimmings, have emerged as an important complementary diagnostic of coronal mass depletion associated with CMEs \citep{2021NatAs...5..697V,2022ApJ...936..170L,2024ApJ...961...23N}. At the same time, 
recent MHz-band radio observations have identified drifting radio bursts from an active M dwarf consistent with solar type II bursts, indicating propagating shock waves most plausibly driven by CMEs and providing strong evidence that such eruptions can propagate into circumstellar space \citep{Callingham2025,2025A&A...703A.198K}. Type IV-like events, which may trace energetic particles trapped in expanding magnetic structures, have also been reported for nearby active M dwarfs \citep{2020ApJ...905...23Z,2024A&A...686A..51M}. 
Taken together, these developments suggest that stellar CMEs likely occur and that, as in the Sun, they can increasingly be diagnosed through a combination of complementary observational techniques.

Current measurements likely probe only part of the overall CMEs, rather than the full picture. 
As a result, CME velocities, masses, and kinetic energies remain much more poorly constrained than flare properties, and the fraction of flares that produce successful mass ejections is still uncertain and under debates \citep{2013ApJ...764..170D,2018ApJ...862...93A}. This uncertainty is astrophysically important not only because CMEs may dominate extreme particle fluences, magnetospheric compression, and episodes of atmospheric erosion in exoplanetary systems, but also because bulk plasma ejections have been proposed to contribute to the long-term evolution of stellar mass and angular momentum \citep{2012ApJ...760....9A,2015ApJ...809...79O,2025ApJ...993...80N}. One of the most promising paths forward is coordinated multi-wavelength monitoring, including magnetic field measurements, to clarify the global picture of stellar CME propagation \citep[e.g.,][]{2024ApJ...961...23N,2026NatAs..10...64N}. 
Such efforts will be essential not only for confirming stellar CMEs more robustly, but also for determining how CME properties depend on stellar mass, rotation, magnetic topology, and age.

\subsection{Long-term variability and activity cycles}

Dynamo observables are crucial in parametrization of different dynamo models, providing reliable observational constraints for stellar and solar dynamo theories. Key stellar dynamo observables include long-term activity cycles \citep{olah2009}, 
persistent active longitudes---such as the ``flip-flop" phenomenon with two permanent active longitudes \citep{Jetsu1991LNP...380..381J}---surface differential rotation revealed using Doppler mapping \citep{Kriskovics2023A&A...674A.143K}
and Fourier spectra of photometric measurements \citep{Olah2003A&A...410..685O, Vida2015A&A...580A..64V}, and, potentially, meridional flows indicated by Doppler mapping \citep{Kovari2007A&A...474..165K, Vida2007AN....328.1078V}. Of these methods, long-term activity cycles are the ones that are studied for the longest time.

The discovery of solar cycle periodicities began in the 19th century with the work of \cite{Schwabe1844AN.....21..233S}, who first identified a repeating pattern in sunspot numbers, followed by \cite{Wolf1852AN.....35..369W}, 
who determined the cycle length to be approximately 11.1 years. 
This 11-year cycle, along with its magnetic 22-year counterpart, is known as the Schwabe/Hale cycle, most clearly observed in the variation in the number of sunspots. Longer-term variations in solar activity have also been identified, including the 70–100 year Gleissberg cycle \citep{gleissbergLongperiodicFluctuationSunspot1939}, also revealed through cosmogenic isotope records such as $^{14}$C and $^{10}$Be in ice cores and tree rings \citep{Usoskin2012ApJ...757...92U}; the $\approx$200-year Suess-cycle detected in radiocarbon proxies \citep{suessSecularVariationsCosmicRayProduced1965}; the possible $\approx$2400-year Hallstatt cycle; and even longer millennium scale variations on the order of $\approx$6000 years \citep[see][and references therein]{usoskinHistorySolarActivity2013}. 
Furthermore, the solar cycle manifested a critical suppression in several proxies during the Maunder Minimum (c. 1645-–1715), a period of sustained magnetic quiescence that serves as the benchmark for Grand Minima in stellar dynamo studies \citep{Usoskin2008LRSP....5....3U}.

Tracers of the solar cycle span a wide range of observational domains. 
Data on sunspots are recorded basically since the advent of the telescope in 1609, but regularly collected measurements are available since the late 19th century.
Photospheric activity can be tracked through the sunspot number, average spot latitudes, and the total sunspot area. Together, these reveal the spatial and temporal progression of active regions.
Chromospheric activity is monitored via the coverage of bright plage regions, through spectroscopic measurements of key emission lines, such as Ca {\sc ii} H\&K. 
Solar flares (both frequency and intensity), X-ray emission, and the 10.7\,cm radio flux all vary with the activity cycle, tracing high-energy processes in the solar corona (Sect.~\ref{ssec:flare}).
Further indicators of the solar cycle are measurements of the solar magnetic field; solar wind parameters; and terrestrial tracers (such as cosmogenic isotope variations in ice cores and tree rings), which provide insight into the solar cycle on large spatio-temporal scales.
 Of these tracers, only a handful are available for other stars, and even less are those, that are recorded for at least several decades.
However, to study stellar activity cycles, systematic observations sustained over decades are needed--improved instrumentation alone cannot replace the need for long-term data. Unfortunately, securing the commitment of funding agencies and institutions for such projects is difficult, as the results are expected typically to fall beyond the timescales of a PhD or typical grant cycle.

The first such long-term endeavour was the Mt. Wilson HK Project (1966--2002) that monitored stellar Ca {\sc ii} H\&K line variations using the S-index 
(the ratio of the flux in the H and K lines and the neighboring continuum on the violet and red side), 
tracking possible activity cycles across different stars over decades. 
The data showed that stellar activity decreases with age and slower rotation, with older stars exhibiting more stable cycles. 
Typical detected cycle lengths were about 10 years, although this could be partly biased by the length of the observations  \citep{Baliunas1995ApJ...438..269B}.  
The results also revealed a connection between the cycle period and the Rossby number, with $P_{\text{cyc}} \propto Ro^{2.0}$ 
\citep{Ossendrijver1997A&A...323..151O}.

\begin{figure}[h!]
    \centering
    \includegraphics[width=0.7\linewidth]{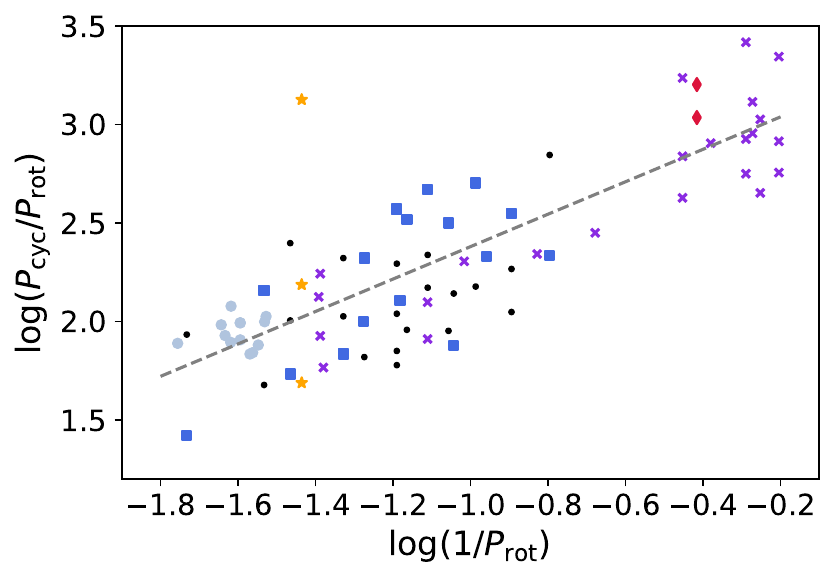}
    \caption{The relation between rotational and cycle periods from 
    {\cite{GorgeiEKDra}}. The cycles obtained from the long-term photometry data for EK\,Dra are marked by red diamonds. The cycles for stars denoted by purple are from \citet{olah2009}. The rest are adopted from \citet{olah2016} where the authors differentiated between simple cycles (light blue circles), complex cycles (dark blue squares) and additional cycles (black dots) to the complex ones. The solar cycles (Gleissberg, Schwabe, and 3–4-yr cycles) are marked in orange. Figure adapted from \citet{GorgeiEKDra}.}
\label{fig:rot_cyc}
\end{figure}

There are multiple databases that provide often decade-long photometric observations of various targets--ASAS, OGLE, different APTs (Automatic Photoelectric Telescopes) and even collection of scanned photographic glass plates (e.g., Harvard's DASCH database) that allow researchers to search for long-term variations.
These observations revealed that stars not only have activity cycles, similar to the Sun, but they have multi-periodic variations, and the length of these cycles seem to change over time \citep{olahMultiperiodicLightVariations2000, olahMultipleChangingCycles2009}. The cycle period shows some correlation with the rotation period, suggesting a relation to the dynamo number (see Sect.~\ref{ssec:sun-in-time}). Figure \ref{fig:rot_cyc}) indicates such a relationship, where the shortening in the cycle period is not overcompensated by the increase in the rotation rate. It is also known that fast rotators often have more irregular cycles, such that stochastic fluctuations in activity proxies leading to spurious cycle periods can affect this relationship. In addition, there could be a branch of low-mass stars that display a more random, irregular magnetic activity generated by a different type of dynamo than that of solar-like stars \citep{savanovActivityCyclesDwarfs2012}. Nevertheless, during the last decade evidences of activity cycles from early to late M dwarfs were reported \citep{IbanezBustos19a,Wargelin2024,Oviedo26}.

Beyond the cycle period itself, the \textit{variability} of stellar cycles
also depends on rotation. \citet{Garg+25} examined 81 Mount Wilson stars
and found that the cycle-to-cycle variance decreases with increasing
rotation rate, consistent with the expectation that fast rotators
operate at large dynamo numbers and recover quickly from weak phases,
while slow rotators with smaller dynamo numbers are more prone to
extended low-activity excursions and grand-minima-like episodes.
Their dynamo-model counterparts \citep{Vashishth+23} reproduce the same
trend, with the empirical scaling $D \propto Ro^{-0.6}$ revising the
$Ro^{-2}$ prediction of linear $\alpha\Omega$ theory and suggesting
that the rotation dependence of the dynamo number is weaker than
classical estimates -- a useful constraint for interpreting the
rotation--variability relation in exoplanet host stars.

Light curves can reveal the variations in the total spotted area, but there are other possible methods for detecting cycles.
These are of special interest in the case of space telescopes, where light curves contain significant instrumental trends and jumps between observing runs, therefore, searching long-term variations in a similar way as in ground-based light curves is not reliable. In these cases a possibility is to look for temporal variations in the standard deviation of the light curve (a proxy for light curve amplitude) -- basically using a measure for the change in spottedness \citep{mathurPhotometricMagneticactivityMetrics2014}. 
Another possibility is to look for changes in the stellar rotation period: in this case the observed rotation period shifts over the activity cycle due to the differential rotation of the stellar surface together with the changing emergence latitude of the active regions, as seen in the butterfly diagram on the Sun \citep{vidaQuestActivityCycles2013}.

The cyclic behavior that produces the butterfly diagram can be used in other ways to detect activity cycles: using Doppler-imaging to trace the variations of spot latitudes \citep{berdyuginaButterflyDiagramActivity2007a}, or even asteroseismology to find out the variations in the rotation period \citep{bazotButterflyDiagramSunlike2018}.

In the case of the Sun, we know that the solar p-mode oscillations vary with the solar cycle: during activity maximum the frequencies of acoustic modes increase and the amplitudes decrease. Such variations were indeed found -- first on an F-type star using CoRoT data \citep{garciaCoRoTRevealsMagnetic2010}, and later in larger number of solar-like stars in Kepler data \citep{kieferStellarMagneticActivity2017, santosSignaturesMagneticActivity2018} -- this can be an interesting alternative of detecting activity cycles on these objects. Unfortunately this method is currently not applicable to later type stars, like M-dwarfs: with their expected p-mode periods being in the range of 20 minutes to 3 hours and estimated amplitudes of just a few $\mu$mag, the detection of this phenomenon is yet out of reach for current instruments \citep{RodriguezLopez2019FrASS...6...76R}.

Magnetic cycles can also be characterized through starspot analysis using transit photometry. \cite{Estrela2016} derived cycle periods of about 1.1–1.3 year for the solar-type stars Kepler-17 and Kepler-63 using independent spot-based methods,
demonstrating the robustness of transit mapping for probing short stellar magnetic activity cycles.

Over the period of a solar cycle, not just the levels of solar radiation and the number/size of sunspots change, but the number of solar flares as well. Interestingly, while the variations in flare activity seems like an obvious way to trace stellar activity cycles as well, the efforts proved yet inconclusive
\citep{davenport10YearsStellar2020, feinsteinEvolutionFlareActivity2024, wainerSearchingStellarActivity2024}.

Using spectropolarimetric observations and reconstruction of the large-scale magnetic field \citep{borosaikiaSolarlikeMagneticCycle2016, 2021MNRAS.500.1243L}, researchers can directly trace polarity reversals and cyclic changes in the global stellar dynamo, providing another complementary method to photometric and chromospheric activity diagnostics in the search for stellar magnetic cycles. For a comprehensive review of recent developments on stellar cycles, we refer the reader to \citet{Jeffers2023SSRv..219...54J}.

\subsection{Modelling magnetic activity diagnostics}

\subsubsection{Photospheric diagnostics}
\label{sssec:photos_diag}

Physics-based forward modeling of photospheric magnetic activity offers a route from stellar MHD to observable diagnostics, complementing the observational measurements (Sect.~\ref{ssec:rotmod}) or inferences 
(Sect.~\ref{sec:mapping}). As of this writing, the primary framework for solar-type stars consists of coupled flux emergence and surface flux transport (SFT) models, which translate a rotation-dependent emergence rate and the dynamics of rising flux tubes into time-evolving surface magnetic flux maps. \cite{Isik18} constructed such a framework, called the Flux Emergence and Transport (FEAT) model, covering rotation rates from solar to eight times the solar value, showing that polar spot formation sets in at around four times the solar rate through the poleward transport of trailing-polarity flux from tilted bipolar regions emerging at mid-latitudes. \cite{Isik11} had earlier shown using a coupled dynamo -- SFT chain that enhanced cycle overlap and large tilt angles at intermediate rotation rates can sustain an unsigned-flux balance between polar caps and low latitudes modulated in anti-phase, providing a natural mechanism for the observed population of moderately active, non-cycling stars. 

Forward models also enable the synthesis of disc-integrated and spatially resolved diagnostics. \citet{Lehmann+17,2019MNRAS.483.5246L} decomposed SFT flux maps into spherical harmonics to synthesize Stokes profiles (Sect.~\ref{sssec:zdivszeeman}), finding that ZDI systematically overestimates contributions from axisymmetric and toroidal field components, particularly at low axial inclinations -- an important caveat when using ZDI maps of exoplanet hosts as boundary conditions for wind or coronal models (see Sect.~\ref{sec:impact}). 
\citet{Senavci21} used FEAT snapshots to synthesize Doppler images of the young solar analogue EK Draconis (Sect.~\ref{sssec:ekdra}), demonstrating that apparent low-latitude spots in observed maps can arise as artifacts of mid-latitude activity projected at the stellar inclination. 
On the rotational timescale, \citet{Isik20} showed that the anomalously large photometric variability amplitudes seen in some near-solar rotators in the Kepler sample can be reproduced by a moderate increase in emergence frequency combined with a high degree of active-region nesting (up to 90\%), without requiring a fundamentally different dynamo. 
\citet{Nemec+22} used SFT simulations to explain the observed transition from faculae-dominated brightening to spot-dominated dimming with increasing chromospheric activity level, attributing it to flux cancellation in the network field of more active stars. Developing a synthetic light-curve generator based on the FEAT platform, \citet{Nemec+23} extended this to synthetic Kepler light curves across a range of rotation rates, reproducing several characteristics of the observed variability pattern distribution. \citet{Sowmya21,Sowmya+22} further used FEAT-based surface distributions to predict astrometric jitter amplitudes for Sun-like and faster-rotating stars, providing model-based noise floors relevant for astrometric planet searches. Taken together, these studies establish that a single self-consistent flux emergence and transport framework can simultaneously constrain spot and facular filling factors, photometric variability, Doppler-imaging reconstruction biases, and astrometric signals. Such tools are versatile for the multi-diagnostic characterization of exoplanet host stars.

Most recently, \citet{Deagan+26} showed analytically that astrometric
photocentre variations probe odd-degree spherical harmonic modes of the
surface brightness distribution that are inaccessible to disc-integrated
photometry. These modes encode north--south asymmetries in active-region
distributions. The authors proposed that the activity-induced jitter traditionally
treated as noise in exoplanet studies (see Sect.~\ref{sssec:astrometry}) can be reframed as a complementary
surface-structure diagnostic: combined photometric and astrometric observations
recover a larger space of surface modes than either technique alone (see also Sect.~\ref{sssec:di} for a similar combination).

\subsubsection{Activity diagnostics and exoplanet signals}
\label{sssec:exosignals}


The photospheric magnetism of exoplanet host stars manifests itself in several observables that directly impact planet detection and their atmospheres characterization. A class of disc-integrated stellar models has been developed specifically to quantify these effects, complementing the physics-based flux emergence and transport approach of Sect.~\ref{sssec:photos_diag} with a focus on the stellar signal as seen by an exoplanet observer.

The \texttt{ECLIPSE} code \citep{Silva2003} is a numerical tool developed to model planetary transit light curves including the effects of starspots (or faculae) occulted by the transiting planet. It was the first code specifically designed for transit mapping, enabling the reconstruction of the spatial distribution and physical properties of stellar active regions on a 2D stellar disc from high-precision photometry. By fitting the small anomalies (seen as “bumps”) on a transit light curve and produced when a planet crosses over starspots, \texttt{ECLIPSE} retrieves parameters such as spot filling factor, intensity (or temperature), and longitude along the transit chord. The code incorporates detailed limb darkening and system geometry, and more recent implementations include multiwavelength capability for application to contamination on planetary transmission spectra \citep{Sumida2026}. 

The Spot Oscillation And Planet ({\tt SOAP}) tool \citep{2012A&A...545A.109B} models the imprint of spots and plages on disc-integrated radial velocity (RV) curves and photometric time series, enabling quantitative predictions of activity-induced RV jitter as a function of spot filling factor, contrast, and configuration. The substantially extended {\tt SOAP} 2.0 \citet{Dumusque14} incorporated the suppression of convective blue-shift by magnetic plages, which is the dominant contributor to long-term RV noise in solar-type stars. Another iteration included the {\tt SOAP-T} extension \citep{2013A&A...549A..35O} for modelling spot-crossing anomalies during planetary transits. These tools established a useful theoretical basis for disentangling stellar and planetary signals in precision RV and transit observations. 

Extending activity correction from analytical forward models to tomographic
surface reconstruction, \citet{Klein+25} applied Doppler imaging to solar
observations collected with HARPS-N between 2022 and 2024, retrieving
brightness maps with $\sim$36$^\circ$ angular resolution whose disc-integrated
imprint matches the observed RV variations at epochs of elevated activity.
Residual RV scatter after Doppler imaging correction reached $\sim$0.6\,m\,s$^{-1}$,
comparable to state-of-the-art activity filtering methods. Injection-recovery
tests further showed that Doppler imaging can support blind planet searches for
orbital periods $\gtrsim$100\,d around Sun-like stars observed with
new-generation highly stable spectrographs, providing a data-driven complement
to parametric tools such as {\tt SOAP}.

A comprehensive multi-diagnostic simulation framework for slowly rotating FGK stars was introduced by \citet{Meunier+19a}, simultaneously modelling RV, astrometry, photometry, and chromospheric emission as functions of stellar properties and activity level. Their work quantified the relative contributions of spots, faculae, and convective blueshift suppression to the total RV budget and planet detection capabilities \citep{Meunier+19b}. The team showed that astrometric jitter from activity cycles could in principle be detected by Gaia \citep{Meunier+20}, and demonstrated the importance of spatial coverage and axial inclination in shaping these signals. This framework is directly relevant for assessing false-positive planet detections and RV noise floors in current and future surveys.

The {\tt StarSim} code \citep{Herrero16} goes a level deeper by computing high-resolution synthetic stellar spectra as a function of the surface distribution of spots, faculae, and quiet photosphere, weighted by their respective model atmospheres. {\tt StarSim} predicts the wavelength-dependent contrast of active regions across a broad spectral range, making it particularly suited to accurate modelling of transit contamination. False positive signals can emerge on transmission spectra when the transit chord samples a stellar disc that is not uniform in surface temperature \citep{Rosich20}. This effect can mimic or mask atmospheric features of exoplanets and constitutes one of the main systematic uncertainties in atmospheric characterization from transmission spectroscopy.

Stellar activity introduces significant biases in the retrieval of exoplanet parameters derived from transit observations. Surface inhomogeneities such as starspots and faculae modify the local stellar flux and, consequently, the disk-integrated observed transit depth. When active regions are occulted during transit, they produce anomalies in the light curve that distort its shape and can bias the determination of key parameters such as the planetary radius, impact parameter, and even the scaled semi-major axis through their influence on transit duration and ingress/egress slopes. These effects have been extensively discussed in the literature \citep{Silva2003, Oshagh2013, Morris2018}, highlighting the need to properly model stellar activity when extracting accurate planetary properties.

These effects become even more critical in transmission spectroscopy, where the planetary radius is measured as a function of wavelength to infer atmospheric composition. Stellar heterogeneities give rise to the “transit light source effect” (TLSE), in which the spectrum of the occulted stellar disk differs from the disk-integrated stellar spectrum \citep{Rackham2018}. As a result, unocculted spots and faculae introduce wavelength-dependent signals that can mimic or obscure atmospheric features. Recent work by \cite{Sumida2026} shows that commonly used simplified contamination models (such as the TLSE approximation) can introduce systematic, wavelength-dependent biases in transmission spectra, leading to incorrect inferences of atmospheric properties. These biases can be comparable in amplitude to real atmospheric signatures, particularly for active stars, complicating the interpretation of spectral features.

Recent studies have therefore emphasized that neglecting stellar activity in atmospheric retrieval frameworks can lead to incorrect estimates of atmospheric composition, temperature structure, and cloud properties \citep{Barstow2015, Rackham2018, Pinhas2018}. In particular, stellar contamination can produce spectral slopes that may be misinterpreted as Rayleigh scattering or haze signatures. Moreover, the impact of stellar activity is inherently time-dependent and wavelength-dependent, as active regions evolve and rotate across the stellar disk \citep{Zellem2017, Morris2018}. Consequently, accurate modeling of stellar surface heterogeneities (e.g., forward models in Sect.~\ref{sssec:photos_diag}) is essential for robust exoplanet characterization, especially in the context of high-precision observations with facilities such as JWST.

\subsubsection{Astrometric exoplanet detectability amid magnetic activity}
\label{sssec:astrometry}

Astrometric planet detection is, in principle, less susceptible to stellar
activity contamination than radial velocity, because fewer physical processes
displace the photocenter of a stellar disc than perturb its disc-integrated
Doppler signal. \citet{Lanza+08} derived a general scaling between the surface
filling factor of active regions and the maximum photocenter excursion for
solar-like stars, identifying spurious planetary detections as a potential
hazard if activity is not monitored simultaneously. Using the observed sunspot
and plage record over one solar cycle, \citet{Lagrange+11} showed that the
Sun's activity-induced astrometric signal at 10\,pc is $\lesssim$0.2\,$\mu$as
(rms $\approx$0.07\,$\mu$as), well below the $\sim$0.3\,$\mu$as wobble
expected from an Earth-mass habitable-zone planet. For near-solar activity
levels, detectability is therefore limited by instrumental precision rather than stellar
noise. \citet{Morris+18} confirmed with simulated Gaia data that centroid
jitter from solar-like spot distributions would remain undetectable, while
highly active stars could produce
measurable signals, and that activity cycles are unlikely to be recovered from
Gaia astrometry alone. 

Building on the SATIRE photometric framework
\citep{Shapiro+14}, the three-paper series of
\citet{Shapiro+21,Sowmya21,Sowmya+22} provided quantitative
predictions directly relevant to exoplanet host stars (activity-jitter
amplitudes reviewed in Sect.~\ref{sssec:photos_diag}): the solar photocenter
displacement of $\sim$0.5\,$\mu$as at 10\,pc in the Gaia~$G$ band falls below
the planetary signal for Earth-mass targets, yet stars observed at
non-equatorial inclinations develop systematic photocenter drifts over the
activity cycle that can confound orbital solutions, active-region nesting can
amplify excursion amplitudes to Gaia-detectable levels, and rapidly rotating
host stars produce progressively spot-dominated, steeply rising astrometric
jitter that makes astrometric variation an activity diagnostic rather than a planet detection method. 

Dedicated mission-oriented studies have assessed the practical consequences
for current and planned astrometric surveys. For the THEIA target sample of
55 nearby FGK stars, \citet{Meunier+22_detection} found detection limits in the
Earth-mass regime with a low false-positive rate from realistic stellar-activity
simulations, with stellar noise playing a significant role only for the nearest
targets (e.g., $\alpha$\,Cen\,A and~B), where the habitable zone is close
enough that the activity-to-planet photocenter ratio becomes unfavourable.
\citet{KaplanLipkin+22} demonstrated that the wavelength dependence of
activity-induced jitter enables multi-passband mitigation that
reduces the effective noise floor by up to a factor of ten, reaching a
best-case six-sigma detection limit of $\sim$0.005\,$M_\oplus$ at 1\,au for
a solar analog with an ideal telescope, and recommending that future astrometry
missions adopt two or more simultaneous passbands. 

In the context of the
Closeby Habitable Exoplanet Survey (CHES), \citet{Bao+24} found that over
90\% of the primary target stars exhibit photocenter jitters below
1\,$\mu$as, with planetary detection efficiencies exceeding 80\% for
about 95\% of the sample. Exceptions are predominantly cool stars whose
habitable zones lie sufficiently close to the host that the jitter-to-planet
signal ratio degrades. For more massive planets, \citet{Sozzetti+23} showed
that combining Gaia astrometry with ground-based radial-velocity campaigns can
constrain transit windows for cold transiting Jupiters to within a few weeks,
demonstrating a complementary demographic access to the long-period giant
planet population at mass scales where stellar activity contamination is
considerably less critical. 

Collectively, these studies establish that
astrometric habitable-zone planet detection around solar-type stars is
generally robust to magnetic activity at near-solar levels, while younger,
faster-rotating, or very nearby host stars enter a regime in which
physics-based photocenter forward modeling (Sect.~\ref{sssec:photos_diag})
becomes an integral component of the detection pipeline.

\subsubsection{Chromospheric diagnostics}
\label{sssec:chrom_diag}
Physics-based modelling of the Ca~{\sc ii} H~\&~K S-index has been developed 
in parallel with the photospheric tools described above. The chromospheric 
emission in these lines is driven primarily by plages --- the chromospheric 
counterparts of faculae --- rather than by starspots, because plage regions 
are strongly brightened in the line cores through non-thermal heating. 
\citet{Sowmya21b} constructed a physics-based forward model by synthesising 
Ca~{\sc ii} H~\&~K spectra in non-local thermodynamic equilibrium using
semiempirical atmospheric models for the quiet Sun and faculae, combined 
with solar magnetic feature distributions derived from observations and from 
surface flux transport simulations. Validated against observed solar S-index 
variations over four activity cycles, the model reveals that the amplitude of 
S-index variability is strongly sensitive to the stellar inclination on rotational 
timescales --- decreasing by $\sim$81\% from an equatorial to a pole-on view --- 
while the cycle-amplitude dependence is comparatively mild ($\sim$22\%).
This inclination dependence has direct consequences for interpreting the 
S-index as a rotation tracer in exoplanet host stars: a pole-on system may exhibit 
substantially attenuated chromospheric rotational modulation regardless of the 
intrinsic activity level, a degeneracy invisible to single-epoch measurements. 

A subsequent study revisited the conventional assumption that starspots are 
negligible contributors to S-index variability \citep{Sowmya23}. Using 
high-resolution Ca~{\sc ii} H observations of sunspots with the Swedish 1-m 
Solar Telescope, it was found that sunspot chromospheres are in fact brighter 
than the quiet
surroundings in the line core, with a contrast comparable to that of plage 
regions, causing spots to increase rather than decrease the S-index. While the 
effect is small for solar-like activity levels, it becomes significant for more 
active stars with large spot filling factors, altering the inferred relationships 
between the S-index and surface area coverages.

The multi-diagnostic simulation framework of \citet{Meunier+19a}, discussed 
in the context of RV and photometry in Sect.~\ref{sssec:exosignals}, also 
produces synthetic $\log R'_{HK}$ time series semi-empirically from plage and 
network filling factors across an F6--K4 grid, finding a strong inclination 
dependence of both the time-averaged level and the long-term cycle amplitude. 
A follow-up study \citep{Meunier+19d} used this set of synthetic time series to 
probe a fundamental limitation of the widely-used technique of correcting RV
variations using $\log R'_{HK}$: the RV--$\log R'_{HK}$ correlation is 
imperfect at cycle timescales because the spatio-temporal structure of the 
butterfly diagram and the contrasting
projection geometries of the two signals produce a cycle-phase hysteresis. 
For a given activity level, the RV amplitude differs between the ascending 
and descending phases of the cycle, with the sense of the asymmetry reversing 
at inclinations of about $60^{\circ}$ from
pole-on. This degeneracy limits the effectiveness of chromospheric emission as 
an activity correction term for habitable-zone planet searches and 
motivates physically informed corrections (see also further studies below on 
the impact on RV amplitudes). 

Extending the framework to joint Ca~{\sc ii} and H$\alpha$ modelling,
\citet{Meunier+22_CaII} analysed 441 F-G-K stars from HARPS and computed synthetic 
emission time series for a range of plage and filament properties. They confirm 
that both chromospheric indicators are correlated for the majority of the sample, 
but find that a few percent of stars show anti-correlations between Ca~{\sc ii} and 
H$\alpha$ on rotational timescales --- a behaviour that plage models alone cannot 
reproduce --- and attribute this to the presence of stellar filaments that absorb 
H$\alpha$ emission, effectively decoupling the two proxies.

Beyond integrated flux diagnostics, the line profile shape of Ca~{\sc ii} H~\&~K 
contains information about the composition and geometry of the active surface that 
is inaccessible to the S-index alone. \citet{Cretignier+24a} characterised the 
distinct Ca~{\sc ii} H~\&~K intensity profiles of plages, network, and spots on 
the Sun using resolved Meudon
spectroheliogram datacubes and disc-integrated spectra from the ISS spectrograph 
and HARPS-N, also deriving the centre-to-limb variation of each component. 
A three-component decomposition of the solar S-index during cycle~24 yields average contributions of $70\pm12$\% from plages, $26\pm12$\% from network, and $4\pm4$\% 
from spots. \citet{Cretignier+24b} extended this approach to actual stellar 
observations of $\alpha$~Cen~B
with HARPS, applying principal and independent component analyses to the 
Ca~{\sc ii} H~\&~K spectral time series. The first extracted component corresponds 
to a denoised S-index, while the second acts as a more powerful proxy for correcting 
the RV contribution from the inhibition of the convective blue-shift. They further 
derived the first principal-component activity profile of Ca~{\sc ii} H~\&~K across 
the spectral type sequence from M1V to F9V, providing a template for future 
high-precision surveys. Taken together, these studies establish that 
Ca~{\sc ii} H~\&~K line profiles are a richer diagnostic resource than
their integrated flux alone, with direct applications to characterising the 
active surfaces of exoplanet host stars 
in the era of extreme-precision RV observations.

\subsubsection{Coronal diagnostics}
\label{sssec:coronal_diag}

High-energy emission from the coronae of cool stars is well-observed in X-ray fluxes and in both their short-term (flaring) \citep[e.g.,][]{Binder2024} and in a few cases their longer-term (stellar cycle) variability \citep[e.g.,][]{Wargelin2017}. However, given the strong correlation between coronal X-ray emission and the underlying stellar magnetic behavior \citep{Pevtsov2003}, there are use cases where modeling the stellar X-ray activity can provide a laboratory to probe magnetic behavior and evolution in cool stars. \citet{Farrish2021} employed a surface flux transport (SFT) treatment of the photospheric magnetic fields of cool stars and produced scaled-up magnetic activity levels representing young, active stars using both empirical and dynamo model-based relationships between stellar properties such as surface flux, rotation period, cycle period, and meridional flows. Employing a scaling based on observations of solar magnetic features between magnetic flux and X-ray luminosity first defined by \citet{Pevtsov2003}, this work produced modeled X-ray activity vs. rotation distributions that matched well observed populations of cool stars, indicating a common relationship between stellar photospheric magnetic activity and coronal X-ray emission for all main sequence cool star spectral types in the unsaturated regime (see further discussion in Sect.~\ref{ssec:model_rotact}). Scatter introduced in the SFT models by the presence of solar cycle-like intrinsic variations in photospheric surface flux also reproduce within statistical significance the spread in the X-ray activity vs. rotation relationship for cool star populations \citep[e.g.,][]{Wright2016,Wright2018}, indicating that for cool stars of moderate age, future long-term monitoring of the variability in their coronal X-ray emission may reveal Sun-like cyclic activity.

Despite the wealth of stellar X-ray observations, challenges are strongly present in monitoring the cooler EUV portion of the coronal emission spectra of other stars, given the extinction of EUV wavelengths by the interstellar medium \citep{Youngblood2019}. This challenge is unfortunate given the importance of EUV information in the modeling of many atmospheric escape processes at exoplanets \citep{GarciaSage2017,Gronoff2020}. A clear scaling law relationship between EUV emission and other properties such as stellar magnetic flux is difficult to define precisely, given the mix of processes that produce coronal EUV emission, including direct heating as well as the cooling of more energetic X-ray emission. Nevertheless, some attempts have been made to reconstruct the EUV portion of stellar spectra by scaling from the more readily-observable X-ray and UV portions of the spectra \citep{Linsky2014}. Attempts have also been made to reconstruct EUV spectra from emission measure distributions for a handful of other stars \citep{SanzForcada2011}. More recently, extensive full-atmosphere models have been developed \citep{Fontenla2016,Tilipman2021}, though this approach is computationally expensive and has thus far only been applied to a handful of stars. The approach of \citet{Peacock2019x}  has similarly utilized full atmospheric modeling and EUV synthesis from the CHIANTI atomic database to produce the PHOENIX grid of reconstructed stellar spectra from X-ray through EUV and UV wavelengths for a variety of cool star spectral types. For a more full review of the challenges inherent to the reconstruction of stellar EUV spectra, see  \citet{Linsky2026}.

\section{Rotation-activity relations}
\label{sec:rotation-activity}
A growing number of studies in the literature have examined empirical trends between stellar rotation and a variety of stellar magnetic activity indicators, comprising the class of rotation-activity relationships. These relationships serve an important function in the field of solar-stellar connections; an understanding of how magnetic activity observables scale with underlying stellar properties (e.g., age and rotation rate) allows us to draw comparisons between the Sun and other stars. Developing and improving upon scaling relationships between intrinsic stellar properties and observable stellar magnetic activity indicators helps to illuminate the underlying physical processes which produce observable magnetic phenomena in other stars. Additionally, models of magnetic activity can be used as laboratories to test how magnetic activity from underlying stellar dynamo action manifests as observable stellar magnetic activity.

\subsection{Photometric trends}

The advent of high-precision, long-baseline space photometry has transformed the study of rotation–activity trends. The NASA missions Kepler \citep{borucki2010} and TESS \citep{Ricker2015} have delivered light curves for hundreds of thousands of stars with photometric precision sufficient to detect rotational modulation at the level of a few hundred parts per million. These datasets enable homogeneous measurements of rotation periods across a wide range of spectral types and evolutionary stages, and allow statistical characterization of variability amplitudes, light curve morphologies, and temporal coherence. As a result, photometry has become a central tool for investigating magnetic activity in large stellar samples.

Photometric variability provides a direct and physically motivated proxy for surface magnetic activity because it traces brightness inhomogeneities associated with starspots and faculae. As active regions rotate in and out of view, they imprint quasi-periodic modulation on the stellar flux. The amplitude and morphology of this modulation encode information about spot filling factor, spatial distribution, contrast, and temporal evolution. Unlike chromospheric or coronal diagnostics, which probe higher atmospheric layers, photometry samples photospheric magnetic manifestations and can be applied uniformly to large stellar populations. While photometric variability is subject to geometric effects (e.g., inclination) and degeneracies between spot properties, it remains one of the most accessible and statistically powerful tracers of stellar magnetic activity.

\subsubsection{Variability Amplitude as an Activity Diagnostic}


Amplitude variability is commonly quantified using peak-to-peak measures or percentile-based ranges that reduce sensitivity to outliers \citep{Basri2011}. Alternatively, root-mean-square (RMS) variability captures the overall dispersion of the light curve and is widely applied in analyses of Kepler data \citep{Basri2013}. In TESS data, \cite{Ponte2023} adopted a percentile-based definition (2.5th–97.5th percentiles normalized by the median flux) to define the rotational photometric amplitude $A_{TESS}$, after careful removal of instrumental systematics and magnitude-dependent noise floors \citep{Ponte2023}. This approach emphasizes the importance of correcting for sector-dependent baselines and photon noise when comparing amplitudes across stars of different brightness.

Empirically, variability amplitude increases, on average, with decreasing rotation period. Rapid rotators exhibit larger modulation amplitudes than slow rotators \citep{McQuillan2014}. Expressed in terms of Rossby number, this behavior mirrors classical rotation–activity relations derived from chromospheric and X-ray diagnostics \citep{Pizzolato2003, 2011ApJ...743...48W}. In the unsaturated regime, amplitude rises with decreasing Rossby number, while at low Rossby numbers it approaches a saturation level. For solar analogs, \cite{Ponte2023} demonstrated a tight correlation between $A_{TESS}$
and the chromospheric index $R'_{HK}$, with a Pearson coefficient $\rho \sim$0.84. This result confirms that photospheric and chromospheric diagnostics trace a common magnetic driver.
A key contribution of \cite{Ponte2023} is the extension of this relation to stellar age in a homogeneous sample of well-characterized solar twins spanning 50 Myr to 8.5 Gyr:
\begin{equation}
  \log t = 12.239 - 0.894 \log A_{TESS},  
\end{equation}
indicating a monotonic decay of photometric amplitude with age and suggesting that variability amplitude can function as a stellar chronometer in the solar-twin regime. Importantly, the Sun, analyzed through TSI data rescaled to the TESS bandpass, lies on the same amplitude–activity–age sequence, reinforcing the physical consistency of the relation. Intriguingly, the scatter in photometric amplitude at a given rotation period is often very large, hinting at strongly differing active-region coverages and/or distributions \citep{Reinhold2020,Isik20}.

Physically, amplitude is governed by the filling factors of faculae and spots including their inhomogeneous contrasts, and the spatial distribution of active regions. A large coverage of axisymmetrically distributed spots may produce weak modulation, whereas concentrated spot groups yield stronger signals, leading to the possibility that more nested active-region emergence can make the star reach a given amplitude at a lower activity level than is required for a random emergence pattern \citep{Isik20}. Rotation-dependent latitudinal distribution and differential rotation affect how surface inhomogeneities evolve over time. The inclination angle of the stellar rotation axis further modulates the observed amplitude, suppressing variability in nearly pole-on systems  \citep{Nemec+23}.

Other limitations are the detection thresholds bias against low-amplitude, slowly rotating stars, and photometric amplitudes that reach instrumental noise levels for old, weakly active stars. For slow- to moderately rotating G stars, for instance, photometric variability transitions from facula-dominated to spot-dominated below near-solar rotation periods \citep{Radick+18}. The physical background of this transition was suggested by \citet{Nemec+22} as magnetic flux cancellation becomes more effective in faculae than in spots. 

Moreover, degeneracies between spot coverage and spot contrast prevent unique inference of surface properties from amplitude alone. Disk-integrated photometry primarily traces the asymmetric component of surface magnetism rather than total magnetic flux.
Nevertheless, variability amplitude scales systematically with rotation and Rossby number. These results strengthen the interpretation of photometric amplitude as a quantitative diagnostic of magnetic surface structure and its rotational evolution.

\subsubsection{Light Curve Morphology and Magnetic Topology}

Beyond variability amplitude, the morphology of stellar light curves encodes information about the spatial distribution and topology of magnetic regions. Stars may exhibit single-dip or double-dip rotational profiles, depending on whether one dominant active longitude or two approximately antipodal spot groups shape the modulation \citep{Basri2011, Reinhold2015}. For slow- to moderate rotation rates, stable, near-sinusoidal light curves typically indicate long-lived, longitudinally concentrated active regions, whereas more complex and evolving patterns reflect multiple spot groups with different contrasts and locations \citep{McQuillan2013}. For rapidly rotating active stars where spots dominate the light curve, moderate nesting or even random emergence may lead to single- or double-dip light curves that look highly periodic and regular \citep{Nemec+23}. Multi-periodic signals in the power spectrum have been interpreted as signatures of surface differential rotation, as spots at different latitudes trace distinct angular velocities \citep{Reinhold2013, Kuker2011}. Thus, light curve morphology provides constraints on magnetic topology that extend beyond amplitude-based diagnostics.

The temporal evolution of light curve shape reveals the dynamical behavior of stellar magnetic fields. Growth and decay of active regions modify both the symmetry and stability of rotational modulation on timescales of days to months \citep{Giles2017}. Phase drifts between successive rotations may arise from differential rotation or from the emergence of new spot groups or active nests at different longitudes \citep{Ozavci18,Breton24}. Persistent active longitudes and flip–flop phenomena, characterized by abrupt switches in dominant spot coverage, have been reported in active stars and interpreted as manifestations of non-axisymmetric dynamo modes \citep{Jetsu1993, Berdyugina2005}. These behaviors connect surface photometric patterns to large-scale magnetic field organization and the underlying hydromagnetic dynamo processes.

Quantitative diagnostics enable systematic characterization of morphological complexity. The autocorrelation function (ACF) provides a robust estimate of rotation period and coherence time of active regions \citep{McQuillan2013}. Lomb–Scargle periodograms identify dominant periodicities and harmonic structure in unevenly sampled data \citep{Lomb1976, Scargle1982}, while secondary peaks can indicate multi-periodicity linked to differential rotation \citep{Reinhold2015}. Wavelet analysis tracks the temporal evolution of periodic signals, revealing transient or migrating spot patterns \citep{Mathur2010}. Additional metrics, such as phase coherence indices or entropy-based measures of light curve complexity 
\citep[e.g.,][]{Basri2013}, quantify the stability of modulation over time. Together, these approaches demonstrate that light curve morphology serves as a diagnostic of magnetic structure and evolution, complementing amplitude-based rotation–activity relations.

\subsubsection{Transit mapping as a photometric activity diagnostic}
\label{sssec:transit-rot}

Planetary transits provide a geometrically anchored complement to disk-integrated rotational modulation: when the planet occults a starspot or facular region, it imprints a localized anomaly on the transit light curve whose amplitude and timing constrain the position and filling factor of the occulted region along the transit chord \citep{Silva2003}.
The spot covering fraction inferred from repeated transit crossings can therefore be compared directly with the photometric amplitude of out-of-transit rotational modulation, testing consistency between spatially resolved and disk-averaged activity diagnostics across a
range of stellar rotation rates. Stellar activity affects planetary parameter retrieval as well:
unocculted spots and faculae bias estimates of planetary radius and atmospheric transmission spectra \citep{Oshagh2013, Rackham2018, Sumida2026}, making activity characterization essential for exoplanet studies.
A full description of the transit mapping technique, the modelling codes, and its applications to differential rotation and magnetic cycle detection is given in Sect.~\ref{sec:transit-mapping}.

\subsection{Chromospheric relationships}

Stellar rotation in cool stars plays a key role in magnetic activity,  which is observed through a complex set of phenomena across the stellar atmosphere (spots, plages, filaments, faculae, etc.). Chromospheric activity proxies, derived from a diverse suite of spectral lines along different wavelengths, are excellent tools to map magnetic heating in the outer stellar atmosphere \citep{SchrijverZwaan00}. The most extended activity indicators reside in the optical range  and  are the Ca {\sc ii} H and K lines (3968.470 Å and 3933.661 Å, respectively) together with  the associated classical Mount Wilson $S$-index, usually employed to perform long-term activity studies \citep{Baliunas1995ApJ...438..269B} and the popular activity proxy log$R'_{HK}$, which is derived by correcting the $S$-index for colour dependence and the underlying photospheric contribution, thereby enabling efficient comparison of activity levels across various spectral types.  

\cite{1984ApJ...279..763N} demonstrated the existence of a fundamental correlation between the chromospheric activity index, $\log R'_{HK}$, and stellar rotation. Furthermore, they established that this relationship is significantly more robust when activity is parameterized taking in consideration the Rossby number $Ro$ (Eq.~\ref{eq:Ro}), 
which encapsulates both rotation and convection zone geometry. Increasing rotation rate and increasing fractional convection zone depth (decreasing $T_{\rm eff}$) both lead to decreasing $Ro$. 

\begin{figure}
    \centering
    \includegraphics[width=\linewidth]{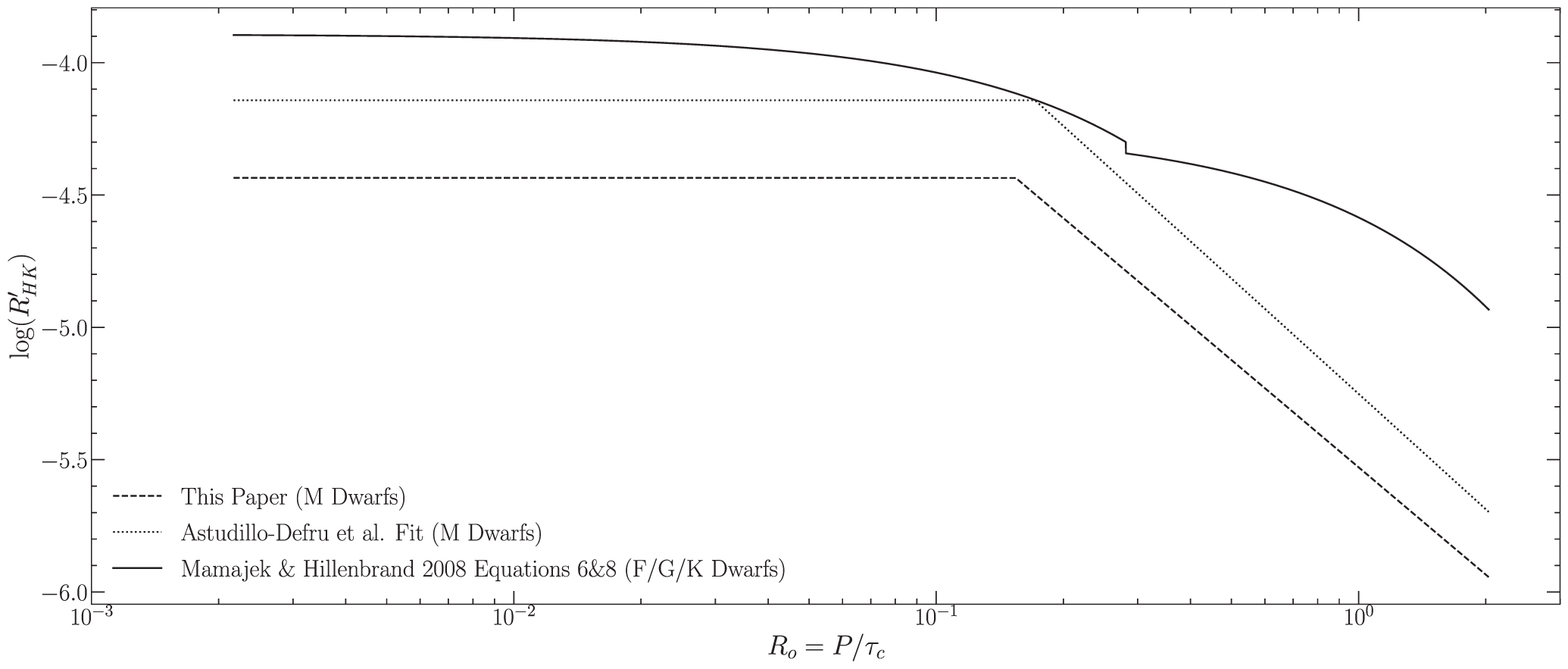}
    \caption{Derived rotation–activity curves from this work, \citet{AstudilloDefru2017}, and \cite{Mamajek2008}. Figure adapted from \citet{Boudreaux2022}.}
    \label{fig:act-rotationHHK}
\end{figure}

Extensions of this pioneer study have demonstrated that the relation between $\log R'_{HK}$ vs. $Ro$ presents two regimes:  the \textit{active} regime ($ \log R'_{HK} <-4.35$), where activity decreases as $Ro$ increases, and the \textit{saturated} regime ($\log R'_{HK} > -4.35$), where chromospheric emission remains constant and becomes independent of $Ro$ \citep{Mamajek2008}, the limit between both sequences being $Ro\sim 0.1$. This empirical relation is also observed in M stars (\citealt{AstudilloDefru2017,Boudreaux2022}, see Fig. \ref{fig:act-rotationHHK}). The rotation-activity decay could be explained by the angular momentum loss with stellar age, first invoked in  \cite{Skumanich1972} and more recently debated in \cite{vanSaders2016} for those stars that are more than halfway through their main-sequence lives, where magnetic breaking is weakened, presumably due to the dynamo approaching a shutdown.

Other popular activity indicators in the optical range include the H$\alpha$ line (at 6562.8 Å) and the Na {\sc i} D resonance lines located at 5895.92 Å (D1 line) and 5889.95 Å (D2 line),  where the H$\alpha$ line is sensitive to magnetic processes in the upper chromosphere and the Na {\sc i} D lines probe the middle-to-lower chromosphere.

\begin{figure}
    \centering
    \includegraphics[width=0.7\linewidth]{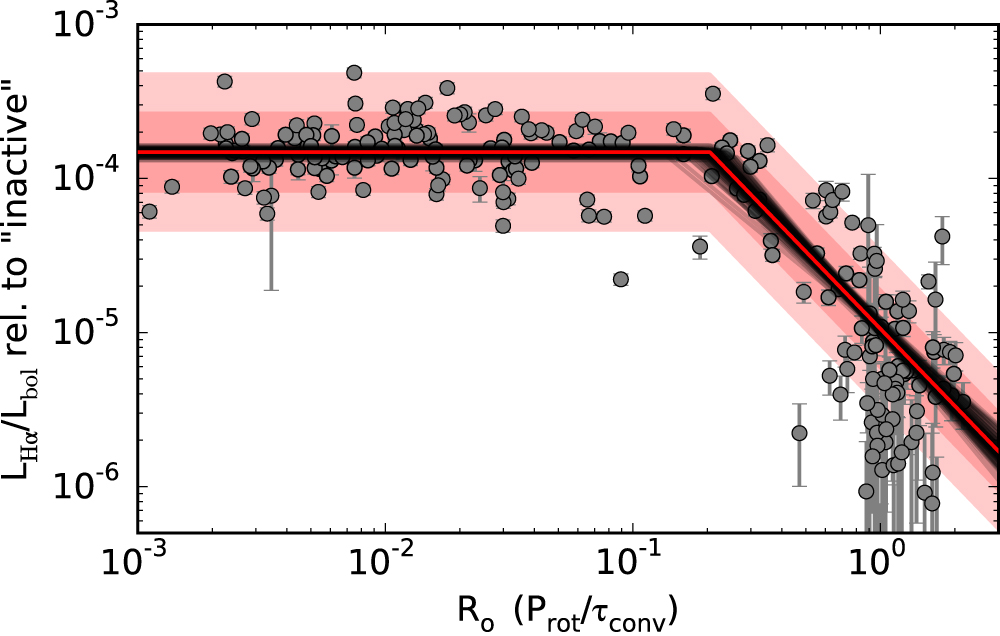}
    \caption{$L_{H\alpha}/_{Lbol}$ as a function of the Rossby number ($Ro$). In solid black line, the fit  of the canonical rotation–activity relation to the data. In  solid red line, the best fit obtained by \cite{Newton2017}, with 100 random draws from the posterior distribution shown in black. The intrinsic scatter is shown by the 1 $\sigma$ and 2$\sigma$ contours marked by the dark and light shaded red regions. Figure adapted from \citet{Newton2017}.}
    \label{fig:halpharot}
\end{figure}

Although the correlation between the Ca {\sc ii} H\&K and H$\alpha$ line-core fluxes is always  positive in the solar case \citep{Livingston2007}, the correlation between both proxies changes along the solar cycle \citep{Meunier2009}: it is enhanced during phases of higher activity \citep{Maldonado19}. This variability underscores that the relationship between different chromospheric layers is dynamic and influenced by the evolution of magnetic features, such as plages and filaments, across the stellar disk. Furthermore, the positive  correlation between H$\alpha$ and Ca {\sc ii} H\&K fluxes observed in the solar case is not universaly valid \citep{Cincunegui2007,GomesdaSilva2014,IbanezBustos2023}. Nevertheless, when considering mean values, H$\alpha$ indexes follow the same empirical  pattern as the $\log R'_{HK}$ with Rossby number (\citealt{Newton2017}, see Fig. \ref{fig:halpharot}).

 One pivotal result on the variability of H$\alpha$ line was provided by  \cite{GomesdaSilva2022}. Using a 20-year baseline from the California Legacy Survey and high-resolution HARPS data of hundreds of FGK stars, they demonstrate that a narrow 0.6 Å bandwidth for H$\alpha$ maximizes the positive correlation with Ca {\sc ii} activity signals, whereas the standard 1.6 Å bandwidth often degrades the signal by incorporating the line wings. In many inactive stars, as activity increases, the H$\alpha$ line core flux rises while the wings simultaneously broaden and decrease in flux. Therefore, \cite{GomesdaSilva2022} suggested to employ this narrower window to measure rotation periods for FGK stars. Furthermore, H$\alpha$ is a good diagnostic to measure stellar rotation period in M stars \citep{Newton2017}.

On the other hand, empirical studies demonstrate that the Na {\sc i} D and Ca {\sc ii} indices are generally also not strictly correlated due to their reliance on different atmospheric sources, illustrating that correlations among these various indices are not ubiquitous and depend heavily on the star's activity level or spectral type, with M dwarfs often diverging from general trends \citep{Meunier2024}.  In this regard, the Na {\sc i} D doublet has been shown to be inadequate for measuring stellar rotation periods \citep{Fuhrmeister2019}.

During the last decades, motivated by the fact that terrestrial habitable-zone planets are easily detected around M dwarf stars due to their low stellar masses, a new generation of high-resolution infrared spectrographs, such as SPIRou \citep{Moutou2015}, CARMENES \citep{Quirrenbach2014}, and NIRPS \citep{Bouchy2025}, has emerged to optimize the signal-to-noise ratio for radial velocity (RV) techniques . With this new instrumentation, longer wavelength stellar activity proxies have become accessible, most notably the He {\sc i} triplet (transitions at 10832.1 Å, 10833.2 Å, and 10833.3 Å), which serves as a key tracer for evaporating exoplanet atmospheres \citep[e.g.,][]{Spake2018}. In the solar case, the He {\sc i} triplet typically exhibits absorption in the chromospheric network but manifests in emission during flares and prominences \citep{Penn2014}. Recent analysis of multi-epoch solar observations by \citet{Mercier2025} reveals that while short-term He {\sc i} variability is often linked to telluric contamination, long-term variations correlate strongly with the $R'_{HK}$ activity index, reflecting rotational modulation by chromospheric plage regions. Nevertheless, the authors show that this variability has a negligible impact on the planetary parameters retrieved in the study of atmospheric escape.

In M dwarfs, the He {\sc i} triplet  and the Paschen series, including Pa $\beta$ at 12821.6 Å, Pa $\gamma$ at 10941.1 Å, and Pa $\delta$ at 10052.1 Å, show no clear monotonic correlation with standard tracers like H$\alpha$, suggesting they probe distinct physical conditions or atmospheric heights \citep{Schofer2019}. In addition to their sensitivity to flares \citep{Fuhrmeister2023},  Paschen lines offer insights into the spatial distribution of activity; for example, in AU Mic, He {\sc i} triplet emission is concentrated at equatorial latitudes, while Pa $\beta$ originates primarily from polar regions \citep{Klein2021}. Complementing these near infrared proxies are the Ca {\sc ii} infrared triplet (IRT) at 8500.4 Å, 8544.4 Å, and 8664.5 Å, which serves as a robust NIR alternative to the Ca {\sc ii} H \& K lines and correlates strongly with H$\alpha$ \citep{Schofer2019}.

One established approach to analyzing chromospheric activity diagnostics is based on the calculation of excess fluxes, obtained by subtracting the underlying photospheric contribution from the total observed flux in the chromospheric emission lines. This spectral subtraction technique traditionally utilizes a template spectrum from a non-active `reference' star of similar spectral type and luminosity class -- or, alternatively, a synthetic stellar model -- to isolate the purely chromospheric component \citep{Frasca1994, Montes1995}. This process is particularly critical for diagnostics such as H$\alpha$ and the Ca {\sc ii} infrared triplet, where the chromospheric emission is often insufficient to produce a distinct emission peak \citep{Busa2007,MartinezArnaiz2010}.  In this context, the {\textsf{iSTARMOD}} tool provides a Python-based implementation of the spectral subtraction technique designed to isolate excess chromospheric emission from the dense absorption lines of late-type stars \citep{Labarga2026} .

On the near-ultraviolet (NUV) side of the electromagnetic spectrum, a good chromospheric activity diagnostic is the Mg {\sc ii} h\&k lines, which are analogous to Ca {\sc ii} H\&K lines and show higher contrast given the fainter NUV background \citep{Schrijver1992}. Both activity proxies seem to follow a tight correlation in FGK stars \citep{Buccino2008}. The NUV chromospheric features are particularly crucial because they act as proxies for the high-energy radiation emitted by the star. Specifically, UV emission from stellar chromospheres drives photo-chemistry in exoplanet atmospheres, making the flux and ratio of specific UV wavebands useful for atmospheric studies \citep{France2016}. Research confirms that the relationship between UV emission line fluxes and optical indices such as $\log R'_{HK}$ can be employed to  predict the average UV emission of an M star to within a factor of 2–4 when direct UV observations are unavailable \citep{Youngblood2017}. Furthermore, while tight correlations between chromospheric (Mg {\sc ii}/FUV) and coronal (X-ray) fluxes exist for F, G, and K dwarfs, the correlations found for M dwarfs show greater scatter or follow different trend lines \citep{Linsky2017}.

\begin{figure}[htb!]
    \centering
    \includegraphics[width=0.9\linewidth]{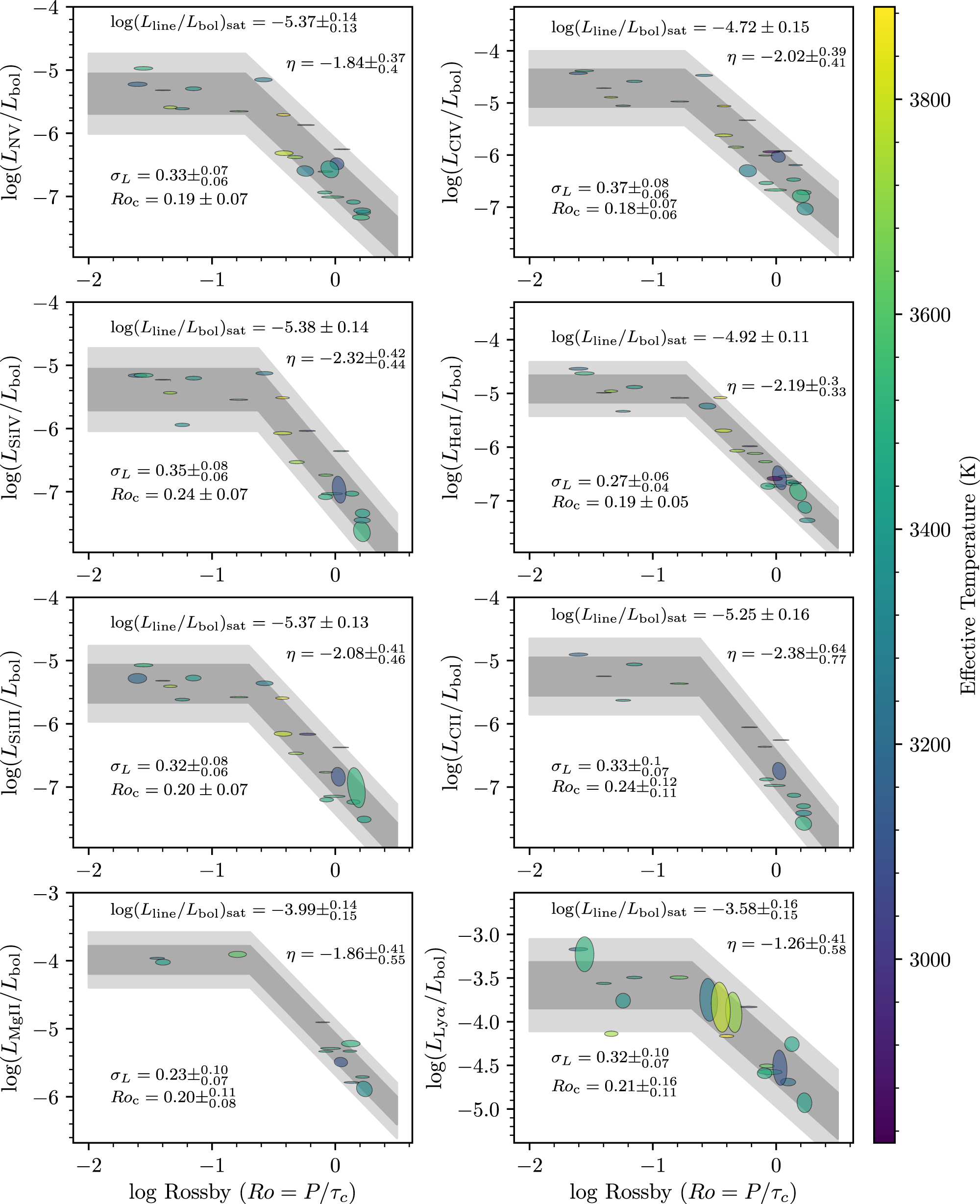}
    \caption{Rotation–activity correlations across all of the UV emission features analyzed by \citet{Pineda2021}. Individual data points are represented by 1$\sigma$ error
ellipses, shaded to indicate each star’s effective temperature. Figure adapted from \cite{Pineda2021}.}
    \label{fig:uvrot}
\end{figure}

\cite{Pineda2021} established that the rotational evolution of far ultraviolet (FUV) emission lines in M dwarfs is well-characterized by a broken power law, exhibiting a saturated regime for fast rotators ($Ro\le 0.2$) and a subsequent decay for slower rotators (see Fig. \ref{fig:uvrot}). This study also demonstrated that FUV spectroscopic features are tightly correlated, allowing for the prediction of line luminosities from known rotation periods to within an intrinsic scatter of $\sim$0.3 dex. These results imply that planets in the habitable zones of early-to-mid M dwarfs accumulate up to 10--20 times more EUV energy relative to modern Earth, with the majority of this exposure occurring during the first gigayear of the star's lifetime. Nevertheless, while these power-law relations are useful as a predictive tool, the significant scatter observed among stars with similar rotation periods suggests that direct observations of host stars remain the gold standard for characterizing the atmospheric evolution of orbiting planets (see Sect.~\ref{sec:impact}).

\subsection{Coronal relationships}

Stellar coronal emission in X-rays is a direct tracer of stellar magnetic activity for cool stars and is closely related to stellar rotation. Various X-ray observatories have measured the X-ray fluxes and luminosities of samples of cool stars to relate them to stellar rotation, where the most relevant X-ray band is soft X-rays, which make up the bulk of coronal emission from cool stars. The exact energy bands used in various studies depend on instrumental characteristics of the telescopes used, but the community standard clusters around 0.2-2~keV.

Which stellar rotation-dependent quantity is best to use for X-ray activity-rotation studies is a matter of ongoing discussion. Most commonly used is the Rossby number $Ro= P_{\rm rot}/\tau_c$ (Sect.~\ref{sec:sun}, Eq.~\ref{eq:Ro}).
While the rotation period can be measured from stellar light curves, the convective turnover time is not directly observable and is either theoretically modelled (see for example \citealt{1996ApJ...457..340K, 2010A&A...510A..46L}) or empirically defined in order to reduce the observed scatter in X-ray-rotation relationships (see for example \citealt{1984ApJ...279..763N, 1994A&A...292..191S, 2011ApJ...743...48W}). An alternative way to study the X-ray-rotation relationship is to use the rotation period (or a simple function of it) on the x axis \citep[see][]{2014ApJ...794..144R}.

The most commonly used X-ray quantity is the so-called fractional X-ray luminosity $R_X = L_X/L_{bol}$, i.e.\ the X-ray luminosity $L_X$ divided by the bolometric luminosity $L_{bol}$ of the star. However, some studies also discuss the diagnostic value of using plain $L_X$ \citep{2020A&A...638A..20M} or the X-ray flux through the stellar surface $F_{X,\,surf}$ for stellar activity characterization (see for example \citealt{1997A&A...318..215S, 2004A&A...417..651S, 2017MNRAS.471.1012B}).
Both the Rossby number and the fractional X-ray luminosity are tools to make stars of different mass and therefore surface area and dynamo efficiency comparable to each other. For this reason, they are widely used as the main quantities for such samples.

\begin{figure}
    \centering
    \includegraphics[width=0.7\linewidth]{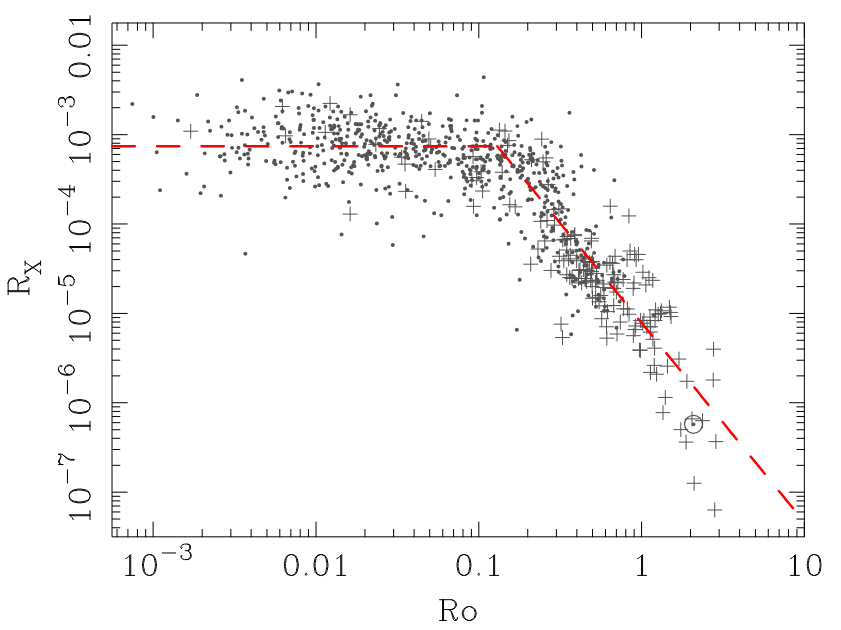}
    \caption{The coronal rotation-activity relationship in the $R_X$ versus $Ro$ form. Stars known to be binaries are shown as plus symbols, and the Sun is indicated with a solar symbol. The best-fitting saturated and non-saturated activity–rotation relations described in \cite{2011ApJ...743...48W} are shown as a dashed red line. Figure adapted from \cite{2011ApJ...743...48W}.}
\label{fig:wright2011}
\end{figure}

The coronal rotation-activity relationship in its classical form of $R_X$ versus $Ro$ displays several regimes (see \citealt{2003A&A...397..147P, 2011ApJ...743...48W, 2018MNRAS.479.2351W, 2020A&A...638A..20M}). As shown in Fig.~\ref{fig:wright2011}, the so-called unsaturated regime is found at $Ro\gtrsim 0.1$, where slower rotation corresponds to lower X-ray activity. The saturated regime is found at $Ro\lesssim 0.1$, where the X-ray activity of the stars displays a plateau and faster rotation does not yield an increase in X-ray activity. 

A historically much-discussed sub-regime is the so-called super-saturation regime of very fast rotation, corresponding to the $Ro\lesssim 0.003$ range. Observationally, not many stars are observed here, but there is a statistically meaningful slight reduction in X-ray luminosity of the stars in that regime compared to the regular saturated regime. The physical reason behind supersaturation has been much debated, with centrifugal stripping \citep{1999A&A...346..883J} or filling factor limitations \citep{2001A&A...370..157S} being the most discussed scenarios. 

Recent advances in measuring magnetic fields in stars have shown that surface-average magnetic fields of low-mass stars show a tight correlation with stellar rotation (see Sect.~\ref{ssec:rotmag}) and X-ray luminosity \cite{Reiners22}. This demonstrates the validity of the central assumption of the coronal rotation-activity relationship, namely that the dynamo itself and not just its stellar atmospheric manifestation depend on the stellar rotation rate.

\subsection{Modelling rotation-activity relationships}
\label{ssec:model_rotact}
The rotation-activity relationship of cool stars shows a power-law in X-ray luminosity in the form $R_X \propto {\rm Ro}^{-p}$ with $2<p<3$ for the unsaturated part (${\rm Ro} \lesssim 0.1$), until it saturates to a nearly constant value for ${\rm Ro}\gtrsim 0.1$. The physical mechanisms responsible for this difference in behavior between the saturated and unsaturated regimes are not yet clear. \citet{Blackman15} interpreted the change from unsaturated to the saturated regime as a change in the differential rotation timescale normalized to the convective eddy correlation time: for fast rotators, the differential rotation shears convective eddies on a timescale shorter than the convective turnover time, potentially explaining the transition to rotation-independent activity. 

\begin{figure}
    \centering
    \includegraphics[width=0.65\linewidth]{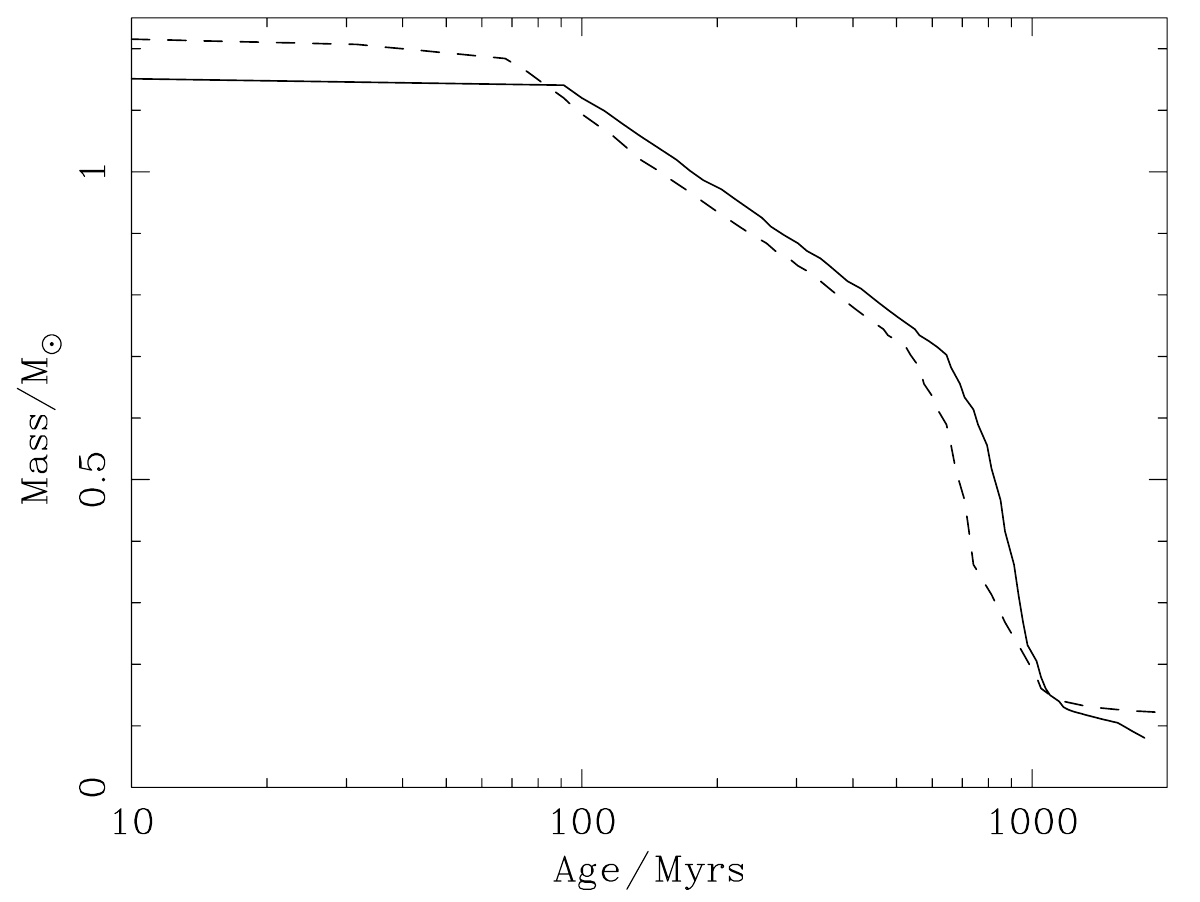}
    \caption{Rotation-activity saturation threshold $Ro < 0.13$ on the 
    mass-age plane (solid line), in comparison to the age when stars transition 
    from the rotational fully-convective sequence to the radiative-convective 
    interface sequence (dashed line). Figure adapted from \cite{2011ApJ...743...48W}.}
    \label{fig:Wright-transition}
\end{figure}
\citet{2011ApJ...743...48W} and references therein posit several possible explanations for the saturation of the $R_X-Ro$ relationship, namely that either the interior dynamo itself saturates, the surface of the stars saturate in filling factor or coverage of active regions \citep{1984A&A...133..117V}, or the corona becomes destabilized by fast rotation and undergoes centrifugal stripping \citep{1999A&A...346..883J}. \citet{2011ApJ...743...48W} provides an additional plausible explanation via the \citet{2003ApJ...586..464B} spin-down formulation whereby stars transition from a convective to an interface dynamo at a critical Rossby number. This work notes that in stellar mass-age space, the Ro = 0.13 saturation threshold and the convective-interface dynamo boundary nearly overlap (shown in Fig.~\ref{fig:Wright-transition}), implying that the spin-down transition from one dynamo type to the other can provide a physical explanation for the critical $Ro = 0.13$ value of the evolution from the saturated to unsaturated branch as cool stars spin down with age. 

A key question in understanding the relationships between underlying dynamo action and observable stellar magnetic activity is whether the same activity-rotation relationships hold for both partially-convective (late-F through early-M) stars and fully-convective (late-M) stars. Additional observations in \citet{Wright2016} and \citet{Wright2018} found that fully-convective M dwarfs generally lie along the same $R_X-Ro$ activity trends in the unsaturated regime as partially-convective stars. This suggests that either the manifestation of stellar dynamo activity as observable stellar magnetic activity is agnostic to the underlying stellar dynamo structure in the stellar interior (perhaps being more reliant on the convective processes near the stellar surface than on the generation of the magnetic field deeper in the interior), or that the dynamo processes in both fully- and partially-convective stars may be more similar than previously thought. More advanced dynamo modeling with connections to real stellar observations is needed to explore this question thoroughly, though dynamo modeling as a field faces challenges such as the prohibitively high computational expense of complex dynamics over very small scale heights near the photosphere  \citep{Yadav15,Brun+Browning17,Kapyla+23}.

\section{Surface reconstructions of magnetic activity}
\label{sec:mapping}

\subsection{Inferred spatio-temporal distributions}
\subsubsection{Photometry}
\label{sect:modelling-photometry}
Starspots are one of the most accessible diagnostic features of stellar magnetic activity: it only needs photometric observations, that can be done using small-sized, ground-based telescopes and space observatories as well. Starspot temperatures can be recovered -- although with high error margins -- using multi-colour photometry \citep[see, e.g.][]{olahTimeseriesPhotometricSPOT1997}. 
In photometric analyses based solely on out-of-transit modulation, spot longitudes can be recovered with reasonable accuracy, whereas spot latitudes remain somewhat harder to constrain,
because they are encoded primarily through visibility and limb-darkening effects \citep{Tuomi2026}. 
In contrast, spot transit mapping \citep{Silva2003} allows the latitude of the occulted spots to be retrieved more directly from the geometry of the planet’s transit chord, substantially improving the latitude determination for the subset of spots crossed by the planet. 

The spot configuration of the stellar surface can be recovered either by analytical models 
\citep{buddingInterpretationCyclicalPhotometric1977,dorrenNewFormulationStarspot1987} or with continuous spot distribution (a.k.a. the disco ball model), similar to those considered for Doppler imaging \citep{lanzaLongtermStarspotEvolution1998b, lanzaLongtermStarspotEvolution2006,roettenbacherImagingStarspotEvolution2013a}. 
The main drawback of photometric spot modelling is the degeneracy of spot contrast, size, and latitude. This is especially true for ground-based photometry, where the data typically have temporal gaps and larger scatter. In such cases, a simple two-spot model is usually sufficient to reproduce the observations within the uncertainties (see, e.g., the tests by \citealt{kovariTestingStabilityReliability1997}).

Due to their higher precision, in the case of space-borne observations, more detailed models are required. However, because these models involve a large number of parameters, an MCMC (Markov Chain Monte Carlo) approach is typically used to find the optimal solution \citep{mosserShortlivedSpotsSolarlike2009, frohlichMagneticActivityDifferential2012}.
There are more detailed analytical models available that include e.g. umbra/penumbra effect as well \citep{kippingAnalyticModelRotational2012a}, however, these features are rarely used in practice, as these increase the number of parameters drastically.
Longer term variations of the light curve due to spot evolution can be modelled by either analytical models by splitting the light curve to overlapping windows and tracing the evolution of the parameters 
\citep{vidaFourcolourPhotometryEY2010, strassmeierBinaryinducedMagneticActivity2011,Lehtinen2011}, or, more recently with Gaussian processes (GPs, see e.g., \citealt{lugerMappingStellarSurfaces2021}). 
Despite the latitude-size degeneracy, employing simultaneous space-borne photometry in Doppler imaging has been shown to enhance latitudinal accuracy of the resulting maps, owing primarily to latitude-dependent visibility times \citep{Lee+26}. 

\subsubsection{Doppler imaging}
\label{sssec:di}
Photospheric line profiles are sensitive to brightness inhomogeneities on stellar surfaces. As the star rotates, starspots cross the visible hemisphere of the star, leaving localized deviations from the average line profile at various wavelength differences from the line center, depending on their relative Doppler shifts. The associated inverse problem is to find the surface-brightness (or temperature) map that would lead to the observed series of line profiles at various rotational phases, called Doppler imaging (DI). 
The first idea of DI states back to the 50's 
\citep{1958IAUS....6..209D}
and application to cool stars was first implemented in the 1980's
\citep[e.g.][]{1983PASP...95..565V,1987ApJ...321..496V}.
Since then several authors and groups have tackled this problem. In early days, especially in case of temperature mapping, the DI was based on one (such as Fe\,{\sc{i}} $\lambda$6546.24\,\AA) or a few spectral lines that were preferably close to unblended. 
\citep[e.g.][]{1990A&A...230..363P}. 
By using a set of lines with different temperature sensitivity,
e.g., including ionised lines, the absolute temperature scale
could be set more accurately \citep[see, e.g.,][]{hackman2001}.

However, this kind of approach was very sensitive to the quality of the spectra. To this purpose, most commonly a least-squares-deconvolution 
\citep[LSD;][]{1997MNRAS.291..658D} 
line profile that represents a selected set of (often hundreds to thousands of) photospheric lines is generated, effectively enhancing the S/N ratio. The Singular Value Decomposition technique provides an alternative framework for the same type of reconstruction 
\citep[SVD;][]{2012A&A...548A..95C}.
This approach is very similar to that of the principle component analysis (PCA) and in it the similarity of the individual Stokes~$V$ profiles allows one to describe the most coherent and systematic features present in all spectral line profiles as a projection
on to a small number of eigenprofiles.
The optimization of a surface brightness map is then carried out an LSD- or SVD-profile synthesis, by minimizing the absolute deviations from the observed profile, where regularization is applied following the maximum entropy minimization
\citep{1987ApJ...321..496V}, 
the Tikhonov criterion
\citep{1977SvAL....3..147G,1991LNP...380..309P}, 
or the Occamian Approach
\citep{1998A&A...338...97B}. 
During the years some alternative LSD codes have also been proposed, like
selective LSD
\citep[sLSD;][]{2005A&A...435..261W}. 

Using LSD or similar techniques, has allowed for implementing DI on fainter star and/or using shorter exposure times. The trade-off is that when combining the information from a large number of absorption lines into one ``mean" profile, information about the temperature behaviour of individual lines is lost.


The Doppler images are mostly based on spectra observed in high-resolution, meaning resolution $R$ between 65\,000
\citep[ESPaDOnS;][]{2006ASPC..358..362D}
and 115\,000
\citep[HARPS-Pol;][]{2008SPIE.7014E..0OS},
although Doppler imaging is possible with moderate resolutions ($R\approx 20\,000$) as well \citep{Kriskovics2023A&A...674A.143K,GorgeiEKDra}.
Currently, only one instrument, namely the Potsdam Echelle Polarimetric and Spectroscopic Instrument
\citep[PEPSI;][]{2015AN....336..324S}
at the 2$\times$8.4\,m (effective aperture of 11.8\,m) Large Binocular
Telescope (LBT) offers significantly higher resolution mode $R>$200\,000. However, this highest resolution mode has been used to map a surface of only one star, the young solar analogue EK\,Dra
\citep{2018A&A...620A.162J}.
This study provided the first possible evidence for starspot penumbrae on a star other than the Sun. The more general result was that the relatively small line broadening together with the only moderately high spectral resolutions previously available appear to be among the main contributors to the lower-than-expected spot contrasts when comparing to the Sun. However, the gain from very high resolution is realized only when the line broadening is dominated by Doppler broadening.

Although DI is a powerful method, its ill-posedness will lead to artifacts in the inversion. These are, of course, more apparent when the quality of the data is poorer. With insufficient information provided by the observations, e.g., the regularisation will influence the solution significantly. 
It is known that DI suffers particularly from uneven and insufficient phase sampling. This can easily be seen in long-term studies of particular targets based on observations with varying phase coverage \citep{hackman2019}. Furthermore, typical artifacts related to both lower S/N ratio and insufficient phase coverage are errors in spot latitudes, ``shadows" of spots of the opposite hemisphere, and vertical stripes or arches in the temperature maps \citep[see, e.g.,][]{hackman2001}.

\citet{Bahar24} introduced SpotDIPy, an open-source Doppler imaging code that can be used to simultaneously invert atomic line profiles and TiO band profiles, demonstrating that molecular bands provide crucial constraints on spot coverage and temperature distributions. SpotDIPy validated its accuracy against DoTS \citep{Cameron92_DOTS}, while offering enhanced features such as automatic limb- and gravity darkening, in addition to improved handling of macroturbulence effects. 

\begin{figure}
    \centering
    \includegraphics[width=\linewidth]{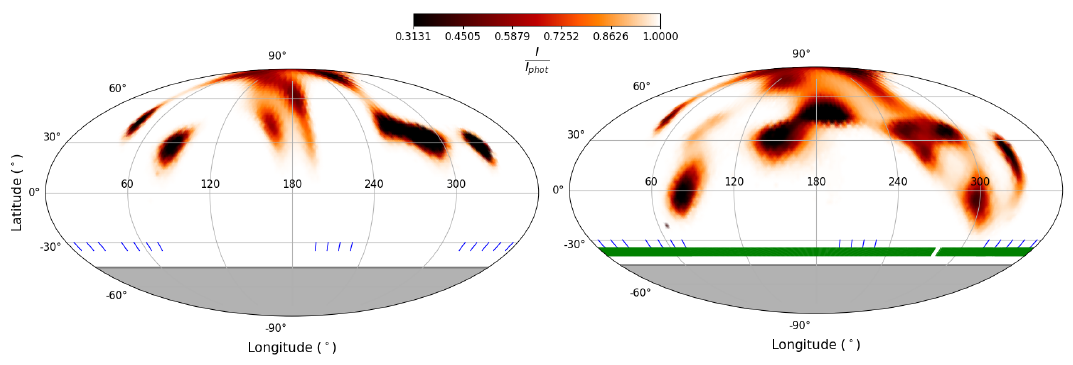}
    \caption{Conventional Doppler imaging (left) compared to a combined reconstruction 
    using Doppler imaging and contemporaneous TESS photometry (right) of PW And, an 
    active K2 dwarf star. Sampled rotational phases are shown with blue (spectroscopy) 
    and green (photometry) ticks. Figure adapted from \citet{Lee+26}.}
    \label{fig:lee26-DI-LCI}
\end{figure}

In addition to high resolution spectroscopy, contemporaneous photometry can be used as a constraint in DI \citep{hackman2001}. A similar approach has been employed in Zeeman-Doppler imaging (ZDI) by \citet{Finociety+21,Finociety+23}, who found better reconstruction performance at low latitudes thanks to TESS photometry in parallel to ZDI. This capacity is also introduced in
the new version of SpotDIPy \citep{Lee+26},
using the same test case PW And (K2V, $P_{\rm rot}=1.76$~d) 
with \citet{Bahar24}. The authors combined high-resolution spectroscopy from the 3.8-m Seimei telescope with high-cadence TESS photometry for PW And. This hybrid approach revealed that densely sampled photometry critically compensates for the inherent limitations of spectroscopy-only methods, particularly in recovering low-latitude and southern hemisphere features that remain poorly constrained by Doppler imaging alone due to inclination effects and visibility biases. The DI-only reconstruction is compared to the combined Doppler and light-curve imaging in Fig.~\ref{fig:lee26-DI-LCI}.  The updated SpotDIPy code advanced from two-temperature to three-temperature approximations (photosphere, cool spots, hot spots) and demonstrated improved reconstruction accuracy under realistic observational conditions with incomplete phase coverage and moderate signal-to-noise ratios. 

These studies establish that combining spectroscopic line profile distortions with photometric brightness variations provides complementary constraints that yield more reliable surface reconstructions, improved latitude determination, and better spot filling factor estimates than either technique alone.

\subsubsection{Magnetic field mapping}
\label{sssec:zdi}
The most commonly used method to map stellar magnetic fields is 
Zeeman-Doppler imaging (ZDI). This
is very similar to DI but it uses in addition polarized spectra to map the magnetic fields of the stars. The basic principles of the technique were given by
\citet{1989A&A...225..456S}
and the first results based on stellar data were published by
\citet{1996IAUS..176...53D}.
Since then, the most cool star ZDI studies have used only circular polarization data \citep[e.g.][]{Petit2004MNRAS.348.1175P,Marsden2006MNRAS.370..468M,Carroll2012A&A...548A..95C,2022A&A...659A..71W}.

The ZDI technique has been tested in various ways. Both \cite{2019MNRAS.483.5246L} and \cite{hackman2024} applied ZDI reconstruction on surface magnetic field maps from simulations. Both cases show, that using Stokes-V profiles, the stellar radial field is better reconstructed than the azimuthal, and especially meridional field. Owing to cancellation of Stokes-V signals from opposite-polarity flux elements, small scale structures are naturally lost in the reconstruction, with significantly reduced total magnetic field as a result. Furthermore, \cite{hackman2024} showed the axisymmetry of the reconstruction is overestimated, and that stellar parameters such as the inclination of the rotation axis will also strongly influence the result. Still, these tests show a surprisingly good performance of the ZDI methods.

For ZDI the multi-line approach is even more crucial than for DI. This is because the Stokes-V signal of individual stellar photospheric absorption lines will not reach above the noise level.
In addition to the most commonly used LSD methods (see section \ref{sssec:di}), several variations of it have been proposed such as nonlinear deconvolution with 
deblending
\citep[NDD;][]{2009A&A...507.1711S},
Zeeman component decomposition
\citep[ZCD;][]{2010A&A...522A..57S}
and improved LSD 
\citep[iLSD;][]{2010A&A...524A...5K}.

The first ever ZDI map of a late-type star using all four Stokes parameters was published by \cite{Rosen2015ApJ...805..169R} for II\,Peg. The main reason for the lack of such ZDI maps is the fact that Zeeman linear polarization signatures in spectral lines are up to 10 times weaker than circular polarization, thus making them more difficult to detect. Furthermore, interpretation of full Stokes vector spectroscopic observations requires a detailed polarized radiative transfer modeling. The main difference between maps using only Stokes~$IV$ and full Stokes is that the recovered radial, meridional, and azimuthal field components become stronger when all four Stokes parameters are used for the inversions. Furthermore, especially the meridional field is insufficiently reconstructed using just Stokes $IV$.

The field topology also becomes much more complex, some individual surface features become stronger, and the total magnetic field energy increases by a factor of 2.1--3.5. 
The study by \cite{Rosen2015ApJ...805..169R} also revealed that the extended field topology obtained from potential field extrapolation from the ZDI results is noticeably affected by the difference between the radial field component recovered in the Stokes~$IV$ and $IQUV$
inversions.

Recently \citet{Donati2025A&A...700A.122D} mapped two M dwarfs, AU\,Mic and EV\,Lac using also full Stokes, but now near-infrared observations were used for mapping instead of visual wavelengths. They also concluded that by using Stokes~$QU$ Zeeman signatures one can reconstruct stellar magnetic fields more reliably. Furthermore, including Stokes~$QU$  is especially useful for slowly rotating stars with more complex fields. 

Comparing ZDI maps with other ways of measuring stellar surface magnetic fields show, that most of the field is hidden in small structures. In these opposite polarities of small scale magnetic fields cancel each other, leaving no disc integrated sign in the Stokes V profiles.
A recent method to measure this effect was introduced by \cite{Kochukhov20}. Here the total field from ZDI maps was compared to the Zeeman line broadening and intensification, which
 is sensitive to the total (scalar) strength of the field since both polarities will contribute. 
 
This method offers a possibility to separate large- and small-scale structures.
In the study by \citep{kochukhov2023} a Doppler imaging method based on Zeeman intensification and broadening ({\it Zeeman Intensification Mapping}) was introduced and applied to the young solar analogue LQ Hya. In order to separate the effects of the magnetic field from the temperature (cool spots), lines from the same multiplet, but with different Land\'e factors 
were used. This enabled comparing a map of the total magnetic field strength, dominated by small scale complex structures, to the large scale field derived by ZDI.
One could be tempted to interpret these as manifestations of the large- and small-scale dynamo. However, since the spatial resolution provided by ZDI based on Stokes $V$ spectropolarimetry depends on both the instrument, $S/N$, rotation phase coverage and a number of stellar parameters, this is not so straightforward.

The surface magnetic field topology of stellar ZDI maps can be described in terms of poloidal vs. toroidal, and axisymmetric vs. non-axisymmetric field components. Using a spherical harmonics decomposition these are easily derivable from the coefficients $\alpha_{\ell,m}$, $\beta_{\ell,m}$
and $\gamma_{\ell,m}$. (For a standard definition of the spherical harmonic decomposition of the stellar magnetic field, see, e.g., \cite{kochukhov2014}). An unfortunate complication is that two different definitions of axisymmetry are used in ZDI studies. On one hand this is defined as the summed magnetic energy in modes $m=0$, on the other hand as all modes with $|m| < \ell/2$.

A substantial body of ZDI work has been carried out within the 
BCool collaboration, which together have mapped the large-scale
surface fields of large stellar samples and characterised how their 
strength and geometry vary with stellar parameters. 
\citet{2008MNRAS.388...80P} showed
that a progressively larger fraction of the surface magnetic energy 
is stored at large scales as rotation increases, with the field 
becoming predominantly poloidal in slow rotators and developing a 
strong large-scale toroidal component once 
$P_{\rm rot}\lesssim 12$~d. The BCool snapshot survey of
$\sim 170$ solar-type stars \citep{BCoolSnapshots} extended these 
detections across mass and rotation, while 
\citet{2014MNRAS.441.2361V} compiled
large-scale field measurements for a heterogeneous sample of F-M 
stars and found the mean large-scale field $\langle B_V\rangle$ to 
decline with age and increase toward smaller Rossby number. The 
TOUPIES survey traced the evolution of the large-scale field from 
the pre-main sequence to the early main sequence, where it weakens 
as stars spin down
\citep{2016MNRAS.457..580F,2018MNRAS.474.4956F}. More recently,
\citet{2025A&A...700A.282B} characterised the large-scale 
field of eleven Sun-like stars spanning $0.2$--$6$~Gyr, deriving 
poloidal/toroidal and axisymmetric energy fractions suitable as 
boundary conditions for wind modelling. The relationship between 
this large-scale (Stokes-$V$) field
$\langle B_V\rangle$ and the total unsigned field 
$\langle B\rangle$ is discussed further in 
Sect.~\ref{sssec:zdivszeeman} and Sect.~\ref{sec:impact}.

\begin{figure}
    \centering
    \includegraphics[width=\textwidth]{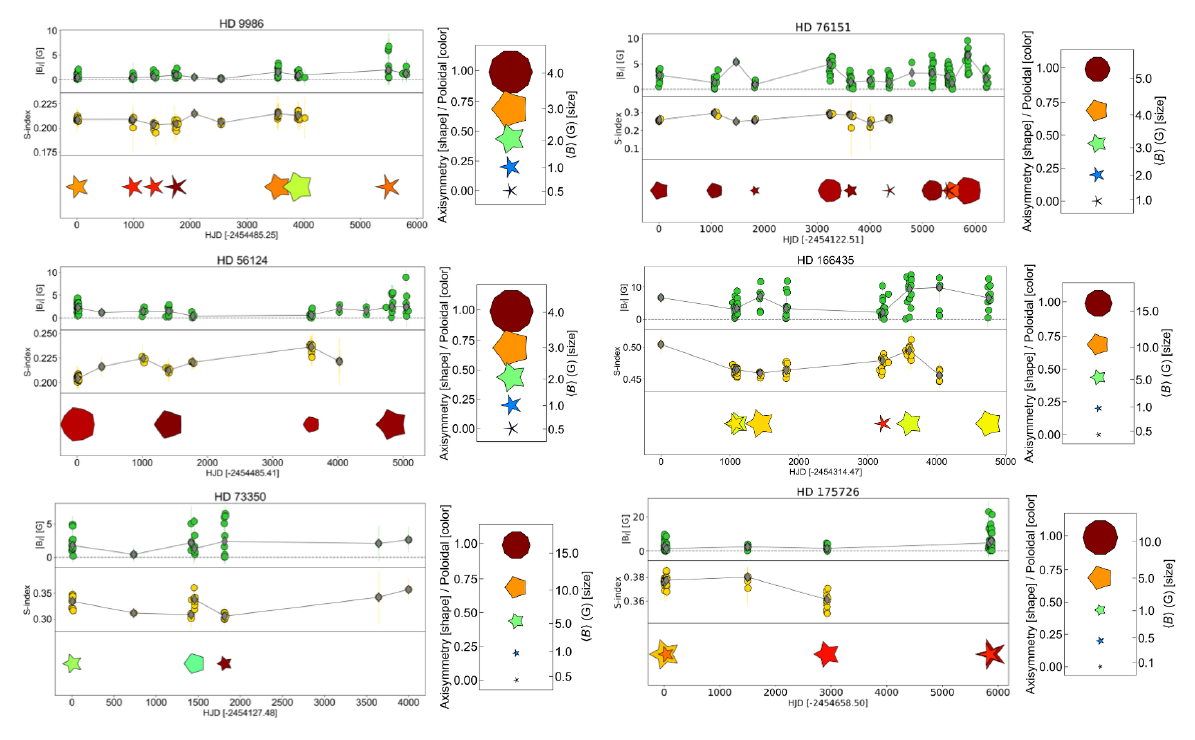}
    \caption{The large-scale
    magnetic field topology of a solar-type stellar sample in time, compared to mean large-scale field and S-index variations. For each star, the
    fractional magnetic energy in the poloidal component is shown against
    the degree of axisymmetry; symbol size scales with the mean large-scale
    (Stokes-$V$) field strength, and symbol colour/shape encode additional
    properties (e.g.\ epoch or activity level). 
    Figure adapted from~\citet{2025A&A...693A.269B}.}
    \label{fig:confusogram}
\end{figure}

The results of ZDI studies is often displayed as ``snapshots" of 
magnetic field properties of a 
sample of stars, the so-called {\it confusograms}. 
An example from the BCool survey is shown in Fig.~\ref{fig:confusogram}, 
in which each star (or epoch) is placed according to the fraction of 
magnetic energy in poloidal versus toroidal field and in axisymmetric versus
non-axisymmetric modes, with the symbol size scaling with the reconstructed
large-scale field strength \citep{2025A&A...693A.269B}. The stars in this sample 
have near-solar temperatures (up to 230 K hotter) with the following rotation 
periods: HD~9986 21.03~d; HD~56124 (see Sect.~\ref{sssec:hd56124}) 20.70~d; 
HD~73350 12.27~d; HD~76151 17.47~d; HD~166435 3.48~d; HD~175726 4.12~d. 
Such diagrams compactly display how the field geometry varies across a sample 
and, for repeated observations, how it evolves in time.
It is clear that the magnetic field topology of individual stars change. In at least some cases, this change could be periodical and connected to stellar magnetic cycles. Thus, the long-term evolution of particular stars can be depicted in terms of axisymmetry vs. non-axisymmetry and poloidality vs. toroidality \citep[see also][]{Kochukhov2013}.

\begin{figure}
    \centering
    \includegraphics[width=\linewidth]{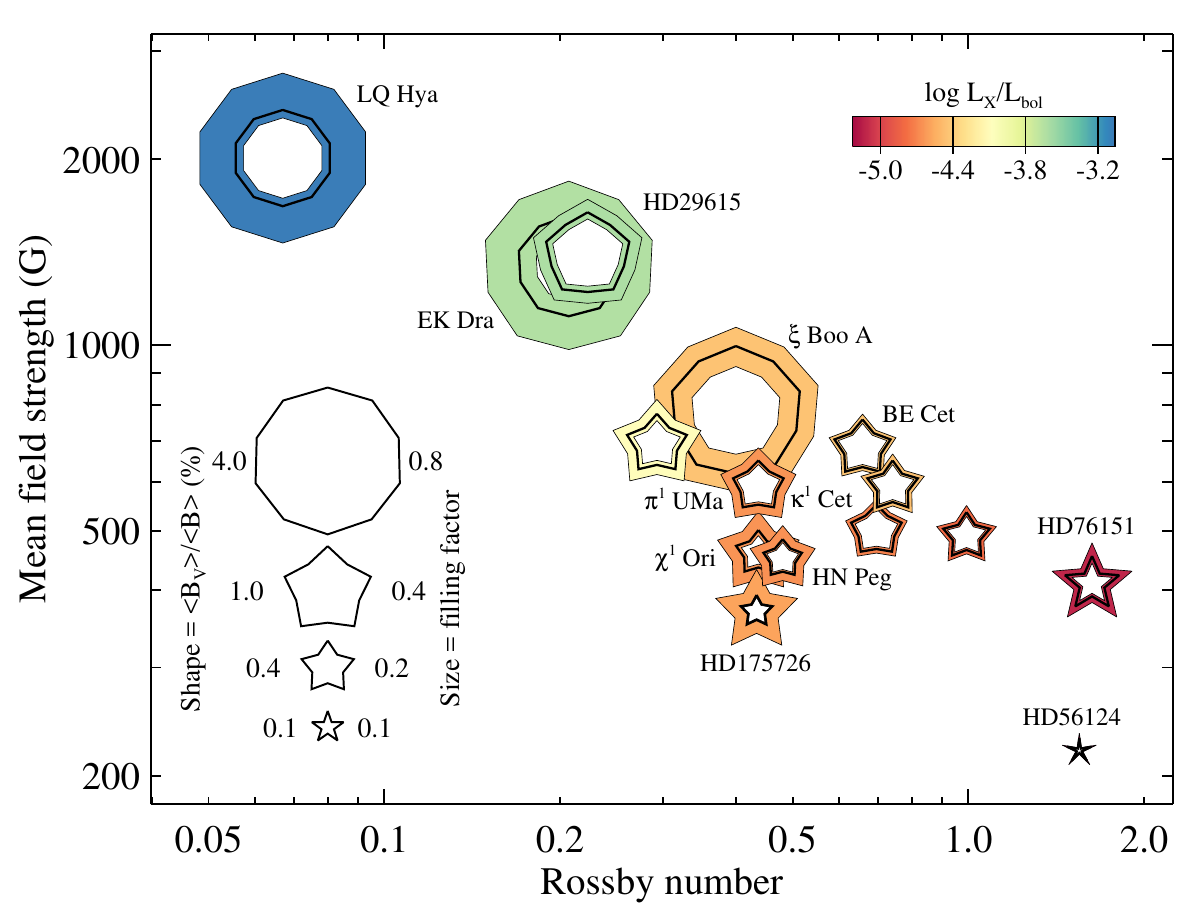}
    \caption{Mean total unsigned magnetic field $\langle B\rangle$ as a
    function of Rossby number for the young solar analogues of
    \citet{Kochukhov20}. Symbol size increases with the magnetic filling
    factor $f$, symbol shape encodes the ratio of the large-scale (ZDI) 
    to total field $\langle B_V\rangle/\langle B\rangle$ (ranging from
    $\lesssim 0.1$\% to a few per cent), and symbol colour indicates 
    the coronal activity level $\log L_{\rm X}/L_{\rm bol}$. The 
    diagram combines the information from Figs.~7--10 of the source 
    paper. Figure adapted from:~\citet{Kochukhov20}.}
    \label{fig:K20Bvsro}
\end{figure}

A complementary view of how the large-scale field relates to the total 
field across the rotation sequence is provided by \citet{Kochukhov20}, 
who combined Zeeman intensification (total unsigned field 
$\langle B\rangle$) and ZDI
(large-scale field $\langle B_V\rangle$) measurements for a sample of 
young solar analogues. Their concluding figure 
(Fig.~\ref{fig:K20Bvsro}) shows that, towards smaller Rossby number, 
the total field strength, the magnetic
filling factor, and the fraction $\langle B_V\rangle/\langle B\rangle$
recovered by ZDI all increase, the latter rising from below $0.1$\% 
in the least active stars to a few per cent in the most active ones. 
This is consistent with the trend found by \citet{See19}, that ZDI 
recovers a larger fraction of the total magnetic flux in more active 
stars (Sect.~\ref{sssec:zdivszeeman}).

This behaviour offers an observational counterpart to the 
rotation-magnetism modelling of Sect.~\ref{ssec:rotmag}. In the 
framework of \citet{Isik26}, the active-region-driven field processed 
by surface flux transport (SFT) grows steeply with rotation, so that its 
contribution to the total unsigned field rises from $\sim 8$\% at 
solar rotation to as much as $\sim 82$\% in the fastest rotators. 
To the extent that the large-scale, organised field recovered by 
ZDI partly traces this SFT-processed field driven by active-region 
emergence, the increase of $\langle B_V\rangle/\langle B\rangle$ 
towards fast rotators reported by 
\citet{2008MNRAS.388...80P}, \citet{Kochukhov20} and \citet{See19} can be
read as the observational signature of the growing active-region fraction, 
reproduced by the model.

\subsubsection{Transit mapping}\label{sec:transit-mapping}
The transit mapping technique is based on the crossing of the stellar disk by an exoplanet. During its transit, the planet may occult surface inhomogeneities such as starspots or faculae. These occultations imprint localized anomalies on the transit light curve, typically short brightenings when the planet covers a darker spot or short dimmings when bright faculae are occulted \citep{Silva2003}. The morphology, timing, and duration of these anomalies encode spatial information about the stellar surface along the transit chord. By modeling these features, one can retrieve the properties of the occulted active regions, including their radii, contrasts, and location (latitudes and longitudes). Because the planet’s projected trajectory across the star is fixed by the orbital geometry, the latitude of the occulted spots is well constrained, while the longitude is determined from the phase of the anomaly within the transit \citep{Silva2003, 2008ApJ...683L.179S}.

The \texttt{ECLIPSE}\footnote{\url{https://github.com/Transit-Model-CRAAM/pipelineMCMC/tree/feat/new-transit-method}} code \citep{Silva2003} was the first developed specifically to model transits across spotted stellar photospheres. The code synthesizes a limb-darkened stellar disk populated with circular spots of adjustable size, contrast, and location, and simulates the passage of a planet with known orbital parameters. By comparing the modeled and observed light curves, the spot parameters responsible for the anomalies can be retrieved. \texttt{ECLIPSE} has been applied to Kepler and CoRoT data to estimate spot magnetic fields \citep{Menezes2024}, stellar rotation periods \citep{2008ApJ...683L.179S}, differential rotation \citep{2011A&A...529A..36S, 2017ApJ...835..294V, Valio2024}, stellar activity cycles \citep{Estrela2016}, surface area coverage \citep{Araujo2025}, and the morphology of active regions \citep{2025A&A...702A.227Z}.
Since then several codes have implemented similar physics for modeling spot occultations during planetary transits, for example, SPOTROD \citep{2014MNRAS.442.3686B}, SOAP and SOAP-T \citep{2012A&A...545A.109B, 2013A&A...549A..35O}, STSP \citep{2017ApJ...846...99M}, 
SpotMCMC \citep{haris2025},
to cite a few.

Transit mapping has been extensively used to study stellar magnetic activity. The method also complements classical out-of-transit rotational modulation studies by providing direct, geometric constraints on the latitudes of individual active regions. With high-precision photometry from CoRoT,  Kepler, K2, and TESS, transit mapping has become a key tool for understanding the magnetic behavior of exoplanet host stars.

By modelling spot-crossing anomalies, it is possible to estimate spot size, intensity contrast, and therefore temperature difference relative to the photosphere \citep{Silva-Valio2010, Sanchis-Ojeda2011,
2017ApJ...846...99M}.
When multiple transits are observed, recurrent anomalies allow the determination of spot longitudinal evolution, as well as constraints on stellar inclination and spin--orbit obliquity when combined with orbital geometry \citep{
Nutzman2011, Sanchis-Ojeda2012, Dai2017, Valio2022}.
Repeated occultations of the same active region provide lower
limits on spot lifetimes and reveal whether active regions persist over several stellar rotations.
In this way, transit mapping yields direct spatial and temporal constraints on magnetic surface structure.

Planetary transit observations also enable measurements of differential
rotation.
If a spot-crossing anomaly recurs in successive transits but at shifted phases, the displacement reflects the difference between the stellar rotation rate at the spot latitude and the orbital period of the planet \citep{2011A&A...529A..36S,
2017ApJ...835..294V, Valio2024}.
From such phase drifts, the latitudinal shear $\Delta\Omega$ can be inferred and compared with rotation periods derived from out-of-transit photometric modulation, linking local,
latitude-specific rotation rates to global rotation diagnostics and providing insight into surface shear and dynamo regimes 

Moreover, transit mapping enables the estimation of stellar surface
differential rotation through different geometric configurations.
In systems with a single transiting planet on a well-aligned
orbit, the transit chord probes essentially one stellar latitude; the rotation period at that latitude is measured from the recurrence of spot-crossing anomalies, and a solar-like differential rotation law is then assumed to extrapolate the global latitudinal shear \citep{2011A&A...529A..36S}. When more than one planet transits at different orbital inclinations, each planet samples a distinct latitude band, allowing a direct estimate of $\Delta\Omega$ without assuming a specific functional form, as demonstrated for the Kepler-411 system \citep{AraujoValio2021}. Lastly, in systems with a significantly oblique orbit, a single planet can probe a broad range of stellar latitudes; recurrent spot-crossing events at different projected latitudes then yield the differential rotation gradient directly, as shown for Kepler-63 \citep{NettoValio2020} and Kepler-210 \citep{Valio2022}.

Finally, transit mapping connects spatially resolved spot properties with global variability measures. In particular, the spot filling factor inferred from the number and size of occulted regions can be quantified and directly compared with the overall photometric amplitude of rotational modulation. Using this approach, \cite{Araujo2025} found spot coverage fractions ranging from a few percent up to nearly 30\%, with a strong anti-correlation with stellar age and a moderate dependence on effective temperature, supporting the use of spot filling factor as a proxy for stellar magnetic activity. Furthermore, transit mapping enables the direct measurement of stellar differential rotation from spots at different latitudes, providing constraints on rotational shear and its dependence on stellar parameters, such as rotation period and effective temperature \citep{Araujo2023}. In addition, it allows estimates of starspot magnetic field strengths from the spot flux deficit, indicating typical values of a few kG and suggesting that stellar magnetic evolution is primarily driven by variations in spot filling factor rather than intrinsic field intensity \citep{Menezes2024}. By complementing disk-integrated photometry with spatially resolved constraints, transit mapping provides a direct link between stellar magnetic topology and rotational evolution.

Stellar activity also affects planetary parameter retrieval: unocculted spots and faculae can bias estimates of planetary radius and atmospheric transmission spectra \citep{Oshagh2013, Rackham2018, Sumida2026}. In active systems, transit mapping therefore plays a dual role, both refining stellar magnetic characterization and mitigating stellar contamination in exoplanet studies. By constraining the distribution and properties of surface inhomogeneities, transit mapping becomes essential for interpreting transmission spectra and improving the reliability of atmospheric inferences in exoplanets.

\subsection{Detecting magnetic cycles on cool stars}
Signs of magnetic cycles based on ZDI studies have been reported for a large number of stars. The most common of these are polarity reversals of the radial field at the visible rotational pole \citep[see. e.g.,][and references therein]{Jeffers2023SSRv..219...54J}. However, in most cases these are not strictly analogue to the solar magnetic cycle. In many cases the polarity reversals of the radial field can be connected to spot variability \citep{borosaikiaSolarlikeMagneticCycle2016,jeffers2018, lehtinen2022}. But there are fewer examples, where solar-like polarity reversals in the azimuthal field have been observed. The case of $\iota$ Hor (Sect.~\ref{sssec:iotaHor}) may be an exception \citep{2025A&A...704A..68A}. One problem is, of course, the short time scale and inadequate accuracy of stellar observations.
Still, the question whether the solar magnetic cycle can be used as a model for other stars, remains open.

Spot transit mapping of a long enough light curve may also infer short activity cycles.
\cite{Estrela2016} analyzed Kepler light curves of active stars (Kepler-17 and Kepler-63) to investigate magnetic-cycle signatures through long-term modulations in spot coverage during transits. By applying transit mapping techniques, the authors derived time series of starspot area coverage and examined their temporal evolution. The study demonstrated that photometric data from space missions can reveal magnetic-cycle behavior in Sun-like and active stars similar to the quasi bi-annual cycles identified on the Sun. This
 highlighted the potential of starspot-based diagnostics for probing stellar dynamos in active exoplanet host stars.

\subsubsection{A cycling young Sun-like star in focus: $\iota$~Horologii}
\label{sssec:iotaHor}

The Far Beyond the Sun programme consists of a coordinated investigation of the magnetic activity of the young solar-like star $\iota$~Horologii ($\iota$~Hor) from the photosphere to the corona, including in a multi-year spectropolarimetric monitoring and mapping of its global magnetic field using ZDI. 
The star has solar-like parameter values, listed in Table~\ref{tab:sun-in-time} and briefly discussed in , Sect.~\ref{sec:Horologii}: its spectral type is F9/G0, putting its convective turnover time somewhat shorter than solar, but it also has a higher metallicity than the Sun (Fe/H$\approx 0.18$), potentially enhancing its magnetic activity level despite the star being hotter (see Sect.~\ref{ssec:rotmag}). At a rotation period of $7.73$~d, the star thus presents as an interesting case of a young planet-hosting star slightly more massive than the Sun. 

Three papers provide successive steps in the analysis, beginning with the identification of two superposed chromospheric cycles, their connection with the reported coronal cycle and the long-term radial velocity behavior (\citealt{2018MNRAS.473.4326A}; Paper I). The study was followed by a multi-wavelength characterization and extension of the monitoring in \citet{2023MNRAS.524.5725A}, refining the activity cycle time-scales and obtaining multi-layer activity diagnostics for this star (Paper II). The third paper on the series contains the reconstruction of 18 ZDI surface magnetic field maps and the first stellar magnetic butterfly diagrams (\citealt{2025A&A...704A..68A}; Paper III). Together these form one of the most complete observational record of magnetic cycle evolution for any known young Sun-like star and represent an exceptional data set for testing the solar-stellar analogy.

\begin{figure*}[!t]
\centering
\includegraphics[trim=0.0cm 0.0cm 0.0cm 0.0cm, clip=true, width=0.99\textwidth]{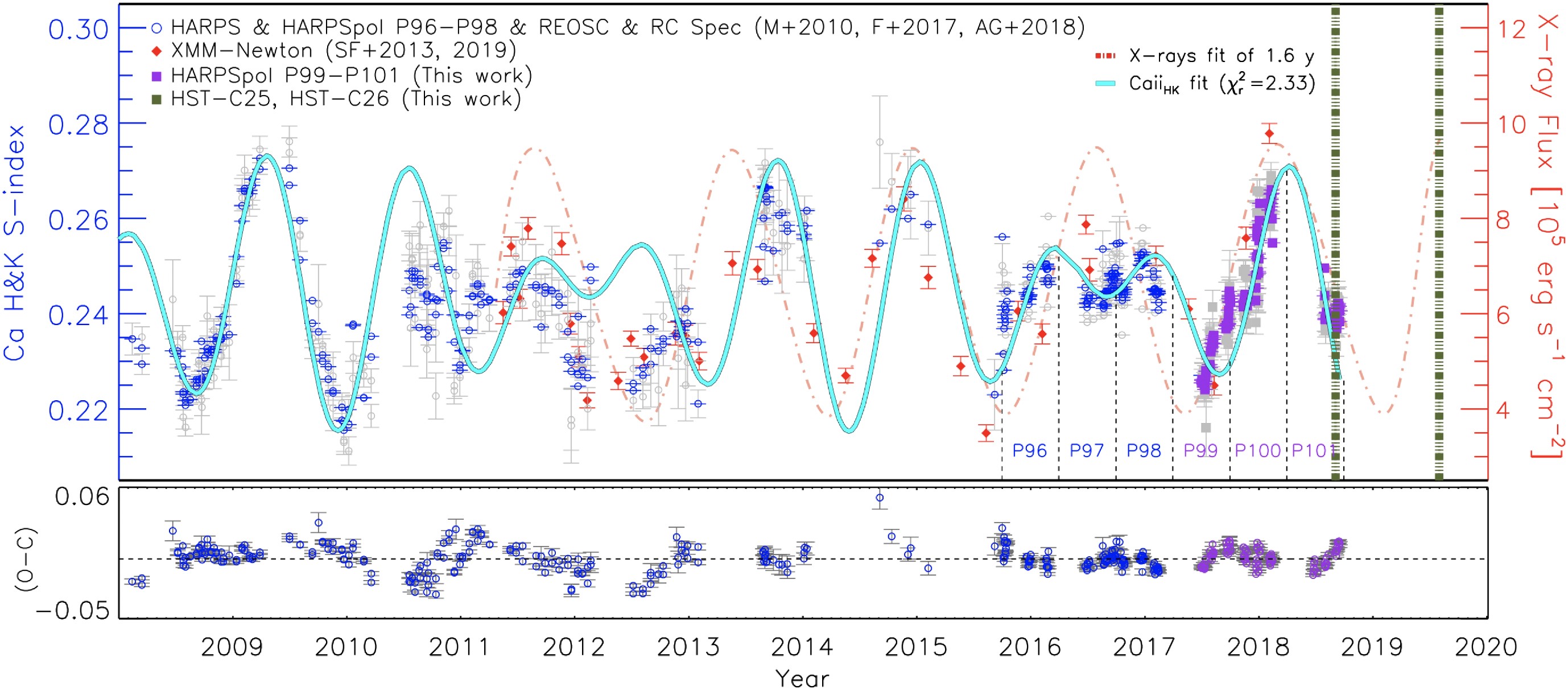}
\caption{Top panel: Long-term evolution of the $S_{HK}$ index derived from the Ca\,II H\&K core emission of $\iota$ Hor, illustrating the presence of two periodic components in the chromospheric activity cycle. The gray points show the unprocessed measurements prior to averaging over the modulation associated with stellar rotation. The blue points represent the data analysed by \citealt{2010ApJ...723L.213M} (M+ 2010), \citealt{2017MNRAS.464.4299F} (F+ 2017), and \citealt{2018MNRAS.473.4326A} (AG+ 2018). The right-hand axis displays the corresponding X ray fluxes (red filled diamonds) reported by \citealt{2013A&A...553L...6S, 2019A&A...631A..45S} (SF+ 2013, 2019). The compilation by \citet{2023MNRAS.524.5725A} incorporates all previous measurements in addition to new HARPSpol points (purple filled squares) and epochs of contemporaneous TESS and HST coverage (olive dashed lines). Bottom panel: residuals from the double period model, shown as observed minus calculated values.
Figure adapted from:~\citet{2023MNRAS.524.5725A}.}\label{Fig:Amazo-Gomez2023_iHor_Cycles}
\label{fig:iHor1}
\end{figure*}

Paper I establishes the basic activity phenomenology of $\iota$\,Hor based on the first half of a long-term HARPSpol spectropolarimetric campaign. The observations span three ESO periods and include circularly polarized spectra processed following the ratio method and Least Squares Deconvolution \citep{1997MNRAS.291..658D, 2009PASP..121..993B}. The star shows the shortest coronal activity cycle known at the time with a period of approximately 1.6\,yr \citep{2013A&A...553L...6S, 2019A&A...631A..45S}. The Ca\,II H and K S-index and the H$\alpha$ index reveal a double periodicity in the long-term magnetic activity, suggesting the beating of two cycles with periods around 1.1-1.5\,yr, shown in Fig.~\ref{fig:iHor1} along with the X-ray flux variation. The authors emphasize that the star is the youngest and most active planet-hosting system with well-sampled chromospheric and coronal variability, and that the short magnetic cycle provides a unique opportunity to test the solar activity paradigm in a regime where the cycle period is about one-tenth of the solar cycle period. 

In addition to activity indices, Paper I measures the longitudinal magnetic field using the LSD Stokes V profiles following \citet{2009ARA&A..47..333D}. The rotation period and radial velocity are examined to verify instrument stability. The radial velocity measurements are consistent with previous orbital determinations \citep{2013A&A...552A..78Z} and show that stellar activity contributes significantly to the residuals, complicating the detection of additional planets. 

Paper II expands the observational baseline by adding three additional semesters of spectropolarimetry, yielding a total of 199 data points \citep{2023MNRAS.524.5725A}. It incorporates visible spectropolarimetry from HARPSpol, space-based photometry from TESS \citep{2015JATIS...1a4003R}, and ultraviolet spectroscopy from HST STIS \citep{1998PASP..110.1183W}. This work presents simultaneous observations across photospheric, chromospheric, and coronal transition region diagnostics and applies a multi-technique analysis including spectropolarimetry, photometry, and ultraviolet spectroscopy of key lines such as C\,III, C\,IV, Si\,IV, and O\,IV. The authors construct a global picture of stellar activity by combining information from the S-index, H$\alpha$ index, longitudinal magnetic field measurements, and UV emission. They show that simultaneous tracing of these diagnostics provides evidence for complex magnetic variability across atmospheric layers. The correlation analysis demonstrates that radial velocity is weakly correlated with magnetic or chromospheric indices, supporting the interpretation that the dominant radial velocity signature continues to be controlled by the known planet rather than magnetic modulation.

Paper II also quantifies the spot and facula contribution to photometric variability using the Gradient of the Power Spectrum (GPS) method and shows that $\iota$\,Hor lies in the spot-dominated branch when compared with a distribution of solar and stellar light curves \citep{2020A&A...633A..32S, 2020A&A...636A..69A, 2020A&A...642A.225A}. The multiwavelength observations permit a connection from the photosphere through the chromosphere to the transition region and corona. Importantly, Paper II delivers the first ZDI reconstruction of the star, acquired simultaneously with TESS photometry as well as two HST NUV/FUV visits. The retrieved map revealed a predominantly negative polarity region at the visible pole during the observed epoch together with evolving mixed polarity patterns at lower latitudes. The authors point out that this particular observation (i.e. ZDI maps with simultaneous multi-wavelength activity diagnostics) can provide robust input and constraints for furture 3D MHD modelling of the corona and stellar wind in this system (planned for Paper IV of the series; Chebly et al. in prep).

\begin{figure*}[!t]
\centering
\includegraphics[trim=0.0cm 0.0cm 0.0cm 0.0cm, clip=true, width=0.99\textwidth]{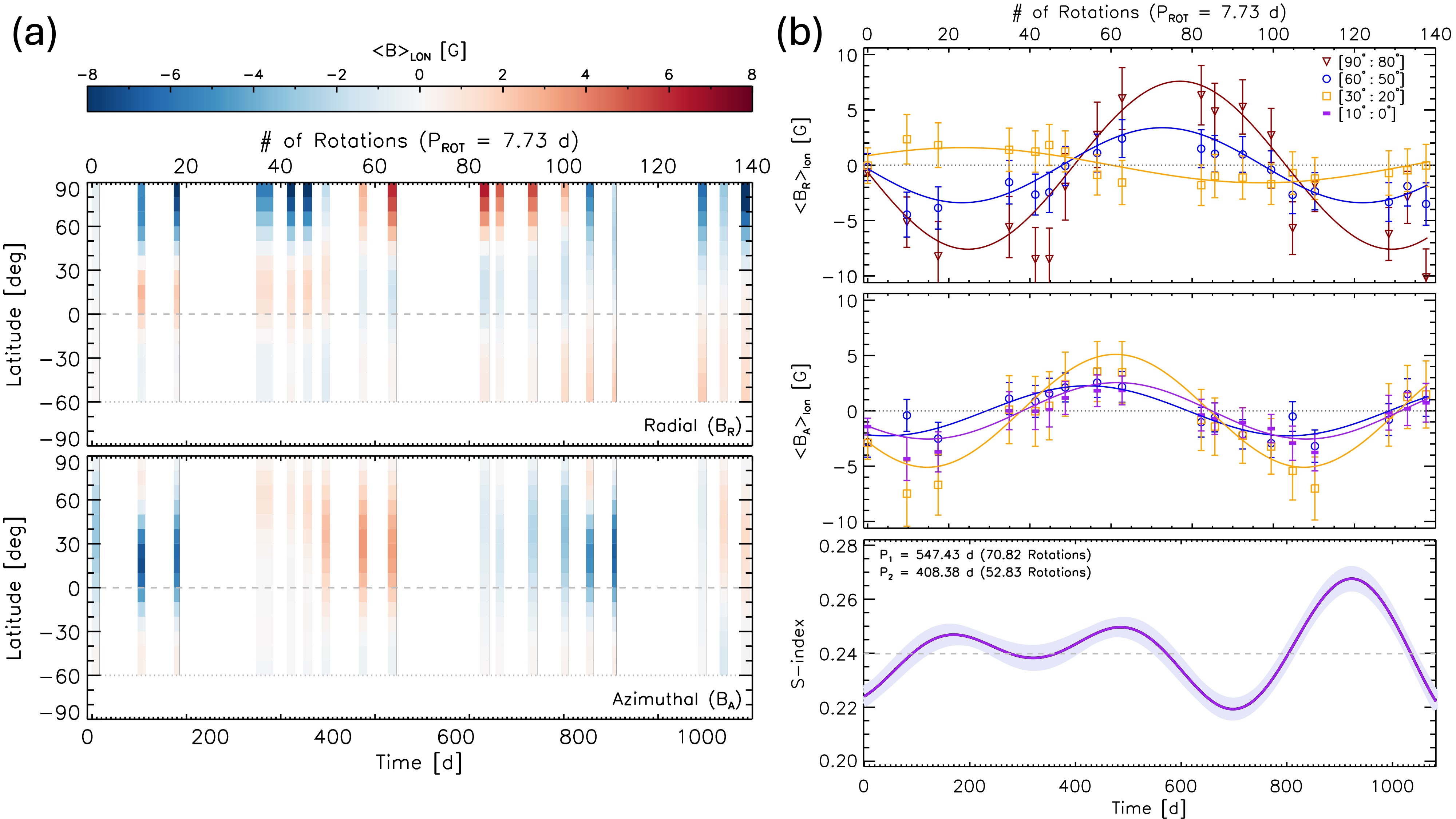}
\caption{(a) Butterfly diagrams of the radial (top) and azimuthal (bottom) large-scale magnetic field of $\iota$ Hor, showing longitudinally averaged field strengths (color scale in gauss) extracted in $10^{\circ}$ latitude bands on each of the 18 ZDI epochs. Day zero corresponds to BJD = 2457300.78580 (2015-10-05 06:51:33.12 UTC), with a secondary axis marking the number of stellar rotations covered by the campaign ($P_{\rm ROT} = 7.73$~d). (b) Temporal evolution of the longitudinally-averaged radial (top) and azimuthal (middle) field components derived from Panel (a), where colored symbols denote latitude ranges. Error bars reflect the standard deviation within each latitude band and across longitude. Solid lines show the corresponding single periodic fits for each latitude bin. The lower panel presents the two period S-index model from \citet{2023MNRAS.524.5725A}, with the shaded envelope indicating the 1$\sigma$ uncertainties in period and amplitude. Figure adapted from~\citet{2025A&A...704A..68A}.}\label{Fig:AG2025_Bfly_iHor}
\label{fig:iHor2}
\end{figure*}

Paper III completes the study progression by performing an intensive ZDI investigation focused entirely on the magnetic cycle of the star \citep{2025A&A...704A..68A}. The authors analyse the complete set of circularly polarised spectra acquired over roughly 3 years and reconstruct 18 separate ZDI maps, thereby tracing the evolution of the large-scale surface field across approximately 139 stellar rotations. The ZDI procedure employs a spherical harmonic description of the field \citep{2002ApJ...575.1078H, 2021MNRAS.500.1243L} and separates the poloidal and toroidal components in order to examine their temporal behavior. The analysis uncovers pronounced evolution including multiple polarity reversals, changes in field strength, and changes in geometry. The toroidal component is found to vary strongly in concert with the chromospheric activity. Most notably, the authors construct stellar magnetic butterfly diagrams for the first time in a young solar analog and use them to trace both the migration of magnetic features and the polarity reversal timescale which is measured to be $\sim100$ rotations ($\approx 773$~days). This provides an empirical estimate of a stellar magnetic cycle based entirely on direct surface magnetic mapping (Fig.~\ref{fig:iHor2}).

Furthermore, Paper III uses field-weighted latitudinal positions extracted from the ZDI maps to obtain the first estimates of large-scale flow properties on a star other than the Sun. By fitting radial magnetic field features at different latitudes and comparing model choices using the Bayesian information criterion \citep{1978AnSta...6..461S}, the authors infer possible poleward and equatorward drift speeds for different field polarities. These flows connect directly to dynamo processes and provide a rare empirical measurement of magnetic transport in a young solar-type star. The results thereby link large-scale magnetic field evolution to stellar dynamo theory and offer a direct comparison with the solar magnetic cycle.

Taken together, the Far Beyond the Sun campaign establishes $\iota$~Hor as a benchmark for magnetic activity in young solar type stars. Across the programme the authors show that intensive spectropolarimetric monitoring can resolve large-scale magnetic field evolution in stars younger and more active than the Sun, providing critical constraints for models of stellar dynamos, stellar wind evolution, and star-planet interactions.

\section{Impact on stellar environments and exoplanets}\label{sec:impact}

In this section, we briefly highlight the importance of stellar magnetism for star-planet interactions, considering both large-scale magnetism and stellar magnetic activity, and reflect on the their consequences for exoplanets, their atmospheric evolution and habitability.

\subsection{Magnetism sculpts the stellar environment}

\subsubsection{Large-scale magnetic fields dictate the environment of stars}

\begin{figure}[!htbp]
    \centering
    \includegraphics[width=0.75\linewidth]{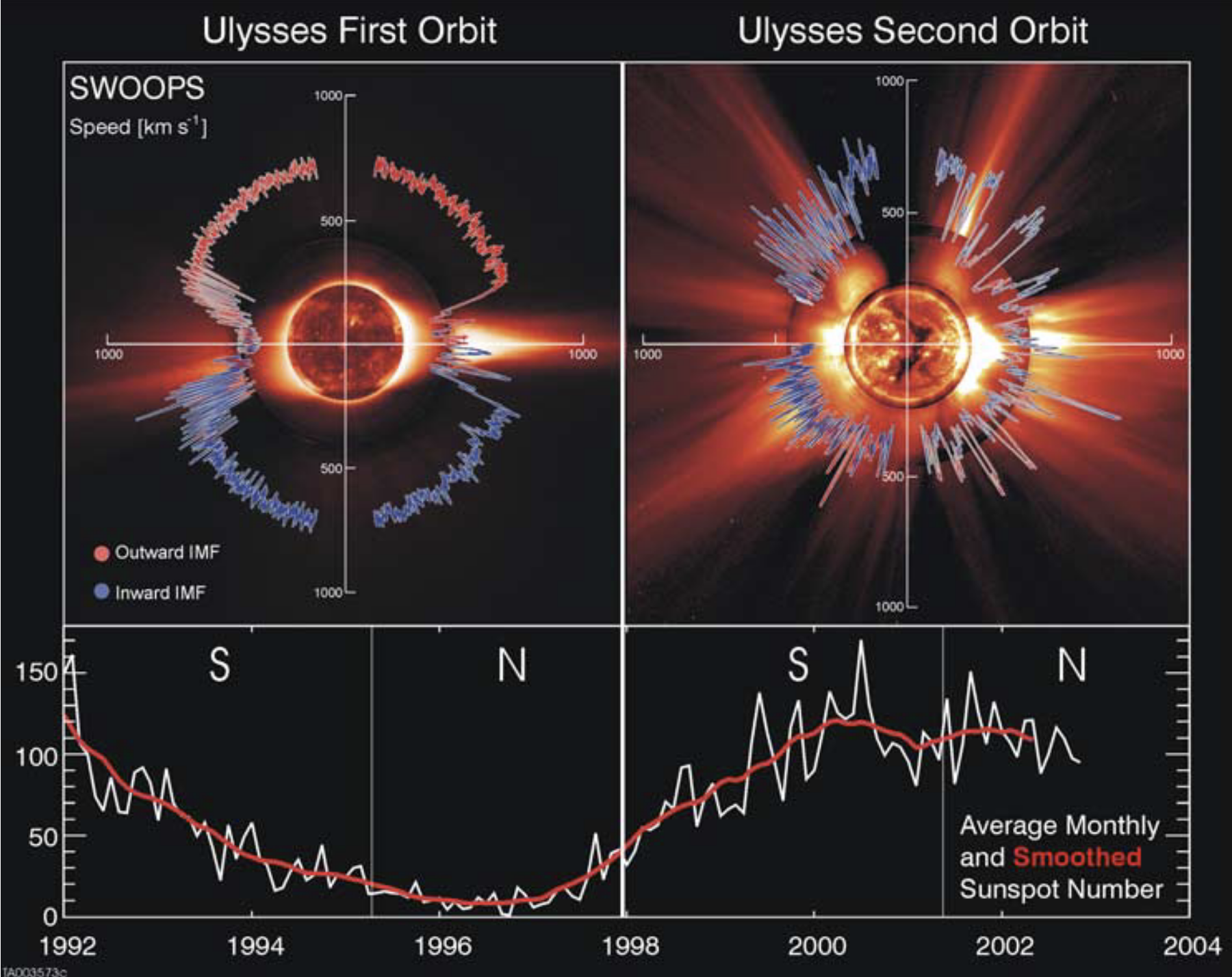}
    \caption{Solar wind speeds as measured by the Ulysses spacecraft, sweeping heliographic latitudes during solar minimum (top left) and solar maximum (top right) as Ulysses made two orbits (see \citealt{2003GeoRL..30.1517M} for more details). The wind speed shown by the red and blue curves is highly modulated by the large-scale magnetic field of the Sun at each epoch. The bottom panel shows the sunspot number during that period. Figure adapted from \citet{2003GeoRL..30.1517M}.}
    \label{fig:McComas2003}
\end{figure}

The magnetic field of a star plays the role of a skeleton sculpting its environment. During the solar minimum, the solar magnetic field is dominated by the dipole, therefore leading to a dipolar structure in its corona. During solar maximum, the magnetic field is notably more complex with multiple smaller scale structures leading to a highly complex and non-axisymmetric coronal structure. These propagate from the low corona of the Sun to far out in the solar system \citep{2020A&A...642A...2R}, as remarkably shown by the solar wind properties between 1 AU and 5 AU (see \citealt{2000JGR...10510419M} and Fig. \ref{fig:McComas2003}). The mechanisms behind such variability will be discussed in Chapter 2. 

In the past decade, significant progress has been made in understanding the magnetic connectivity between objects in the solar system and the Sun, mostly thanks to the space mission Parker Solar Probe and Solar Orbiter. 
It is interesting to note that Parker Solar Probe closest perihelion reached a distance of 6.9 $\times 10^3$ Mm to the center of the Sun. This corresponds to roughly 0.046 AU and 9.86 $R_\odot$. Such a distance corresponds, for an exoplanet circularly orbiting a solar twin, to an orbital period of 3.6 days. With such a close perihelion, Parker Solar Probe crossed at multiple times the alfvénic transition in the solar wind, where the velocity of the accelerating wind becomes larger than the local Alfvén speed \citep{2021PhRvL.127y5101K}. \citet{2025A&A...702A.252F} further compiled all crossings, leading to an estimate of the average Alfvén surface to be between 11 and 16 $R_\odot$ depending on the solar activity level. For exoplanets around solar twins, these distances correspond to orbital periods of 4.2 and 7.4 days, respectively. As such, the in-situ measurements of Parker Solar Probe and Solar Orbiter and the modelling efforts to assess their magnetic connectivity give invaluable context to assess the local environment of close-in exoplanets. 

Nevertheless, extrapolating this knowledge to exoplanets presents numerous challenges. It is generally assumed that all cool stars form thermal winds similarly to what occurs for the Sun (see e.g. \citealt{2020A&A...635A.170A,2023MNRAS.524.5060C}). Mass-loss rate estimates for cool stars support this idea, with a mass-loss observed to increase with X-ray luminosity \citep{2021ApJ...915...37W}, albeit with a large spread that could come from various origins (rotation regimes, stellar metallicity, magnetic cycles, to cite only a few). The application of solar wind driving mechanisms to these stars raise nevertheless concerns when stars rotate much faster than the Sun (e.g. \citealt{2015ApJ...814...99R,2019MNRAS.482.2853J}), or when the large-scale magnetic of the star is very strong (e.g. dipolar fields above 1 kG for some M-dwarfs, see \citealt{2010MNRAS.407.2269M}). In the latter case, it is important to realize that state-of-the-art stellar wind models predict Alfvén surfaces as large as about 100 $R_\star$ for the most magnetized M-dwarfs \citep{2022MNRAS.514..675K,2024CRPhy..24S.138S}.

Another strong uncertainty comes in how to properly extrapolate magnetic field in a stellar corona. In the Sun, the potential-field source surface (PFSS) method \citep{1969SoPh....6..442S} based on the radial magnetic field in the photosphere generally gives a relatively good first-order estimate of the magnetic connectivity and on the structuring in the low corona \citep{2021FrASS...8...84P}. The value of the source-surface radius can affect the exact magnetic connectivity in the corona \citep{2023JSWSC..13...11R}, and is often taken to a canonical value of 2.5 $R_\odot$ for the Sun. For cool stars an optimal source-surface radius can be obtained based on constraints about mass and angular momentum loss rates \citep{2015ApJ...814...99R}. 
It must be noted, in addition, that most MHD models (see Chapter~2) in the literature rely heavily on PFSS extrapolation: they are initialized with PFSS, and then the full MHD solution adjusts slightly the 3D solution, affecting only mildly the magnetic connectivity and the structure of the corona \citep{2015ApJ...814...99R}. Nevertheless, ZDI maps \citep{Donati2025A&A...700A.122D} from distant stars sometimes exhibit a dominant toroidal component \citep{2025MNRAS.542.1318S}, which questions the correctness of using standard PFSS extrapolations for modelling the magnetic corona of such stars. When the magnetic field is mostly toroidal, non-potential extrapolations are needed, as shown in e.g. \citet{2019MNRAS.483.5246L,2022MNRAS.512.4556S}. Such extrapolations add other tunable parameters, and we have yet as a community to correctly assess how to get rid of such tunable parameter to properly model the magnetic environment of such stars.


\subsubsection{Transient events and stellar environment variability}

On top of the global structuring provided by the large-scale magnetic field, it is well known in the solar environment that transient events provide drastic changes in the local properties (magnetic field intensity and direction, density, temperature, flow speed). For instance, energetic particles can be accelerated during flares \citep{2011SSRv..159..357Z} and CME propagation \citep[e.g.][]{2022RvMPP...6....8S,2023FrASS..1054266R}, dramatically and temporarily changing the population of fast particles around the star. These particles strongly affect the atmosphere of planets in the Solar System (see hapter~2, Section 5). In addition, the amount of galactic cosmic rays received at Earth is also modulated by the solar cycle \citep{2013LRSP...10....3P}, showing the strong importance of the large-scale magnetic field of the central star in mediating the environment of exoplanets. 

\begin{figure}[!htbp]
    \centering
    \includegraphics[width=\linewidth]{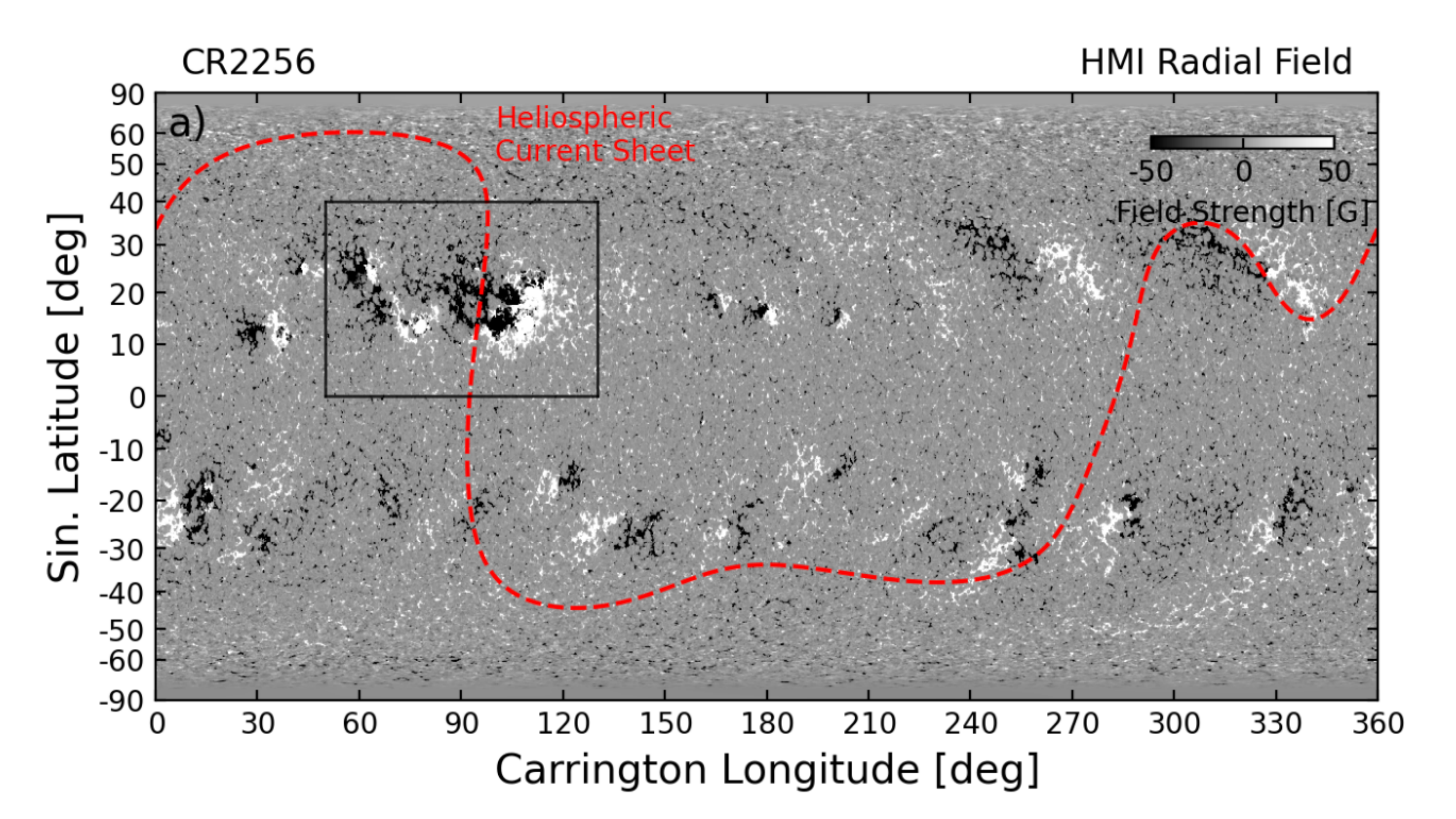}
    \caption{Example of heliospheric current sheet (red dashed line) projected on top of the photospheric magnetic field (in black and white). The nested active region is highlighted by the black box and anchors the heliospheric current sheet above it to be oriented in the latitudinal direction at this particular Carrington longitude. Figure adapted from \citet{2024A&A...692A..29F}.}
    \label{fig:Finley2024}
\end{figure}

To be more specific, we can first consider magnetic flux emergence which is a transient phenomenon that affects significantly the surrounding of a cool star. Taking the Sun as an example, flux emergence is mediated by the solar cycle and appears to emerge with specific patterns. Strong emergence at the solar photosphere (from $10^{16}$ to $10^{20}$ Mx/h within solar active regions, see \citealt{SchrijverZwaan00,vanDriel-Gesztelyi15}) occurs within the butterfly equatorial bands as the cycle progresses, and is also found to emerge at preferred longitude, a phenomenon sometimes called `active longitudes' \citep{2003A&A...405.1121B} or activity nesting (see Sect.~\ref{ssec:solar}; \citealt{vanDriel-Gesztelyi15,Finley25} and references therein). These nests of emergence have been shown to have a strong impact on the location and shape of the current sheet in the Solar System (see Figure \ref{fig:Finley2024} and \citealt{2024A&A...692A..29F}), thereby directly influencing the position of magnetic sectors seen by orbiting planets. It is worth noting that the nesting of emergence and activity is seen on the Sun, but its broad applicability to cool stars remains an open question as of today (see, e.g., \citealt{NettoValio2020,2020ApJ...901L..12I}). 

Transients structures in the corona and wind of cools stars take also the form of coronal mass ejections (CMEs) and sub-structures within the stellar wind like co-rotating interaction regions (CIRs) and stream interactions regions (SIRs) \citep{2021LRSP...18....4T}. All of them ultimately stem from stellar magnetic fields: CMEs are powered by the release of free magnetic energy through magnetic reconnection, and CIRs and SIRs originate in the structuring of the rotating wind leading to the existence of different wind streams reaching different velocities at the same distance to the star (see, e.g., fast solar wind catching up slow solar wind as shown in Figure \ref{fig:act-rotationHHK}). 
The observation of such transients in stellar environment is nowadays in its infancy, with for instance a remarkable new recent detection of stellar CMEs through a radio type II burst by \citet{Callingham2025}. The list of transient events permeating the stellar environment is developed in more details in Chapter~2, which we defer the reader to. We now turn to the specific impact magnetism on close-in objects in a stellar system.

\subsection{Impact of magnetism on close-in objects}


\subsubsection{Magnetic coupling between exoplanets and their environment}

\begin{figure}[!htbp]
    \centering
    \includegraphics[width=\linewidth]{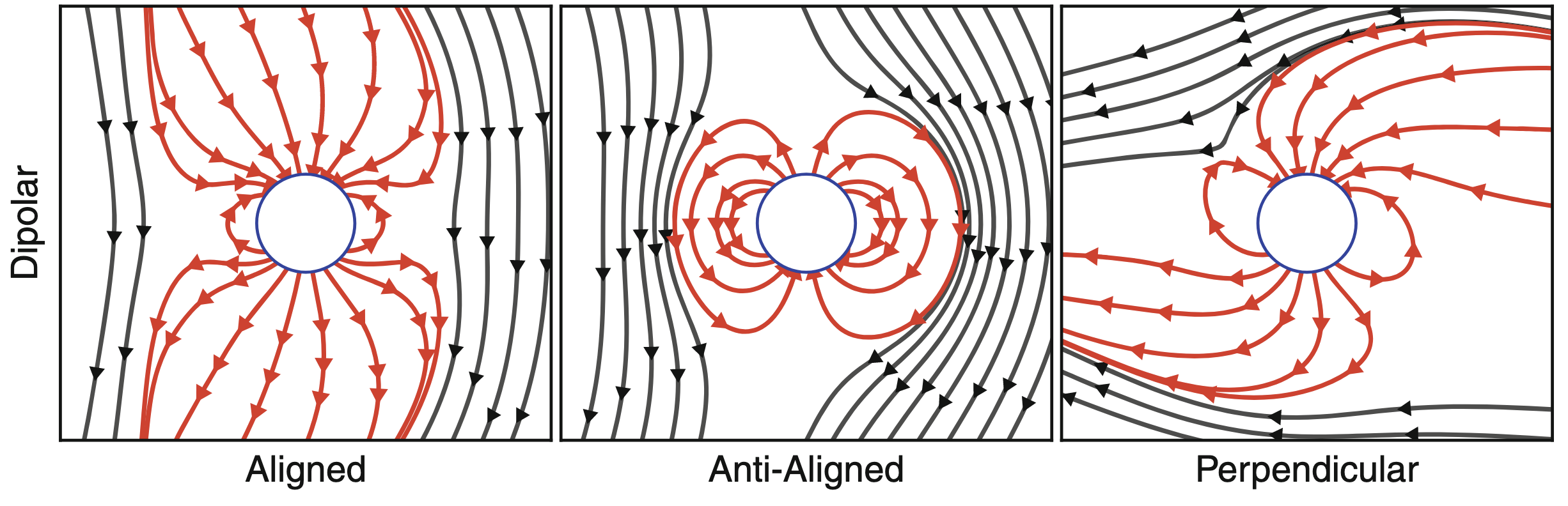}
    \caption{Example of three different magnetosphere-stellar wind magnetic configurations, adapted from \citet{2018haex.bookE..25S}.}
    \label{fig:Strugarek2018}
\end{figure}

The magnetic structuring of the stellar environment leads a to highly variable magnetic environment for orbiting planets. Taking the iconic example of the Earth, it is well known that the orientation of the interplanetary magnetic field (IMF) has a drastic impact on how it couples to the Earth magnetosphere \citep{2015SSRv..188..251E}. For instance, the orientation of the IMF affects how the magnetosphere reacts to interplanetary CMEs \citep{2017LRSP...14....5K} and the current system in the Earth magnetosphere \citep{2017SSRv..206..547M}. A simple example of how a planetary magnetosphere couples to its environment can be found in Figure~\ref{fig:Strugarek2018} for three different ambient magnetic field directions \citep{2018haex.bookE..25S}. One can immediately remark that the magnetosphere can be open (first panel), closed (second panel), or connect differently its northern and southern hemisphere to the environment (third panel). For the exoplanets orbiting close-in to their hosts, the magnetic environment of the planets can strongly change along their orbit when the star harbors complex magnetic field, leading to strong modulations of the wind-magnetosphere coupling. Their proximity to their host makes them interact with CMEs at the beginning of their expansion phase, which can have a severe impact on the CMEs themselves \citep{2011ApJ...738..166C}. 

In addition to the magnetic geometry,
the coupling can operate in different regimes.
The coupling regime can be simply approximated by the alfv\'enic Mach number $M_a=v/v_a$ regime, 
where $v_a$ is the local Alfv\'en speed at the planet position and $v$ is the modulus of the relative speed between the planet and its surrounding
\cite{2001Ap&SS.277..293Z,2004ApJ...602L..53I,2007P&SS...55..598Z}. In general, the relative speed between the planet and its environment is controlled by two aspects: the local wind speed, which stems from both the rotation of the star and the stellar wind accelerating away from the star on one side, and the orbital motion of the planet on the other side \citep{2023spi..conf....1S}. 
In the star-planet direction, the relative speed is dominated by the accelerating wind speed. As a result, the stellar wind Alfvén surface (the surface delimiting sub- and super-alfvénic stellar wind) is traditionally seen as the delimiting surface within which a planet can be magnetically connected to its star. Outside of this Alfvén surface, which is the case of all the Solar System planets, the planet intercepts a super-alfvénic wind, leading to the existence of a shock in front of it. Inside the Alfvén surface, subtantial power can be channeled towards the star in the form of Alfvén waves (typically up to a few 10$^{20}$ W, see \citealt{2007P&SS...55..598Z,2012A&A...544A..23L,2013A&A...552A.119S,2015ApJ...815..111S,2026A&A...705A..12P}). Nevertheless, it is also important to realize that the orbital motion (relative to the rotation of the medium at the orbit of the planet) can itself be super-alfvénic as well \citep{2010ApJ...722L.168V}. In that case, a bow shock is created in front of the planet in the direction of the orbit, without preventing the planet to magnetically connect to its host \citep{2023spi..conf....1S}. These aspects will be discussed in many more details in Chapter 3.

Finally, let us note that in general, the wind speed and the large-scale magnetic field of individual stars is expected to decrease over time during the main sequence lifetime \citep{Vidotto21LRSP}. The magnetic coupling between a planet and its environment will therefore also change significantly over secular timescales (and the planet dynamo can also change on those timescales). \citet{2011JGRA..116.1217S} studied how smaller the Earth magnetosphere could have been 3.5 Gyr ago and \citet{2023MNRAS.525.4008V} studied the wind-Earth magnetosphere coupling during the main sequence. \citet{Fionnagain2018,Carolan2019} studied the evolution of the solar wind in time and deduced two possible regimes for the Earth magnetospheric size, following $r_M\propto \Omega_\star^{-0.27}$ in the saturated regime and $r_M\propto \Omega_\star^{-2}$ in the unsaturated regime. For close-in planets, the evolution of the stellar magnetic field and rotation leads to changes in the Alfv\'en surface, effectively changing their migration rate over secular timescales (see \citealt{Ahuir2021} and Chapter 3).


\subsubsection{Influence on exoplanet evolution, atmospheric retention and habitability}

The evolution of magnetic activity over time also has profound effects on planets and on their atmosphere. For instance, the consequences of the young Sun's intense radiation and particle environment for the early Solar System were profound. The Sun's magnetic activity steadily declined during its main-sequence. While the solar photospheric luminosity was approximately (30\%) lower compared to present-day levels, its faster rotation and active dynamo generated magnetic heating processes in the chromosphere, transition region, and corona that induced UV, EUV, and X-ray emission roughly 10, 100, and 1000 times the present-day levels, respectively, as inferred from young solar analogs (\citealt{2007LRSP....4....3G}). This enhanced radiative and particle flux would have played a critical role in  shaping the early atmospheres of Venus, Earth, and Mars through photoionization, sputtering, and thermal escape processes. These effects likely shaped as well the Neptunian desert and savanna \citep{2021A&A...647A..40A,2023MNRAS.520.3749L,2025A&A...702A.132A}. These aspects will be discussed in many details for both short and long timescales in Chapter 4.

In addition, the aforementioned magnetic coupling can also affect the atmospheric retention of exoplanets. Two competing effects come into play: on one side the planet magnetic field can divert high energy particles around the planetary atmosphere (effectively shielding it from interactions), on the other side a large magnetosphere  leads to a larger interaction region with the environment with possibly larger energy depositions (see \citealt{2018MNRAS.481.5146B,2025ARA&A..63..299V} and references therein). For small magnetospheres (for instance for planets around young, active stars), the interplay between the magnetic field and the escaping material is still being actively investigated \citep[e.g.][]{Das2019,2021MNRAS.507.3626K,Gupta2023}. The escape mechanisms have different interplay with magnetic fields \citep{2020JGRA..12527639G}. Even though planets with and without an intrinsic magnetosphere are expected to interact differently with their surroundings \citep{Basak2021}, the example of Mars and the Earth strikingly show very comparable escape rates despite one having an intrinsic magnetosphere and not the other. The reason here likely lies in both the escape process under scrutiny and the relative importance of shielding (e.g. \citealt{2019MNRAS.488.2108E}). These dynamic interactions between the host star activity and planetary atmospheres have important implications for atmospheric escape and the habitability of exoplanets. 





\section{Open questions and outlook}
\label{sec:outlook}

The breadth of phenomena and diagnostics surveyed in this chapter naturally
opens a wide horizon of unresolved questions, several of which sit at the
interface of stellar physics and exoplanet science. We collect below a set of
open questions, each accompanied by a brief context, together with promising 
paths for progress.

\begin{enumerate}
    \item {\bf How typical is the Sun among other early G stars?} 
    Curated samples spanning ages of $\sim 0.1$--$8$~Gyr
    (Sect.~\ref{ssec:solar-analogs}) reveal substantial diversity in
    activity level, cycle behaviour, and magnetic geometry at comparable
    mass, rotation, and metallicity. The Sun appears photometrically quiet
    relative to some similarly rotating {\it Kepler} stars \citep{Reinhold2020}, 
    and recent spectropolarimetric evidence \citep{Metcalfe25} suggests this
    may reflect a real, physical departure from the cyclic regime sampled
    by faster rotators. Disentangling whether the Sun is genuinely
    atypical, or simply observed in a particular phase of a bi-stable
    near-critical dynamo (Sect.~\ref{ssec:sun-in-time}), requires sustained
    chromospheric, spectropolarimetric, and photometric monitoring of
    narrowly defined solar twins.

    \item {\bf Magnetic parity transitions and the global state of the
    cyclic dynamo.}
    Dynamo modelling now suggests that the equilibrium parity of
    solar-type stars is itself rotation-dependent, with rapid rotators
    favouring quadrupolar configurations driven by high-latitude flux
    emergence (Sect.~\ref{ssec:solar}). How frequently solar analogs
    cross between dipolar and quadrupolar states, and on what timescale,
    remains essentially unconstrained observationally. Such transitions
    reshape the heliospheric current sheet, the stellar wind, and the
    propagation of CMEs, and hence the radiative/particle environment of
    close-in planets (Sect.~\ref{sec:impact}). Observational signatures
    that could indicate a transition in progress -- including
    butterfly-diagram asymmetries derived from ZDI time series
    \citep{2025A&A...704A..68A} -- are a natural target for the coming
    decade.

    \item {\bf Subcritical dynamo, weakened magnetic braking, and the
    nature of low-activity states.}
    Three previously puzzling observations -- the breakdown of
    gyrochronology beyond mid main sequence \citep{vanSaders2016}, the low
    present-day activity of the Sun \citep{Reinhold2020}, and the
    occurrence of grand minima -- can now be addressed within the unified
    framework of a subcritical dynamo near solar age
    (Sect.~\ref{ssec:sun-in-time}; \citealt{Tripathi2021,Metcalfe25}).
    Yet several questions remain open: how typical are episodes of broad
    activity minima or full Grand Minima among solar-like stars; whether
    stars in a broad minimum follow the same activity--rotation scaling as
    cycling stars; and whether the distinction between a ``broad minimum''
    (an extended low-activity excursion within a regular cycle) and a true
    Grand Minimum (like the Maunder Minimum) reflects a fundamental change
    in dynamo mode or merely the surface manifestation of submerged flux.
    Long photometric baselines (decadal to century scale) combined with
    chromospheric monitoring and large-scale-field reconstructions are the
    most direct route to constrain transition probabilities and recovery
    timescales, building on existing evidence for cycle 
    min-max variability and rotation rate \citep[e.g.,][]{Garg+25}. 

    \item {\bf Rotation--magnetism scaling and the role of complementary
    field diagnostics.}
    Direct Zeeman broadening (sensitive to the total unsigned field,
    including the small-scale dynamo and mixed-polarity flux) and Stokes-V
    ZDI (sensitive to the large-scale organised field) give complementary
    but partly orthogonal pictures of stellar magnetism
    (Sect.~\ref{sssec:zdivszeeman}). Reconciling these is essential to
    constrain how steeply the magnetic flux emergence rate must scale with
    rotation -- recent modelling indicates at least quadratic scaling in the
    unsaturated regime, considerably steeper than earlier assumptions
    \citep{Isik26}. Independent confirmation will require homogeneous
    samples controlled for metallicity and effective temperature
    \citep[e.g.,][]{CarvalhoSilva25,Pezzotti25}, expanded Stokes-V monitoring 
    across the Rossby-number axis, and forward modelling that bridges Zeeman
    broadening and Stokes-V observables on the same surface mapping.

    \item {\bf Active nests, active longitudes.}
    On the Sun, magnetic flux preferentially emerges in spatial clusters 
    \citep[e.g.,][]{Csaszar25,Karapinar26}
    whose physical origin is not yet well-understood \citep[e.g.,][]{Raphaldini23}. 
    Active nests anchor the heliospheric
    current sheet over preferred longitudes (Sect.~\ref{sec:impact};
    \citealt{2024A&A...692A..29F}), and an increased degree of nesting can
    reproduce the elevated photometric variability and light curve 
    morphologies of some near-solar rotators \citep{Isik20}. To what
    extent active nests and longitudes are a generic feature of cool
    stars, how they depend on rotation, age, and metallicity, and whether
    they can be reliably diagnosed from disc-integrated time series all
    remain open. Combined transit mapping, photometric morphology
    analysis, and Doppler imaging of well-monitored systems are the most
    promising routes.

    \item {\bf The relation between starspots and stellar flares.}
    Solar studies indicate that disc-integrated active-region magnetic
    flux is a poor flare predictor; what controls flare energetics is
    the spatial complexity of the small fraction of flux that actually
    reconnects -- characterised by polarity-inversion-line gradients
    \citep{Dhakal+24} and by ribbon reconnection fluxes
    \citep{Kazachenko+17} -- which can plausibly power flares up to the
    lower end of the stellar superflare regime \citep{Krivova+26}.
    This naturally explains why disc-integrated stellar studies find
    mixed correlations between spot occupancy and flare occurrence
    (Sect.~\ref{ssec:flare};
    \citealt{Roettenbacher2018ApJ...868....3R,Zhang2025arXiv251201051Z,Lee+26}),
    while transit-mapping studies that probe smaller spatial scales do
    reveal positive trends with active-region area \citep{Araujo2021}.
    Progress will require coordinated photometric, spectroscopic, and
    Doppler-imaging campaigns \citep[e.g.,][]{2024ApJ...961...23N} 
    capable of distinguishing the
    disc-integrated spot \textit{coverage} from the small-scale magnetic
    complexity that the Sun teaches us is the actual flare driver.

    \item {\bf Stellar mass ejections and the energy partitioning of
    transients.}
    Stellar flares are now well characterised statistically (see, e.g., \citealt{2025A&A...694A.161S}), but their
    association with mass ejections, the fraction that escape as \textit{bona-fide}
    CMEs, and the partitioning of energy across the optical--UV--X-ray
    bands remain only loosely constrained (Sect.~\ref{ssec:eruptions}).
    Anomalies relative to the solar paradigm include $(i)$ the apparent 
    absence of a Neupert effect in some events 
    \citep{2023ApJ...951...33T} and $(ii)$ the
    unexpectedly large NUV continuum contribution in M-dwarf megaflares
    \citep{2025ApJ...978...81K}. They highlight how poorly the spectral 
    energy distribution of stellar transients is currently mapped. 
    Coordinated optical--UV--X-ray monitoring \citep[e.g.,][]{2024ApJ...961...23N,2026NatAs..10...64N} and dedicated radio 
    searches for stellar type-II bursts \citep{Callingham2025} should 
    provide the empirical basis
    needed to assess the radiative and particle environment experienced by
    close-in exoplanets.

    \item {\bf What constitutes a stellar activity cycle?}
    The diversity of cyclic behaviour seen across chromospheric S-indices,
    photometric amplitudes, X-ray fluxes, and large-scale-field reversals
    (Sect.~\ref{ssec:rotmod} and~\ref{sssec:iotaHor}) suggests that a
    single, solar-anchored notion of ``activity cycle'' is too narrow for
    the cool-star sample. Some objects show multi-periodic envelopes,
    flip-flop behaviour, or apparent absence of a cycle altogether
    \citep{olahMultiperiodicLightVariations2000,Jeffers2023SSRv..219...54J},
    and the cycle period--rotation period relation itself remains
    scatter-dominated and sensitive to the operational cycle definition.
    Definitions that distinguish polarity-reversal cycles from spot-area
    cycles, and that account for cycle-phase hysteresis between
    diagnostics \citep{Meunier+19d}, would clarify both sample selection
    and dynamo-model comparison.

    \item {\bf ZDI's reach and the meaning of recovered field
    components.}
    Stokes-V ZDI is a powerful but information-limited tool: tests against
    synthetic surface maps show that small-scale and mixed-polarity flux
    is lost to polarity cancellation, axisymmetric and toroidal components
    are systematically biased, and the results depend on inclination,
    signal-to-noise ratio, and phase coverage (Sect.~\ref{sssec:zdi};
    \citealt{2019MNRAS.483.5246L,hackman2024}). What ``toroidal field''
    actually means physically, and to what extent recovered topologies can
    serve as boundary conditions for wind and coronal models of exoplanet
    hosts (Sect.~\ref{sec:impact}), therefore remain open. Full Stokes
    $IQUV$ inversions \citep{Rosen2015ApJ...805..169R,
    Donati2025A&A...700A.122D}, Zeeman-intensification mapping
    \citep{kochukhov2023}, and direct comparison with
    surface-flux-transport-based forward models are the most promising
    routes to interpretive clarity. 
    One challenging task where Zeeman-intensification mapping is crucial is the investigation of (the lack of) spatial correlations between Doppler-imaged cool spots and strong 
    mixed-polarity magnetic flux producing the Zeeman signal. 

\item {\bf Spot characterization from transit mapping: breaking the intensity–size degeneracy and constraining spot lifetimes.}
Spot transit mapping directly characterises individual starspots:  location, size, and intensity, by modeling brightness bumps in planetary transit light curves. However, a fundamental degeneracy limits single-band photometry: a small, colder spot produces the same flux deficit as a larger, warmer spot. Breaking this requires simultaneous multiwavelength observations, as demonstrated by \cite{Valio2025} using the SPARC4 instrument on CoRoT-2. Another  challenge is that spot lifetimes are poorly constrained and vary with spectral type: on solar-type stars, lifetimes range from 10 to 350 days depending on spot area \cite{Namekata2019}, while cooler stars harbour spots that persist considerably longer, particularly at longer rotation periods, consistent with diffusive decay mechanisms \citep{Giles2017}.  Since spot evolution between successive transits undermines cross-identification of the same feature, establishing empirical lifetime distributions as a function of spectral type is essential for transit mapping to become a reliable surface diagnostic across the cool-star sequence.
    
    \item {\bf Multi-diagnostic surface mapping: photometry, Doppler
    imaging, transit mapping, and astrometry.}
    Each surface-reconstruction technique probes only a partial and
    complementary subset of the brightness and magnetic-field distribution:
    photometry constrains longitudinal asymmetries but suffers from
    latitude--size--contrast degeneracies
    (Sect.~\ref{sect:modelling-photometry}); Doppler imaging adds
    rotational-phase information and constrains spot latitudes within its
    own biases (Sect.~\ref{sssec:di}); transit mapping yields tight
    latitudinal localisation along a single chord but contends with
    spot-size/intensity degeneracies (Sect.~\ref{sec:transit-mapping});
    and astrometric photocentre variations probe odd spherical-harmonic
    modes inaccessible to disc-integrated photometry \citep{Deagan+26}.
    Joint inversions that combine two or more of these diagnostics --
    particularly simultaneous space-borne photometry alongside
    spectroscopic monitoring \citep[e.g.,][]{Lee+26} -- promise to break
    long-standing degeneracies and provide the most reliable surface maps
    for exoplanet host stars.

    \item {\bf Alternative activity proxies for the exoplanet era.}
    The classical Mt.\ Wilson Ca\,{\sc ii} H\&K $S$-index is being
    complemented by a growing set of proxies sensitive to different
    atmospheric layers and surface components: the He~{\sc i} 10830~\AA{}
    triplet, the Mg~{\sc ii} h\&k lines and FUV emission lines, the
    Paschen series and the Ca~{\sc ii} infrared triplet, narrow-band
    H$\alpha$ \citep{GomesdaSilva2022}, multi-colour photometric
    amplitudes, and asteroseismic p-mode frequency shifts
    \citep{garciaCoRoTRevealsMagnetic2010,kieferStellarMagneticActivity2017}.
    Line-profile decompositions of Ca~{\sc ii} H\&K
    \citep{Cretignier+24a,Cretignier+24b} unlock information lost in
    integrated-flux indices. Establishing the inter-calibration of these
    proxies across spectral types -- and ideally their behaviour through 
    full activity cycles -- is essential for high-precision radial-velocity 
    and atmospheric characterisation surveys.

    \item {\bf Saturation, supersaturation, and the partial- to
    fully-convective transition.}
    The physical origin of the $R_X$--$Ro$ saturation near $Ro\simeq 0.1$
    remains unresolved: candidate mechanisms include intrinsic dynamo
    saturation, filling-factor saturation of the surface, centrifugal
    stripping of the corona, and a transition from convective to interface
    dynamo (Sect.~\ref{ssec:model_rotact};
    \citealt{2011ApJ...743...48W,Blackman15}). Equally 
    interesting
    is the
    apparent continuity of fully-convective and partially-convective stars
    on the same rotation--activity sequence \citep{Wright2016,Wright2018}:
    either the observable activity is largely set by near-surface
    convective processes regardless of interior dynamo structure, or the
    dynamo mechanisms themselves are more similar than 
  often
  thought.
    Targeted 3D MHD modelling, EUV and X-ray monitoring at the very-fast-
    rotator end, and large homogeneous samples spanning the convective
    transition are needed to test these alternatives.

    \item {\bf From surface fields to stellar wind, EUV, and exoplanet
    environments.}
    Translating ZDI maps into reliable wind, EUV, and astrospheric
    current-sheet predictions for exoplanet hosts requires assumptions --
    a PFSS source-surface radius, a wind-driving prescription, an EUV
    reconstruction method -- whose validity outside the solar regime is
    uncertain (Sect.~\ref{sec:impact}; see also
    \citealt{Linsky2014,SanzForcada2011,Peacock2019x,Linsky2026}).
    Pinning down EUV spectra is especially urgent because ISM extinction
    precludes direct measurement for almost all stars, yet EUV flux drives
    atmospheric escape from close-in planets. Cross-validating
    extrapolation methods on the Sun viewed as a star, extending
    stellar wind models to fast rotators and strongly magnetised
    M dwarfs, and quantifying the impact of cycle phase, active nests, and
    transients on the local exoplanetary environment will be central to
    the coming decade of exoplanet astrophysics.
    
    
\end{enumerate}

The open questions collected above, diverse as they are, converge on two
structural pillars and one critical constraint for future progress.
The first is coordinated multi-wavelength and multi-diagnostic observations:
activity manifests differently at every atmospheric layer, and the same
magnetic feature leaves complementary signatures across diagnostics 
(e.g., photometry, line profiles, polarimetry, astrometry) so that
consistency between techniques acquired contemporaneously is the optimal
route to resolving the degeneracies surveyed throughout this chapter.
The second is forward modelling: physically self-consistent models that
predict synthetic observables from realistic surface configurations,
spanning flux transport, coronal structure, and stellar wind, allow
techniques to be compared on equal terms and place stellar characterisation
on a predictive rather than purely empirical footing. Underpinning both 
pillars is a prerequisite that recurs throughout this chapter: stellar samples 
must be homogeneous in the parameters that otherwise
confound activity trends -- principally mass, effective temperature, and
metallicity -- so that observed scaling relations isolate the variable
of interest and remain physically interpretable. Together, these three 
elements are what is needed to ensure progress in understanding magnetic
activity as a powerful tool for characterising exoplanets and
probing the environments in which planets exist and evolve.

\backmatter


\bmhead{Acknowledgements}



\renewcommand{\thefootnote}{\fnsymbol{footnote}}

This article was written following the workshop ``Magnetic Activity and its Impact 
on (Exo)planets", hosted and supported by the International Space Science Institute 
(ISSI) in Bern, Switzerland. The authors wish to express their thanks to ISSI 
for their financial and logistical support.
KP acknowledges support from the European Research Council (ERC) under grant agreement 101170037 (ERC-CoG Evaporator).\footnotemark[2]
AS acknowledges support from the European Research Council (ERC) under grant agreement 101125367 (ERC-CoG ExoMagnets).\footnotemark[2]
KV acknowledges funding provided by the Hungarian National Research, Development and Innovation Office grant \'Elvonal KKP-143986. 
EI and KN acknowledge support from the Japanese-Turkish bilateral programme by JSPS/T\"UB{\.I}TAK under grant agreements JPJSBP120269403/225N340. 
PF acknowledges financial support from the Severo Ochoa grant CEX2021-001131-S funded by MCIN/AEI/10.13039/501100011033. PF is also funded by the European Union (ERC, THIRSTEE, 101164189). Views and opinions expressed are however those of the author(s) only and do not necessarily reflect those of the European Union or the European Research Council. Neither the European Union nor the granting authority can be held responsible for them.

\footnotetext[2]{Views and opinions expressed are however those of the author(s) only and do not necessarily reflect those of the European Union or the European Research Council Executive Agency. Neither the European Union nor the granting authority can be held responsible for them.}

\section*{Declarations}


{\bf Conflict of interest}~~The authors have no conflicts of interest to declare that 
are relevant to the content of this article.

\bibliography{Ch1}

@book{Schussler2025book,
  editor    = {Sch{\"u}ssler, Manfred and Cameron, Robert and
               Charbonneau, Paul and Dikpati, Mausumi and
               Hotta, Hideyuki and Kitchatinov, Leonid},
  title     = {Solar and Stellar Dynamos: {A} New Era},
  series    = {Space Sciences Series of ISSI},
  volume    = {90},
  publisher = {Springer},
  address   = {Dordrecht},
  year      = {2025},
  isbn      = {978-94-024-2261-0},
  doi       = {10.1007/978-94-024-2260-3},
}

@ARTICLE{Brun+Browning17,
       author = {{Brun}, Allan Sacha and {Browning}, Matthew K.},
        title = "{Magnetism, dynamo action and the solar-stellar connection}",
      journal = {Living Reviews in Solar Physics},
         year = 2017,
        month = dec,
       volume = {14},
       number = {1},
          eid = {4},
        pages = {4},
          doi = {10.1007/s41116-017-0007-8},
       adsurl = {https://ui.adsabs.harvard.edu/abs/2017LRSP...14....4B}
}

@BOOK{Engvold+19book,
       author = {{Engvold}, Oddbj{\o}rn and {Vial}, Jean-Claude and {Skumanich}, Andrew},
        title = "{The Sun as a Guide to Stellar Physics}",
         year = 2019,
          doi = {10.1016/C2017-0-01365-4},
       adsurl = {https://ui.adsabs.harvard.edu/abs/2019sgsp.book.....E}
}

@ARTICLE{Kapyla+23,
       author = {{K{\"a}pyl{\"a}}, Petri J. and {Browning}, Matthew K. and {Brun}, Allan Sacha and {Guerrero}, Gustavo and {Warnecke}, J{\"o}rn},
        title = "{Simulations of Solar and Stellar Dynamos and Their Theoretical Interpretation}",
      journal = {\ssr},
         year = 2023,
        month = oct,
       volume = {219},
       number = {7},
          eid = {58},
        pages = {58},
          doi = {10.1007/s11214-023-01005-6},
archivePrefix = {arXiv},
       eprint = {2305.16790},
 primaryClass = {astro-ph.SR},
       adsurl = {https://ui.adsabs.harvard.edu/abs/2023SSRv..219...58K}
}

@ARTICLE{2015SSRv..188..251E,
       author = {{Eastwood}, J.~P. and {Hietala}, H. and {Toth}, G. and {Phan}, T.~D. and {Fujimoto}, M.},
        title = "{What Controls the Structure and Dynamics of Earth's Magnetosphere?}",
      journal = {\ssr},
         year = 2015,
        month = may,
       volume = {188},
       number = {1-4},
        pages = {251-286},
          doi = {10.1007/s11214-014-0050-x},
       adsurl = {https://ui.adsabs.harvard.edu/abs/2015SSRv..188..251E}
}

@ARTICLE{2017LRSP...14....5K,
       author = {{Kilpua}, Emilia and {Koskinen}, Hannu E.~J. and {Pulkkinen}, Tuija I.},
        title = "{Coronal mass ejections and their sheath regions in interplanetary space}",
      journal = {Living Reviews in Solar Physics},
         year = 2017,
        month = dec,
       volume = {14},
       number = {1},
          eid = {5},
        pages = {5},
          doi = {10.1007/s41116-017-0009-6},
       adsurl = {https://ui.adsabs.harvard.edu/abs/2017LRSP...14....5K}
}

@INCOLLECTION{2018haex.bookE..25S,
       author = {{Strugarek}, Antoine},
        title = "{Models of Star-Planet Magnetic Interaction}",
    booktitle = {Handbook of Exoplanets},
         year = 2018,
       editor = {{Deeg}, Hans J. and {Belmonte}, Juan Antonio},
          eid = {25},
        pages = {25},
          doi = {10.1007/978-3-319-55333-7_25},
       adsurl = {https://ui.adsabs.harvard.edu/abs/2018haex.bookE..25S}
}

@ARTICLE{2017SSRv..206..547M,
       author = {{Milan}, S.~E. and {Clausen}, L.~B.~N. and {Coxon}, J.~C. and {Carter}, J.~A. and {Walach}, M.-T. and {Laundal}, K. and {{\O}stgaard}, N. and {Tenfjord}, P. and {Reistad}, J. and {Snekvik}, K. and {Korth}, H. and {Anderson}, B.~J.},
        title = "{Overview of Solar Wind-Magnetosphere-Ionosphere-Atmosphere Coupling and the Generation of Magnetospheric Currents}",
      journal = {\ssr},
         year = 2017,
        month = mar,
       volume = {206},
       number = {1-4},
        pages = {547-573},
          doi = {10.1007/s11214-017-0333-0},
       adsurl = {https://ui.adsabs.harvard.edu/abs/2017SSRv..206..547M}
}

@ARTICLE{2025ApJ...993...80N,
       author = {{Namekata}, Kosuke and {Maehara}, Hiroyuki and {Notsu}, Yuta and {Honda}, Satoshi and {Ikuta}, Kai and {Nogami}, Daisaku and {Shibata}, Kazunari},
        title = "{Do Young Suns Produce Frequent, Massive CMEs? Results from Five-year Dedicated Optical Observations of EK Draconis and V889 Hercules}",
      journal = {\apj},
         year = 2025,
        month = nov,
       volume = {993},
       number = {1},
          eid = {80},
        pages = {80},
          doi = {10.3847/1538-4357/adfe70},
archivePrefix = {arXiv},
       eprint = {2510.22111},
 primaryClass = {astro-ph.SR},
       adsurl = {https://ui.adsabs.harvard.edu/abs/2025ApJ...993...80N}
}

@ARTICLE{2015ApJ...809...79O,
       author = {{Osten}, Rachel A. and {Wolk}, Scott J.},
        title = "{Connecting Flares and Transient Mass-loss Events in Magnetically Active Stars}",
      journal = {\apj},
         year = 2015,
        month = aug,
       volume = {809},
       number = {1},
          eid = {79},
        pages = {79},
          doi = {10.1088/0004-637X/809/1/79},
archivePrefix = {arXiv},
       eprint = {1506.04994},
 primaryClass = {astro-ph.SR},
       adsurl = {https://ui.adsabs.harvard.edu/abs/2015ApJ...809...79O}
}

@ARTICLE{2018ApJ...862...93A,
       author = {{Alvarado-G{\'o}mez}, Juli{\'a}n D. and {Drake}, Jeremy J. and {Cohen}, Ofer and {Moschou}, Sofia P. and {Garraffo}, Cecilia},
        title = "{Suppression of Coronal Mass Ejections in Active Stars by an Overlying Large-scale Magnetic Field: A Numerical Study}",
      journal = {\apj},
         year = 2018,
        month = aug,
       volume = {862},
       number = {2},
          eid = {93},
        pages = {93},
          doi = {10.3847/1538-4357/aacb7f},
archivePrefix = {arXiv},
       eprint = {1806.02828},
 primaryClass = {astro-ph.SR},
       adsurl = {https://ui.adsabs.harvard.edu/abs/2018ApJ...862...93A}
}

@ARTICLE{2013ApJ...764..170D,
       author = {{Drake}, Jeremy J. and {Cohen}, Ofer and {Yashiro}, Seiji and {Gopalswamy}, Nat},
        title = "{Implications of Mass and Energy Loss due to Coronal Mass Ejections on Magnetically Active Stars}",
      journal = {\apj},
         year = 2013,
        month = feb,
       volume = {764},
       number = {2},
          eid = {170},
        pages = {170},
          doi = {10.1088/0004-637X/764/2/170},
archivePrefix = {arXiv},
       eprint = {1302.1136},
 primaryClass = {astro-ph.SR},
       adsurl = {https://ui.adsabs.harvard.edu/abs/2013ApJ...764..170D}
}

@ARTICLE{2012ApJ...760....9A,
       author = {{Aarnio}, Alicia N. and {Matt}, Sean P. and {Stassun}, Keivan G.},
        title = "{Mass Loss in Pre-main-sequence Stars via Coronal Mass Ejections and Implications for Angular Momentum Loss}",
      journal = {\apj},
         year = 2012,
        month = nov,
       volume = {760},
       number = {1},
          eid = {9},
        pages = {9},
          doi = {10.1088/0004-637X/760/1/9},
archivePrefix = {arXiv},
       eprint = {1209.6410},
 primaryClass = {astro-ph.SR},
       adsurl = {https://ui.adsabs.harvard.edu/abs/2012ApJ...760....9A}
}

@ARTICLE{2026NatAs..10...64N,
       author = {{Namekata}, Kosuke and {France}, Kevin and {Chae}, Jongchul and {Airapetian}, Vladimir S. and {Kowalski}, Adam and {Notsu}, Yuta and {Young}, Peter R. and {Honda}, Satoshi and {Kang}, Soosang and {Kang}, Juhyung and {Lee}, Kyeore and {Maehara}, Hiroyuki and {Lee}, Kyoung-Sun and {Tamburri}, Cole and {Ohshima}, Tomohito and {Takayama}, Masaki and {Shibata}, Kazunari},
        title = "{Discovery of multi-temperature coronal mass ejection signatures from a young solar analogue}",
      journal = {Nature Astronomy},
         year = 2026,
        month = jan,
       volume = {10},
        pages = {64-75},
          doi = {10.1038/s41550-025-02691-8},
archivePrefix = {arXiv},
       eprint = {2510.22110},
 primaryClass = {astro-ph.SR},
       adsurl = {https://ui.adsabs.harvard.edu/abs/2026NatAs..10...64N}
}

@ARTICLE{2024A&A...686A..51M,
       author = {{Mohan}, Atul and {Mondal}, Surajit and {Wedemeyer}, Sven and {Gopalswamy}, Natchimuthuk},
        title = "{Energetic particle activity in AD Leo: Detection of a solar-like type-IV burst}",
      journal = {\aap},
         year = 2024,
        month = jun,
       volume = {686},
          eid = {A51},
        pages = {A51},
          doi = {10.1051/0004-6361/202347924},
archivePrefix = {arXiv},
       eprint = {2402.00185},
 primaryClass = {astro-ph.SR},
       adsurl = {https://ui.adsabs.harvard.edu/abs/2024A&A...686A..51M}
}

@ARTICLE{2020ApJ...905...23Z,
       author = {{Zic}, Andrew and {Murphy}, Tara and {Lynch}, Christene and {Heald}, George and {Lenc}, Emil and {Kaplan}, David L. and {Cairns}, Iver H. and {Coward}, David and {Gendre}, Bruce and {Johnston}, Helen and {MacGregor}, Meredith and {Price}, Danny C. and {Wheatland}, Michael S.},
        title = "{A Flare-type IV Burst Event from Proxima Centauri and Implications for Space Weather}",
      journal = {\apj},
         year = 2020,
        month = dec,
       volume = {905},
       number = {1},
          eid = {23},
        pages = {23},
          doi = {10.3847/1538-4357/abca90},
archivePrefix = {arXiv},
       eprint = {2012.04642},
 primaryClass = {astro-ph.SR},
       adsurl = {https://ui.adsabs.harvard.edu/abs/2020ApJ...905...23Z}
}

@ARTICLE{2025A&A...703A.198K,
       author = {{Konijn}, David C. and {Vedantham}, Harish K. and {Tasse}, Cyril and {Shimwell}, Timothy W. and {Hardcastle}, Martin J. and {Callingham}, Joseph R. and {Ilin}, Ekaterina and {Drabent}, Alexander and {Zarka}, Philippe and {van der Tak}, Floris F.~S. and {Bloot}, Sanne},
        title = "{Occurrence rate of stellar Type II radio bursts from a 100 star-year search for coronal mass ejections}",
      journal = {\aap},
         year = 2025,
        month = nov,
       volume = {703},
          eid = {A198},
        pages = {A198},
          doi = {10.1051/0004-6361/202554317},
archivePrefix = {arXiv},
       eprint = {2511.09296},
 primaryClass = {astro-ph.SR},
       adsurl = {https://ui.adsabs.harvard.edu/abs/2025A&A...703A.198K}
}

@ARTICLE{2022ApJ...936..170L,
       author = {{Loyd}, R.~O. Parke and {Mason}, James Paul and {Jin}, Meng and {Shkolnik}, Evgenya L. and {France}, Kevin and {Youngblood}, Allison and {Villadsen}, Jackie and {Schneider}, Christian and {Schneider}, Adam C. and {Llama}, Joe and {Ramiaramanantsoa}, Tahina and {Richey-Yowell}, Tyler},
        title = "{Constraining the Physical Properties of Stellar Coronal Mass Ejections with Coronal Dimming: Application to Far-ultraviolet Data of $\epsilon$ Eridani}",
      journal = {\apj},
         year = 2022,
        month = sep,
       volume = {936},
       number = {2},
          eid = {170},
        pages = {170},
          doi = {10.3847/1538-4357/ac80c1},
archivePrefix = {arXiv},
       eprint = {2207.05115},
 primaryClass = {astro-ph.SR},
       adsurl = {https://ui.adsabs.harvard.edu/abs/2022ApJ...936..170L}
}

@ARTICLE{2021NatAs...5..697V,
       author = {{Veronig}, Astrid M. and {Odert}, Petra and {Leitzinger}, Martin and {Dissauer}, Karin and {Fleck}, Nikolaus C. and {Hudson}, Hugh S.},
        title = "{Indications of stellar coronal mass ejections through coronal dimmings}",
      journal = {Nature Astronomy},
         year = 2021,
        month = jan,
       volume = {5},
        pages = {697-706},
          doi = {10.1038/s41550-021-01345-9},
archivePrefix = {arXiv},
       eprint = {2110.12029},
 primaryClass = {astro-ph.SR},
       adsurl = {https://ui.adsabs.harvard.edu/abs/2021NatAs...5..697V}
}

@ARTICLE{1990A&A...238..249H,
       author = {{Houdebine}, E.~R. and {Foing}, B.~H. and {Rodono}, M.},
        title = "{Dynamics of flares on late-type dMe stars. I. Flare mass ejections and stellar evolution.}",
      journal = {\aap},
         year = 1990,
        month = nov,
       volume = {238},
        pages = {249},
       adsurl = {https://ui.adsabs.harvard.edu/abs/1990A&A...238..249H}
}

@ARTICLE{2022NatAs...6..241N,
       author = {{Namekata}, Kosuke and {Maehara}, Hiroyuki and {Honda}, Satoshi and {Notsu}, Yuta and {Okamoto}, Soshi and {Takahashi}, Jun and {Takayama}, Masaki and {Ohshima}, Tomohito and {Saito}, Tomoki and {Katoh}, Noriyuki and {Tozuka}, Miyako and {Murata}, Katsuhiro L. and {Ogawa}, Futa and {Niwano}, Masafumi and {Adachi}, Ryo and {Oeda}, Motoki and {Shiraishi}, Kazuki and {Isogai}, Keisuke and {Seki}, Daikichi and {Ishii}, Takako T. and {Ichimoto}, Kiyoshi and {Nogami}, Daisaku and {Shibata}, Kazunari},
        title = "{Probable detection of an eruptive filament from a superflare on a solar-type star}",
      journal = {Nature Astronomy},
         year = 2022,
        month = jan,
       volume = {6},
        pages = {241-248},
          doi = {10.1038/s41550-021-01532-8},
archivePrefix = {arXiv},
       eprint = {2112.04808},
 primaryClass = {astro-ph.SR},
       adsurl = {https://ui.adsabs.harvard.edu/abs/2022NatAs...6..241N}
}

@ARTICLE{2025LRSP...22....2V,
       author = {{Veronig}, Astrid M. and {Dissauer}, Karin and {Kliem}, Bernhard and {Downs}, Cooper and {Hudson}, Hugh S. and {Jin}, Meng and {Osten}, Rachel and {Podladchikova}, Tatiana and {Prasad}, Avijeet and {Qiu}, Jiong and {Thompson}, Barbara and {Tian}, Hui and {Vourlidas}, Angelos},
        title = "{Coronal dimmings and what they tell us about solar and stellar coronal mass ejections}",
      journal = {Living Reviews in Solar Physics},
         year = 2025,
        month = jul,
       volume = {22},
       number = {1},
          eid = {2},
        pages = {2},
          doi = {10.1007/s41116-025-00041-4},
archivePrefix = {arXiv},
       eprint = {2505.19228},
 primaryClass = {astro-ph.SR},
       adsurl = {https://ui.adsabs.harvard.edu/abs/2025LRSP...22....2V}
}

@ARTICLE{2016A&A...590A..11V,
       author = {{Vida}, K. and {Kriskovics}, L. and {Ol{\'a}h}, K. and {Leitzinger}, M. and {Odert}, P. and {K{\H{o}}v{\'a}ri}, Zs. and {Korhonen}, H. and {Greimel}, R. and {Robb}, R. and {Cs{\'a}k}, B. and {Kov{\'a}cs}, J.},
        title = "{Investigating magnetic activity in very stable stellar magnetic fields. Long-term photometric and spectroscopic study of the fully convective M4 dwarf V374 Pegasi}",
      journal = {\aap},
         year = 2016,
        month = may,
       volume = {590},
          eid = {A11},
        pages = {A11},
          doi = {10.1051/0004-6361/201527925},
archivePrefix = {arXiv},
       eprint = {1603.00867},
 primaryClass = {astro-ph.SR},
       adsurl = {https://ui.adsabs.harvard.edu/abs/2016A&A...590A..11V}
}

@ARTICLE{2024Univ...10..313V,
       author = {{Vida}, Kriszti{\'a}n and {K{\H{o}}v{\'a}ri}, Zsolt and {Leitzinger}, Martin and {Odert}, Petra and {Ol{\'a}h}, Katalin and {Seli}, B{\'a}lint and {Kriskovics}, Levente and {Greimel}, Robert and {G{\"o}rgei}, Anna M{\'a}ria},
        title = "{Stellar Flares, Superflares, and Coronal Mass Ejections{\textemdash}Entering the Big Data Era}",
      journal = {Universe},
         year = 2024,
        month = jul,
       volume = {10},
       number = {8},
          eid = {313},
        pages = {313},
          doi = {10.3390/universe10080313},
archivePrefix = {arXiv},
       eprint = {2407.16446},
 primaryClass = {astro-ph.SR},
       adsurl = {https://ui.adsabs.harvard.edu/abs/2024Univ...10..313V}
}

@ARTICLE{2022SerAJ.205....1L,
       author = {{Leitzinger}, M. and {Odert}, P.},
        title = "{Stellar Coronal Mass Ejections}",
      journal = {Serbian Astronomical Journal},
         year = 2022,
        month = dec,
       volume = {205},
        pages = {1-22},
          doi = {10.2298/SAJ2205001L},
archivePrefix = {arXiv},
       eprint = {2212.09079},
 primaryClass = {astro-ph.SR},
       adsurl = {https://ui.adsabs.harvard.edu/abs/2022SerAJ.205....1L}
}

@ARTICLE{2022arXiv221105506N,
       author = {{Namekata}, Kosuke and {Maehara}, Hiroyuki and {Honda}, Satoshi and {Notsu}, Yuta and {Nogami}, Daisaku and {Shibata}, Kazunari},
        title = "{Hunting for stellar coronal mass ejections}",
      journal = {arXiv e-prints},
         year = 2022,
        month = nov,
          eid = {arXiv:2211.05506},
        pages = {arXiv:2211.05506},
          doi = {10.48550/arXiv.2211.05506},
archivePrefix = {arXiv},
       eprint = {2211.05506},
 primaryClass = {astro-ph.SR},
       adsurl = {https://ui.adsabs.harvard.edu/abs/2022arXiv221105506N}
}

@INPROCEEDINGS{2017IAUS..328..243O,
       author = {{Osten}, Rachel A. and {Wolk}, Scott J.},
        title = "{A Framework for Finding and Interpreting Stellar CMEs}",
    booktitle = {Living Around Active Stars},
         year = 2017,
       editor = {{Nandy}, D. and {Valio}, A. and {Petit}, P.},
       volume = {328},
        month = oct,
        pages = {243-251},
          doi = {10.1017/S1743921317004252},
       adsurl = {https://ui.adsabs.harvard.edu/abs/2017IAUS..328..243O}
}

@ARTICLE{2022AA...667L...9S,
       author = {{Stelzer}, B. and {Caramazza}, M. and {Raetz}, S. and {Argiroffi}, C. and {Coffaro}, M.},
        title = "{The Great Flare of 2021 November 19 on AD Leonis. Simultaneous XMM-Newton and TESS observations}",
      journal = {\aap},
         year = 2022,
        month = nov,
       volume = {667},
          eid = {L9},
        pages = {L9},
          doi = {10.1051/0004-6361/202244642},
archivePrefix = {arXiv},
       eprint = {2209.05068},
 primaryClass = {astro-ph.SR},
       adsurl = {https://ui.adsabs.harvard.edu/abs/2022A&A...667L...9S}
}

@ARTICLE{2023ApJ...951...33T,
       author = {{Tristan}, Isaiah I. and {Notsu}, Yuta and {Kowalski}, Adam F. and {Brown}, Alexander and {Wisniewski}, John P. and {Osten}, Rachel A. and {Vrijmoet}, Eliot H. and {White}, Graeme L. and {Carter}, Brad D. and {Grady}, Carol A. and {Henry}, Todd J. and {Hinojosa}, Rodrigo H. and {Lomax}, Jamie R. and {Neff}, James E. and {Paredes}, Leonardo A. and {Soutter}, Jack},
        title = "{A 7 Day Multiwavelength Flare Campaign on AU Mic. I. High-time-resolution Light Curves and the Thermal Empirical Neupert Effect}",
      journal = {\apj},
         year = 2023,
        month = jul,
       volume = {951},
       number = {1},
          eid = {33},
        pages = {33},
          doi = {10.3847/1538-4357/acc94f},
archivePrefix = {arXiv},
       eprint = {2304.05692},
 primaryClass = {astro-ph.SR},
       adsurl = {https://ui.adsabs.harvard.edu/abs/2023ApJ...951...33T}
}

@ARTICLE{2026ApJ...996...13O,
       author = {{Osten}, Rachel A. and {Kowalski}, Adam F. and {Hawley}, Suzanne and {Tristan}, Isaiah I. and {Schmidt}, Sarah J. and {Tofflemire}, Ben and {Hilton}, Eric},
        title = "{Radio and Optical Flares on the dMe Flare Star EV Lac}",
      journal = {\apj},
         year = 2026,
        month = jan,
       volume = {996},
       number = {1},
          eid = {13},
        pages = {13},
          doi = {10.3847/1538-4357/ae1a74},
archivePrefix = {arXiv},
       eprint = {2511.02719},
 primaryClass = {astro-ph.SR},
       adsurl = {https://ui.adsabs.harvard.edu/abs/2026ApJ...996...13O}
}

@ARTICLE{2024ApJ...961..189N,
       author = {{Notsu}, Yuta and {Kowalski}, Adam F. and {Maehara}, Hiroyuki and {Namekata}, Kosuke and {Hamaguchi}, Kenji and {Enoto}, Teruaki and {Tristan}, Isaiah I. and {Hawley}, Suzanne L. and {Davenport}, James R.~A. and {Honda}, Satoshi and {Ikuta}, Kai and {Inoue}, Shun and {Namizaki}, Keiichi and {Nogami}, Daisaku and {Shibata}, Kazunari},
        title = "{Apache Point Observatory (APO)/SMARTS Flare Star Campaign Observations. I. Blue Wing Asymmetries in Chromospheric Lines during Mid-M-Dwarf Flares from Simultaneous Spectroscopic and Photometric Observation Data}",
      journal = {\apj},
         year = 2024,
        month = feb,
       volume = {961},
       number = {2},
          eid = {189},
        pages = {189},
          doi = {10.3847/1538-4357/ad062f},
archivePrefix = {arXiv},
       eprint = {2310.02450},
 primaryClass = {astro-ph.SR},
       adsurl = {https://ui.adsabs.harvard.edu/abs/2024ApJ...961..189N}
}

@ARTICLE{2025ApJ...978...81K,
       author = {{Kowalski}, Adam F. and {Osten}, Rachel A. and {Notsu}, Yuta and {Tristan}, Isaiah I. and {Segura}, Antigona and {Maehara}, Hiroyuki and {Namekata}, Kosuke and {Inoue}, Shun},
        title = "{Rising Near-ultraviolet Spectra in Stellar Megaflares}",
      journal = {\apj},
         year = 2025,
        month = jan,
       volume = {978},
       number = {1},
          eid = {81},
        pages = {81},
          doi = {10.3847/1538-4357/ad9395},
archivePrefix = {arXiv},
       eprint = {2411.07913},
 primaryClass = {astro-ph.SR},
       adsurl = {https://ui.adsabs.harvard.edu/abs/2025ApJ...978...81K}
}

@ARTICLE{2021ApJ...911L..25M,
       author = {{MacGregor}, Meredith A. and {Weinberger}, Alycia J. and {Loyd}, R.~O. Parke and {Shkolnik}, Evgenya and {Barclay}, Thomas and {Howard}, Ward S. and {Zic}, Andrew and {Osten}, Rachel A. and {Cranmer}, Steven R. and {Kowalski}, Adam F. and {Lenc}, Emil and {Youngblood}, Allison and {Estes}, Anna and {Wilner}, David J. and {Forbrich}, Jan and {Hughes}, Anna and {Law}, Nicholas M. and {Murphy}, Tara and {Boley}, Aaron and {Matthews}, Jaymie},
        title = "{Discovery of an Extremely Short Duration Flare from Proxima Centauri Using Millimeter through Far-ultraviolet Observations}",
      journal = {\apjl},
         year = 2021,
        month = apr,
       volume = {911},
       number = {2},
          eid = {L25},
        pages = {L25},
          doi = {10.3847/2041-8213/abf14c},
archivePrefix = {arXiv},
       eprint = {2104.09519},
 primaryClass = {astro-ph.SR},
       adsurl = {https://ui.adsabs.harvard.edu/abs/2021ApJ...911L..25M}
}

@ARTICLE{2024ApJ...961...23N,
       author = {{Namekata}, Kosuke and {Airapetian}, Vladimir S. and {Petit}, Pascal and {Maehara}, Hiroyuki and {Ikuta}, Kai and {Inoue}, Shun and {Notsu}, Yuta and {Paudel}, Rishi R. and {Arzoumanian}, Zaven and {Avramova-Boncheva}, Antoaneta A. and {Gendreau}, Keith and {Jeffers}, Sandra V. and {Marsden}, Stephen and {Morin}, Julien and {Neiner}, Coralie and {Vidotto}, Aline A. and {Shibata}, Kazunari},
        title = "{Multiwavelength Campaign Observations of a Young Solar-type Star, EK Draconis. I. Discovery of Prominence Eruptions Associated with Superflares}",
      journal = {\apj},
         year = 2024,
        month = jan,
       volume = {961},
       number = {1},
          eid = {23},
        pages = {23},
          doi = {10.3847/1538-4357/ad0b7c},
archivePrefix = {arXiv},
       eprint = {2311.07380},
 primaryClass = {astro-ph.SR},
       adsurl = {https://ui.adsabs.harvard.edu/abs/2024ApJ...961...23N}
}

@ARTICLE{2024Sci...386.1301V,
       author = {{Vasilyev}, Valeriy and {Reinhold}, Timo and {Shapiro}, Alexander I. and {Usoskin}, Ilya and {Krivova}, Natalie A. and {Maehara}, Hiroyuki and {Notsu}, Yuta and {Brun}, Allan Sacha and {Solanki}, Sami K. and {Gizon}, Laurent},
        title = "{Sun-like stars produce superflares roughly once per century}",
      journal = {Science},
         year = 2024,
        month = dec,
       volume = {386},
       number = {6727},
        pages = {1301-1305},
          doi = {10.1126/science.adl5441},
archivePrefix = {arXiv},
       eprint = {2412.12265},
 primaryClass = {astro-ph.SR},
       adsurl = {https://ui.adsabs.harvard.edu/abs/2024Sci...386.1301V}
}

@ARTICLE{2013ApJS..209....5S,
       author = {{Shibayama}, Takuya and {Maehara}, Hiroyuki and {Notsu}, Shota and {Notsu}, Yuta and {Nagao}, Takashi and {Honda}, Satoshi and {Ishii}, Takako T. and {Nogami}, Daisaku and {Shibata}, Kazunari},
        title = "{Superflares on Solar-type Stars Observed with Kepler. I. Statistical Properties of Superflares}",
      journal = {\apjs},
         year = 2013,
        month = nov,
       volume = {209},
       number = {1},
          eid = {5},
        pages = {5},
          doi = {10.1088/0067-0049/209/1/5},
archivePrefix = {arXiv},
       eprint = {1308.1480},
 primaryClass = {astro-ph.SR},
       adsurl = {https://ui.adsabs.harvard.edu/abs/2013ApJS..209....5S}
}

@ARTICLE{2020AJ....160..219F,
       author = {{Feinstein}, Adina D. and {Montet}, Benjamin T. and {Ansdell}, Megan and {Nord}, Brian and {Bean}, Jacob L. and {G{\"u}nther}, Maximilian N. and {Gully-Santiago}, Michael A. and {Schlieder}, Joshua E.},
        title = "{Flare Statistics for Young Stars from a Convolutional Neural Network Analysis of TESS Data}",
      journal = {\aj},
         year = 2020,
        month = nov,
       volume = {160},
       number = {5},
          eid = {219},
        pages = {219},
          doi = {10.3847/1538-3881/abac0a},
archivePrefix = {arXiv},
       eprint = {2005.07710},
 primaryClass = {astro-ph.SR},
       adsurl = {https://ui.adsabs.harvard.edu/abs/2020AJ....160..219F}
}

@ARTICLE{2021A&A...645A..42I,
       author = {{Ilin}, Ekaterina and {Schmidt}, Sarah J. and {Poppenh{\"a}ger}, Katja and {Davenport}, James R.~A. and {Kristiansen}, Martti H. and {Omohundro}, Mark},
        title = "{Flares in open clusters with K2. II. Pleiades, Hyades, Praesepe, Ruprecht 147, and M 67}",
      journal = {\aap},
         year = 2021,
        month = jan,
       volume = {645},
          eid = {A42},
        pages = {A42},
          doi = {10.1051/0004-6361/202039198},
archivePrefix = {arXiv},
       eprint = {2010.05576},
 primaryClass = {astro-ph.SR},
       adsurl = {https://ui.adsabs.harvard.edu/abs/2021A&A...645A..42I}
}

@ARTICLE{2019A&A...622A.133I,
       author = {{Ilin}, Ekaterina and {Schmidt}, Sarah J. and {Davenport}, James R.~A. and {Strassmeier}, Klaus G.},
        title = "{Flares in open clusters with K2 . I. M 45 (Pleiades), M 44 (Praesepe), and M 67}",
      journal = {\aap},
         year = 2019,
        month = feb,
       volume = {622},
          eid = {A133},
        pages = {A133},
          doi = {10.1051/0004-6361/201834400},
archivePrefix = {arXiv},
       eprint = {1812.06725},
 primaryClass = {astro-ph.SR},
       adsurl = {https://ui.adsabs.harvard.edu/abs/2019A&A...622A.133I}
}

@ARTICLE{2019ApJ...871..241D,
       author = {{Davenport}, James R.~A. and {Covey}, Kevin R. and {Clarke}, Riley W. and {Boeck}, Austin C. and {Cornet}, Jonathan and {Hawley}, Suzanne L.},
        title = "{The Evolution of Flare Activity with Stellar Age}",
      journal = {\apj},
         year = 2019,
        month = feb,
       volume = {871},
       number = {2},
          eid = {241},
        pages = {241},
          doi = {10.3847/1538-4357/aafb76},
archivePrefix = {arXiv},
       eprint = {1901.00890},
 primaryClass = {astro-ph.SR},
       adsurl = {https://ui.adsabs.harvard.edu/abs/2019ApJ...871..241D}
}

@ARTICLE{2012Natur.485..478M,
       author = {{Maehara}, Hiroyuki and {Shibayama}, Takuya and {Notsu}, Shota and {Notsu}, Yuta and {Nagao}, Takashi and {Kusaba}, Satoshi and {Honda}, Satoshi and {Nogami}, Daisaku and {Shibata}, Kazunari},
        title = "{Superflares on solar-type stars}",
      journal = {Nature},
         year = 2012,
        month = may,
       volume = {485},
       number = {7399},
        pages = {478-481},
          doi = {10.1038/nature11063},
       adsurl = {https://ui.adsabs.harvard.edu/abs/2012Natur.485..478M}
}

@ARTICLE{2020arXiv201102117O,
       author = {{Okamoto}, Soshi and {Notsu}, Yuta and {Maehara}, Hiroyuki and {Namekata}, Kosuke and {Honda}, Satoshi and {Ikuta}, Kai and {Nogami}, Daisaku and {Shibata}, Kazunari},
        title = "{Statistical Properties of Superflares on Solar-type Stars: Results Using All of the Kepler Primary Mission Data}",
      journal = {\apj},
         year = 2021,
        month = jan,
       volume = {906},
       number = {2},
          eid = {72},
        pages = {72},
          doi = {10.3847/1538-4357/abc8f5},
archivePrefix = {arXiv},
       eprint = {2011.02117},
 primaryClass = {astro-ph.SR},
       adsurl = {https://ui.adsabs.harvard.edu/abs/2021ApJ...906...72O}
}

@ARTICLE{2019ApJ...876...58N,
       author = {{Notsu}, Yuta and {Maehara}, Hiroyuki and {Honda}, Satoshi and {Hawley}, Suzanne L. and {Davenport}, James R.~A. and {Namekata}, Kosuke and {Notsu}, Shota and {Ikuta}, Kai and {Nogami}, Daisaku and {Shibata}, Kazunari},
        title = "{Do Kepler Superflare Stars Really Include Slowly Rotating Sun-like Stars?{\textemdash}Results Using APO 3.5 m Telescope Spectroscopic Observations and Gaia-DR2 Data}",
      journal = {\apj},
         year = 2019,
        month = may,
       volume = {876},
       number = {1},
          eid = {58},
        pages = {58},
          doi = {10.3847/1538-4357/ab14e6},
archivePrefix = {arXiv},
       eprint = {1904.00142},
 primaryClass = {astro-ph.SR},
       adsurl = {https://ui.adsabs.harvard.edu/abs/2019ApJ...876...58N}
}

@ARTICLE{2016ApJ...829...23D,
       author = {{Davenport}, James R.~A.},
        title = "{The Kepler Catalog of Stellar Flares}",
      journal = {\apj},
         year = 2016,
        month = sep,
       volume = {829},
       number = {1},
          eid = {23},
        pages = {23},
          doi = {10.3847/0004-637X/829/1/23},
archivePrefix = {arXiv},
       eprint = {1607.03494},
 primaryClass = {astro-ph.SR},
       adsurl = {https://ui.adsabs.harvard.edu/abs/2016ApJ...829...23D}
}

@ARTICLE{1991ApJ...378..725H,
       author = {{Hawley}, Suzanne L. and {Pettersen}, Bjorn R.},
        title = "{The Great Flare of 1985 April 12 on AD Leonis}",
      journal = {\apj},
         year = 1991,
        month = sep,
       volume = {378},
        pages = {725},
          doi = {10.1086/170474},
       adsurl = {https://ui.adsabs.harvard.edu/abs/1991ApJ...378..725H}
}

@ARTICLE{2005ApJ...621..398O,
       author = {{Osten}, Rachel A. and {Hawley}, Suzanne L. and {Allred}, Joel C. and {Johns-Krull}, Christopher M. and {Roark}, Christine},
        title = "{From Radio to X-Ray: Flares on the dMe Flare Star EV Lacertae}",
      journal = {\apj},
         year = 2005,
        month = mar,
       volume = {621},
       number = {1},
        pages = {398-416},
          doi = {10.1086/427275},
archivePrefix = {arXiv},
       eprint = {astro-ph/0411236},
 primaryClass = {astro-ph},
       adsurl = {https://ui.adsabs.harvard.edu/abs/2005ApJ...621..398O}
}

@ARTICLE{2007LRSP....4....3G,
       author = {{G{\"u}del}, Manuel},
        title = "{The Sun in Time: Activity and Environment}",
      journal = {Living Reviews in Solar Physics},
         year = 2007,
        month = dec,
       volume = {4},
       number = {1},
          eid = {3},
        pages = {3},
          doi = {10.12942/lrsp-2007-3},
archivePrefix = {arXiv},
       eprint = {0712.1763},
 primaryClass = {astro-ph},
       adsurl = {https://ui.adsabs.harvard.edu/abs/2007LRSP....4....3G}
}

@ARTICLE{Gudel2004A&ARv,
       author = {{G{\"u}del}, Manuel},
        title = "{X-ray astronomy of stellar coronae}",
      journal = {\aapr},
         year = 2004,
        month = sep,
       volume = {12},
       number = {2-3},
        pages = {71-237},
          doi = {10.1007/s00159-004-0023-2},
archivePrefix = {arXiv},
       eprint = {astro-ph/0406661},
 primaryClass = {astro-ph},
       adsurl = {https://ui.adsabs.harvard.edu/abs/2004A&ARv..12...71G}
}

@ARTICLE{Benz2017LRSP,
       author = {{Benz}, Arnold O.},
        title = "{Flare Observations}",
      journal = {Living Reviews in Solar Physics},
         year = 2017,
        month = dec,
       volume = {14},
       number = {1},
          eid = {2},
        pages = {2},
          doi = {10.1007/s41116-016-0004-3},
       adsurl = {https://ui.adsabs.harvard.edu/abs/2017LRSP...14....2B}
}

@ARTICLE{2024LRSP...21....1K,
       author = {{Kowalski}, Adam F.},
        title = "{Stellar flares}",
      journal = {Living Reviews in Solar Physics},
         year = 2024,
        month = dec,
       volume = {21},
       number = {1},
          eid = {1},
        pages = {1},
          doi = {10.1007/s41116-024-00039-4},
archivePrefix = {arXiv},
       eprint = {2402.07885},
 primaryClass = {astro-ph.SR},
       adsurl = {https://ui.adsabs.harvard.edu/abs/2024LRSP...21....1K}
}

@ARTICLE{Vidotto21LRSP,
       author = {{Vidotto}, Aline A.},
        title = "{The evolution of the solar wind}",
      journal = {Living Reviews in Solar Physics},
         year = 2021,
        month = dec,
       volume = {18},
       number = {1},
          eid = {3},
        pages = {3},
          doi = {10.1007/s41116-021-00029-w},
archivePrefix = {arXiv},
       eprint = {2103.15748},
 primaryClass = {astro-ph.SR},
       adsurl = {https://ui.adsabs.harvard.edu/abs/2021LRSP...18....3V}
}

@ARTICLE{2023JSWSC..13...11R,
       author = {{R{\'e}ville}, Victor and {Poirier}, Nicolas and {Kouloumvakos}, Athanasios and {Rouillard}, Alexis Paul and {Ferreira Pinto}, Rui and {Fargette}, Na{\"\i}s and {Indurain}, Mikel and {Fournon}, Rapha{\"e}l and {James}, Th{\'e}o and {Pobeda}, Rapha{\"e}l and {Scoul}, Cyril},
        title = "{HelioCast: heliospheric forecasting based on white-light observations of the solar corona}",
      journal = {Journal of Space Weather and Space Climate},
         year = 2023,
        month = mar,
       volume = {13},
          eid = {11},
        pages = {11},
          doi = {10.1051/swsc/2023008},
archivePrefix = {arXiv},
       eprint = {2303.14972},
 primaryClass = {astro-ph.SR},
       adsurl = {https://ui.adsabs.harvard.edu/abs/2023JSWSC..13...11R}
}

@ARTICLE{2021LRSP...18....4T,
       author = {{Temmer}, Manuela},
        title = "{Space weather: the solar perspective: An update to Schwenn (2006)}",
      journal = {Living Reviews in Solar Physics},
         year = 2021,
        month = dec,
       volume = {18},
       number = {1},
          eid = {4},
        pages = {4},
          doi = {10.1007/s41116-021-00030-3},
archivePrefix = {arXiv},
       eprint = {2104.04261},
 primaryClass = {astro-ph.SR},
       adsurl = {https://ui.adsabs.harvard.edu/abs/2021LRSP...18....4T}
}

@ARTICLE{2021FrASS...8...84P,
       author = {{Poirier}, Nicolas and {Rouillard}, Alexis P. and {Kouloumvakos}, Athanasios and {Przybylak}, Alexis and {Fargette}, Na{\"\i}s and {Pobeda}, Rapha{\"e}l and {R{\'e}ville}, Victor and {Pinto}, Rui F. and {Indurain}, Mikel and {Alexandre}, Matthieu},
        title = "{Exploiting White-Light Observations to Improve Estimates of Magnetic Connectivity}",
      journal = {Frontiers in Astronomy and Space Sciences},
         year = 2021,
        month = may,
       volume = {8},
          eid = {84},
        pages = {84},
          doi = {10.3389/fspas.2021.684734},
       adsurl = {https://ui.adsabs.harvard.edu/abs/2021FrASS...8...84P}
}

@ARTICLE{2003GeoRL..30.1517M,
       author = {{McComas}, D.~J. and {Elliott}, H.~A. and {Schwadron}, N.~A. and {Gosling}, J.~T. and {Skoug}, R.~M. and {Goldstein}, B.~E.},
        title = "{The three-dimensional solar wind around solar maximum}",
      journal = {\grl},
         year = 2003,
        month = may,
       volume = {30},
       number = {10},
          eid = {1517},
        pages = {1517},
          doi = {10.1029/2003GL017136},
       adsurl = {https://ui.adsabs.harvard.edu/abs/2003GeoRL..30.1517M}
}

@ARTICLE{2020ApJ...901L..12I,
       author = {{I{\textcommabelow s}{\i}k}, Emre and {Shapiro}, Alexander I. and {Solanki}, Sami K. and {Krivova}, Natalie A.},
        title = "{Amplification of Brightness Variability by Active-region Nesting in Solar-like Stars}",
      journal = {\apjl},
         year = 2020,
        month = sep,
       volume = {901},
       number = {1},
          eid = {L12},
        pages = {L12},
          doi = {10.3847/2041-8213/abb409},
archivePrefix = {arXiv},
       eprint = {2009.00692},
 primaryClass = {astro-ph.SR},
       adsurl = {https://ui.adsabs.harvard.edu/abs/2020ApJ...901L..12I}
}

@ARTICLE{2024A&A...692A..29F,
       author = {{Finley}, A.~J.},
        title = "{Nested active regions anchor the heliospheric current sheet and stall the reversal of the coronal magnetic field}",
      journal = {\aap},
         year = 2024,
        month = dec,
       volume = {692},
          eid = {A29},
        pages = {A29},
          doi = {10.1051/0004-6361/202451896},
archivePrefix = {arXiv},
       eprint = {2410.18244},
 primaryClass = {astro-ph.SR},
       adsurl = {https://ui.adsabs.harvard.edu/abs/2024A&A...692A..29F}
}

@ARTICLE{2003A&A...405.1121B,
       author = {{Berdyugina}, S.~V. and {Usoskin}, I.~G.},
        title = "{Active longitudes in sunspot activity: Century scale persistence}",
      journal = {\aap},
         year = 2003,
        month = jul,
       volume = {405},
        pages = {1121-1128},
          doi = {10.1051/0004-6361:20030748},
       adsurl = {https://ui.adsabs.harvard.edu/abs/2003A&A...405.1121B}
}

@ARTICLE{2013LRSP...10....3P,
       author = {{Potgieter}, Marius S.},
        title = "{Solar Modulation of Cosmic Rays}",
      journal = {Living Reviews in Solar Physics},
         year = 2013,
        month = dec,
       volume = {10},
       number = {1},
          eid = {3},
        pages = {3},
          doi = {10.12942/lrsp-2013-3},
archivePrefix = {arXiv},
       eprint = {1306.4421},
 primaryClass = {physics.space-ph},
       adsurl = {https://ui.adsabs.harvard.edu/abs/2013LRSP...10....3P}
}

@ARTICLE{2022RvMPP...6....8S,
       author = {{Shen}, Fang and {Shen}, Chenglong and {Xu}, Mengjiao and {Liu}, Yousheng and {Feng}, Xueshang and {Wang}, Yuming},
        title = "{Propagation characteristics of coronal mass ejections (CMEs) in the corona and interplanetary space}",
      journal = {Reviews of Modern Plasma Physics},
         year = 2022,
        month = dec,
       volume = {6},
       number = {1},
          eid = {8},
        pages = {8},
          doi = {10.1007/s41614-022-00069-1},
       adsurl = {https://ui.adsabs.harvard.edu/abs/2022RvMPP...6....8S}
}

@ARTICLE{2023FrASS..1054266R,
       author = {{Reames}, Donald V.},
        title = "{Review and outlook of solar energetic particle measurements on multispacecraft missions}",
      journal = {Frontiers in Astronomy and Space Sciences},
         year = 2023,
        month = aug,
       volume = {10},
          eid = {1254266},
        pages = {1254266},
          doi = {10.3389/fspas.2023.1254266},
archivePrefix = {arXiv},
       eprint = {2307.04182},
 primaryClass = {astro-ph.SR},
       adsurl = {https://ui.adsabs.harvard.edu/abs/2023FrASS..1054266R}
}

@ARTICLE{2011SSRv..159..357Z,
       author = {{Zharkova}, V.~V. and {Arzner}, K. and {Benz}, A.~O. and {Browning}, P. and {Dauphin}, C. and {Emslie}, A.~G. and {Fletcher}, L. and {Kontar}, E.~P. and {Mann}, G. and {Onofri}, M. and {Petrosian}, V. and {Turkmani}, R. and {Vilmer}, N. and {Vlahos}, L.},
        title = "{Recent Advances in Understanding Particle Acceleration Processes in Solar Flares}",
      journal = {\ssr},
         year = 2011,
        month = sep,
       volume = {159},
       number = {1-4},
        pages = {357-420},
          doi = {10.1007/s11214-011-9803-y},
archivePrefix = {arXiv},
       eprint = {1110.2359},
 primaryClass = {astro-ph.SR},
       adsurl = {https://ui.adsabs.harvard.edu/abs/2011SSRv..159..357Z}
}

@ARTICLE{2015ApJ...814...99R,
       author = {{R{\'e}ville}, Victor and {Brun}, Allan Sacha and {Strugarek}, Antoine and {Matt}, Sean P. and {Bouvier}, J{\'e}r{\^o}me and {Folsom}, Colin P. and {Petit}, Pascal},
        title = "{From Solar to Stellar Corona: The Role of Wind, Rotation, and Magnetism}",
      journal = {\apj},
         year = 2015,
        month = dec,
       volume = {814},
       number = {2},
          eid = {99},
        pages = {99},
          doi = {10.1088/0004-637X/814/2/99},
archivePrefix = {arXiv},
       eprint = {1509.06982},
 primaryClass = {astro-ph.SR},
       adsurl = {https://ui.adsabs.harvard.edu/abs/2015ApJ...814...99R}
}

@ARTICLE{kochukhov2014,
       author = {{Kochukhov}, O. and {L{\"u}ftinger}, T. and {Neiner}, C. and {Alecian}, E. and {MiMeS Collaboration}},
        title = "{Magnetic field topology of the unique chemically peculiar star CU Virginis}",
      journal = {\aap},
         year = 2014,
        month = may,
       volume = {565},
          eid = {A83},
        pages = {A83},
          doi = {10.1051/0004-6361/201423472},
archivePrefix = {arXiv},
       eprint = {1404.2645},
 primaryClass = {astro-ph.SR},
       adsurl = {https://ui.adsabs.harvard.edu/abs/2014A&A...565A..83K}
}

@ARTICLE{Schofer2019,
       author = {{Sch{\"o}fer}, P. and {Jeffers}, S.~V. and {Reiners}, A. and {Shulyak}, D. and {Fuhrmeister}, B. and {Johnson}, E.~N. and {Zechmeister}, M. and {Ribas}, I. and {Quirrenbach}, A. and {Amado}, P.~J. and {Caballero}, J.~A. and {Anglada-Escud{\'e}}, G. and {Bauer}, F.~F. and {B{\'e}jar}, V.~J.~S. and {Cort{\'e}s-Contreras}, M. and {Dreizler}, S. and {Guenther}, E.~W. and {Kaminski}, A. and {K{\"u}rster}, M. and {Lafarga}, M. and {Montes}, D. and {Morales}, J.~C. and {Pedraz}, S. and {Tal-Or}, L.},
        title = "{The CARMENES search for exoplanets around M dwarfs. Activity indicators at visible and near-infrared wavelengths}",
      journal = {\aap},
         year = 2019,
        month = mar,
       volume = {623},
          eid = {A44},
        pages = {A44},
          doi = {10.1051/0004-6361/201834114},
archivePrefix = {arXiv},
       eprint = {1901.08861},
 primaryClass = {astro-ph.SR},
       adsurl = {https://ui.adsabs.harvard.edu/abs/2019A&A...623A..44S}
}

@ARTICLE{Bouchy2025,
       author = {{Bouchy}, Fran{\c{c}}ois and {Doyon}, Ren{\'e} and {Pepe}, Francesco and {Melo}, Claudio and {Artigau}, {\'E}tienne and {Malo}, Lison and {Wildi}, Fran{\c{c}}ois and {Baron}, Fr{\'e}d{\'e}rique and {Delfosse}, Xavier and {De Medeiros}, Jose Renan and {Rebolo}, Rafael and {Santos}, Nuno C. and {Wade}, Gregg and {Allart}, Romain and {Al Moulla}, Khaled and {Blind}, Nicolas and {Cadieux}, Charles and {Canto Martins}, Bruno L. and {Cook}, Neil J. and {Dumusque}, Xavier and {Frensch}, Yolanda and {Genest}, Fr{\'e}d{\'e}ric and {Gonz{\'a}lez Hern{\'a}ndez}, Jonay I. and {Grieves}, Nolan and {Lo Curto}, Gaspare and {Lovis}, Christophe and {Mignon}, Lucile and {Nielsen}, Louise D. and {Poulin-Girard}, Anne-Sophie and {Rasilla}, Jos{\'e} Luis and {Reshetov}, Vladimir and {Sosnowska}, Danuta and {Sordet}, Michael and {Saint-Antoine}, Jonathan and {Su{\'a}rez Mascare{\~n}o}, Alejandro and {Thibault}, Simon and {Vall{\'e}e}, Philippe and {Vandal}, Thomas and {Abreu}, Manuel and {Aguiar}, Jos{\'e} L.~A. and {Allain}, Guillaume and {Arial}, Tomy and {Auger}, Hugues and {Barros}, Susana C.~C. and {Bazinet}, Luc and {Benneke}, Bj{\"o}rn and {Bonfils}, Xavier and {Boucher}, Anne and {Bourrier}, Vincent and {Bovay}, S{\'e}bastien and {Broeg}, Christopher and {Brousseau}, Denis and {Bruniquel}, Vincent and {Bryan}, Marta and {Cabral}, Alexandre and {Carmona}, Andres and {Carteret}, Yann and {Challita}, Zalpha and {Chazelas}, Bruno and {Cloutier}, Ryan and {Coelho}, Jo{\~a}o and {Cointepas}, Marion and {Conod}, Uriel and {Cowan}, Nicolas B. and {Cristo}, Eduardo and {Gomes da Silva}, Jo{\~a}o and {Dauplaise}, Laurie and {Darveau-Bernier}, Antoine and {de Lima Gomes}, Roseane and {de Freitas}, Daniel Brito and {Delgado-Mena}, Elisa and {Delisle}, Jean-Baptiste and {Ehrenreich}, David and {Faria}, Jo{\~a}o and {Figueira}, Pedro and {Fontinele}, Dasaev O. and {Forveille}, Thierry and {Gagn{\'e}}, Jonathan and {Genolet}, Ludovic and {T{\'e}mich}, F{\'e}lix Gracia and {Hernandez}, Olivier and {Hobson}, Melissa J. and {Hoeijmakers}, Jens and {Hubin}, Norbert and {Jahandar}, Farbod and {Jayawardhana}, Ray and {K{\"a}ufl}, Hans-Ulrich and {Kerley}, Dan and {Kolb}, Johann and {Krishnamurthy}, Vigneshwaran and {Lafreni{\`e}re}, David and {Lamontagne}, Pierrot and {Larue}, Pierre and {Leath}, Henry and {L'Heureux}, Alexandrine and {de Castro Le{\~a}o}, Izan and {Lim}, Olivia and {Martins}, Allan M. and {Matthews}, Jaymie and {Mayer}, Jean-S{\'e}bastien and {Messias}, Yuri S. and {Metchev}, Stan and {Moranta}, Leslie and {Mordasini}, Christoph and {Mounzer}, Dany and {Nari}, Nicola and {Osborn}, Ares and {Ouellet}, Mathieu and {Otegi}, Jon and {Parc}, L{\'e}na and {Pasquini}, Luca and {Passegger}, Vera M. and {Pelletier}, Stefan and {Peroux}, C{\'e}line and {Piaulet-Ghorayeb}, Caroline and {Plotnykov}, Mykhaylo and {Pompei}, Emanuela and {Rowe}, Jason and {Sarajlic}, Mirsad and {Segovia}, Alex anhttps://wiki.helsinki.fi/xwiki/bin/view/SMA/d {Seidel}, Julia and {S{\'e}gransan}, Damien and {Schnell}, Robin and {Costa Silva}, Ana Rita and {Srivastava}, Avidaan and {Stefanov}, Atanas K. and {Teixeira}, M{\'a}rcio A. and {Udry}, St{\'e}phane and {Valencia}, Diana and {Vaulato}, Valentina and {Wardenier}, Joost P. and {Wehbe}, Bachar and {Weisserman}, Drew and {Wevers}, Ivan and {Yariv}, Vincent and {Zins}, G{\'e}rard},
        title = "{NIRPS joining HARPS at ESO 3.6 m: On-sky performance and science objectives}",
      journal = {\aap},
         year = 2025,
        month = aug,
       volume = {700},
          eid = {A10},
        pages = {A10},
          doi = {10.1051/0004-6361/202453341},
archivePrefix = {arXiv},
       eprint = {2507.21767},
 primaryClass = {astro-ph.IM},
       adsurl = {https://ui.adsabs.harvard.edu/abs/2025A&A...700A..10B}
}

@INPROCEEDINGS{Quirrenbach2014,
       author = {{Quirrenbach}, A. and {Amado}, P.~J. and {Caballero}, J.~A. and {Mundt}, R. and {Reiners}, A. and {Ribas}, I. and {Seifert}, W. and {Abril}, M. and {Aceituno}, J. and {Alonso-Floriano}, F.~J. and {Ammler-von Eiff}, M. and {Antona Jim{\'e}nez}, R. and {Anwand-Heerwart}, H. and {Azzaro}, M. and {Bauer}, F. and {Barrado}, D. and {Becerril}, S. and {B{\'e}jar}, V.~J.~S. and {Ben{\'\i}tez}, D. and {Berdi{\~n}as}, Z.~M. and {C{\'a}rdenas}, M.~C. and {Casal}, E. and {Claret}, A. and {Colom{\'e}}, J. and {Cort{\'e}s-Contreras}, M. and {Czesla}, S. and {Doellinger}, M. and {Dreizler}, S. and {Feiz}, C. and {Fern{\'a}ndez}, M. and {Galad{\'\i}}, D. and {G{\'a}lvez-Ortiz}, M.~C. and {Garc{\'\i}a-Piquer}, A. and {Garc{\'\i}a-Vargas}, M.~L. and {Garrido}, R. and {Gesa}, L. and {G{\'o}mez Galera}, V. and {Gonz{\'a}lez {\'A}lvarez}, E. and {Gonz{\'a}lez Hern{\'a}ndez}, J.~I. and {Gr{\"o}zinger}, U. and {Gu{\`a}rdia}, J. and {Guenther}, E.~W. and {de Guindos}, E. and {Guti{\'e}rrez-Soto}, J. and {Hagen}, H.-J. and {Hatzes}, A.~P. and {Hauschildt}, P.~H. and {Helmling}, J. and {Henning}, T. and {Hermann}, D. and {Hern{\'a}ndez Casta{\~n}o}, L. and {Herrero}, E. and {Hidalgo}, D. and {Holgado}, G. and {Huber}, A. and {Huber}, K.~F. and {Jeffers}, S. and {Joergens}, V. and {de Juan}, E. and {Kehr}, M. and {Klein}, R. and {K{\"u}rster}, M. and {Lamert}, A. and {Lalitha}, S. and {Laun}, W. and {Lemke}, U. and {Lenzen}, R. and {L{\'o}pez del Fresno}, Mauro and {L{\'o}pez Mart{\'\i}}, B. and {L{\'o}pez-Santiago}, J. and {Mall}, U. and {Mandel}, H. and {Mart{\'\i}n}, E.~L. and {Mart{\'\i}n-Ruiz}, S. and {Mart{\'\i}nez-Rodr{\'\i}guez}, H. and {Marvin}, C.~J. and {Mathar}, R.~J. and {Mirabet}, E. and {Montes}, D. and {Morales Mu{\~n}oz}, R. and {Moya}, A. and {Naranjo}, V. and {Ofir}, A. and {Oreiro}, R. and {Pall{\'e}}, E. and {Panduro}, J. and {Passegger}, V.-M. and {P{\'e}rez-Calpena}, A. and {P{\'e}rez Medialdea}, D. and {Perger}, M. and {Pluto}, M. and {Ram{\'o}n}, A. and {Rebolo}, R. and {Redondo}, P. and {Reffert}, S. and {Reinhardt}, S. and {Rhode}, P. and {Rix}, H.-W. and {Rodler}, F. and {Rodr{\'\i}guez}, E. and {Rodr{\'\i}guez-L{\'o}pez}, C. and {Rodr{\'\i}guez-P{\'e}rez}, E. and {Rohloff}, R.-R. and {Rosich}, A. and {S{\'a}nchez-Blanco}, E. and {S{\'a}nchez Carrasco}, M.~A. and {Sanz-Forcada}, J. and {Sarmiento}, L.~F. and {Sch{\"a}fer}, S. and {Schiller}, J. and {Schmidt}, C. and {Schmitt}, J.~H.~M.~M. and {Solano}, E. and {Stahl}, O. and {Storz}, C. and {St{\"u}rmer}, J. and {Su{\'a}rez}, J.~C. and {Ulbrich}, R.~G. and {Veredas}, G. and {Wagner}, K. and {Winkler}, J. and {Zapatero Osorio}, M.~R. and {Zechmeister}, M. and {Abell{\'a}n de Paco}, F.~J. and {Anglada-Escud{\'e}}, G. and {del Burgo}, C. and {Klutsch}, A. and {Lizon}, J.~L. and {L{\'o}pez-Morales}, M. and {Morales}, J.~C. and {Perryman}, M.~A.~C. and {Tulloch}, S.~M. and {Xu}, W.},
        title = "{CARMENES instrument overview}",
    booktitle = {Ground-based and Airborne Instrumentation for Astronomy V},
         year = 2014,
       editor = {{Ramsay}, Suzanne K. and {McLean}, Ian S. and {Takami}, Hideki},
       series = {Society of Photo-Optical Instrumentation Engineers (SPIE) Conference Series},
       volume = {9147},
        month = jul,
          eid = {91471F},
        pages = {91471F},
          doi = {10.1117/12.2056453},
       adsurl = {https://ui.adsabs.harvard.edu/abs/2014SPIE.9147E..1FQ}
}

@ARTICLE{Penn2014,
       author = {{Penn}, Matthew J.},
        title = "{Infrared Solar Physics}",
      journal = {Living Reviews in Solar Physics},
         year = 2014,
        month = dec,
       volume = {11},
       number = {1},
          eid = {2},
        pages = {2},
          doi = {10.12942/lrsp-2014-2},
       adsurl = {https://ui.adsabs.harvard.edu/abs/2014LRSP...11....2P}
}

@ARTICLE{IbanezBustos2023,
       author = {{Iba{\~n}ez Bustos}, R.~V. and {Buccino}, A.~P. and {Flores}, M. and {Martinez}, C.~F. and {Mauas}, P.~J.~D.},
        title = "{Correlation between activity indicators: H{\ensuremath{\alpha}} and Ca II lines in M-dwarf stars}",
      journal = {\aap},
         year = 2023,
        month = apr,
       volume = {672},
          eid = {A37},
        pages = {A37},
          doi = {10.1051/0004-6361/202245352},
archivePrefix = {arXiv},
       eprint = {2303.17237},
 primaryClass = {astro-ph.SR},
       adsurl = {https://ui.adsabs.harvard.edu/abs/2023A&A...672A..37I}
}

@ARTICLE{Maldonado19,
       author = {{Maldonado}, J. and {Phillips}, D.~F. and {Dumusque}, X. and
         {Collier Cameron}, A. and {Haywood}, R.~D. and {Lanza}, A.~F. and
         {Micela}, G. and {Mortier}, A. and {Saar}, S.~H. and {Sozzetti}, A. and
         {Rice}, K. and {Milbourne}, T. and {Cecconi}, M. and {Cegla}, H.~M. and
         {Cosentino}, R. and {Costes}, J. and {Ghedina}, A. and {Gonzalez}, M. and
         {Guerra}, J. and {Hern{\'a}ndez}, N. and {Li}, C. -H. and {Lodi}, M. and
         {Malavolta}, L. and {Molinari}, E. and {Pepe}, F. and {Piotto}, G. and
         {Poretti}, E. and {Sasselov}, D. and {San Juan}, J. and {Thompson}, S. and
         {Udry}, S. and {Watson}, C.},
        title = "{Temporal evolution and correlations of optical activity indicators measured in Sun-as-a-star observations}",
      journal = {\aap},
         year = "2019",
        month = "Jul",
       volume = {627},
          eid = {A118},
        pages = {A118},
          doi = {10.1051/0004-6361/201935233},
archivePrefix = {arXiv},
       eprint = {1906.03002},
 primaryClass = {astro-ph.SR},
       adsurl = {https://ui.adsabs.harvard.edu/abs/2019A&A...627A.118M}
}

@ARTICLE{Mamajek2008,
       author = {{Mamajek}, Eric E. and {Hillenbrand}, Lynne A.},
        title = "{Improved Age Estimation for Solar-Type Dwarfs Using Activity-Rotation Diagnostics}",
      journal = {\apj},
         year = 2008,
        month = nov,
       volume = {687},
       number = {2},
        pages = {1264-1293},
          doi = {10.1086/591785},
archivePrefix = {arXiv},
       eprint = {0807.1686},
 primaryClass = {astro-ph},
       adsurl = {https://ui.adsabs.harvard.edu/abs/2008ApJ...687.1264M}
}

@ARTICLE{Busa2007,
       author = {{Bus{\`a}}, I. and {Aznar Cuadrado}, R. and {Terranegra}, L. and {Andretta}, V. and {Gomez}, M.~T.},
        title = "{The Ca II infrared triplet as a stellar activity diagnostic. II. Test and calibration with high resolution observations}",
      journal = {\aap},
         year = 2007,
        month = may,
       volume = {466},
       number = {3},
        pages = {1089-1098},
          doi = {10.1051/0004-6361:20065588},
       adsurl = {https://ui.adsabs.harvard.edu/abs/2007A&A...466.1089B}
}

@ARTICLE{Cincunegui2007,
       author = {{Cincunegui}, C. and {D{\'\i}az}, R.~F. and {Mauas}, P.~J.~D.},
        title = "{H{\ensuremath{\alpha}} and the Ca II H and K lines as activity proxies for late-type stars}",
      journal = {\aap},
         year = 2007,
        month = jul,
       volume = {469},
       number = {1},
        pages = {309-317},
          doi = {10.1051/0004-6361:20066503},
archivePrefix = {arXiv},
       eprint = {astro-ph/0703511},
 primaryClass = {astro-ph},
       adsurl = {https://ui.adsabs.harvard.edu/abs/2007A&A...469..309C}
}

@ARTICLE{Frasca1994,
       author = {{Frasca}, A. and {Catalano}, S.},
        title = "{H{\ensuremath{\alpha}} survey of late-type active binaries.}",
      journal = {\aap},
         year = 1994,
        month = apr,
       volume = {284},
        pages = {883-899},
       adsurl = {https://ui.adsabs.harvard.edu/abs/1994A&A...284..883F}
}

@ARTICLE{GomesdaSilva2014,
   author = {{Gomes da Silva}, J. and {Santos}, N.~C. and {Boisse}, I. and 
	{Dumusque}, X. and {Lovis}, C.},
    title = "{On the long-term correlation between the flux in the Ca ii H \& K and H{$\alpha$} lines for FGK stars}",
  journal = {\aap},
archivePrefix = "arXiv",
   eprint = {1311.6642},
 primaryClass = "astro-ph.SR",
     year = 2014,
    month = jun,
   volume = 566,
      eid = {A66},
    pages = {A66},
      doi = {10.1051/0004-6361/201322697},
   adsurl = {http://adsabs.harvard.edu/abs/2014A%26A...566A..66G}
}

@ARTICLE{MartinezArnaiz2010,
       author = {{Mart{\'\i}nez-Arn{\'a}iz}, R. and {Maldonado}, J. and {Montes}, D. and {Eiroa}, C. and {Montesinos}, B.},
        title = "{Chromospheric activity and rotation of FGK stars in the solar vicinity. An estimation of the radial velocity jitter}",
      journal = {\aap},
         year = 2010,
        month = sep,
       volume = {520},
          eid = {A79},
        pages = {A79},
          doi = {10.1051/0004-6361/200913725},
archivePrefix = {arXiv},
       eprint = {1002.4391},
 primaryClass = {astro-ph.SR},
       adsurl = {https://ui.adsabs.harvard.edu/abs/2010A&A...520A..79M}
}

@ARTICLE{Meunier2024,
       author = {{Meunier}, N. and {Mignon}, L. and {Kretzschmar}, M. and {Delfosse}, X.},
        title = "{Characterisation of the stellar activity of M dwarfs. II. Relationship between Ca, H{\ensuremath{\alpha}}, and Na chromospheric emissions}",
      journal = {\aap},
         year = 2024,
        month = apr,
       volume = {684},
          eid = {A106},
        pages = {A106},
          doi = {10.1051/0004-6361/202347362},
archivePrefix = {arXiv},
       eprint = {2403.01790},
 primaryClass = {astro-ph.SR},
       adsurl = {https://ui.adsabs.harvard.edu/abs/2024A&A...684A.106M}
}

@ARTICLE{Montes1995,
       author = {{Montes}, D. and {de Castro}, E. and {Fernandez-Figueroa}, M.~J. and {Cornide}, M.},
        title = "{Application of the spectral subtraction technique to the CA II H \& K and H{\ensuremath{\in}} lines in a sample of chromospherically active binaries.}",
      journal = {\aaps},
         year = 1995,
        month = dec,
       volume = {114},
        pages = {287},
       adsurl = {https://ui.adsabs.harvard.edu/abs/1995A&AS..114..287M}
}

@ARTICLE{vanDriel-Gesztelyi15,
       author = {{van Driel-Gesztelyi}, Lidia and {Green}, Lucie May},
        title = "{Evolution of Active Regions}",
      journal = {Living Reviews in Solar Physics},
         year = 2015,
        month = dec,
       volume = {12},
       number = {1},
          eid = {1},
        pages = {1},
          doi = {10.1007/lrsp-2015-1},
       adsurl = {https://ui.adsabs.harvard.edu/abs/2015LRSP...12....1V}
}

@ARTICLE{Livingston2007,
       author = {{Livingston}, W. and {Wallace}, L. and {White}, O.~R. and {Giampapa}, M.~S.},
        title = "{Sun-as-a-Star Spectrum Variations 1974-2006}",
      journal = {\apj},
         year = 2007,
        month = mar,
       volume = {657},
       number = {2},
        pages = {1137-1149},
          doi = {10.1086/511127},
archivePrefix = {arXiv},
       eprint = {astro-ph/0612554},
 primaryClass = {astro-ph},
       adsurl = {https://ui.adsabs.harvard.edu/abs/2007ApJ...657.1137L}
}

@ARTICLE{Meunier2009,
       author = {{Meunier}, N. and {Delfosse}, X.},
        title = "{On the correlation between Ca and H{\ensuremath{\alpha}} solar emission and consequences for stellar activity observations}",
      journal = {\aap},
         year = 2009,
        month = jul,
       volume = {501},
       number = {3},
        pages = {1103-1112},
          doi = {10.1051/0004-6361/200911823},
archivePrefix = {arXiv},
       eprint = {0905.4037},
 primaryClass = {astro-ph.SR},
       adsurl = {https://ui.adsabs.harvard.edu/abs/2009A&A...501.1103M}
}

@ARTICLE{Hathaway15,
       author = {{Hathaway}, David H.},
        title = "{The Solar Cycle}",
      journal = {Living Reviews in Solar Physics},
         year = 2015,
        month = dec,
       volume = {12},
       number = {1},
          eid = {4},
        pages = {4},
          doi = {10.1007/lrsp-2015-4},
archivePrefix = {arXiv},
       eprint = {1502.07020},
 primaryClass = {astro-ph.SR},
       adsurl = {https://ui.adsabs.harvard.edu/abs/2015LRSP...12....4H}
}

@ARTICLE{Cameron23,
       author = {{Cameron}, Robert H. and {Sch{\"u}ssler}, Manfred},
        title = "{Observationally Guided Models for the Solar Dynamo and the Role of the Surface Field}",
      journal = {\ssr},
         year = 2023,
        month = oct,
       volume = {219},
       number = {7},
          eid = {60},
        pages = {60},
          doi = {10.1007/s11214-023-01004-7},
archivePrefix = {arXiv},
       eprint = {2305.02253},
 primaryClass = {astro-ph.SR},
       adsurl = {https://ui.adsabs.harvard.edu/abs/2023SSRv..219...60C}
}

@ARTICLE{Zhang_Zhebin+24,
       author = {{Zhang}, Zebin and {Jiang}, Jie and {Kitchatinov}, Leonid},
        title = "{Modeling the effects of starspots on stellar magnetic cycles}",
      journal = {\aap},
         year = 2024,
        month = jun,
       volume = {686},
          eid = {A90},
        pages = {A90},
          doi = {10.1051/0004-6361/202348201},
archivePrefix = {arXiv},
       eprint = {2402.17449},
 primaryClass = {astro-ph.SR},
       adsurl = {https://ui.adsabs.harvard.edu/abs/2024A&A...686A..90Z}
}

@ARTICLE{Karak+14,
       author = {{Karak}, Bidya Binay and {Kitchatinov}, Leonid L. and {Choudhuri}, Arnab Rai},
        title = "{A Dynamo Model of Magnetic Activity in Solar-like Stars with Different Rotational Velocities}",
      journal = {\apj},
         year = 2014,
        month = aug,
       volume = {791},
       number = {1},
          eid = {59},
        pages = {59},
          doi = {10.1088/0004-637X/791/1/59},
archivePrefix = {arXiv},
       eprint = {1402.1874},
 primaryClass = {astro-ph.SR},
       adsurl = {https://ui.adsabs.harvard.edu/abs/2014ApJ...791...59K}
}

@ARTICLE{Vashishth+23,
       author = {{Vashishth}, Vindya and {Karak}, Bidya Binay and {Kitchatinov}, Leonid},
        title = "{Dynamo modelling for cycle variability and occurrence of grand minima in Sun-like stars: rotation rate dependence}",
      journal = {\mnras},
         year = 2023,
        month = jun,
       volume = {522},
       number = {2},
        pages = {2601-2610},
          doi = {10.1093/mnras/stad1105},
archivePrefix = {arXiv},
       eprint = {2304.05819},
 primaryClass = {astro-ph.SR},
       adsurl = {https://ui.adsabs.harvard.edu/abs/2023MNRAS.522.2601V}
}

@ARTICLE{Vashishth+26,
       author = {{Vashishth}, Vindya and {Karak}, Bidya Binay},
        title = "{Does the Babcock--Leighton dynamo operate in rapidly rotating solar-type stars? Exploration using a 3D dynamo model at different rotation rates}",
      journal = {\mnras},
         year = 2026,
        month = jan,
       volume = {545},
       number = {3},
          eid = {staf2214},
        pages = {staf2214},
          doi = {10.1093/mnras/staf2214},
archivePrefix = {arXiv},
       eprint = {2512.10508},
 primaryClass = {astro-ph.SR},
       adsurl = {https://ui.adsabs.harvard.edu/abs/2026MNRAS.545f2214V}
}

@ARTICLE{Garg+25,
       author = {{Garg}, Suyog and {Mandrai}, Rohan B. and {Karak}, Bidya Binay},
        title = "{Stellar Cycle Variability in Mount Wilson Stars and Dynamo Models: Rotation Rate and Dynamo Number Dependency}",
      journal = {\apj},
         year = 2025,
        month = dec,
       volume = {995},
       number = {2},
          eid = {194},
        pages = {194},
          doi = {10.3847/1538-4357/ae1fdd},
archivePrefix = {arXiv},
       eprint = {2511.08481},
 primaryClass = {astro-ph.SR},
       adsurl = {https://ui.adsabs.harvard.edu/abs/2025ApJ...995..194G}
}

@ARTICLE{Castenmiller86,
       author = {{Castenmiller}, M.~J.~M. and {Zwaan}, C. and {van der Zalm}, E.~B.~J.},
        title = "{Sunspot Nests - Manifestations of Sequences in Magnetic Activity}",
      journal = {\solphys},
         year = 1986,
        month = jun,
       volume = {105},
       number = {2},
        pages = {237-255},
          doi = {10.1007/BF00172045},
       adsurl = {https://ui.adsabs.harvard.edu/abs/1986SoPh..105..237C}
}

@ARTICLE{Brouwer90,
       author = {{Brouwer}, M.~P. and {Zwaan}, C.},
        title = "{Sunspot Nests as Traced by a Cluster Analysis}",
      journal = {\solphys},
         year = 1990,
        month = oct,
       volume = {129},
       number = {2},
        pages = {221-246},
          doi = {10.1007/BF00159038},
       adsurl = {https://ui.adsabs.harvard.edu/abs/1990SoPh..129..221B}
}

@ARTICLE{Csaszar25,
       author = {{Cs{\'a}sz{\'a}r}, K. and {Kors{\'o}s}, M.~B. and {So{\'o}s}, Sz. and {Erd{\'e}lyi}, R.},
        title = "{Exploring spatial and temporal patterns across solar cycles: Focusing on active longitudes}",
      journal = {\aap},
         year = 2026,
        month = jan,
       volume = {},
        pages = {},
          doi = {10.1051/0004-6361/202556599},
       adsurl = {}
}

@ARTICLE{Karapinar26,
       author = {{Karap{\i}nar}, Nurdan and {I{\textcommabelow s}{\i}k}, Emre and {Krivova}, Natalie A. and {{\c{S}}enavc{\i}}, Hakan V.},
        title = "{Quantifying Sunspot Group Nesting with Density-Based Unsupervised Clustering}",
      journal = {\solphys},
         year = 2026,
        month = feb,
       volume = {301},
       number = {3},
          eid = {34},
        pages = {34},
          doi = {10.1007/s11207-026-02632-2},
archivePrefix = {arXiv},
       eprint = {2512.17364},
 primaryClass = {astro-ph.SR},
       adsurl = {https://ui.adsabs.harvard.edu/abs/2026SoPh..301...34K}
}

@ARTICLE{Usoskin07,
       author = {{Usoskin}, I.~G. and {Berdyugina}, S.~V. and {Moss}, D. and {Sokoloff}, D.~D.},
        title = "{Long-term persistence of solar active longitudes and its implications for the solar dynamo theory}",
      journal = {Advances in Space Research},
         year = 2007,
        month = jan,
       volume = {40},
       number = {7},
        pages = {951-958},
          doi = {10.1016/j.asr.2006.12.050},
       adsurl = {https://ui.adsabs.harvard.edu/abs/2007AdSpR..40..951U}
}

@ARTICLE{Raphaldini23,
       author = {{Raphaldini}, Breno and {Dikpati}, Mausumi and {McIntosh}, Scott W.},
        title = "{Information-theoretic Analysis of Longitude Distribution of Photospheric Magnetic Fields from MDI/HMI Synoptic Maps: Evidence for Rossby Waves}",
      journal = {\apj},
         year = 2023,
        month = aug,
       volume = {953},
       number = {2},
          eid = {156},
        pages = {156},
          doi = {10.3847/1538-4357/ace320},
       adsurl = {https://ui.adsabs.harvard.edu/abs/2023ApJ...953..156R}
}

@ARTICLE{Finley25,
       author = {{Finley}, A.~J. and {Brun}, A.~S. and {Strugarek}, A. and {Perri}, B.},
        title = "{A prolific solar flare factory: Nearly continuous monitoring of an active region nest with Solar Orbiter}",
      journal = {\aap},
         year = 2025,
        month = may,
       volume = {697},
          eid = {A217},
        pages = {A217},
          doi = {10.1051/0004-6361/202554323},
archivePrefix = {arXiv},
       eprint = {2504.06345},
 primaryClass = {astro-ph.SR},
       adsurl = {https://ui.adsabs.harvard.edu/abs/2025A&A...697A.217F}
}

@ARTICLE{Dhakal+24,
       author = {{Dhakal}, Suman K. and {Zhang}, Jie},
        title = "{What Are the Causes of Super Activity of Solar Active Regions?}",
      journal = {\apj},
         year = 2024,
        month = jan,
       volume = {960},
       number = {1},
          eid = {36},
        pages = {36},
          doi = {10.3847/1538-4357/ad07d2},
archivePrefix = {arXiv},
       eprint = {2312.15083},
 primaryClass = {astro-ph.SR},
       adsurl = {https://ui.adsabs.harvard.edu/abs/2024ApJ...960...36D}
}

@ARTICLE{Krivova+26,
       author = {{Krivova}, N.~A. and {Chatzistergos}, T. and {Kazachenko}, M. and {Isik}, E.},
        title = "{Empirical flare energy limits for the largest historical sunspots}",
      journal = {Phil. Trans. Roy. Soc., in press; arXiv e-prints},
         year = 2026,
        month = mar,
          eid = {arXiv:2603.09474},
        pages = {arXiv:2603.09474},
          doi = {10.48550/arXiv.2603.09474},
archivePrefix = {arXiv},
       eprint = {2603.09474},
 primaryClass = {astro-ph.SR},
       adsurl = {https://ui.adsabs.harvard.edu/abs/2026arXiv260309474K}
}

@ARTICLE{Kazachenko+17,
       author = {{Kazachenko}, Maria D. and {Lynch}, Benjamin J. and {Welsch}, Brian T. and {Sun}, Xudong},
        title = "{A Database of Flare Ribbon Properties from the Solar Dynamics Observatory. I. Reconnection Flux}",
      journal = {\apj},
         year = 2017,
        month = aug,
       volume = {845},
       number = {1},
          eid = {49},
        pages = {49},
          doi = {10.3847/1538-4357/aa7ed6},
archivePrefix = {arXiv},
       eprint = {1704.05097},
 primaryClass = {astro-ph.SR},
       adsurl = {https://ui.adsabs.harvard.edu/abs/2017ApJ...845...49K}
}

@ARTICLE{IbanezBustos19a,
       author = {{Iba{\~n}ez Bustos}, R.~V. and {Buccino}, A.~P. and {Flores}, M. and {Mauas}, P.~J.~D.},
        title = "{Ross 128 - GL 447. A possible activity cycle for a slow-rotating fully convective star}",
      journal = {\aap},
         year = 2019,
        month = aug,
       volume = {628},
          eid = {L1},
        pages = {L1},
          doi = {10.1051/0004-6361/201936030},
archivePrefix = {arXiv},
       eprint = {1907.05728},
 primaryClass = {astro-ph.SR},
       adsurl = {https://ui.adsabs.harvard.edu/abs/2019A&A...628L...1I}
}

@ARTICLE{Isik20,
       author = {{I{\c{s}}{\i}k}, Emre and {Shapiro}, Alexander I. and {Solanki}, Sami K. and {Krivova}, Natalie A.},
        title = "{Amplification of Brightness Variability by Active-region Nesting in Solar-like Stars}",
      journal = {\apjl},
         year = 2020,
        month = sep,
       volume = {901},
       number = {1},
          eid = {L12},
        pages = {L12},
          doi = {10.3847/2041-8213/abb409},
archivePrefix = {arXiv},
       eprint = {2009.00692},
 primaryClass = {astro-ph.SR},
       adsurl = {https://ui.adsabs.harvard.edu/abs/2020ApJ...901L..12I}
}

@ARTICLE{Ozavci18,
       author = {{{\"O}zavc{\i}}, I. and {{\c{S}}enavc{\i}}, H.~V. and {I{\c{s}}{\i}k}, E. and {Hussain}, G.~A.~J. and {O'Neal}, D. and {Y{\i}lmaz}, M. and {Selam}, S.~O.},
        title = "{Recurrent star-spot activity and differential rotation in KIC 11560447}",
      journal = {\mnras},
         year = 2018,
        month = mar,
       volume = {474},
       number = {4},
        pages = {5534-5548},
          doi = {10.1093/mnras/stx3053},
archivePrefix = {arXiv},
       eprint = {1711.08949},
 primaryClass = {astro-ph.SR},
       adsurl = {https://ui.adsabs.harvard.edu/abs/2018MNRAS.474.5534O}
}

@ARTICLE{Breton24,
       author = {{Breton}, S.~N. and {Lanza}, A.~F. and {Messina}, S.},
        title = "{Tracking active nests in solar-type pulsators: Ensemble starspot modelling of Kepler asteroseismic targets}",
      journal = {\aap},
         year = 2024,
        month = feb,
       volume = {682},
          eid = {A67},
        pages = {A67},
          doi = {10.1051/0004-6361/202348298},
archivePrefix = {arXiv},
       eprint = {2312.02811},
 primaryClass = {astro-ph.SR},
       adsurl = {https://ui.adsabs.harvard.edu/abs/2024A&A...682A..67B}
}

@ARTICLE{Isik18,
       author = {{I{\c{s}}{\i}k}, E. and {Solanki}, S.~K. and {Krivova}, N.~A. and {Shapiro}, A.~I.},
        title = "{Forward modelling of brightness variations in Sun-like stars. I. Emergence and surface transport of magnetic flux}",
      journal = {\aap},
         year = 2018,
        month = dec,
       volume = {620},
          eid = {A177},
        pages = {A177},
          doi = {10.1051/0004-6361/201833393},
archivePrefix = {arXiv},
       eprint = {1810.06728},
 primaryClass = {astro-ph.SR},
       adsurl = {https://ui.adsabs.harvard.edu/abs/2018A&A...620A.177I}
}

@ARTICLE{Isik23,
       author = {{I{\c{s}}{\i}k}, Emre and {van Saders}, Jennifer L. and {Reiners}, Ansgar and {Metcalfe}, Travis S.},
        title = "{Scaling and Evolution of Stellar Magnetic Activity}",
      journal = {\ssr},
         year = 2023,
        month = dec,
       volume = {219},
       number = {8},
          eid = {70},
        pages = {70},
          doi = {10.1007/s11214-023-01016-3},
archivePrefix = {arXiv},
       eprint = {2310.09515},
 primaryClass = {astro-ph.SR},
       adsurl = {https://ui.adsabs.harvard.edu/abs/2023SSRv..219...70I}
}

@BOOK{SchrijverZwaan00,
       author = {{Schrijver}, Carolus J. and {Zwaan}, Cornelis},
        title = "{Solar and Stellar Magnetic Activity}",
         year = 2000,
    publisher = "{Cambridge University Press}",
       adsurl = {https://ui.adsabs.harvard.edu/abs/2000ssma.book.....S}
}

@ARTICLE{Schrijver1992,
       author = {{Schrijver}, C.~J. and {Dobson}, A.~K. and {Radick}, R.~R.},
        title = "{Nearly simultaneous observations of chromospheric and coronal radiative losses of cool stars.}",
      journal = {\aap},
         year = 1992,
        month = may,
       volume = {258},
        pages = {432-448},
       adsurl = {https://ui.adsabs.harvard.edu/abs/1992A&A...258..432S}
}

@INPROCEEDINGS{Saar96,
       author = {{Saar}, S.~H.},
        title = "{Recent magnetic fields measurements of stellar magnetic fields}",
    booktitle = {Stellar Surface Structure},
         year = 1996,
       editor = {{Strassmeier}, Klaus G. and {Linsky}, Jeffrey L.},
       series = {IAU Symposium},
       volume = {176},
        month = jan,
        pages = {237},
       adsurl = {https://ui.adsabs.harvard.edu/abs/1996IAUS..176..237S}
}

@ARTICLE{Kochukhov20,
       author = {{Kochukhov}, O. and {Hackman}, T. and {Lehtinen}, J.~J. and {Wehrhahn}, A.},
        title = "{Hidden magnetic fields of young suns}",
      journal = {\aap},
         year = 2020,
        month = mar,
       volume = {635},
          eid = {A142},
        pages = {A142},
          doi = {10.1051/0004-6361/201937185},
archivePrefix = {arXiv},
       eprint = {2002.10469},
 primaryClass = {astro-ph.SR},
       adsurl = {https://ui.adsabs.harvard.edu/abs/2020A&A...635A.142K}
}

@ARTICLE{Reiners22,
       author = {{Reiners}, A. and {Shulyak}, D. and {K{\"a}pyl{\"a}}, P.~J. and {Ribas}, I. and {Nagel}, E. and {Zechmeister}, M. and {Caballero}, J.~A. and {Shan}, Y. and {Fuhrmeister}, B. and {Quirrenbach}, A. and {Amado}, P.~J. and {Montes}, D. and {Jeffers}, S.~V. and {Azzaro}, M. and {B{\'e}jar}, V.~J.~S. and {Chaturvedi}, P. and {Henning}, Th. and {K{\"u}rster}, M. and {Pall{\'e}}, E.},
        title = "{Magnetism, rotation, and nonthermal emission in cool stars. Average magnetic field measurements in 292 M dwarfs}",
      journal = {\aap},
         year = 2022,
        month = jun,
       volume = {662},
          eid = {A41},
        pages = {A41},
          doi = {10.1051/0004-6361/202243251},
archivePrefix = {arXiv},
       eprint = {2204.00342},
 primaryClass = {astro-ph.SR},
       adsurl = {https://ui.adsabs.harvard.edu/abs/2022A&A...662A..41R}
}

@ARTICLE{Chen25,
       author = {{Chen}, Yue-Hong and {Alvarado-G{\'o}mez}, Juli{\'a}n D. and {Cheng}, Xin and {Dai}, Yu and {Shi}, Tong and {Poppenh{\"a}ger}, Katja and {Xing}, Chen and {Inoue}, Shun and {Warnecke}, J{\"o}rn and {Korpi-Lagg}, Maarit J. and {Ding}, Mingde},
        title = "{High-Resolution Modeling of Coronae and Winds in Solar-Type Stars with Varying Rotation Rates. I. X-Ray Coronae}",
      journal = {\apj},
         year = 2025,
        month = dec,
       volume = {995},
       number = {1},
          eid = {83},
        pages = {83},
          doi = {10.3847/1538-4357/ae1697},
archivePrefix = {arXiv},
       eprint = {2510.12969},
 primaryClass = {astro-ph.SR},
       adsurl = {https://ui.adsabs.harvard.edu/abs/2025ApJ...995...83C}
}

@ARTICLE{Viviani18,
       author = {{Viviani}, M. and {Warnecke}, J. and {K{\"a}pyl{\"a}}, M.~J. and {K{\"a}pyl{\"a}}, P.~J. and {Olspert}, N. and {Cole-Kodikara}, E.~M. and {Lehtinen}, J.~J. and {Brandenburg}, A.},
        title = "{Transition from axi- to nonaxisymmetric dynamo modes in spherical convection models of solar-like stars}",
      journal = {\aap},
         year = 2018,
        month = aug,
       volume = {616},
          eid = {A160},
        pages = {A160},
          doi = {10.1051/0004-6361/201732191},
archivePrefix = {arXiv},
       eprint = {1710.10222},
 primaryClass = {astro-ph.SR},
       adsurl = {https://ui.adsabs.harvard.edu/abs/2018A&A...616A.160V}
}

@ARTICLE{Karoff18,
       author = {{Karoff}, Christoffer and {Metcalfe}, Travis S. and {Santos}, {\^A}ngela R.~G. and {Montet}, Benjamin T. and {Isaacson}, Howard and {Witzke}, Veronika and {Shapiro}, Alexander I. and {Mathur}, Savita and {Davies}, Guy R. and {Lund}, Mikkel N. and {Garcia}, Rafael A. and {Brun}, Allan S. and {Salabert}, David and {Avelino}, Pedro P. and {van Saders}, Jennifer and {Egeland}, Ricky and {Cunha}, Margarida S. and {Campante}, Tiago L. and {Chaplin}, William J. and {Krivova}, Natalie and {Solanki}, Sami K. and {Stritzinger}, Maximilian and {Knudsen}, Mads F.},
        title = "{The Influence of Metallicity on Stellar Differential Rotation and Magnetic Activity}",
      journal = {\apj},
         year = 2018,
        month = jan,
       volume = {852},
       number = {1},
          eid = {46},
        pages = {46},
          doi = {10.3847/1538-4357/aaa026},
archivePrefix = {arXiv},
       eprint = {1711.07716},
 primaryClass = {astro-ph.SR},
       adsurl = {https://ui.adsabs.harvard.edu/abs/2018ApJ...852...46K}
}

@ARTICLE{See21,
       author = {{See}, Victor and {Roquette}, Julia and {Amard}, Louis and {Matt}, Sean P.},
        title = "{Photometric Variability as a Proxy for Magnetic Activity and Its Dependence on Metallicity}",
      journal = {\apj},
         year = 2021,
        month = may,
       volume = {912},
       number = {2},
          eid = {127},
        pages = {127},
          doi = {10.3847/1538-4357/abed47},
archivePrefix = {arXiv},
       eprint = {2103.05675},
 primaryClass = {astro-ph.SR},
       adsurl = {https://ui.adsabs.harvard.edu/abs/2021ApJ...912..127S}
}

@ARTICLE{See23,
       author = {{See}, Victor and {Roquette}, Julia and {Amard}, Louis and {Matt}, Sean},
        title = "{Further evidence of the link between activity and metallicity using the flaring properties of stars in the Kepler field}",
      journal = {\mnras},
         year = 2023,
        month = oct,
       volume = {524},
       number = {4},
        pages = {5781-5786},
          doi = {10.1093/mnras/stad2020},
archivePrefix = {arXiv},
       eprint = {2307.01688},
 primaryClass = {astro-ph.SR},
       adsurl = {https://ui.adsabs.harvard.edu/abs/2023MNRAS.524.5781S}
}

@ARTICLE{See24,
       author = {{See}, Victor and {Lu}, Yuxi (Lucy) and {Amard}, Louis and {Roquette}, Julia},
        title = "{The impact of stellar metallicity on rotation and activity evolution in the Kepler field using gyro-kinematic ages}",
      journal = {\mnras},
         year = 2024,
        month = sep,
       volume = {533},
       number = {2},
        pages = {1290-1299},
          doi = {10.1093/mnras/stae1828},
archivePrefix = {arXiv},
       eprint = {2405.00779},
 primaryClass = {astro-ph.SR},
       adsurl = {https://ui.adsabs.harvard.edu/abs/2024MNRAS.533.1290S}
}

@ARTICLE{Pezzotti25,
       author = {{Pezzotti}, C. and {B{\'e}trisey}, J. and {Buldgen}, G. and {Gilfanov}, M. and {Bikmaev}, I. and {Sunyaev}, R. and {I{\textcommabelow s}{\i}k}, E. and {Gosset}, E. and {Wright}, N.~J.},
        title = "{The stellar activity-rotation-age relationship under the lens of asteroseismology}",
      journal = {\aap},
         year = 2026,
        month = feb,
       volume = {706},
          eid = {A257},
        pages = {A257},
          doi = {10.1051/0004-6361/202557390},
archivePrefix = {arXiv},
       eprint = {2512.14517},
 primaryClass = {astro-ph.SR},
       adsurl = {https://ui.adsabs.harvard.edu/abs/2026A&A...706A.257P}
}

@ARTICLE{Isik26,
       author = {{I\c{s}{\i}k}, Emre and {Solanki}, Sami K. and {Krivova}, Natalie A. and {Shapiro}, Alexander I.},
        title = "{The rotation-magnetism relationship in solar-type stars. Constraining magnetic flux emergence rates}",
      journal = {\aap},
         year = 2026,
        month = feb,
       volume = {711},
          eid = {A48},
        pages = {A48},
          doi = {10.1051/0004-6361/202558620},
archivePrefix = {arXiv},
       eprint = {2512.18095},
 primaryClass = {astro-ph.SR},
       adsurl = {https://ui.adsabs.harvard.edu/abs/2025arXiv251218095I}
}

@ARTICLE{Witzke23,
       author = {{Witzke}, V. and {Duehnen}, H.~B. and {Shapiro}, A.~I. and {Przybylski}, D. and {Bhatia}, T.~S. and {Cameron}, R. and {Solanki}, S.~K.},
        title = "{Small-scale dynamo in cool stars. II. The effect of metallicity}",
      journal = {\aap},
         year = 2023,
        month = jan,
       volume = {669},
          eid = {A157},
        pages = {A157},
          doi = {10.1051/0004-6361/202244771},
archivePrefix = {arXiv},
       eprint = {2211.02722},
 primaryClass = {astro-ph.SR},
       adsurl = {https://ui.adsabs.harvard.edu/abs/2023A&A...669A.157W}
}

@ARTICLE{Metcalfe25,
       author = {{Metcalfe}, Travis S. and {van Saders}, Jennifer L. and {Pinsonneault}, Marc H. and {Ayres}, Thomas R. and {Kochukhov}, Oleg and {Stassun}, Keivan G. and {Finley}, Adam J. and {See}, Victor and {Ilyin}, Ilya V. and {Strassmeier}, Klaus G.},
        title = "{Weakened Magnetic Braking Signals the Collapse of the Global Stellar Dynamo}",
      journal = {\apjl},
         year = 2025,
        month = sep,
       volume = {991},
       number = {1},
          eid = {L17},
        pages = {L17},
          doi = {10.3847/2041-8213/ae03bc},
archivePrefix = {arXiv},
       eprint = {2509.03717},
 primaryClass = {astro-ph.SR},
       adsurl = {https://ui.adsabs.harvard.edu/abs/2025ApJ...991L..17M}
}

@ARTICLE{Cameron17,
       author = {{Cameron}, R.~H. and {Sch{\"u}ssler}, M.},
        title = "{Understanding Solar Cycle Variability}",
      journal = {\apj},
         year = 2017,
        month = jul,
       volume = {843},
       number = {2},
          eid = {111},
        pages = {111},
          doi = {10.3847/1538-4357/aa767a},
archivePrefix = {arXiv},
       eprint = {1705.10746},
 primaryClass = {astro-ph.SR},
       adsurl = {https://ui.adsabs.harvard.edu/abs/2017ApJ...843..111C}
}

@ARTICLE{Vidotto16,
       author = {{Vidotto}, A.~A.},
        title = "{The magnetic field vector of the Sun-as-a-star}",
      journal = {\mnras},
         year = 2016,
        month = jun,
       volume = {459},
       number = {2},
        pages = {1533-1542},
          doi = {10.1093/mnras/stw758},
archivePrefix = {arXiv},
       eprint = {1603.09226},
 primaryClass = {astro-ph.SR},
       adsurl = {https://ui.adsabs.harvard.edu/abs/2016MNRAS.459.1533V}
}

@ARTICLE{See19,
       author = {{See}, Victor and {Matt}, Sean P. and {Folsom}, Colin P. and {Boro Saikia}, Sudeshna and {Donati}, Jean-Francois and {Fares}, Rim and {Finley}, Adam J. and {H{\'e}brard}, {\'E}lodie M. and {Jardine}, Moira M. and {Jeffers}, Sandra V. and {Lehmann}, Lisa T. and {Marsden}, Stephen C. and {Mengel}, Matthew W. and {Morin}, Julien and {Petit}, Pascal and {Vidotto}, Aline A. and {Waite}, Ian A. and {BCool Collaboration}},
        title = "{Estimating Magnetic Filling Factors from Zeeman-Doppler Magnetograms}",
      journal = {\apj},
         year = 2019,
        month = may,
       volume = {876},
       number = {2},
          eid = {118},
        pages = {118},
          doi = {10.3847/1538-4357/ab1096},
archivePrefix = {arXiv},
       eprint = {1903.05595},
 primaryClass = {astro-ph.SR},
       adsurl = {https://ui.adsabs.harvard.edu/abs/2019ApJ...876..118S}
}

@ARTICLE{Isik11,
       author = {{I{\c{s}}{\i}k}, E. and {Schmitt}, D. and {Sch{\"u}ssler}, M.},
        title = "{Magnetic flux generation and transport in cool stars}",
      journal = {\aap},
         year = 2011,
        month = apr,
       volume = {528},
          eid = {A135},
        pages = {A135},
          doi = {10.1051/0004-6361/201014501},
archivePrefix = {arXiv},
       eprint = {1102.0569},
 primaryClass = {astro-ph.SR},
       adsurl = {https://ui.adsabs.harvard.edu/abs/2011A&A...528A.135I}
}

@ARTICLE{Radick+18,
       author = {{Radick}, Richard R. and {Lockwood}, G.~W. and {Skiff}, B.~A. and {Baliunas}, S.~L.},
        title = "{Patterns of Variation among Sun-like Stars}",
      journal = {\apjs},
         year = 1998,
        month = sep,
       volume = {118},
       number = {1},
        pages = {239-258},
          doi = {10.1086/313135},
       adsurl = {https://ui.adsabs.harvard.edu/abs/1998ApJS..118..239R}
}

@ARTICLE{Nemec+22,
       author = {{N{\`e}mec}, N. -E. and {Shapiro}, A.~I. and {I{\c{s}}{\i}k}, E. and {Sowmya}, K. and {Solanki}, S.~K. and {Krivova}, N.~A. and {Cameron}, R.~H. and {Gizon}, L.},
        title = "{Faculae Cancel out on the Surfaces of Active Suns}",
      journal = {\apjl},
         year = 2022,
        month = aug,
       volume = {934},
       number = {2},
          eid = {L23},
        pages = {L23},
          doi = {10.3847/2041-8213/ac8155},
archivePrefix = {arXiv},
       eprint = {2207.06816},
 primaryClass = {astro-ph.SR},
       adsurl = {https://ui.adsabs.harvard.edu/abs/2022ApJ...934L..23N}
}

@ARTICLE{Nemec+23,
       author = {{N{\`e}mec}, N. -E. and {Shapiro}, A.~I. and {I{\c{s}}{\i}k}, E. and {Solanki}, S.~K. and {Reinhold}, T.},
        title = "{Forward modelling of brightness variations in Sun-like stars. II. Light curves and variability}",
      journal = {\aap},
         year = 2023,
        month = apr,
       volume = {672},
          eid = {A138},
        pages = {A138},
          doi = {10.1051/0004-6361/202244412},
archivePrefix = {arXiv},
       eprint = {2303.03040},
 primaryClass = {astro-ph.SR},
       adsurl = {https://ui.adsabs.harvard.edu/abs/2023A&A...672A.138N}
}

@ARTICLE{Lanza+08,
       author = {{Lanza}, A.~F. and {De Martino}, C. and {Rodon{\`o}}, M.},
        title = "{Astrometric effects of solar-like magnetic activity in late-type stars and their relevance for the detection of extrasolar planets}",
      journal = {New Astronomy},
         year = 2008,
        month = feb,
       volume = {13},
       number = {2},
        pages = {77-84},
          doi = {10.1016/j.newast.2007.06.009},
archivePrefix = {arXiv},
       eprint = {0706.2942},
 primaryClass = {astro-ph},
       adsurl = {https://ui.adsabs.harvard.edu/abs/2008NewA...13...77L}
}

@ARTICLE{Lagrange+11,
       author = {{Lagrange}, A.-M. and {Meunier}, N. and {Desort}, M. and {Malbet}, F.},
        title = "{Using the Sun to estimate Earth-like planets detection capabilities . III. Impact of spots and plages on astrometric detection}",
      journal = {\aap},
         year = 2011,
        month = apr,
       volume = {528},
          eid = {L9},
        pages = {L9},
          doi = {10.1051/0004-6361/201016354},
archivePrefix = {arXiv},
       eprint = {1101.2512},
 primaryClass = {astro-ph.SR},
       adsurl = {https://ui.adsabs.harvard.edu/abs/2011A&A...528L...9L}
}

@ARTICLE{Morris+18,
       author = {{Morris}, Brett M. and {Agol}, Eric and {Davenport}, James R.~A. and {Hawley}, Suzanne L.},
        title = "{Spotting stellar activity cycles in Gaia astrometry}",
      journal = {\mnras},
         year = 2018,
        month = jun,
       volume = {476},
       number = {4},
        pages = {5408-5416},
          doi = {10.1093/mnras/sty568},
archivePrefix = {arXiv},
       eprint = {1802.09943},
 primaryClass = {astro-ph.SR},
       adsurl = {https://ui.adsabs.harvard.edu/abs/2018MNRAS.476.5408M}
}

@ARTICLE{Shapiro+14,
       author = {{Shapiro}, A.~I. and {Solanki}, S.~K. and {Krivova}, N.~A. and {Schmutz}, W.~K. and {Ball}, W.~T. and {Knaack}, R. and {Rozanov}, E.~V. and {Unruh}, Y.~C.},
        title = "{Variability of Sun-like stars: reproducing observed photometric trends}",
      journal = {\aap},
         year = 2014,
        month = sep,
       volume = {569},
          eid = {A38},
        pages = {A38},
          doi = {10.1051/0004-6361/201323086},
archivePrefix = {arXiv},
       eprint = {1406.2383},
 primaryClass = {astro-ph.SR},
       adsurl = {https://ui.adsabs.harvard.edu/abs/2014A&A...569A..38S}
}

@ARTICLE{Shapiro+21,
       author = {{Shapiro}, Alexander I. and {Solanki}, Sami K. and {Krivova}, Natalie A.},
        title = "{Predictions of Astrometric Jitter for Sun-like Stars. I. The Model and Its Application to the Sun as Seen from the Ecliptic}",
      journal = {\apj},
         year = 2021,
        month = feb,
       volume = {908},
       number = {2},
          eid = {223},
        pages = {223},
          doi = {10.3847/1538-4357/abd630},
archivePrefix = {arXiv},
       eprint = {2012.12312},
 primaryClass = {astro-ph.SR},
       adsurl = {https://ui.adsabs.harvard.edu/abs/2021ApJ...908..223S}
}

@ARTICLE{Sowmya21,
       author = {{Sowmya}, K. and {N{\`e}mec}, N. -E. and {Shapiro}, A.~I. and {I{\c{s}}{\i}k}, E. and {Witzke}, V. and {Mints}, A. and {Krivova}, N.~A. and {Solanki}, S.~K.},
        title = "{Predictions of Astrometric Jitter for Sun-like Stars. II. Dependence on Inclination, Metallicity, and Active-region Nesting}",
      journal = {\apj},
         year = 2021,
        month = oct,
       volume = {919},
       number = {2},
          eid = {94},
        pages = {94},
          doi = {10.3847/1538-4357/ac111b},
archivePrefix = {arXiv},
       eprint = {2107.01493},
 primaryClass = {astro-ph.SR},
       adsurl = {https://ui.adsabs.harvard.edu/abs/2021ApJ...919...94S}
}

@ARTICLE{Sowmya+22,
       author = {{Sowmya}, K. and {N{\`e}mec}, N. -E. and {Shapiro}, A.~I. and {I{\c{s}}{\i}k}, E. and {Krivova}, N.~A. and {Solanki}, S.~K.},
        title = "{Predictions of Astrometric Jitter for Sun-like Stars. III. Fast Rotators}",
      journal = {\apj},
         year = 2022,
        month = aug,
       volume = {934},
       number = {2},
          eid = {146},
        pages = {146},
          doi = {10.3847/1538-4357/ac79b3},
archivePrefix = {arXiv},
       eprint = {2206.07702},
 primaryClass = {astro-ph.SR},
       adsurl = {https://ui.adsabs.harvard.edu/abs/2022ApJ...934..146S}
}

@ARTICLE{Meunier+22_detection,
       author = {{Meunier}, N. and {Lagrange}, A.-M.},
        title = "{A new estimation of astrometric exoplanet detection limits in the habitable zone around nearby stars}",
      journal = {\aap},
         year = 2022,
        month = mar,
       volume = {659},
          eid = {A104},
        pages = {A104},
          doi = {10.1051/0004-6361/202142702},
archivePrefix = {arXiv},
       eprint = {2202.06301},
 primaryClass = {astro-ph.EP},
       adsurl = {https://ui.adsabs.harvard.edu/abs/2022A&A...659A.104M}
}

@ARTICLE{KaplanLipkin+22,
       author = {{Kaplan-Lipkin}, Avi and {Macintosh}, Bruce and {Madurowicz}, Alexander and {Sowmya}, Krishnamurthy and {Shapiro}, Alexander and {Krivova}, Natalie and {Solanki}, Sami K.},
        title = "{Multiwavelength Mitigation of Stellar Activity in Astrometric Planet Detection}",
      journal = {\aj},
         year = 2022,
        month = may,
       volume = {163},
       number = {5},
          eid = {205},
        pages = {205},
          doi = {10.3847/1538-3881/ac56e0},
archivePrefix = {arXiv},
       eprint = {2112.06383},
 primaryClass = {astro-ph.EP},
       adsurl = {https://ui.adsabs.harvard.edu/abs/2022AJ....163..205K}
}

@ARTICLE{Bao+24,
       author = {{Bao}, Chunhui and {Ji}, Jianghui and {Tan}, Dongjie and {Chen}, Guo and {Huang}, Xiumin and {Wang}, Su and {Dong}, Yao},
        title = "{Closeby Habitable Exoplanet Survey (CHES). I. Astrometric Noise and Planetary Detection Efficiency Due to Stellar Spots and Faculae}",
      journal = {\aj},
         year = 2024,
        month = jun,
       volume = {167},
       number = {6},
          eid = {286},
        pages = {286},
          doi = {10.3847/1538-3881/ad4031},
archivePrefix = {arXiv},
       eprint = {2404.11210},
 primaryClass = {astro-ph.EP},
       adsurl = {https://ui.adsabs.harvard.edu/abs/2024AJ....167..286B}
}

@ARTICLE{Sozzetti+23,
       author = {{Sozzetti}, A. and {Giacobbe}, P. and {Lattanzi}, M.~G. and {Pinamonti}, M.},
        title = "{On the follow-up efforts of long-period transiting planet candidates detected with Gaia astrometry}",
      journal = {\mnras},
         year = 2023,
        month = apr,
       volume = {520},
       number = {2},
        pages = {1748-1756},
          doi = {10.1093/mnras/stad253},
archivePrefix = {arXiv},
       eprint = {2302.00420},
 primaryClass = {astro-ph.EP},
       adsurl = {https://ui.adsabs.harvard.edu/abs/2023MNRAS.520.1748S}
}

@ARTICLE{Deagan+26,
       author = {{Deagan}, C. and {Montet}, Benjamin T.},
        title = "{Inferring hemispheric asymmetries of stellar active regions through the information content of astrometric signals}",
      journal = {\mnras},
         year = 2026,
        month = may,
       volume = {548},
       number = {2},
          eid = {stag675},
        pages = {stag675},
          doi = {10.1093/mnras/stag675},
archivePrefix = {arXiv},
       eprint = {2601.11707},
 primaryClass = {astro-ph.SR},
       adsurl = {https://ui.adsabs.harvard.edu/abs/2026MNRAS.548ag675D}
}

@ARTICLE{Sowmya21b,
       author = {{Sowmya}, K. and {Shapiro}, A.~I. and {Witzke}, V. and {N{\`e}mec}, N.-E. and {Chatzistergos}, T. and {Yeo}, K.~L. and {Krivova}, N.~A. and {Solanki}, S.~K.},
        title = "{Modeling Stellar Ca II H and K Emission Variations. I. Effect of Inclination on the S-index}",
      journal = {\apj},
         year = 2021,
        month = jun,
       volume = {914},
       number = {1},
          eid = {21},
        pages = {21},
          doi = {10.3847/1538-4357/abf247},
archivePrefix = {arXiv},
       eprint = {2103.13893},
 primaryClass = {astro-ph.SR},
       adsurl = {https://ui.adsabs.harvard.edu/abs/2021ApJ...914...21S}
}

@ARTICLE{Sowmya23,
       author = {{Sowmya}, K. and {Shapiro}, A.~I. and {Rouppe van der Voort}, L.~H.~M. and {Krivova}, N.~A. and {Solanki}, S.~K.},
        title = "{Modeling Stellar Ca II H and K Emission Variations: Spot Contribution to the S-index}",
      journal = {\apjl},
         year = 2023,
        month = oct,
       volume = {956},
       number = {1},
          eid = {L10},
        pages = {L10},
          doi = {10.3847/2041-8213/acf92a},
archivePrefix = {arXiv},
       eprint = {2309.03690},
 primaryClass = {astro-ph.SR},
       adsurl = {https://ui.adsabs.harvard.edu/abs/2023ApJ...956L..10S}
}

@ARTICLE{Solanki13,
       author = {{Solanki}, Sami K. and {Krivova}, Natalie A. and {Haigh}, Joanna D.},
        title = "{Solar Irradiance Variability and Climate}",
      journal = {\araa},
         year = 2013,
        month = aug,
       volume = {51},
       number = {1},
        pages = {311-351},
          doi = {10.1146/annurev-astro-082812-141007},
archivePrefix = {arXiv},
       eprint = {1306.2770},
 primaryClass = {astro-ph.SR},
       adsurl = {https://ui.adsabs.harvard.edu/abs/2013ARA&A..51..311S}
}

@ARTICLE{Dumusque14,
       author = {{Dumusque}, X. and {Boisse}, I. and {Santos}, N.~C.},
        title = "{SOAP 2.0: A Tool to Estimate the Photometric and Radial Velocity Variations Induced by Stellar Spots and Plages}",
      journal = {\apj},
         year = 2014,
        month = dec,
       volume = {796},
       number = {2},
          eid = {132},
        pages = {132},
          doi = {10.1088/0004-637X/796/2/132},
archivePrefix = {arXiv},
       eprint = {1409.3594},
 primaryClass = {astro-ph.SR},
       adsurl = {https://ui.adsabs.harvard.edu/abs/2014ApJ...796..132D}
}

@ARTICLE{Meunier+20,
       author = {{Meunier}, N. and {Lagrange}, A. -M. and {Borgniet}, S.},
        title = "{Activity time series of old stars from late F to early K. V. Effect on exoplanet detectability with high-precision astrometry}",
      journal = {\aap},
         year = 2020,
        month = dec,
       volume = {644},
          eid = {A77},
        pages = {A77},
          doi = {10.1051/0004-6361/202038710},
archivePrefix = {arXiv},
       eprint = {2011.02158},
 primaryClass = {astro-ph.SR},
       adsurl = {https://ui.adsabs.harvard.edu/abs/2020A&A...644A..77M}
}

@ARTICLE{Meunier+19a,
       author = {{Meunier}, N. and {Lagrange}, A. -M. and {Boulet}, T. and {Borgniet}, S.},
        title = "{Activity time series of old stars from late F to early K. I. Simulating radial velocity, astrometry, photometry, and chromospheric emission}",
      journal = {\aap},
         year = 2019,
        month = jul,
       volume = {627},
          eid = {A56},
        pages = {A56},
          doi = {10.1051/0004-6361/201834796},
archivePrefix = {arXiv},
       eprint = {1904.01437},
 primaryClass = {astro-ph.SR},
       adsurl = {https://ui.adsabs.harvard.edu/abs/2019A&A...627A..56M}
}

@ARTICLE{Meunier+19b,
       author = {{Meunier}, N. and {Lagrange}, A. -M.},
        title = "{Activity time series of old stars from late F to early K. II. Radial velocity jitter and exoplanet detectability}",
      journal = {\aap},
         year = 2019,
        month = aug,
       volume = {628},
          eid = {A125},
        pages = {A125},
          doi = {10.1051/0004-6361/201935347},
archivePrefix = {arXiv},
       eprint = {1909.02969},
 primaryClass = {astro-ph.SR},
       adsurl = {https://ui.adsabs.harvard.edu/abs/2019A&A...628A.125M}
}

@ARTICLE{Meunier+19d,
       author = {{Meunier}, N. and {Lagrange}, A.-M. and {Cuzacq}, S.},
        title = "{Activity time series of old stars from late F to early K. IV. Limits of the correction of radial velocities using chromospheric emission}",
      journal = {\aap},
         year = 2019,
        month = dec,
       volume = {632},
          eid = {A81},
        pages = {A81},
          doi = {10.1051/0004-6361/201935348},
archivePrefix = {arXiv},
       eprint = {1911.05319},
 primaryClass = {astro-ph.SR},
       adsurl = {https://ui.adsabs.harvard.edu/abs/2019A&A...632A..81M}
}

@ARTICLE{Meunier+22_CaII,
       author = {{Meunier}, N. and {Kretzschmar}, M. and {Gravet}, R. and {Mignon}, L. and {Delfosse}, X.},
        title = "{Relationship between Ca and H{\ensuremath{\alpha}} chromospheric emission in F-G-K stars: Indication of stellar filaments?}",
      journal = {\aap},
         year = 2022,
        month = feb,
       volume = {658},
          eid = {A57},
        pages = {A57},
          doi = {10.1051/0004-6361/202142120},
archivePrefix = {arXiv},
       eprint = {2201.05492},
 primaryClass = {astro-ph.SR},
       adsurl = {https://ui.adsabs.harvard.edu/abs/2022A&A...658A..57M}
}

@ARTICLE{Cretignier+24a,
       author = {{Cretignier}, M. and {Pietrow}, A.~G.~M. and {Aigrain}, S.},
        title = "{Stellar surface information from the Ca II H\&K lines - I. Intensity profiles of the solar activity components}",
      journal = {\mnras},
         year = 2024,
        month = jan,
       volume = {527},
       number = {2},
        pages = {2940-2962},
          doi = {10.1093/mnras/stad3292},
archivePrefix = {arXiv},
       eprint = {2310.15926},
 primaryClass = {astro-ph.SR},
       adsurl = {https://ui.adsabs.harvard.edu/abs/2024MNRAS.527.2940C}
}

@ARTICLE{Cretignier+24b,
       author = {{Cretignier}, M. and {Hara}, N.~C. and {Pietrow}, A.~G.~M. and {Zhao}, Y. and {Yu}, H. and {Dumusque}, X. and {Sozzetti}, A. and {Lovis}, C. and {Aigrain}, S.},
        title = "{Stellar surface information from the Ca II H\&K lines - II. Defining better activity proxies}",
      journal = {\mnras},
         year = 2024,
        month = dec,
       volume = {535},
       number = {3},
        pages = {2562-2584},
          doi = {10.1093/mnras/stae2508},
archivePrefix = {arXiv},
       eprint = {2411.00557},
 primaryClass = {astro-ph.SR},
       adsurl = {https://ui.adsabs.harvard.edu/abs/2024MNRAS.535.2562C}
}

@ARTICLE{Herrero16,
       author = {{Herrero}, Enrique and {Ribas}, Ignasi and {Jordi}, Carme and {Morales}, Juan Carlos and {Perger}, Manuel and {Rosich}, Albert},
        title = "{Modelling the photosphere of active stars for planet detection and characterization}",
      journal = {\aap},
         year = 2016,
        month = feb,
       volume = {586},
          eid = {A131},
        pages = {A131},
          doi = {10.1051/0004-6361/201425369},
archivePrefix = {arXiv},
       eprint = {1511.06717},
 primaryClass = {astro-ph.EP},
       adsurl = {https://ui.adsabs.harvard.edu/abs/2016A&A...586A.131H}
}

@ARTICLE{Rosich20,
       author = {{Rosich}, A. and {Herrero}, E. and {Mallonn}, M. and {Ribas}, I. and {Morales}, J.~C. and {Perger}, M. and {Anglada-Escud{\'e}}, G. and {Granzer}, T.},
        title = "{Correcting for chromatic stellar activity effects in transits with multiband photometric monitoring: application to WASP-52}",
      journal = {\aap},
         year = 2020,
        month = sep,
       volume = {641},
          eid = {A82},
        pages = {A82},
          doi = {10.1051/0004-6361/202037586},
archivePrefix = {arXiv},
       eprint = {2007.00573},
 primaryClass = {astro-ph.EP},
       adsurl = {https://ui.adsabs.harvard.edu/abs/2020A&A...641A..82R}
}

@ARTICLE{2010A&A...510A..46L,
       author = {{Landin}, N.~R. and {Mendes}, L.~T.~S. and {Vaz}, L.~P.~R.},
        title = "{Theoretical values of convective turnover times and Rossby numbers for solar-like, pre-main sequence stars}",
      journal = {\aap},
         year = 2010,
        month = feb,
       volume = {510},
          eid = {A46},
        pages = {A46},
          doi = {10.1051/0004-6361/200913015},
archivePrefix = {arXiv},
       eprint = {1001.2754},
 primaryClass = {astro-ph.SR},
       adsurl = {https://ui.adsabs.harvard.edu/abs/2010A&A...510A..46L}
}

@ARTICLE{2014ApJ...794..144R,
       author = {{Reiners}, A. and {Sch{\"u}ssler}, M. and {Passegger}, V.~M.},
        title = "{Generalized Investigation of the Rotation-Activity Relation: Favoring Rotation Period instead of Rossby Number}",
      journal = {\apj},
         year = 2014,
        month = oct,
       volume = {794},
       number = {2},
          eid = {144},
        pages = {144},
          doi = {10.1088/0004-637X/794/2/144},
archivePrefix = {arXiv},
       eprint = {1408.6175},
 primaryClass = {astro-ph.SR},
       adsurl = {https://ui.adsabs.harvard.edu/abs/2014ApJ...794..144R}
}

@ARTICLE{1996ApJ...457..340K,
       author = {{Kim}, Yong-Cheol and {Demarque}, Pierre},
        title = "{The Theoretical Calculation of the Rossby Number and the ``Nonlocal'' Convective Overturn Time for Pre-Main-Sequence and Early Post-Main-Sequence Stars}",
      journal = {\apj},
         year = 1996,
        month = jan,
       volume = {457},
        pages = {340},
          doi = {10.1086/176733},
       adsurl = {https://ui.adsabs.harvard.edu/abs/1996ApJ...457..340K}
}

@ARTICLE{1999A&A...346..883J,
       author = {{Jardine}, M. and {Unruh}, Y.~C.},
        title = "{Coronal emission and dynamo saturation}",
      journal = {\aap},
         year = 1999,
        month = jun,
       volume = {346},
        pages = {883-891},
       adsurl = {https://ui.adsabs.harvard.edu/abs/1999A&A...346..883J}
}

@ARTICLE{2001A&A...370..157S,
       author = {{St{\c{e}}pie{\'n}}, K. and {Schmitt}, J.~H.~M.~M. and {Voges}, W.},
        title = "{ROSAT all-sky survey of W Ursae Majoris stars and the problem of supersaturation}",
      journal = {\aap},
         year = 2001,
        month = apr,
       volume = {370},
   
pages = {157-169},
          doi = {10.1051/0004-6361:20010197},
       adsurl = {https://ui.adsabs.harvard.edu/abs/2001A&A...370..157S}
}

@ARTICLE{1994A&A...292..191S,
       author = {{Stepien}, K.},
        title = "{Applicability of the Rossby number in activity-rotation relations for dwarfs and giants.}",
      journal = {\aap},
         year = 1994,
        month = dec,
       volume = {292},
        pages = {191-207},
       adsurl = {https://ui.adsabs.harvard.edu/abs/1994A&A...292..191S}
}

@ARTICLE{1984ApJ...279..763N,
       author = {{Noyes}, R.~W. and {Hartmann}, L.~W. and {Baliunas}, S.~L. and {Duncan}, D.~K. and {Vaughan}, A.~H.},
        title = "{Rotation, convection, and magnetic activity in lower main-sequence stars.}",
      journal = {\apj},
         year = 1984,
        month = apr,
       volume = {279},
        pages = {763-777},
          doi = {10.1086/161945},
       adsurl = {https://ui.adsabs.harvard.edu/abs/1984ApJ...279..763N}
}

@ARTICLE{2004A&A...417..651S,
       author = {{Schmitt}, J.~H.~M.~M. and {Liefke}, C.},
        title = "{NEXXUS: A comprehensive ROSAT survey of coronal X-ray emission among nearby solar-like stars}",
      journal = {\aap},
         year = 2004,
        month = apr,
       volume = {417},
        pages = {651-665},
          doi = {10.1051/0004-6361:20030495},
archivePrefix = {arXiv},
       eprint = {astro-ph/0308510},
 primaryClass = {astro-ph},
       adsurl = {https://ui.adsabs.harvard.edu/abs/2004A&A...417..651S}
}

@ARTICLE{1997A&A...318..215S,
       author = {{Schmitt}, J.~H.~M.~M.},
        title = "{Coronae on solar-like stars.}",
      journal = {\aap},
         year = 1997,
        month = feb,
       volume = {318},
        pages = {215-230},
       adsurl = {https://ui.adsabs.harvard.edu/abs/1997A&A...318..215S}
}

@ARTICLE{2003A&A...397..147P,
       author = {{Pizzolato}, N. and {Maggio}, A. and {Micela}, G. and {Sciortino}, S. and {Ventura}, P.},
        title = "{The stellar activity-rotation relationship revisited: Dependence of saturated and non-saturated X-ray emission regimes on stellar mass for late-type dwarfs}",
      journal = {\aap},
         year = 2003,
        month = jan,
       volume = {397},
        pages = {147-157},
          doi = {10.1051/0004-6361:20021560},
       adsurl = {https://ui.adsabs.harvard.edu/abs/2003A&A...397..147P}
}

@ARTICLE{2018MNRAS.479.2351W,
       author = {{Wright}, Nicholas J. and {Newton}, Elisabeth R. and {Williams}, Peter K.~G. and {Drake}, Jeremy J. and {Yadav}, Rakesh K.},
        title = "{The stellar rotation-activity relationship in fully convective M dwarfs}",
      journal = {\mnras},
         year = 2018,
        month = sep,
       volume = {479},
       number = {2},
        pages = {2351-2360},
          doi = {10.1093/mnras/sty1670},
archivePrefix = {arXiv},
       eprint = {1807.03304},
 primaryClass = {astro-ph.SR},
       adsurl = {https://ui.adsabs.harvard.edu/abs/2018MNRAS.479.2351W}
}

@ARTICLE{Yadav15,
       author = {{Yadav}, Rakesh K. and {Christensen}, Ulrich R. and {Morin}, Julien and {Gastine}, Thomas and {Reiners}, Ansgar and {Poppenhaeger}, Katja and {Wolk}, Scott J.},
        title = "{Explaining the Coexistence of Large-scale and Small-scale Magnetic Fields in Fully Convective Stars}",
      journal = {\apjl},
         year = 2015,
        month = nov,
       volume = {813},
       number = {2},
          eid = {L31},
        pages = {L31},
          doi = {10.1088/2041-8205/813/2/L31},
archivePrefix = {arXiv},
       eprint = {1510.05541},
 primaryClass = {astro-ph.SR},
       adsurl = {https://ui.adsabs.harvard.edu/abs/2015ApJ...813L..31Y}
}

@ARTICLE{2017MNRAS.471.1012B,
       author = {{Booth}, R.~S. and {Poppenhaeger}, K. and {Watson}, C.~A. and {Silva Aguirre}, V. and {Wolk}, S.~J.},
        title = "{An improved age-activity relationship for cool stars older than a gigayear}",
      journal = {\mnras},
         year = 2017,
        month = oct,
       volume = {471},
       number = {1},
        pages = {1012-1025},
          doi = {10.1093/mnras/stx1630},
archivePrefix = {arXiv},
       eprint = {1706.08979},
 primaryClass = {astro-ph.SR},
       adsurl = {https://ui.adsabs.harvard.edu/abs/2017MNRAS.471.1012B}
}

@ARTICLE{2020A&A...638A..20M,
       author = {{Magaudda}, E. and {Stelzer}, B. and {Covey}, K.~R. and {Raetz}, St. and {Matt}, S.~P. and {Scholz}, A.},
        title = "{Relation of X-ray activity and rotation in M dwarfs and predicted time-evolution of the X-ray luminosity}",
      journal = {\aap},
         year = 2020,
        month = jun,
       volume = {638},
          eid = {A20},
        pages = {A20},
          doi = {10.1051/0004-6361/201937408},
archivePrefix = {arXiv},
       eprint = {2004.02904},
 primaryClass = {astro-ph.SR},
       adsurl = {https://ui.adsabs.harvard.edu/abs/2020A&A...638A..20M}
}

@ARTICLE{Fuhrmeister2023,
       author = {{Fuhrmeister}, B. and {Czesla}, S. and {Schmitt}, J.~H.~M.~M. and {Schneider}, P.~C. and {Caballero}, J.~A. and {Jeffers}, S.~V. and {Nagel}, E. and {Montes}, D. and {G{\'a}lvez Ortiz}, M.~C. and {Reiners}, A. and {Ribas}, I. and {Quirrenbach}, A. and {Amado}, P.~J. and {Henning}, Th. and {Lodieu}, N. and {Mart{\'\i}n-Fern{\'a}ndez}, P. and {Morales}, J.~C. and {Sch{\"o}fer}, P. and {Seifert}, W. and {Zechmeister}, M.},
        title = "{The CARMENES search for exoplanets around M dwarfs. Behaviour of the Paschen lines during flares and quiescence}",
      journal = {\aap},
         year = 2023,
        month = oct,
       volume = {678},
          eid = {A1},
        pages = {A1},
          doi = {10.1051/0004-6361/202347161},
archivePrefix = {arXiv},
       eprint = {2308.07685},
 primaryClass = {astro-ph.SR},
       adsurl = {https://ui.adsabs.harvard.edu/abs/2023A&A...678A...1F}
}

@ARTICLE{2011ApJ...743...48W,
       author = {{Wright}, Nicholas J. and {Drake}, Jeremy J. and {Mamajek}, Eric E. and {Henry}, Gregory W.},
        title = "{The Stellar-activity-Rotation Relationship and the Evolution of Stellar Dynamos}",
      journal = {\apj},
         year = 2011,
        month = dec,
       volume = {743},
       number = {1},
          eid = {48},
        pages = {48},
          doi = {10.1088/0004-637X/743/1/48},
archivePrefix = {arXiv},
       eprint = {1109.4634},
 primaryClass = {astro-ph.SR},
       adsurl = {https://ui.adsabs.harvard.edu/abs/2011ApJ...743...48W}
}

@ARTICLE{Blackman15,
       author = {{Blackman}, E.~G. and {Thomas}, J.~H.},
        title = "{Explaining the observed relation between stellar activity and rotation.}",
      journal = {\mnras},
         year = 2015,
        month = jan,
       volume = {446},
        pages = {L51-L55},
          doi = {10.1093/mnrasl/slu163},
archivePrefix = {arXiv},
       eprint = {1407.8500},
 primaryClass = {astro-ph.SR},
       adsurl = {https://ui.adsabs.harvard.edu/abs/2015MNRAS.446L..51B}
}

@ARTICLE{Weber2023,
       author = {{Weber}, Maria A. and {Schunker}, Hannah and {Jouve}, Laur{\`e}ne and {I{\c{s}}{\i}k}, Emre},
        title = "{Understanding Active Region Origins and Emergence on the Sun and Other Cool Stars}",
      journal = {\ssr},
         year = 2023,
        month = dec,
       volume = {219},
       number = {8},
          eid = {63},
        pages = {63},
          doi = {10.1007/s11214-023-01006-5},
archivePrefix = {arXiv},
       eprint = {2306.06536},
 primaryClass = {astro-ph.SR},
       adsurl = {https://ui.adsabs.harvard.edu/abs/2023SSRv..219...63W}
}

@ARTICLE{Yeates2023,
       author = {{Yeates}, Anthony R. and {Cheung}, Mark C.~M. and {Jiang}, Jie and {Petrovay}, Kristof and {Wang}, Yi-Ming},
        title = "{Surface Flux Transport on the Sun}",
      journal = {\ssr},
         year = 2023,
        month = jun,
       volume = {219},
       number = {4},
          eid = {31},
        pages = {31},
          doi = {10.1007/s11214-023-00978-8},
archivePrefix = {arXiv},
       eprint = {2303.01209},
 primaryClass = {astro-ph.SR},
       adsurl = {https://ui.adsabs.harvard.edu/abs/2023SSRv..219...31Y}
}

@ARTICLE{CharbonneauSokoloff2023,
       author = {{Charbonneau}, Paul and {Sokoloff}, Dmitry},
        title = "{Evolution of Solar and Stellar Dynamo Theory}",
      journal = {\ssr},
         year = 2023,
        month = aug,
       volume = {219},
       number = {5},
          eid = {35},
        pages = {35},
          doi = {10.1007/s11214-023-00980-0},
archivePrefix = {arXiv},
       eprint = {2305.16553},
 primaryClass = {astro-ph.SR},
       adsurl = {https://ui.adsabs.harvard.edu/abs/2023SSRv..219...35C}
}

@ARTICLE{Schwabe1844AN.....21..233S,
       author = {{Schwabe}, Heinrich},
        title = "{Sonnenbeobachtungen im Jahre 1843. Von Herrn Hofrath Schwabe in Dessau}",
      journal = {Astronomische Nachrichten},
         year = 1844,
        month = feb,
       volume = {21},
       number = {15},
        pages = {233},
          doi = {10.1002/asna.18440211505},
       adsurl = {https://ui.adsabs.harvard.edu/abs/1844AN.....21..233S}
}

@ARTICLE{Wolf1852AN.....35..369W,
       author = {{Wolf}, Rudolf},
        title = "{Bericht {\"u}ber neue Untersuchungen {\"u}ber die Periode der Sonnenflecken und ihrer Bedeutung von Herrn Prof. Wolf}",
      journal = {Astronomische Nachrichten},
         year = 1852,
        month = dec,
       volume = {35},
       number = {25},
        pages = {369},
          doi = {10.1002/asna.18530352504},
       adsurl = {https://ui.adsabs.harvard.edu/abs/1852AN.....35..369W}
}

@ARTICLE{Usoskin2012ApJ...757...92U,
       author = {{Usoskin}, Ilya G. and {Kovaltsov}, Gennady A.},
        title = "{Occurrence of Extreme Solar Particle Events: Assessment from Historical Proxy Data}",
      journal = {\apj},
         year = 2012,
        month = sep,
       volume = {757},
       number = {1},
          eid = {92},
        pages = {92},
          doi = {10.1088/0004-637X/757/1/92},
archivePrefix = {arXiv},
       eprint = {1207.5932},
 primaryClass = {astro-ph.SR},
       adsurl = {https://ui.adsabs.harvard.edu/abs/2012ApJ...757...92U}
}

@ARTICLE{Baliunas1995ApJ...438..269B,
       author = {{Baliunas}, S.~L. and {Donahue}, R.~A. and {Soon}, W.~H. and {Horne}, J.~H. and {Frazer}, J. and {Woodard-Eklund}, L. and {Bradford}, M. and {Rao}, L.~M. and {Wilson}, O.~C. and {Zhang}, Q. and {Bennett}, W. and {Briggs}, J. and {Carroll}, S.~M. and {Duncan}, D.~K. and {Figueroa}, D. and {Lanning}, H.~H. and {Misch}, T. and {Mueller}, J. and {Noyes}, R.~W. and {Poppe}, D. and {Porter}, A.~C. and {Robinson}, C.~R. and {Russell}, J. and {Shelton}, J.~C. and {Soyumer}, T. and {Vaughan}, A.~H. and {Whitney}, J.~H.},
        title = "{Chromospheric Variations in Main-Sequence Stars. II.}",
      journal = {\apj},
         year = 1995,
        month = jan,
       volume = {438},
        pages = {269},
          doi = {10.1086/175072},
       adsurl = {https://ui.adsabs.harvard.edu/abs/1995ApJ...438..269B}
}

@ARTICLE{Ossendrijver1997A&A...323..151O,
       author = {{Ossendrijver}, A.~J.~H.},
        title = "{On the cycle periods of stellar dynamos.}",
      journal = {\aap},
         year = 1997,
        month = jul,
       volume = {323},
        pages = {151-157},
       adsurl = {https://ui.adsabs.harvard.edu/abs/1997A&A...323..151O}
}

@ARTICLE{RodriguezLopez2019FrASS...6...76R,
       author = {{Rodr{\'\i}guez-L{\'o}pez}, Cristina},
        title = "{The quest for pulsating M dwarf stars}",
      journal = {Frontiers in Astronomy and Space Sciences},
         year = 2019,
        month = dec,
       volume = {6},
          eid = {76},
        pages = {76},
          doi = {10.3389/fspas.2019.00076},
       adsurl = {https://ui.adsabs.harvard.edu/abs/2019FrASS...6...76R}
}

@ARTICLE{Jeffers2023SSRv..219...54J,
       author = {{Jeffers}, Sandra V. and {Kiefer}, Ren{\'e} and {Metcalfe}, Travis S.},
        title = "{Stellar Activity Cycles}",
      journal = {\ssr},
         year = 2023,
        month = oct,
       volume = {219},
       number = {7},
          eid = {54},
        pages = {54},
          doi = {10.1007/s11214-023-01000-x},
archivePrefix = {arXiv},
       eprint = {2309.14138},
 primaryClass = {astro-ph.SR},
       adsurl = {https://ui.adsabs.harvard.edu/abs/2023SSRv..219...54J}
}

@ARTICLE{Rosen2015ApJ...805..169R,
       author = {{Ros{\'e}n}, L. and {Kochukhov}, O. and {Wade}, G.~A.},
        title = "{First Zeeman Doppler Imaging of a Cool Star Using all Four Stokes Parameters}",
      journal = {\apj},
         year = 2015,
        month = jun,
       volume = {805},
       number = {2},
          eid = {169},
        pages = {169},
          doi = {10.1088/0004-637X/805/2/169},
archivePrefix = {arXiv},
       eprint = {1504.00176},
 primaryClass = {astro-ph.SR},
       adsurl = {https://ui.adsabs.harvard.edu/abs/2015ApJ...805..169R}
}

@ARTICLE{Petit2004MNRAS.348.1175P,
       author = {{Petit}, P. and {Donati}, J. -F. and {Wade}, G.~A. and {Landstreet}, J.~D. and {Bagnulo}, S. and {L{\"u}ftinger}, T. and {Sigut}, T.~A.~A. and {Shorlin}, S.~L.~S. and {Strasser}, S. and {Auri{\`e}re}, M. and {Oliveira}, J.~M.},
        title = "{Magnetic topology and surface differential rotation on the K1 subgiant of the RS CVn system HR 1099}",
      journal = {\mnras},
         year = 2004,
        month = mar,
       volume = {348},
       number = {4},
        pages = {1175-1190},
          doi = {10.1111/j.1365-2966.2004.07420.x},
archivePrefix = {arXiv},
       eprint = {astro-ph/0312238},
 primaryClass = {astro-ph},
       adsurl = {https://ui.adsabs.harvard.edu/abs/2004MNRAS.348.1175P}
}

@ARTICLE{2009PASP..121..993B,
       author = {{Bagnulo}, S. and {Landolfi}, M. and {Landstreet}, J.~D. and {Landi Degl'Innocenti}, E. and {Fossati}, L. and {Sterzik}, M.},
        title = "{Stellar Spectropolarimetry with Retarder Waveplate and Beam Splitter Devices}",
      journal = {\pasp},
         year = 2009,
        month = sep,
       volume = {121},
       number = {883},
        pages = {993},
          doi = {10.1086/605654},
       adsurl = {https://ui.adsabs.harvard.edu/abs/2009PASP..121..993B}
}

@ARTICLE{2013A&A...553L...6S,
       author = {{Sanz-Forcada}, J. and {Stelzer}, B. and {Metcalfe}, T.~S.},
        title = "{{\ensuremath{\i}}Horologi, the first coronal activity cycle in a young solar-like star}",
      journal = {\aap},
         year = 2013,
        month = may,
       volume = {553},
          eid = {L6},
        pages = {L6},
          doi = {10.1051/0004-6361/201321388},
archivePrefix = {arXiv},
       eprint = {1305.1132},
 primaryClass = {astro-ph.SR},
       adsurl = {https://ui.adsabs.harvard.edu/abs/2013A&A...553L...6S}
}

@ARTICLE{2019A&A...631A..45S,
       author = {{Sanz-Forcada}, J. and {Stelzer}, B. and {Coffaro}, M. and {Raetz}, S. and {Alvarado-G{\'o}mez}, J.~D.},
        title = "{Multi-wavelength variability of the young solar analog {\ensuremath{\i}} Horologii. X-ray cycle, star spots, flares, and UV emission}",
      journal = {\aap},
         year = 2019,
        month = nov,
       volume = {631},
          eid = {A45},
        pages = {A45},
          doi = {10.1051/0004-6361/201935703},
archivePrefix = {arXiv},
       eprint = {1909.01320},
 primaryClass = {astro-ph.SR},
       adsurl = {https://ui.adsabs.harvard.edu/abs/2019A&A...631A..45S}
}

@ARTICLE{2018MNRAS.473.4326A,
       author = {{Alvarado-G{\'o}mez}, Juli{\'a}n D. and {Hussain}, Gaitee A.~J. and {Drake}, Jeremy J. and {Donati}, Jean-Fran{\c{c}}ois and {Sanz-Forcada}, Jorge and {Stelzer}, Beate and {Cohen}, Ofer and {Amazo-G{\'o}mez}, Eliana M. and {Grunhut}, Jason H. and {Garraffo}, Cecilia and et al.},
        title = "{Far beyond the Sun - I. The beating magnetic heart in Horologium}",
      journal = {\mnras},
         year = 2018,
        month = feb,
       volume = {473},
       number = {4},
        pages = {4326-4338},
          doi = {10.1093/mnras/stx2642},
archivePrefix = {arXiv},
       eprint = {1710.02438},
 primaryClass = {astro-ph.SR},
       adsurl = {https://ui.adsabs.harvard.edu/abs/2018MNRAS.473.4326A}
}

@ARTICLE{2023MNRAS.524.5725A,
       author = {{Amazo-G{\'o}mez}, E.~M. and {Alvarado-G{\'o}mez}, J.~D. and {Poppenh{\"a}ger}, K. and {Hussain}, G.~A.~J. and {Wood}, B.~E. and {Drake}, J.~J. and {do Nascimento}, J.-D. and {Anthony}, F. and {Sanz-Forcada}, J. and {Stelzer}, B. and et al.},
        title = "{Far beyond the Sun - II. Probing the stellar magnetism of the young Sun {\ensuremath{\i}} Horologii from the photosphere to its corona}",
      journal = {\mnras},
         year = 2023,
        month = oct,
       volume = {524},
       number = {4},
        pages = {5725-5748},
          doi = {10.1093/mnras/stad2086},
archivePrefix = {arXiv},
       eprint = {2307.01744},
 primaryClass = {astro-ph.SR},
       adsurl = {https://ui.adsabs.harvard.edu/abs/2023MNRAS.524.5725A}
}

@ARTICLE{2025A&A...704A..68A,
       author = {{Alvarado-G{\'o}mez}, J.~D. and {Hussain}, G.~A.~J. and {Amazo-G{\'o}mez}, E.~M. and {Xu}, Y. and {Poppenh{\"a}ger}, K. and {Chebly}, J. and {Donati}, J.-F. and {Stelzer}, B. and {Sanz-Forcada}, J.},
        title = "{Far beyond the Sun: III. The magnetic cycle of {\ensuremath{\i}} Horologii}",
      journal = {\aap},
         year = 2025,
        month = dec,
       volume = {704},
          eid = {A68},
        pages = {A68},
          doi = {10.1051/0004-6361/202555349},
archivePrefix = {arXiv},
       eprint = {2510.03146},
 primaryClass = {astro-ph.SR},
       adsurl = {https://ui.adsabs.harvard.edu/abs/2025A&A...704A..68A}
}

@ARTICLE{2015JATIS...1a4003R,
       author = {{Ricker}, George R. and {Winn}, Joshua N. and {Vanderspek}, Roland and {Latham}, David W. and {Bakos}, G{\'a}sp{\'a}r {\'A}. and {Bean}, Jacob L. and {Berta-Thompson}, Zachory K. and {Brown}, Timothy M. and {Buchhave}, Lars and {Butler}, Nathaniel R. and et al.},
        title = "{Transiting Exoplanet Survey Satellite (TESS)}",
      journal = {Journal of Astronomical Telescopes, Instruments, and Systems},
         year = 2015,
        month = jan,
       volume = {1},
          eid = {014003},
        pages = {014003},
          doi = {10.1117/1.JATIS.1.1.014003},
       adsurl = {https://ui.adsabs.harvard.edu/abs/2015JATIS...1a4003R}
}

@ARTICLE{1998PASP..110.1183W,
       author = {{Woodgate}, B.~E. and {Kimble}, R.~A. and {Bowers}, C.~W. and {Kraemer}, S. and {Kaiser}, M.~E. and {Danks}, A.~C. and {Grady}, J.~F. and {Loiacono}, J.~J. and {Brumfield}, M. and {Feinberg}, L. and et al.},
        title = "{The Space Telescope Imaging Spectrograph Design}",
      journal = {\pasp},
         year = 1998,
        month = oct,
       volume = {110},
       number = {752},
        pages = {1183-1204},
          doi = {10.1086/316243},
       adsurl = {https://ui.adsabs.harvard.edu/abs/1998PASP..110.1183W}
}

@ARTICLE{2020A&A...633A..32S,
       author = {{Shapiro}, A.~I. and {Amazo-G{\'o}mez}, E.~M. and {Krivova}, N.~A. and {Solanki}, S.~K.},
        title = "{Inflection point in the power spectrum of stellar brightness variations. I. The model}",
      journal = {\aap},
         year = 2020,
        month = jan,
       volume = {633},
          eid = {A32},
        pages = {A32},
          doi = {10.1051/0004-6361/201936018},
archivePrefix = {arXiv},
       eprint = {1910.08351},
 primaryClass = {astro-ph.SR},
       adsurl = {https://ui.adsabs.harvard.edu/abs/2020A&A...633A..32S}
}

@ARTICLE{2011ApJ...738..166C,
       author = {{Cohen}, O. and {Kashyap}, V.~L. and {Drake}, J.~J. and {Sokolov}, I.~V. and {Gombosi}, T.~I.},
        title = "{The Dynamics of Stellar Coronae Harboring Hot Jupiters. II. A Space Weather Event on a Hot Jupiter}",
      journal = {\apj},
         year = 2011,
        month = sep,
       volume = {738},
       number = {2},
          eid = {166},
        pages = {166},
          doi = {10.1088/0004-637X/738/2/166},
archivePrefix = {arXiv},
       eprint = {1102.4125},
 primaryClass = {astro-ph.SR},
       adsurl = {https://ui.adsabs.harvard.edu/abs/2011ApJ...738..166C}
}

@ARTICLE{2004ApJ...602L..53I,
       author = {{Ip}, Wing-Huen and {Kopp}, Andreas and {Hu}, Juei-Hwa},
        title = "{On the Star-Magnetosphere Interaction of Close-in Exoplanets}",
      journal = {\apjl},
         year = 2004,
        month = feb,
       volume = {602},
       number = {1},
        pages = {L53-L56},
          doi = {10.1086/382274},
       adsurl = {https://ui.adsabs.harvard.edu/abs/2004ApJ...602L..53I}
}

@INPROCEEDINGS{2023spi..conf....1S,
       author = {{Strugarek}, Antoine},
        title = "{Physics of star-planet magnetic interactions}",
    booktitle = {Star-Planet Interactions},
         year = 2023,
        month = feb,
        pages = {1},
          doi = {10.48550/arXiv.2104.05968},
archivePrefix = {arXiv},
       eprint = {2104.05968},
 primaryClass = {astro-ph.SR},
       adsurl = {https://ui.adsabs.harvard.edu/abs/2023spi..conf....1S}
}

@ARTICLE{2010ApJ...722L.168V,
       author = {{Vidotto}, A.~A. and {Jardine}, M. and {Helling}, Ch.},
        title = "{Early UV Ingress in WASP-12b: Measuring Planetary Magnetic Fields}",
      journal = {\apjl},
         year = 2010,
        month = oct,
       volume = {722},
       number = {2},
        pages = {L168-L172},
          doi = {10.1088/2041-8205/722/2/L168},
archivePrefix = {arXiv},
       eprint = {1009.5947},
 primaryClass = {astro-ph.EP},
       adsurl = {https://ui.adsabs.harvard.edu/abs/2010ApJ...722L.168V}
}

@ARTICLE{2012A&A...544A..23L,
       author = {{Lanza}, A.~F.},
        title = "{Star-planet magnetic interaction and activity in late-type stars with close-in planets}",
      journal = {\aap},
         year = 2012,
        month = aug,
       volume = {544},
          eid = {A23},
        pages = {A23},
          doi = {10.1051/0004-6361/201219002},
archivePrefix = {arXiv},
       eprint = {1206.5893},
 primaryClass = {astro-ph.EP},
       adsurl = {https://ui.adsabs.harvard.edu/abs/2012A&A...544A..23L}
}

@ARTICLE{2026A&A...705A..12P,
       author = {{Paul}, Arghyadeep and {Strugarek}, Antoine},
        title = "{Energetics of star--planet magnetic interactions: Novel insights from 3D modelling}",
      journal = {\aap},
         year = 2026,
        month = jan,
       volume = {705},
          eid = {A12},
        pages = {A12},
          doi = {10.1051/0004-6361/202556212},
archivePrefix = {arXiv},
       eprint = {2510.23277},
 primaryClass = {astro-ph.SR},
       adsurl = {https://ui.adsabs.harvard.edu/abs/2026A&A...705A..12P}
}

@ARTICLE{2015ApJ...815..111S,
       author = {{Strugarek}, A. and {Brun}, A.~S. and {Matt}, S.~P. and {R{\'e}ville}, V.},
        title = "{Magnetic Games between a Planet and Its Host Star: The Key Role of Topology}",
      journal = {\apj},
         year = 2015,
        month = dec,
       volume = {815},
       number = {2},
          eid = {111},
        pages = {111},
          doi = {10.1088/0004-637X/815/2/111},
archivePrefix = {arXiv},
       eprint = {1511.02837},
 primaryClass = {astro-ph.EP},
       adsurl = {https://ui.adsabs.harvard.edu/abs/2015ApJ...815..111S}
}

@ARTICLE{2013A&A...552A.119S,
       author = {{Saur}, J. and {Grambusch}, T. and {Duling}, S. and {Neubauer}, F.~M. and {Simon}, S.},
        title = "{Magnetic energy fluxes in sub-Alfv{\'e}nic planet star and moon planet interactions}",
      journal = {\aap},
         year = 2013,
        month = apr,
       volume = {552},
          eid = {A119},
        pages = {A119},
          doi = {10.1051/0004-6361/201118179},
       adsurl = {https://ui.adsabs.harvard.edu/abs/2013A&A...552A.119S}
}

@ARTICLE{2001Ap&SS.277..293Z,
       author = {{Zarka}, Philippe and {Treumann}, Rudolf A. and {Ryabov}, Boris P. and {Ryabov}, Vladimir B.},
        title = "{Magnetically-Driven Planetary Radio Emissions and Application to Extrasolar Planets}",
      journal = {\apss},
         year = 2001,
        month = jun,
       volume = {277},
        pages = {293-300},
          doi = {10.1023/A:1012221527425},
       adsurl = {https://ui.adsabs.harvard.edu/abs/2001Ap&SS.277..293Z}
}

@ARTICLE{2007P&SS...55..598Z,
       author = {{Zarka}, Philippe},
        title = "{Plasma interactions of exoplanets with their parent star and associated radio emissions}",
      journal = {\planss},
         year = 2007,
        month = apr,
       volume = {55},
       number = {5},
        pages = {598-617},
          doi = {10.1016/j.pss.2006.05.045},
       adsurl = {https://ui.adsabs.harvard.edu/abs/2007P&SS...55..598Z}
}

@ARTICLE{2020A&A...636A..69A,
       author = {{Amazo-G{\'o}mez}, E.~M. and {Shapiro}, A.~I. and {Solanki}, S.~K. and {Krivova}, N.~A. and {Kopp}, G. and {Reinhold}, T. and {Oshagh}, M. and {Reiners}, A.},
        title = "{Inflection point in the power spectrum of stellar brightness variations. II. The Sun}",
      journal = {\aap},
         year = 2020,
        month = apr,
       volume = {636},
          eid = {A69},
        pages = {A69},
          doi = {10.1051/0004-6361/201936925},
archivePrefix = {arXiv},
       eprint = {2002.03455},
 primaryClass = {astro-ph.SR},
       adsurl = {https://ui.adsabs.harvard.edu/abs/2020A&A...636A..69A}
}

@ARTICLE{2020A&A...642A.225A,
       author = {{Amazo-G{\'o}mez}, E.~M. and {Shapiro}, A.~I. and {Solanki}, S.~K. and {Kopp}, G. and {Oshagh}, M. and {Reinhold}, T. and {Reiners}, A.},
        title = "{Inflection point in the power spectrum of stellar brightness variations. III. Facular versus spot dominance on stars with known rotation periods}",
      journal = {\aap},
         year = 2020,
        month = oct,
       volume = {642},
          eid = {A225},
        pages = {A225},
          doi = {10.1051/0004-6361/202038926},
archivePrefix = {arXiv},
       eprint = {2008.11492},
 primaryClass = {astro-ph.SR},
       adsurl = {https://ui.adsabs.harvard.edu/abs/2020A&A...642A.225A}
}

@ARTICLE{Marsden2006MNRAS.370..468M,
       author = {{Marsden}, S.~C. and {Donati}, J. -F. and {Semel}, M. and {Petit}, P. and {Carter}, B.~D.},
        title = "{Surface differential rotation and photospheric magnetic field of the young solar-type star HD 171488 (V889 Her)}",
      journal = {\mnras},
         year = 2006,
        month = jul,
       volume = {370},
       number = {1},
        pages = {468-476},
          doi = {10.1111/j.1365-2966.2006.10503.x},
       adsurl = {https://ui.adsabs.harvard.edu/abs/2006MNRAS.370..468M}
}

@ARTICLE{Carroll2012A&A...548A..95C,
       author = {{Carroll}, T.~A. and {Strassmeier}, K.~G. and {Rice}, J.~B. and {K{\"u}nstler}, A.},
        title = "{The magnetic field topology of the weak-lined T Tauri star V410 Tauri. New strategies for Zeeman-Doppler imaging}",
      journal = {\aap},
         year = 2012,
        month = dec,
       volume = {548},
          eid = {A95},
        pages = {A95},
          doi = {10.1051/0004-6361/201220215},
archivePrefix = {arXiv},
       eprint = {1211.2720},
 primaryClass = {astro-ph.SR},
       adsurl = {https://ui.adsabs.harvard.edu/abs/2012A&A...548A..95C}
}

@ARTICLE{Donati2025A&A...700A.122D,
       author = {{Donati}, J. -F. and {Cristofari}, P.~I. and {Klein}, B. and {Finociety}, B. and {Moutou}, C.},
        title = "{Full Stokes magnetometry of the active M dwarfs AU Mic and EV Lac with SPIRou}",
      journal = {\aap},
         year = 2025,
        month = aug,
       volume = {700},
          eid = {A122},
        pages = {A122},
          doi = {10.1051/0004-6361/202555428},
archivePrefix = {arXiv},
       eprint = {2507.01754},
 primaryClass = {astro-ph.SR},
       adsurl = {https://ui.adsabs.harvard.edu/abs/2025A&A...700A.122D}
}

@article{gleissbergLongperiodicFluctuationSunspot1939,
  title = {A Long-Periodic Fluctuation of the Sun-Spot Numbers},
  author = {Gleissberg, W.},
  year = {1939},
  month = jun,
  journal = {The Observatory},
  volume = {62},
  pages = {158--159},
  issn = {0029-7704},
  urldate = {2025-09-11},
}

@article{suessSecularVariationsCosmicRayProduced1965,
  title = {Secular {{Variations}} of the {{Cosmic-Ray-Produced Carbon}} 14 in the {{Atmosphere}} and {{Their Interpretations}}},
  author = {Suess, Hans E.},
  year = {1965},
  month = dec,
  journal = {Journal of Geophysical Research},
  volume = {70},
  pages = {5937--5952},
  issn = {0148-0227},
  doi = {10.1029/JZ070i023p05937},
  urldate = {2025-09-11}
}

@article{usoskinHistorySolarActivity2013,
  title = {A {{History}} of {{Solar Activity}} over {{Millennia}}},
  author = {Usoskin, Ilya G.},
  year = {2013},
  month = mar,
  journal = {Living Reviews in Solar Physics},
  volume = {10},
  number = {1},
  pages = {1},
  issn = {1614-4961},
  doi = {10.12942/lrsp-2013-1},
  urldate = {2025-09-04},
  langid = {english},
}

@ARTICLE{olah2009,
       author = {{Ol{\'a}h}, K. and {Koll{\'a}th}, Z. and {Granzer}, T. and {Strassmeier}, K.~G. and {Lanza}, A.~F. and {J{\"a}rvinen}, S. and {Korhonen}, H. and {Baliunas}, S.~L. and {Soon}, W. and {Messina}, S. and {Cutispoto}, G.},
        title = "{Multiple and changing cycles of active stars. II. Results}",
      journal = {\aap},
         year = 2009,
        month = jul,
       volume = {501},
       number = {2},
        pages = {703-713},
          doi = {10.1051/0004-6361/200811304},
archivePrefix = {arXiv},
       eprint = {0904.1747},
 primaryClass = {astro-ph.SR},
       adsurl = {https://ui.adsabs.harvard.edu/abs/2009A&A...501..703O}
}

@ARTICLE{olah2016,
       author = {{Ol{\'a}h}, K. and {K{\H{o}}v{\'a}ri}, {\protect Zs}. and {Petrovay}, K. and {Soon}, W. and {Baliunas}, S. and {Koll{\'a}th}, Z. and {Vida}, K.},
        title = "{Magnetic cycles at different ages of stars}",
      journal = {\aap},
         year = 2016,
        month = jun,
       volume = {590},
          eid = {A133},
        pages = {A133},
          doi = {10.1051/0004-6361/201628479},
archivePrefix = {arXiv},
       eprint = {1604.06701},
 primaryClass = {astro-ph.SR},
       adsurl = {https://ui.adsabs.harvard.edu/abs/2016A&A...590A.133O}
}

@article{olahMultipleChangingCycles2009,
  title = {Multiple and Changing Cycles of Active Stars {{II}}. {{Results}}},
  author = {Ol{\'a}h, K. and Koll{\'a}th, Z. and Granzer, T. and Strassmeier, K. G. and Lanza, A. F. and J{\"a}rvinen, S. and Korhonen, H. and Baliunas, S. L. and Soon, W. and Messina, S. and Cutispoto, G.},
  year = {2009},
  month = jul,
  journal = {Astronomy and Astrophysics},
  volume = {501},
  number = {2},
  eprint = {0904.1747},
  pages = {703--713},
  issn = {00046361},
  doi = {10.1051/0004-6361/200811304},
  archiveprefix = {arXiv},
}

@techreport{olahMultiperiodicLightVariations2000,
  title = {Multiperiodic Light Variations of Active Stars},
  author = {Ol{\'a}h, K and Koll{\'a}th, Z and Strassmeier, K G},
  year = {2000},
  journal = {Astron. Astrophys},
  volume = {356},
  pages = {643--653},
}

@article{savanovActivityCyclesDwarfs2012,
  title = {Activity Cycles of {{M}} Dwarfs},
  author = {Savanov, I. S.},
  year = {2012},
  month = sep,
  journal = {Astronomy Reports},
  volume = {56},
  number = {9},
  pages = {716--721},
  issn = {10637729},
  doi = {10.1134/S1063772912090077},
}

@article{vidaQuestActivityCycles2013,
  title = {A Quest for Activity Cycles in Low-Mass Stars},
  author = {Vida, K. and Kriskovics, L. and Ol{\'a}h, K.},
  year = {2013},
  month = nov,
  journal = {Astronomische Nachrichten},
  volume = {334},
  number = {9},
  pages = {972--975},
  issn = {00046337},
  doi = {10.1002/asna.201211973},
}

@article{mathurPhotometricMagneticactivityMetrics2014,
  title = {Photometric Magnetic-Activity Metrics Tested with the {{Sun}}: {{Application}} to {{Kepler M}} Dwarfs},
  author = {Mathur, S. and Salabert, D. and Garcia, R. A. and Ceillier, T.},
journal = {Journal of Space Weather and Space Climate},
  year = {2014},
  month = apr,
  eprint = {1404.3076},
  archiveprefix = {arXiv},
}

@article{garciaCoRoTRevealsMagnetic2010,
  title = {{{CoRoT Reveals}} a {{Magnetic Activity Cycle}} in a {{Sun-Like Star}}},
  author = {Garc{\'i}a, Rafael A. and Mathur, Savita and Salabert, David and Ballot, J{\'e}r{\^o}me and R{\'e}gulo, Clara and Metcalfe, Travis S. and Baglin, Annie},
  year = {2010},
  month = aug,
  journal = {Science},
  volume = {329},
  number = {5995},
  pages = {1032--1032},
  publisher = {American Association for the Advancement of Science},
  doi = {10.1126/science.1191064},
  urldate = {2025-09-19},
}

@article{kieferStellarMagneticActivity2017,
  title = {Stellar Magnetic Activity and Variability of Oscillation Parameters: {{An}} Investigation of 24 Solar-like Stars Observed by {{Kepler}}},
  shorttitle = {Stellar Magnetic Activity and Variability of Oscillation Parameters},
  author = {Kiefer, Ren{\'e} and Schad, Ariane and Davies, Guy and Roth, Markus},
  year = {2017},
  month = feb,
  journal = {Astronomy \& Astrophysics},
  volume = {598},
  pages = {A77},
  publisher = {EDP Sciences},
  issn = {0004-6361, 1432-0746},
  doi = {10.1051/0004-6361/201628469},
  urldate = {2025-09-19},
  copyright = {{\copyright} ESO, 2017},
  langid = {english},
}

@article{santosSignaturesMagneticActivity2018,
  title = {Signatures of {{Magnetic Activity}} in the {{Seismic Data}} of {{Solar-type Stars Observed}} by {{Kepler}}},
  author = {Santos, A. R. G. and Campante, T. L. and Chaplin, W. J. and Cunha, M. S. and Lund, M. N. and Kiefer, R. and Salabert, D. and Garc{\'i}a, R. A. and Davies, G. R. and Elsworth, Y. and Howe, R.},
  year = {2018},
  month = jul,
  journal = {The Astrophysical Journal Supplement Series},
  volume = {237},
  number = {1},
  pages = {17},
  publisher = {The American Astronomical Society},
  issn = {0067-0049},
  doi = {10.3847/1538-4365/aac9b6},
  urldate = {2025-09-19},
  langid = {english}
}

@article{davenport10YearsStellar2020,
  title = {10 {{Years}} of {{Stellar Activity}} for {{GJ}} 1243},
  author = {Davenport, {\relax James}. R. A. and Mendoza, Guadalupe Tovar and Hawley, Suzanne L.},
  year = {2020},
  month = jul,
  journal = {The Astronomical Journal},
  volume = {160},
  pages = {36},
  issn = {0004-6256},
  doi = {10.3847/1538-3881/ab9536},
  urldate = {2025-09-23},
}

@article{feinsteinEvolutionFlareActivity2024,
  title = {Evolution of {{Flare Activity}} in {{GKM Stars Younger Than}} 300 {{Myr}} over {{Five Years}} of {{TESS Observations}}},
  author = {Feinstein, Adina D. and Seligman, Darryl Z. and France, Kevin and Gagn{\'e}, Jonathan and Kowalski, Adam},
  year = {2024},
  month = aug,
  journal = {The Astronomical Journal},
  volume = {168},
  pages = {60},
  issn = {0004-6256},
  doi = {10.3847/1538-3881/ad4edf},
  urldate = {2025-09-23},
}

@article{wainerSearchingStellarActivity2024,
  title = {Searching for {{Stellar Activity Cycles Using Flares}}: {{The Short-}} and {{Long-timescale Activity Variations}} of {{TIC-272272592}}},
  shorttitle = {Searching for {{Stellar Activity Cycles Using Flares}}},
  author = {Wainer, Tobin M. and Davenport, James R. A. and Tovar Mendoza, Guadalupe and Feinstein, Adina D. and Wagg, Tom},
  year = {2024},
  month = dec,
  journal = {The Astronomical Journal},
  volume = {168},
  pages = {232},
  issn = {0004-6256},
  doi = {10.3847/1538-3881/ad7bb2},
  urldate = {2025-09-23},
}

@article{berdyuginaButterflyDiagramActivity2007a,
  title = {Butterfly {{Diagram}} and {{Activity Cycles}} in {{HR}} 1099},
  author = {Berdyugina, Svetlana V. and Henry, Gregory W.},
  year = {2007},
  month = mar,
  journal = {The Astrophysical Journal},
  volume = {659},
  number = {2},
  pages = {L157},
  issn = {0004-637X},
  doi = {10.1086/517881},
  urldate = {2025-09-23},
  langid = {english},
}

@article{bazotButterflyDiagramSunlike2018,
  title = {Butterfly Diagram of a {{Sun-like}} Star Observed Using Asteroseismology},
  author = {Bazot, M. and Nielsen, M. B. and Mary, D. and {Christensen-Dalsgaard}, J. and Benomar, O. and Petit, P. and Gizon, L. and Sreenivasan, K. R. and White, T. R.},
  year = {2018},
  month = nov,
  journal = {Astronomy \& Astrophysics},
  volume = {619},
  pages = {L9},
  publisher = {EDP Sciences},
  issn = {0004-6361, 1432-0746},
  doi = {10.1051/0004-6361/201834251},
  urldate = {2025-09-23},
  copyright = {{\copyright} ESO 2018},
  langid = {english},
}

@article{2021MNRAS.500.1243L,
  title = {Identifying Solar-like Magnetic Cycles with {{Zeeman-Doppler-Imaging}}},
  author = {Lehmann, L. T. and Hussain, G. A. J. and Vidotto, A. A. and Jardine, M. M. and Mackay, D. H.},
  year = {2021},
  month = jan,
  journal = {Monthly Notices of the Royal Astronomical Society},
  volume = {500},
  pages = {1243--1260},
  issn = {0035-8711},
  doi = {10.1093/mnras/staa3284},
  urldate = {2025-09-23}
}

@ARTICLE{Labarga2026,
       author = {{Labarga}, Fernando and {Montes}, David},
        title = "{iSTARMOD: A Python Code to Quantify Chromospheric Activity by Using the Spectral Subtraction Technique}",
      journal = {\aj},
         year = 2026,
        month = jan,
       volume = {171},
       number = {1},
          eid = {15},
        pages = {15},
          doi = {10.3847/1538-3881/ae173e},
archivePrefix = {arXiv},
       eprint = {2512.09192},
 primaryClass = {astro-ph.SR},
       adsurl = {https://ui.adsabs.harvard.edu/abs/2026AJ....171...15L}
}

@ARTICLE{Boudreaux2022,
       author = {{Boudreaux}, Emily M. and {Newton}, Elisabeth R. and {Mondrik}, Nicholas and {Charbonneau}, David and {Irwin}, Jonathan},
        title = "{The Ca II H and K Rotation-Activity Relation in 53 Mid-to-late-type M Dwarfs}",
      journal = {\apj},
         year = 2022,
        month = apr,
       volume = {929},
       number = {1},
          eid = {80},
        pages = {80},
          doi = {10.3847/1538-4357/ac5cbf},
archivePrefix = {arXiv},
       eprint = {2203.04999},
 primaryClass = {astro-ph.SR},
       adsurl = {https://ui.adsabs.harvard.edu/abs/2022ApJ...929...80B}
}

@ARTICLE{AstudilloDefru2017,
       author = {{Astudillo-Defru}, N. and {Delfosse}, X. and {Bonfils}, X. and {Forveille}, T. and {Lovis}, C. and {Rameau}, J.},
        title = "{Magnetic activity in the HARPS M dwarf sample. The rotation-activity relationship for very low-mass stars through R'$_{HK}$}",
      journal = {\aap},
         year = 2017,
        month = apr,
       volume = {600},
          eid = {A13},
        pages = {A13},
          doi = {10.1051/0004-6361/201527078},
archivePrefix = {arXiv},
       eprint = {1610.09007},
 primaryClass = {astro-ph.SR},
       adsurl = {https://ui.adsabs.harvard.edu/abs/2017A&A...600A..13A}
}

@ARTICLE{Newton2017,
       author = {{Newton}, Elisabeth R. and {Irwin}, Jonathan and {Charbonneau}, David and {Berlind}, Perry and {Calkins}, Michael L. and {Mink}, Jessica},
        title = "{The H{\ensuremath{\alpha}} Emission of Nearby M Dwarfs and its Relation to Stellar Rotation}",
      journal = {\apj},
         year = 2017,
        month = jan,
       volume = {834},
       number = {1},
          eid = {85},
        pages = {85},
          doi = {10.3847/1538-4357/834/1/85},
archivePrefix = {arXiv},
       eprint = {1611.03509},
 primaryClass = {astro-ph.SR},
       adsurl = {https://ui.adsabs.harvard.edu/abs/2017ApJ...834...85N}
}

@ARTICLE{Mercier2025,
       author = {{Mercier}, Samson J. and {Dumusque}, Xavier and {Bourrier}, Vincent and {Al Moulla}, Khaled and {Cretignier}, Michael and {Dethier}, William and {Lo Curto}, Gaspare and {Figueira}, Pedro and {Lovis}, Christophe and {Pepe}, Francesco and {Santos}, Nuno C. and {Udry}, St{\'e}phane and {Wildi}, Fran{\c{c}}ois and {Allart}, Romain and {Baron}, Fr{\'e}d{\'e}rique and {Bouchy}, Fran{\c{c}}ois and {Carmona}, Andres and {Cointepas}, Marion and {Doyon}, Ren{\'e} and {Frensch}, Yolanda G.~C. and {Grieves}, Nolan and {Mignon}, Lucile and {Nielsen}, Louise D.},
        title = "{Studying the variability of the He triplet to understand the detection limits of evaporating exoplanet atmospheres}",
      journal = {\aap},
         year = 2025,
        month = aug,
       volume = {700},
          eid = {A8},
        pages = {A8},
          doi = {10.1051/0004-6361/202452856},
archivePrefix = {arXiv},
       eprint = {2507.21290},
 primaryClass = {astro-ph.EP},
       adsurl = {https://ui.adsabs.harvard.edu/abs/2025A&A...700A...8M}
}

@ARTICLE{GomesdaSilva2022,
       author = {{Gomes da Silva}, J. and {Bensabat}, A. and {Monteiro}, T. and {Santos}, N.~C.},
        title = "{Optimising the H{\ensuremath{\alpha}} index for the identification of activity signals in FGK stars. Improvement of the correlation between H{\ensuremath{\alpha}} and Ca II H\&K}",
      journal = {\aap},
         year = 2022,
        month = dec,
       volume = {668},
          eid = {A174},
        pages = {A174},
          doi = {10.1051/0004-6361/202244595},
archivePrefix = {arXiv},
       eprint = {2210.06903},
 primaryClass = {astro-ph.SR},
       adsurl = {https://ui.adsabs.harvard.edu/abs/2022A&A...668A.174G}
}

@article{borosaikiaSolarlikeMagneticCycle2016,
  title = {A Solar-like Magnetic Cycle on the Mature {{K-dwarf}} 61 {{Cygni A}} ({{HD}} 201091)},
  author = {Boro Saikia, S. and Jeffers, S. V. and Morin, J. and Petit, P. and Folsom, C. P. and Marsden, S. C. and Donati, J. -F. and Cameron, R. and Hall, J. C. and Perdelwitz, V. and Reiners, A. and Vidotto, A. A.},
  year = {2016},
  month = oct,
  journal = {Astronomy and Astrophysics},
  volume = {594},
  pages = {A29},
  issn = {0004-6361},
  doi = {10.1051/0004-6361/201628262},
  urldate = {2025-09-23},
}

@article{buddingInterpretationCyclicalPhotometric1977,
  title = {The {{Interpretation}} of {{Cyclical Photometric Variations}} in {{Certain Dwarf ME-Type Stars}}},
  author = {Budding, E.},
  year = {1977},
  month = may,
  journal = {Astrophysics and Space Science},
  volume = {48},
  pages = {207--223},
  publisher = {Springer},
  issn = {0004-640X},
  doi = {10.1007/BF00643052},
  urldate = {2025-10-03},
}

@article{dorrenNewFormulationStarspot1987,
  title = {A {{New Formulation}} of the {{Starspot Model}}, and the {{Consequences}} of {{Starspot Structure}}},
  author = {Dorren, J. D.},
  year = {1987},
  month = sep,
  journal = {The Astrophysical Journal},
  volume = {320},
  pages = {756},
  publisher = {IOP},
  issn = {0004-637X},
  doi = {10.1086/165593},
  urldate = {2025-10-03},
}

@article{mosserShortlivedSpotsSolarlike2009,
  title = {Short-Lived Spots in Solar-like Stars as Observed by {{CoRoT}}},
  author = {Mosser, B. and Baudin, F. and Lanza, A. F. and Hulot, J. C. and Catala, C. and Baglin, A. and Auvergne, M.},
  year = {2009},
  month = oct,
  journal = {Astronomy and Astrophysics},
  volume = {506},
  pages = {245--254},
  publisher = {EDP},
  issn = {0004-6361},
  doi = {10.1051/0004-6361/200911942},
  urldate = {2025-10-06},
}

@article{frohlichMagneticActivityDifferential2012,
  title = {Magnetic Activity and Differential Rotation in the Young {{Sun-like}} Stars {{KIC}} 7985370 and {{KIC}} 7765135},
  author = {Fr{\"o}hlich, H. -E. and Frasca, A. and Catanzaro, G. and Bonanno, A. and Corsaro, E. and {Molenda-{\.Z}akowicz}, J. and Klutsch, A. and Montes, D.},
  year = {2012},
  month = jul,
  journal = {Astronomy and Astrophysics},
  volume = {543},
  pages = {A146},
  publisher = {EDP},
  issn = {0004-6361},
  doi = {10.1051/0004-6361/201219167},
  urldate = {2025-10-06},
}

@article{roettenbacherImagingStarspotEvolution2013a,
  title = {Imaging {{Starspot Evolution}} on {{Kepler Target KIC}} 5110407 {{Using Light-Curve Inversion}}},
  author = {Roettenbacher, Rachael M. and Monnier, John D. and Harmon, Robert O. and Barclay, Thomas and Still, Martin},
  year = {2013},
  month = apr,
  journal = {The Astrophysical Journal},
  volume = {767},
  pages = {60},
  publisher = {IOP},
  issn = {0004-637X},
  doi = {10.1088/0004-637X/767/1/60},
  urldate = {2025-10-06},
}

@article{lanzaLongtermStarspotEvolution2006,
  title = {Long-Term Starspot Evolution, Activity Cycle, and Orbital Period Variation of {{V711 Tauri}} ({{HR}} 1099)},
  author = {Lanza, A. F. and Piluso, N. and Rodon{\`o}, M. and Messina, S. and Cutispoto, G.},
  year = {2006},
  month = aug,
  journal = {Astronomy and Astrophysics},
  volume = {455},
  pages = {595--606},
  publisher = {EDP},
  issn = {0004-6361},
  doi = {10.1051/0004-6361:20064847},
  urldate = {2025-10-06},
}

@article{kovariTestingStabilityReliability1997,
  title = {Testing the Stability and Reliability of Starspot Modelling.},
  author = {K\H{o}v\'ari, {\relax Zs}. and Bartus, J.},
  year = {1997},
  month = jul,
  journal = {Astronomy and Astrophysics},
  volume = {323},
  pages = {801--808},
  publisher = {EDP},
  issn = {0004-6361},
  urldate = {2025-10-06},
}

@article{kippingAnalyticModelRotational2012a,
  title = {An Analytic Model for Rotational Modulations in the Photometry of Spotted Stars},
  author = {Kipping, David M.},
  year = {2012},
  month = dec,
  journal = {Monthly Notices of the Royal Astronomical Society},
  volume = {427},
  pages = {2487--2511},
  publisher = {OUP},
  issn = {0035-8711},
  doi = {10.1111/j.1365-2966.2012.22124.x},
  urldate = {2025-10-06},
}

@article{strassmeierBinaryinducedMagneticActivity2011,
  title = {Binary-Induced Magnetic Activity?. {{Time-series}} Echelle Spectroscopy and Photometry of {{HD}} 123351 = {{CZ CVn}}},
  shorttitle = {Binary-Induced Magnetic Activity?},
  author = {Strassmeier, K. G. and Carroll, T. A. and Weber, M. and Granzer, T. and Bartus, J. and Ol{\'a}h, K. and Rice, J. B.},
  year = {2011},
  month = nov,
  journal = {Astronomy and Astrophysics},
  volume = {535},
  pages = {A98},
  publisher = {EDP},
  issn = {0004-6361},
  doi = {10.1051/0004-6361/201117167},
  urldate = {2025-10-06},
}

@article{vidaFourcolourPhotometryEY2010,
  title = {Four-Colour Photometry of {{EY Dra}}: {{A}} Study of an Ultra-Fast Rotating Active {{dM1-2e}} Star},
  author = {Vida, K. and Ol{\'a}h, K. and Kov{\'a}ri, Zs and Jurcsik, J. and S{\'o}dor, {\'A} and V{\'a}radi, M. and Belucz, B. and D{\'e}k{\'a}ny, I. and Hurta, Zs and Nagy, I. and Posztob{\'a}nyi, K.},
  year = {2010},
  month = mar,
  journal = {Astronomische Nachrichten},
  volume = {331},
  number = {3},
  pages = {250--256},
  issn = {00046337},
  doi = {10.1002/asna.200911341},
}

@article{lugerMappingStellarSurfaces2021,
  title = {Mapping {{Stellar Surfaces}}. {{II}}. {{An Interpretable Gaussian Process Model}} for {{Light Curves}}},
  author = {Luger, Rodrigo and {Foreman-Mackey}, Daniel and Hedges, Christina},
  year = {2021},
  month = sep,
  journal = {The Astronomical Journal},
  volume = {162},
  pages = {124},
  publisher = {IOP},
  issn = {0004-6256},
  doi = {10.3847/1538-3881/abfdb9},
  urldate = {2025-10-06},
}

@article{biczUnveilingSpectacular24hour2024,
  title = {Unveiling the Spectacular over 24-Hour Flare of Star {{CD-36}} 3202},
  author = {Bicz, K. and Falewicz, R. and Pietras, M.},
  year = {2024},
  month = feb,
  journal = {Astronomy and Astrophysics},
  volume = {682},
  pages = {A176},
  publisher = {EDP},
  issn = {0004-6361},
  doi = {10.1051/0004-6361/202347901},
  urldate = {2025-10-06},
}

@article{vidaPhotosphericChromosphericActivity2009,
  title = {Photospheric and Chromospheric Activity in {{V405 Andromedae}}. {{An M}} Dwarf Binary with Components on the Two Sides of the Full Convection Limit},
  author = {Vida, K. and Ol{\'a}h, K. and K{\H o}v{\'a}ri, {\relax Zs}. and Korhonen, H. and Bartus, J. and Hurta, {\relax Zs}. and Posztob{\'a}nyi, K.},
  year = {2009},
  month = sep,
  journal = {Astronomy and Astrophysics},
  volume = {504},
  pages = {1021--1029},
  issn = {0004-6361},
  doi = {10.1051/0004-6361/200912326},
  urldate = {2025-09-16},
  copyright = {All rights reserved},
}

@article{olahTimeseriesPhotometricSPOT1997,
  title = {Time-Series Photometric {{SPOT}} Modeling. {{III}}. {{Thirty}} Years in the Life of {{HK Lacertae}}.},
  author = {Ol\'ah, K. and K{\H o}v{\'a}ri, {\relax Zs}. and Bartus, J. and Strassmeier, K. G. and Hall, D. S. and Henry, G. W.},
  year = {1997},
  month = may,
  journal = {Astronomy and Astrophysics},
  volume = {321},
  pages = {811--821},
  publisher = {EDP},
  issn = {0004-6361},
  urldate = {2025-10-06},
}

@article{lanzaLongtermStarspotEvolution1998b,
  title = {Long-Term Starspot Evolution, Activity Cycle and Orbital Period Variation of {{AR Lacertae}}},
  author = {Lanza, A. F. and Catalano, S. and Cutispoto, G. and Pagano, I. and Rodono, M.},
  year = {1998},
  month = apr,
  journal = {Astronomy and Astrophysics},
  volume = {332},
  pages = {541--560},
  publisher = {EDP},
  issn = {0004-6361},
  urldate = {2025-10-07},
}

@ARTICLE{Senavci21,
       author = {{{S}enavc{\i}}, H.~V. and {K{\i}l{\i}{\c{c}}o{\u{g}}lu}, T. and {I{\c{s}}{\i}k}, E. and {Hussain}, G.~A.~J. and {Montes}, D. and {Bahar}, E. and {Solanki}, S.~K.},
        title = "{Observing and modelling the young solar analogue EK Draconis: starspot distribution, elemental abundances, and evolutionary status}",
      journal = {\mnras},
         year = 2021,
        month = apr,
       volume = {502},
       number = {3},
        pages = {3343-3356},
          doi = {10.1093/mnras/stab199},
archivePrefix = {arXiv},
       eprint = {2101.07248},
 primaryClass = {astro-ph.SR},
       adsurl = {https://ui.adsabs.harvard.edu/abs/2021MNRAS.502.3343S}
}

@ARTICLE{Araujo2021,
       author = {{Ara{\'u}jo}, Alexandre and {Valio}, Adriana},
        title = "{Kepler-411 Star Activity: Connection between Starspots and Superflares}",
      journal = {\apjl},
         year = 2021,
        month = dec,
       volume = {922},
       number = {2},
          eid = {L23},
        pages = {L23},
          doi = {10.3847/2041-8213/ac3767},
archivePrefix = {arXiv},
       eprint = {2111.05452},
 primaryClass = {astro-ph.SR},
       adsurl = {https://ui.adsabs.harvard.edu/abs/2021ApJ...922L..23A}
}

@ARTICLE{Estrela2016,
       author = {{Estrela}, Raissa and {Valio}, Adriana},
        title = "{Stellar Magnetic Cycles in the Solar-like Stars Kepler-17 and Kepler-63}",
      journal = {\apj},
         year = 2016,
        month = nov,
       volume = {831},
       number = {1},
          eid = {57},
        pages = {57},
          doi = {10.3847/0004-637X/831/1/57},
       adsurl = {https://ui.adsabs.harvard.edu/abs/2016ApJ...831...57E}
}

@INCOLLECTION{Cameron92_DOTS,
       author = {{Collier Cameron}, A.},
        title = "{Modelling Stellar Photospheric Spots Using Spectroscopy (Invited)}",
    booktitle = {Surface Inhomogeneities on Late-Type Stars},
         year = 1992,
       editor = {{Byrne}, Patrick B. and {Mullan}, Dermott J.},
       volume = {397},
        pages = {33},
          doi = {10.1007/3-540-55310-X_131},
       adsurl = {https://ui.adsabs.harvard.edu/abs/1992LNP...397...33C}
}

@ARTICLE{Bahar24,
       author = {{Bahar}, Engin and {{\c{S}}enavc{\i}}, Hakan V. and {I{\c{s}}{\i}k}, Emre and {Hussain}, Gaitee A.~J. and {Kochukhov}, Oleg and {Montes}, David and {Xiang}, Yue},
        title = "{First Chromospheric Activity and Doppler Imaging Study of PW And Using a New Doppler Imaging Code: SpotDIPy}",
      journal = {\apj},
         year = 2024,
        month = jan,
       volume = {960},
       number = {1},
          eid = {60},
        pages = {60},
          doi = {10.3847/1538-4357/ad055d},
archivePrefix = {arXiv},
       eprint = {2310.14865},
 primaryClass = {astro-ph.SR},
       adsurl = {https://ui.adsabs.harvard.edu/abs/2024ApJ...960...60B}
}

@ARTICLE{Lee+26,
       author = {{Lee}, S. and {Bahar}, E. and {{\c{S}}enavc{\i}}, H.~V. and {I{\textcommabelow s}{\i}k}, E. and {Ikuta}, K. and {Namekata}, K. and {Nagata}, H. and {Kawauchi}, K. and {Omiya}, M. and {Izumiura}, H. and {Tajitsu}, A. and {Sato}, B. and {Honda}, S. and {Nogami}, D.},
        title = "{Doppler imaging combined with high-cadence photometry: I. Revisiting the surface of a pre-main-sequence flare star PW Andromedae}",
      journal = {\aap},
         year = 2026,
        month = mar,
       volume = {707},
          eid = {A24},
        pages = {A24},
          doi = {10.1051/0004-6361/202556616},
archivePrefix = {arXiv},
       eprint = {2511.12190},
 primaryClass = {astro-ph.SR},
       adsurl = {https://ui.adsabs.harvard.edu/abs/2026A&A...707A..24L}
}

@ARTICLE{CarvalhoSilva25,
       author = {{Carvalho-Silva}, Gabriela and {Mel{\'e}ndez}, Jorge and {Rathsam}, Anne and {Shejeelammal}, J. and {Martos}, Giulia and {Lorenzo-Oliveira}, Diego and {Spina}, Lorenzo and {Ribeiro Alves}, D{\'e}bora},
        title = "{A New Age-Activity Relation For Solar Analogs that Accounts for Metallicity}",
      journal = {\apjl},
         year = 2025,
        month = apr,
       volume = {983},
       number = {2},
          eid = {L31},
        pages = {L31},
          doi = {10.3847/2041-8213/adc382},
archivePrefix = {arXiv},
       eprint = {2504.17482},
 primaryClass = {astro-ph.SR},
       adsurl = {https://ui.adsabs.harvard.edu/abs/2025ApJ...983L..31C}
}

@ARTICLE{Finociety+21,
       author = {{Finociety}, B. and {Donati}, J.-F. and {Klein}, B. and {Zaire}, B. and {Lehmann}, L. and {Moutou}, C. and {Bouvier}, J. and {Alencar}, S.~H.~P. and {Yu}, L. and {Grankin}, K. and {Artigau}, {\'E}. and {Doyon}, R. and {Delfosse}, X. and {Fouqu{\'e}}, P. and {H{\'e}brard}, G. and {Jardine}, M. and {K{\'o}sp{\'a}l}, {\'A}. and {M{\'e}nard}, F. and {M{\'e}nard}, F. and {SLS Consortium}},
        title = "{The T Tauri star V410 Tau in the eyes of SPIRou and TESS}",
      journal = {\mnras},
         year = 2021,
        month = dec,
       volume = {508},
       number = {3},
        pages = {3427-3445},
          doi = {10.1093/mnras/stab2778},
archivePrefix = {arXiv},
       eprint = {2109.11755},
 primaryClass = {astro-ph.SR},
       adsurl = {https://ui.adsabs.harvard.edu/abs/2021MNRAS.508.3427F}
}

@ARTICLE{Finociety+23,
       author = {{Finociety}, B. and {Donati}, J.-F. and {Grankin}, K. and {Bouvier}, J. and {Alencar}, S. and {M{\'e}nard}, F. and {Ray}, T.~P. and {K{\'o}sp{\'a}l}, {\'A}. and {consortium}, the S.~L.~S.},
        title = "{The active weak-line T Tauri star LkCa 4 observed with SPIRou and TESS}",
      journal = {\mnras},
         year = 2023,
        month = apr,
       volume = {520},
       number = {2},
        pages = {3049-3065},
          doi = {10.1093/mnras/stad267},
archivePrefix = {arXiv},
       eprint = {2301.09584},
 primaryClass = {astro-ph.SR},
       adsurl = {https://ui.adsabs.harvard.edu/abs/2023MNRAS.520.3049F}
}

@ARTICLE{2002ApJ...575.1078H,
       author = {{Hussain}, G.~A.~J. and {van Ballegooijen}, A.~A. and {Jardine}, M. and {Collier Cameron}, A.},
        title = "{The Coronal Topology of the Rapidly Rotating K0 Dwarf AB Doradus. I. Using Surface Magnetic Field Maps to Model the Structure of the Stellar Corona}",
      journal = {\apj},
         year = 2002,
        month = aug,
       volume = {575},
       number = {2},
        pages = {1078-1086},
          doi = {10.1086/341429},
archivePrefix = {arXiv},
       eprint = {astro-ph/0207452},
 primaryClass = {astro-ph},
       adsurl = {https://ui.adsabs.harvard.edu/abs/2002ApJ...575.1078H}
}

@ARTICLE{2010ApJ...723L.213M,
       author = {{Metcalfe}, T.~S. and {Basu}, S. and {Henry}, T.~J. and {Soderblom}, D.~R. and {Judge}, P.~G. and {Kn{\"o}lker}, M. and {Mathur}, S. and {Rempel}, M.},
        title = "{Discovery of a 1.6 Year Magnetic Activity Cycle in the Exoplanet Host Star {\ensuremath{\i}} Horologii}",
      journal = {\apjl},
         year = 2010,
        month = nov,
       volume = {723},
       number = {2},
        pages = {L213-L217},
          doi = {10.1088/2041-8205/723/2/L213},
archivePrefix = {arXiv},
       eprint = {1009.5399},
 primaryClass = {astro-ph.SR},
       adsurl = {https://ui.adsabs.harvard.edu/abs/2010ApJ...723L.213M}
}

@ARTICLE{Buccino2008,
       author = {{Buccino}, A.~P. and {Mauas}, P.~J.~D.},
        title = "{Mg II h+k emission lines as stellar activity indicators of main sequence F-K stars}",
      journal = {\aap},
         year = 2008,
        month = jun,
       volume = {483},
       number = {3},
        pages = {903-910},
          doi = {10.1051/0004-6361:20078925},
archivePrefix = {arXiv},
       eprint = {0804.1101},
 primaryClass = {astro-ph},
       adsurl = {https://ui.adsabs.harvard.edu/abs/2008A&A...483..903B}
}

@ARTICLE{Spake2018,
       author = {{Spake}, J.~J. and {Sing}, D.~K. and {Evans}, T.~M. and {Oklop{\v{c}}i{\'c}}, A. and {Bourrier}, V. and {Kreidberg}, L. and {Rackham}, B.~V. and {Irwin}, J. and {Ehrenreich}, D. and {Wyttenbach}, A. and {Wakeford}, H.~R. and {Zhou}, Y. and {Chubb}, K.~L. and {Nikolov}, N. and {Goyal}, J.~M. and {Henry}, G.~W. and {Williamson}, M.~H. and {Blumenthal}, S. and {Anderson}, D.~R. and {Hellier}, C. and {Charbonneau}, D. and {Udry}, S. and {Madhusudhan}, N.},
        title = "{Helium in the eroding atmosphere of an exoplanet}",
      journal = {\nat},
         year = 2018,
        month = may,
       volume = {557},
       number = {7703},
        pages = {68-70},
          doi = {10.1038/s41586-018-0067-5},
archivePrefix = {arXiv},
       eprint = {1805.01298},
 primaryClass = {astro-ph.EP},
       adsurl = {https://ui.adsabs.harvard.edu/abs/2018Natur.557...68S}
}

@ARTICLE{2017MNRAS.464.4299F,
       author = {{Flores}, Mat{\'\i}as G. and {Buccino}, Andrea P. and {Saffe}, Carlos E. and {Mauas}, Pablo J.~D.},
        title = "{A possible long-term activity cycle for {\ensuremath{\i}} Horologii: First results from SPI-HK{\ensuremath{\alpha}} project}",
      journal = {\mnras},
         year = 2017,
        month = feb,
       volume = {464},
       number = {4},
        pages = {4299-4305},
          doi = {10.1093/mnras/stw2650},
archivePrefix = {arXiv},
       eprint = {1610.03731},
 primaryClass = {astro-ph.SR},
       adsurl = {https://ui.adsabs.harvard.edu/abs/2017MNRAS.464.4299F}
}

@ARTICLE{1978AnSta...6..461S,
       author = {{Schwarz}, Gideon},
        title = "{Estimating the Dimension of a Model}",
      journal = {Annals of Statistics},
         year = 1978,
        month = jul,
       volume = {6},
       number = {2},
        pages = {461-464},
       adsurl = {https://ui.adsabs.harvard.edu/abs/1978AnSta...6..461S}
}

@ARTICLE{1997MNRAS.291..658D,
       author = {{Donati}, J.-F. and {Semel}, M. and {Carter}, B.~D. and {Rees}, D.~E. and {Collier Cameron}, A.},
        title = "{Spectropolarimetric observations of active stars}",
      journal = {\mnras},
         year = 1997,
        month = nov,
       volume = {291},
       number = {4},
        pages = {658-682},
          doi = {10.1093/mnras/291.4.658},
       adsurl = {https://ui.adsabs.harvard.edu/abs/1997MNRAS.291..658D}
}

@ARTICLE{2012A&A...548A..95C,
       author = {{Carroll}, T.~A. and {Strassmeier}, K.~G. and {Rice}, J.~B. and {K{\"u}nstler}, A.},
        title = "{The magnetic field topology of the weak-lined T Tauri star V410 Tauri. New strategies for Zeeman-Doppler imaging}",
      journal = {\aap},
         year = 2012,
        month = dec,
       volume = {548},
          eid = {A95},
        pages = {A95},
          doi = {10.1051/0004-6361/201220215},
archivePrefix = {arXiv},
       eprint = {1211.2720},
 primaryClass = {astro-ph.SR},
       adsurl = {https://ui.adsabs.harvard.edu/abs/2012A&A...548A..95C}
}

@INPROCEEDINGS{1958IAUS....6..209D,
       author = {{Deutsch}, A.~J.},
        title = "{Harmonic Analysis of the Periodic Spectrum Variables}",
    booktitle = {Electromagnetic Phenomena in Cosmical Physics},
         year = 1958,
       editor = {{Lehnert}, B.},
       series = {IAU Symposium},
       volume = {6},
        month = jan,
        pages = {209},
       adsurl = {https://ui.adsabs.harvard.edu/abs/1958IAUS....6..209D}
}

@ARTICLE{1998A&A...338...97B,
       author = {{Berdyugina}, Svetlana V.},
        title = "{Surface imaging by the Occamian approach. Basic principles, simulations, and tests}",
      journal = {\aap},
         year = 1998,
        month = oct,
       volume = {338},
        pages = {97-105},
       adsurl = {https://ui.adsabs.harvard.edu/abs/1998A&A...338...97B}
}

@ARTICLE{1987ApJ...321..496V,
       author = {{Vogt}, Steven S. and {Penrod}, G. Donald and {Hatzes}, Artie P.},
        title = "{Doppler Images of Rotating Stars Using Maximum Entropy Image Reconstruction}",
      journal = {\apj},
         year = 1987,
        month = oct,
       volume = {321},
        pages = {496},
          doi = {10.1086/165647},
       adsurl = {https://ui.adsabs.harvard.edu/abs/1987ApJ...321..496V}
}

@ARTICLE{1977SvAL....3..147G,
       author = {{Goncharskii}, A.~V. and {Stepanov}, V.~V. and {Kokhlova}, V.~L. and {Yagola}, A.~G.},
        title = "{Reconstruction of local line profiles from those observed in an Ap spectrum.}",
      journal = {Soviet Astronomy Letters},
         year = 1977,
        month = jan,
       volume = {3},
        pages = {147-149},
       adsurl = {https://ui.adsabs.harvard.edu/abs/1977SvAL....3..147G}
}

@ARTICLE{Skumanich1972,
       author = {{Skumanich}, A.},
        title = "{Time Scales for Ca II Emission Decay, Rotational Braking, and Lithium Depletion}",
      journal = {\apj},
         year = 1972,
        month = feb,
       volume = {171},
        pages = {565},
          doi = {10.1086/151310},
       adsurl = {https://ui.adsabs.harvard.edu/abs/1972ApJ...171..565S}
}

@INproceedings{1991LNP...380..309P,
       author = {{Piskunov}, N.~E.},
        title = "{The Art of Surface Imaging}",
    booktitle = {IAU Colloquium 130: The Sun and Cool Stars. Activity, Magnetism, Dynamos},
         year = 1991,
       editor = {{Tuominen}, I. and {Moss}, D. and {R{\"u}diger}, G.},
       volume = {380},
        pages = {309},
          doi = {10.1007/3-540-53955-7_147},
       adsurl = {https://ui.adsabs.harvard.edu/abs/1991LNP...380..309P}
}

@ARTICLE{2005A&A...435..261W,
       author = {{Wolter}, U. and {Schmitt}, J.~H.~M.~M. and {van Wyk}, F.},
        title = "{Doppler imaging of Speedy Mic using the VLT. Fast spot evolution on a young K-dwarf star}",
      journal = {\aap},
         year = 2005,
        month = may,
       volume = {435},
       number = {1},
        pages = {261-273},
          doi = {10.1051/0004-6361:20042239},
archivePrefix = {arXiv},
       eprint = {astro-ph/0504104},
 primaryClass = {astro-ph},
       adsurl = {https://ui.adsabs.harvard.edu/abs/2005A&A...435..261W}
}

@INPROCEEDINGS{2006ASPC..358..362D,
       author = {{Donati}, J.-F. and {Catala}, C. and {Landstreet}, J.~D. and {Petit}, P.},
        title = "{ESPaDOnS: The New Generation Stellar Spectro-Polarimeter. Performances and First Results}",
    booktitle = {Solar Polarization 4},
         year = 2006,
       editor = {{Casini}, R. and {Lites}, B.~W.},
       series = {Astronomical Society of the Pacific Conference Series},
       volume = {358},
        month = dec,
        pages = {362},
       adsurl = {https://ui.adsabs.harvard.edu/abs/2006ASPC..358..362D}
}

@INPROCEEDINGS{2008SPIE.7014E..0OS,
       author = {{Snik}, Frans and {Jeffers}, Sandra and {Keller}, Christoph and {Piskunov}, Nikolai and {Kochukhov}, Oleg and {Valenti}, Jeff and {Johns-Krull}, Christopher},
        title = "{The upgrade of HARPS to a full-Stokes high-resolution spectropolarimeter}",
    booktitle = {Ground-based and Airborne Instrumentation for Astronomy II},
         year = 2008,
       editor = {{McLean}, Ian S. and {Casali}, Mark M.},
       series = {Society of Photo-Optical Instrumentation Engineers (SPIE) Conference Series},
       volume = {7014},
        month = jul,
          eid = {70140O},
        pages = {70140O},
          doi = {10.1117/12.787393},
       adsurl = {https://ui.adsabs.harvard.edu/abs/2008SPIE.7014E..0OS}
}

@ARTICLE{2015AN....336..324S,
       author = {{Strassmeier}, K.~G. and {Ilyin}, I. and {J{\"a}rvinen}, A. and {Weber}, M. and {Woche}, M. and {Barnes}, S.~I. and {Bauer}, S.-M. and {Beckert}, E. and {Bittner}, W. and {Bredthauer}, R. and {Carroll}, T.~A. and {Denker}, C. and {Dionies}, F. and {DiVarano}, I. and {D{\"o}scher}, D. and {Fechner}, T. and {Feuerstein}, D. and {Granzer}, T. and {Hahn}, T. and {Harnisch}, G. and {Hofmann}, A. and {Lesser}, M. and {Paschke}, J. and {Pankratow}, S. and {Plank}, V. and {Pl{\"u}schke}, D. and {Popow}, E. and {Sablowski}, D.},
        title = "{PEPSI: The high-resolution {\'e}chelle spectrograph and polarimeter for the Large Binocular Telescope}",
      journal = {Astronomische Nachrichten},
         year = 2015,
        month = may,
       volume = {336},
       number = {4},
        pages = {324},
          doi = {10.1002/asna.201512172},
archivePrefix = {arXiv},
       eprint = {1505.06492},
 primaryClass = {astro-ph.IM},
       adsurl = {https://ui.adsabs.harvard.edu/abs/2015AN....336..324S}
}

@ARTICLE{2018A&A...620A.162J,
       author = {{J{\"a}rvinen}, S.~P. and {Strassmeier}, K.~G. and {Carroll}, T.~A. and {Ilyin}, I. and {Weber}, M.},
        title = "{Mapping EK Draconis with PEPSI. Possible evidence for starspot penumbrae}",
      journal = {\aap},
         year = 2018,
        month = dec,
       volume = {620},
          eid = {A162},
        pages = {A162},
          doi = {10.1051/0004-6361/201833496},
archivePrefix = {arXiv},
       eprint = {1812.03675},
 primaryClass = {astro-ph.SR},
       adsurl = {https://ui.adsabs.harvard.edu/abs/2018A&A...620A.162J}
}

@ARTICLE{1983PASP...95..565V,
       author = {{Vogt}, S.~S. and {Penrod}, G.~D.},
        title = "{Doppler imaging of spotted stars : application to the RS Canum Venaticorum star HR 1099.}",
      journal = {Publications of the Astronomical Society of the Pacific },
         year = 1983,
        month = sep,
       volume = {95},
        pages = {565-576},
          doi = {10.1086/131208},
       adsurl = {https://ui.adsabs.harvard.edu/abs/1983PASP...95..565V}
}

@ARTICLE{1989A&A...225..456S,
       author = {{Semel}, M.},
        title = "{Zeeman-Doppler imaging of active stars. I - Basic principles.}",
      journal = {\aap},
         year = 1989,
        month = nov,
       volume = {225},
        pages = {456-466},
       adsurl = {https://ui.adsabs.harvard.edu/abs/1989A&A...225..456S}
}

@INPROCEEDINGS{1996IAUS..176...53D,
       author = {{Donati}, J.-F.},
        title = "{Stellar magnetic imaging from spectropolarimetric data}",
    booktitle = {Stellar Surface Structure},
         year = 1996,
       editor = {{Strassmeier}, Klaus G. and {Linsky}, Jeffrey L.},
       series = {IAU Symposium},
       volume = {176},
        month = jan,
        pages = {53},
       adsurl = {https://ui.adsabs.harvard.edu/abs/1996IAUS..176...53D}
}

@ARTICLE{2009ARA&A..47..333D,
       author = {{Donati}, J.-F. and {Landstreet}, J.~D.},
        title = "{Magnetic Fields of Nondegenerate Stars}",
      journal = {\araa},
         year = 2009,
        month = sep,
       volume = {47},
       number = {1},
        pages = {333-370},
          doi = {10.1146/annurev-astro-082708-101833},
archivePrefix = {arXiv},
       eprint = {0904.1938},
 primaryClass = {astro-ph.SR},
       adsurl = {https://ui.adsabs.harvard.edu/abs/2009ARA&A..47..333D}
}

@ARTICLE{2013A&A...552A..78Z,
       author = {{Zechmeister}, M. and {K{\"u}rster}, M. and {Endl}, M. and {Lo Curto}, G. and {Hartman}, H. and {Nilsson}, H. and {Henning}, T. and {Hatzes}, A.~P. and {Cochran}, W.~D.},
        title = "{The planet search programme at the ESO CES and HARPS. IV. The search for Jupiter analogues around solar-like stars}",
      journal = {\aap},
         year = 2013,
        month = apr,
       volume = {552},
          eid = {A78},
        pages = {A78},
          doi = {10.1051/0004-6361/201116551},
archivePrefix = {arXiv},
       eprint = {1211.7263},
 primaryClass = {astro-ph.EP},
       adsurl = {https://ui.adsabs.harvard.edu/abs/2013A&A...552A..78Z}
}

@ARTICLE{1990A&A...230..363P,
       author = {{Piskunov}, N.~E. and {Tuominen}, I. and {Vilhu}, O.},
        title = "{Surface imaging of late-type stars.}",
      journal = {\aap},
         year = 1990,
        month = apr,
       volume = {230},
        pages = {363-370},
       adsurl = {https://ui.adsabs.harvard.edu/abs/1990A&A...230..363P}
}

@ARTICLE{2009A&A...507.1711S,
       author = {{Sennhauser}, C. and {Berdyugina}, S.~V. and {Fluri}, D.~M.},
        title = "{Nonlinear deconvolution with deblending: a new analyzing technique for spectroscopy}",
      journal = {\aap},
         year = 2009,
        month = dec,
       volume = {507},
       number = {3},
        pages = {1711-1718},
          doi = {10.1051/0004-6361/200912467},
       adsurl = {https://ui.adsabs.harvard.edu/abs/2009A&A...507.1711S}
}

@ARTICLE{2010A&A...522A..57S,
       author = {{Sennhauser}, C. and {Berdyugina}, S.~V.},
        title = "{Zeeman component decomposition for recovering common profiles and magnetic fields}",
      journal = {\aap},
         year = 2010,
        month = nov,
       volume = {522},
          eid = {A57},
        pages = {A57},
          doi = {10.1051/0004-6361/201014971},
       adsurl = {https://ui.adsabs.harvard.edu/abs/2010A&A...522A..57S}
}

@ARTICLE{2010A&A...524A...5K,
       author = {{Kochukhov}, O. and {Makaganiuk}, V. and {Piskunov}, N.},
        title = "{Least-squares deconvolution of the stellar intensity and polarization spectra}",
      journal = {\aap},
         year = 2010,
        month = dec,
       volume = {524},
          eid = {A5},
        pages = {A5},
          doi = {10.1051/0004-6361/201015429},
archivePrefix = {arXiv},
       eprint = {1008.5115},
 primaryClass = {astro-ph.SR},
       adsurl = {https://ui.adsabs.harvard.edu/abs/2010A&A...524A...5K}
}

@ARTICLE{GorgeiEKDra,
       author = {{G{\"o}rgei}, A. and {Kriskovics}, L. and {Vida}, K. and {Seli}, B. and {Ol{\'a}h}, K. and {S{\'a}gi}, P. and {B{\'o}di}, A. and {J{\"a}rvinen}, S.~P. and {Strassmeier}, K.~G. and {P{\'a}l}, A. and {K{\H{o}}v{\'a}ri}, Zs.},
        title = "{Magnetic activity on the young Sun: a case study of EK Draconis}",
      journal = {arXiv e-prints},
         year = 2025,
        month = dec,
          eid = {arXiv:2512.03830},
        pages = {arXiv:2512.03830},
          doi = {10.48550/arXiv.2512.03830},
archivePrefix = {arXiv},
       eprint = {2512.03830},
 primaryClass = {astro-ph.SR},
       adsurl = {https://ui.adsabs.harvard.edu/abs/2025arXiv251203830G}
}

@ARTICLE{Kriskovics2023A&A...674A.143K,
       author = {{Kriskovics}, L. and {K{\H{o}}v{\'a}ri}, Zs. and {Seli}, B. and {Ol{\'a}h}, K. and {Vida}, K. and {Henry}, G.~W. and {Granzer}, T. and {G{\"o}rgei}, A.},
        title = "{EI Eridani: A star under the influence. The effect of magnetic activity in the short and long term}",
      journal = {\aap},
         year = 2023,
        month = jun,
       volume = {674},
          eid = {A143},
        pages = {A143},
          doi = {10.1051/0004-6361/202245767},
archivePrefix = {arXiv},
       eprint = {2304.13234},
 primaryClass = {astro-ph.SR},
       adsurl = {https://ui.adsabs.harvard.edu/abs/2023A&A...674A.143K}
}

@inproceedings{johnstoneStellarWindsPlanetary2021,
       author = {{Johnstone}, Colin},
        title = "{Stellar Winds and Planetary Atmospheres}",
    booktitle = {Star-Planet Interactions},
         year = 2023,
       editor = {{Bigot}, Lionel and {Bouvier}, J{\'e}r{\^o}me and {Lebreton}, Yveline and {Chiavassa}, Andrea and {L{\`e}bre}, Agn{\`e}s},
        month = feb,
        pages = {154},
          doi = {10.48550/arXiv.2105.11243},
archivePrefix = {arXiv},
       eprint = {2105.11243},
 primaryClass = {astro-ph.EP},
       adsurl = {https://ui.adsabs.harvard.edu/abs/2023spi..conf..154J}
}

@ARTICLE{2008ApJ...683L.179S,
       author = {{Silva-Valio}, Adriana},
        title = "{Estimating Stellar Rotation from Starspot Detection during Planetary Transits}",
      journal = {\apjl},
         year = 2008,
        month = aug,
       volume = {683},
       number = {2},
        pages = {L179},
          doi = {10.1086/591846},
archivePrefix = {arXiv},
       eprint = {0808.2156},
 primaryClass = {astro-ph},
       adsurl = {https://ui.adsabs.harvard.edu/abs/2008ApJ...683L.179S}
}

@ARTICLE{2011A&A...529A..36S,
       author = {{Silva-Valio}, A. and {Lanza}, A.~F.},
        title = "{Time evolution and rotation of starspots on CoRoT-2 from the modelling of transit photometry}",
      journal = {\aap},
         year = 2011,
        month = may,
       volume = {529},
          eid = {A36},
        pages = {A36},
          doi = {10.1051/0004-6361/201015382},
archivePrefix = {arXiv},
       eprint = {1102.2192},
 primaryClass = {astro-ph.SR},
       adsurl = {https://ui.adsabs.harvard.edu/abs/2011A&A...529A..36S}
}

@ARTICLE{2017ApJ...835..294V,
       author = {{Valio}, Adriana and {Estrela}, Raissa and {Netto}, Yuri and {Bravo}, J.~P. and {de Medeiros}, J.~R.},
        title = "{Activity and Rotation of Kepler-17}",
      journal = {\apj},
         year = 2017,
        month = feb,
       volume = {835},
       number = {2},
          eid = {294},
        pages = {294},
          doi = {10.3847/1538-4357/835/2/294},
archivePrefix = {arXiv},
       eprint = {1702.02213},
 primaryClass = {astro-ph.SR},
       adsurl = {https://ui.adsabs.harvard.edu/abs/2017ApJ...835..294V}
}

@ARTICLE{2025A&A...702A.227Z,
       author = {{Zaleski}, S.~M. and {Valio}, A. and {Carter}, B.},
        title = "{Extracting starspot structures from exoplanet transit photometry}",
      journal = {\aap},
         year = 2025,
        month = oct,
       volume = {702},
          eid = {A227},
        pages = {A227},
          doi = {10.1051/0004-6361/202452779},
       adsurl = {https://ui.adsabs.harvard.edu/abs/2025A&A...702A.227Z}
}

@ARTICLE{2014MNRAS.442.3686B,
       author = {{B{\'e}ky}, Bence and {Kipping}, David M. and {Holman}, Matthew J.},
        title = "{SPOTROD: a semi-analytic model for transits of spotted stars}",
      journal = {\mnras},
         year = 2014,
        month = aug,
       volume = {442},
       number = {4},
        pages = {3686-3699},
          doi = {10.1093/mnras/stu1061},
archivePrefix = {arXiv},
       eprint = {1407.4465},
 primaryClass = {astro-ph.EP},
       adsurl = {https://ui.adsabs.harvard.edu/abs/2014MNRAS.442.3686B}
}

@ARTICLE{2012A&A...545A.109B,
       author = {{Boisse}, I. and {Bonfils}, X. and {Santos}, N.~C.},
        title = "{SOAP. A tool for the fast computation of photometry and radial velocity induced by stellar spots}",
      journal = {\aap},
         year = 2012,
        month = sep,
       volume = {545},
          eid = {A109},
        pages = {A109},
          doi = {10.1051/0004-6361/201219115},
archivePrefix = {arXiv},
       eprint = {1206.5493},
 primaryClass = {astro-ph.IM},
       adsurl = {https://ui.adsabs.harvard.edu/abs/2012A&A...545A.109B}
}

@ARTICLE{2013A&A...549A..35O,
       author = {{Oshagh}, M. and {Boisse}, I. and {Bou{\'e}}, G. and {Montalto}, M. and {Santos}, N.~C. and {Bonfils}, X. and {Haghighipour}, N.},
        title = "{SOAP-T: a tool to study the light curve and radial velocity of a system with a transiting planet and a rotating spotted star}",
      journal = {\aap},
         year = 2013,
        month = jan,
       volume = {549},
          eid = {A35},
        pages = {A35},
          doi = {10.1051/0004-6361/201220173},
archivePrefix = {arXiv},
       eprint = {1211.1311},
 primaryClass = {astro-ph.EP},
       adsurl = {https://ui.adsabs.harvard.edu/abs/2013A&A...549A..35O}
}

@ARTICLE{2017ApJ...846...99M,
       author = {{Morris}, Brett M. and {Hebb}, Leslie and {Davenport}, James R.~A. and {Rohn}, Graeme and {Hawley}, Suzanne L.},
        title = "{The Starspots of HAT-P-11: Evidence for a Solar-like Dynamo}",
      journal = {\apj},
         year = 2017,
        month = sep,
       volume = {846},
       number = {2},
          eid = {99},
        pages = {99},
          doi = {10.3847/1538-4357/aa8555},
archivePrefix = {arXiv},
       eprint = {1708.02583},
 primaryClass = {astro-ph.SR},
       adsurl = {https://ui.adsabs.harvard.edu/abs/2017ApJ...846...99M}
}

@ARTICLE{2005A&A...440..735J,
       author = {{J{\"a}rvinen}, S.~P. and {Berdyugina}, S.~V. and {Strassmeier}, K.~G.},
        title = "{Spots on EK Draconis. Active longitudes and cycles from long-term photometry}",
      journal = {\aap},
         year = 2005,
        month = sep,
       volume = {440},
       number = {2},
        pages = {735-741},
          doi = {10.1051/0004-6361:20053297},
       adsurl = {https://ui.adsabs.harvard.edu/abs/2005A&A...440..735J}
}

@ARTICLE{1998A&A...330..685S,
       author = {{Strassmeier}, K.~G. and {Rice}, J.~B.},
        title = "{Doppler imaging of stellar surface structure. VI. HD 129333 = EK Draconis: a stellar analog of the active young Sun}",
      journal = {\aap},
         year = 1998,
        month = feb,
       volume = {330},
        pages = {685-695},
       adsurl = {https://ui.adsabs.harvard.edu/abs/1998A&A...330..685S}
}

@ARTICLE{2025A&A...702A.252F,
       author = {{Finley}, A.~J.},
        title = "{Reconstructing the Sun's Alfv{\'e}n surface and wind braking torque with Parker Solar Probe}",
      journal = {\aap},
         year = 2025,
        month = oct,
       volume = {702},
          eid = {A252},
        pages = {A252},
          doi = {10.1051/0004-6361/202556391},
archivePrefix = {arXiv},
       eprint = {2509.07088},
 primaryClass = {astro-ph.SR},
       adsurl = {https://ui.adsabs.harvard.edu/abs/2025A&A...702A.252F}
}

@ARTICLE{2025MNRAS.542.1318S,
       author = {{See}, Victor and {Amard}, Louis and {Bellotti}, Stefano and {Boro Saikia}, Sudeshna and {Brown}, Emma L. and {Donati}, Jean-Francois and {Fares}, Rim and {Finley}, Adam J. and {Folsom}, Colin P. and {H{\'e}brard}, {\'E}lodie M. and {Jardine}, Moira M. and {Jeffers}, Sandra V. and {Klein}, Baptiste and {Lehmann}, Lisa T. and {Marsden}, Stephen C. and {Matt}, Sean P. and {Mengel}, Matthew W. and {Morin}, Julien and {Petit}, Pascal and {Smith}, Katelyn and {Vidotto}, Aline A. and {Waite}, Ian A.},
        title = "{The magnetic and spin-down properties of slowly rotating fully convective M dwarfs}",
      journal = {\mnras},
         year = 2025,
        month = sep,
       volume = {542},
       number = {2},
        pages = {1318-1330},
          doi = {10.1093/mnras/staf1197},
archivePrefix = {arXiv},
       eprint = {2507.16986},
 primaryClass = {astro-ph.SR},
       adsurl = {https://ui.adsabs.harvard.edu/abs/2025MNRAS.542.1318S}
}

@ARTICLE{2023MNRAS.524.5060C,
       author = {{Chebly}, Judy J. and {Alvarado-G{\'o}mez}, Juli{\'a}n D. and {Poppenh{\"a}ger}, Katja and {Garraffo}, Cecilia},
        title = "{Numerical quantification of the wind properties of cool main sequence stars}",
      journal = {\mnras},
         year = 2023,
        month = oct,
       volume = {524},
       number = {4},
        pages = {5060-5079},
          doi = {10.1093/mnras/stad2100},
archivePrefix = {arXiv},
       eprint = {2307.04615},
 primaryClass = {astro-ph.SR},
       adsurl = {https://ui.adsabs.harvard.edu/abs/2023MNRAS.524.5060C}
}

@ARTICLE{Klein2021,
       author = {{Klein}, Baptiste and {Donati}, Jean-Fran{\c{c}}ois and {Moutou}, Claire and {Delfosse}, Xavier and {Bonfils}, Xavier and {Martioli}, Eder and {Fouqu{\'e}}, Pascal and {Cloutier}, Ryan and {Artigau}, {\'E}tienne and {Doyon}, Ren{\'e} and {H{\'e}brard}, Guillaume and {Morin}, Julien and {Rameau}, Julien and {Plavchan}, Peter and {Gaidos}, Eric},
        title = "{Investigating the young AU Mic system with SPIRou: large-scale stellar magnetic field and close-in planet mass}",
      journal = {\mnras},
         year = 2021,
        month = mar,
       volume = {502},
       number = {1},
        pages = {188-205},
          doi = {10.1093/mnras/staa3702},
archivePrefix = {arXiv},
       eprint = {2011.13357},
 primaryClass = {astro-ph.EP},
       adsurl = {https://ui.adsabs.harvard.edu/abs/2021MNRAS.502..188K}
}

@ARTICLE{Fuhrmeister2019,
       author = {{Fuhrmeister}, B. and {Czesla}, S. and {Schmitt}, J.~H.~M.~M. and {Johnson}, E.~N. and {Sch{\"o}fer}, P. and {Jeffers}, S.~V. and {Caballero}, J.~A. and {Zechmeister}, M. and {Reiners}, A. and {Ribas}, I. and {Amado}, P.~J. and {Quirrenbach}, A. and {Bauer}, F. and {B{\'e}jar}, V.~J.~S. and {Cort{\'e}s-Contreras}, M. and {D{\'\i}ez Alonso}, E. and {Dreizler}, S. and {Galad{\'\i}-Enr{\'\i}quez}, D. and {Guenther}, E.~W. and {Kaminski}, A. and {K{\"u}rster}, M. and {Lafarga}, M. and {Montes}, D.},
        title = "{The CARMENES search for exoplanets around M dwarfs. Period search in H{\ensuremath{\alpha}}, Na I D, and Ca II IRT lines}",
      journal = {\aap},
         year = 2019,
        month = mar,
       volume = {623},
          eid = {A24},
        pages = {A24},
          doi = {10.1051/0004-6361/201834483},
archivePrefix = {arXiv},
       eprint = {1901.05173},
 primaryClass = {astro-ph.SR},
       adsurl = {https://ui.adsabs.harvard.edu/abs/2019A&A...623A..24F}
}


\end{document}